\documentclass[a4paper,12pt]{dmathesis}

\usepackage{imakeidx}
\makeindex[name=sym,title={List of Symbols}]
\makeindex

\usepackage{import}
\usepackage{imakeidx}
\makeindex

\usepackage{amsmath}
\usepackage{amssymb}
\usepackage{amsthm}
\usepackage{epsfig}
\usepackage{graphicx}

\usepackage[numbers,sort&compress]{natbib}

\usepackage[colorlinks=true,
	urlcolor=blue,
	anchorcolor=blue,
	citecolor=blue,
	filecolor=blue,
	linkcolor=blue,
	menucolor=blue,
	pagecolor=blue,
	linktocpage=true,
	pdfproducer=medialab,
	pdfa=true,
	unicode=true
]{hyperref}

\usepackage{dsfont}
\usepackage{mathtools}
\usepackage{physics}
\usepackage{tikz}
\usepackage{tikz-cd}
\usepackage{slashed}

\usepackage{newpxtext}
\usepackage[euler-digits,euler-hat-accent]{eulervm}

\usepackage{comment}
\specialcomment{comment}{\begingroup \color{red}}{\endgroup}

\allowdisplaybreaks[0] 

\usepackage[numbered]{bookmark}

\let\originalleft\left
\let\originalright\right
\renewcommand{\left}{\mathopen{}\mathclose\bgroup\originalleft}
\renewcommand{\right}{\aftergroup\egroup\originalright}

\DeclareMathOperator{\DD}{D}
\DeclareMathOperator{\sign}{sign}
\DeclareMathOperator{\ch}{ch}
\DeclareMathOperator{\td}{td}
\DeclareMathOperator{\vol}{vol}
\DeclareMathOperator{\diag}{diag}

\newcommand{\sslash}{\mathbin{/\mkern-6mu/}}

\theoremstyle{definition}
\newtheorem{thm}{Theorem}[section]

\newtheorem{defn}{Definition}[section]
\newtheorem{conj}{Conjecture}[section]

\theoremstyle{remark}

\definecolor{phantom}{gray}{0.98}

\newcommand{\id}{\operatorname{id}}
\newcommand{\im}{\operatorname{im}}
\newcommand{\eff}{\mathrm{eff}}
\newcommand{\crit}{\mathrm{crit}}
\newcommand{\JK}[1]{\underset{x=#1}{\operatorname{JK-Res}}}
\newcommand{\pic}[2]{\mathrm{Pic}^{#1}\left(#2\right)}
\newcommand{\sym}[2]{\mathrm{Sym}^{#1}\left(#2\right)}

\makeatletter
\let\save@mathaccent\mathaccent
\newcommand*\if@single[3]{%
  \setbox0\hbox{${\mathaccent"0362{#1}}^H$}%
  \setbox2\hbox{${\mathaccent"0362{\kern0pt#1}}^H$}%
  \ifdim\ht0=\ht2 #3\else #2\fi
  }
\newcommand*\rel@kern[1]{\kern#1\dimexpr\macc@kerna}
\newcommand*\widebar[1]{\@ifnextchar^{{\wide@bar{#1}{0}}}{\wide@bar{#1}{1}}}
\newcommand*\wide@bar[2]{\if@single{#1}{\wide@bar@{#1}{#2}{1}}{\wide@bar@{#1}{#2}{2}}}
\newcommand*\wide@bar@[3]{%
  \begingroup
  \def\mathaccent##1##2{%
    \let\mathaccent\save@mathaccent
    \if#32 \let\macc@nucleus\first@char \fi
    \setbox\z@\hbox{$\macc@style{\macc@nucleus}_{}$}%
    \setbox\tw@\hbox{$\macc@style{\macc@nucleus}{}_{}$}%
    \dimen@\wd\tw@
    \advance\dimen@-\wd\z@
    \divide\dimen@ 3
    \@tempdima\wd\tw@
    \advance\@tempdima-\scriptspace
    \divide\@tempdima 10
    \advance\dimen@-\@tempdima
    \ifdim\dimen@>\z@ \dimen@0pt\fi
    \rel@kern{0.6}\kern-\dimen@
    \if#31
      \overline{\rel@kern{-0.6}\kern\dimen@\macc@nucleus\rel@kern{0.4}\kern\dimen@}%
      \advance\dimen@0.4\dimexpr\macc@kerna
      \let\final@kern#2%
      \ifdim\dimen@<\z@ \let\final@kern1\fi
      \if\final@kern1 \kern-\dimen@\fi
    \else
      \overline{\rel@kern{-0.6}\kern\dimen@#1}%
    \fi
  }%
  \macc@depth\@ne
  \let\math@bgroup\@empty \let\math@egroup\macc@set@skewchar
  \mathsurround\z@ \frozen@everymath{\mathgroup\macc@group\relax}%
  \macc@set@skewchar\relax
  \let\mathaccentV\macc@nested@a
  \if#31
    \macc@nested@a\relax111{#1}%
  \else
    \def\gobble@till@marker##1\endmarker{}%
    \futurelet\first@char\gobble@till@marker#1\endmarker
    \ifcat\noexpand\first@char A\else
      \def\first@char{}%
    \fi
    \macc@nested@a\relax111{\first@char}%
  \fi
  \endgroup
}
\makeatother

\usepackage[british]{babel}
\usepackage[figuresright]{rotating}

\usepackage{csquotes}
\usepackage{booktabs}
\usepackage{xfrac}

\usepackage[margin=15mm,hang]{caption}
\usepackage{subcaption}

\makeatletter
\newcommand{\polytope}{\mathord{\mathpalette\raise@btd\bigtriangledown}}
\newcommand\raise@btd[2]{%
  \raisebox{.9\depth}{$\m@th#1#2$}%
}

\makeatother

\usepackage{listings}

\usepackage{jhepthesis}

\usepackage{setspace}
\definecolor{cardinal}{rgb}{0.6,0,0}
\definecolor{darkgreen}{rgb}{0,0.5,0}
\definecolor{golden}{rgb}{0.92, 0.7, 0}
\definecolor{midnight}{rgb}{0, 0, 0.5}
\definecolor{darkblue}{rgb}{0.2, 0, 0.8}

\definecolor{mylightgrey}{RGB}{230,230,230}
\definecolor{mygrey}{RGB}{190,190,190}
\definecolor{mydarkgrey}{RGB}{110,110,110}
\definecolor{mygreen}{RGB}{120,220,160}
\definecolor{mydarkgreen}{RGB}{60,120,60}
\definecolor{myverydarkgreen}{RGB}{20,60,20}
\definecolor{mydarkred}{RGB}{140,40,40}

\excludecomment{comment}

\begin{document}

\title{An Algebro-Geometric Approach to Twisted Indices of Supersymmetric Gauge Theories}
\author{Guangyu Xu}
\supervisor{Mathew Richard Bullimore}
\date{2022}

\pagenumbering{roman}
\maketitlepage

\begin{abstract}
	 This thesis studies the algebro-geometric aspects of supersymmetric abelian gauge theories in three dimensions. The supersymmetric vacua are demonstrated to exhibit a window phenomenon in Chern-Simons levels, which is analogous to the window phenomenon in quantum K-theory with level structures. This correspondence between three-dimensional gauge theories and quantum K-theory is investigated from the perspectives of semi-classical vacua, twisted chiral rings, and twisted indices. In particular, the twisted index admits an algebro-geometric interpretation as the supersymmetric index of an effective quantum mechanics. Via supersymmetric localisation, the contributions from both topological and vortex saddle points are shown to agree with the Jeffrey-Kirwan contour integral formula. The algebro-geometric construction of Chern-Simons contributions to the twisted index from determinant line bundles provides a natural connection with quantum K-theory.
\end{abstract}

\begin{declaration*}
	The work in this thesis is based on research carried out in the Department of
	Mathematical Sciences at the University of Durham. No part of this thesis has been
	submitted elsewhere for any degree or qualification.
\end{declaration*}

\begin{acknowledgements*}
	I would like to thank the unconditional love from my parents, who have supported me to pursue this passion during my entire academic life.
	
	The companionship from my partner and our cat has given me a warm home at a foreign place, even during the darkest times.
	
	I am indebted to the mentorship provided by my supervisor, who introduced me to this fascinating branch of research.
	
	Lastly I also want to thank the theoretical physics community for providing endless grounds of constant musing.
\end{acknowledgements*}

\begin{epigraph*}
	The scientist does not study nature because it is useful; he studies it because he delights in it, and he delights in it because it is beautiful. If nature were not beautiful, it would not be worth knowing, and if nature were not worth knowing, life would not be worth living.
	\source{Science and Method}{Henri Poincar\'e}
\end{epigraph*}

\setlength{\parskip}{0pt}
\tableofcontents
\setlength{\parskip}{1.6em plus 3pt minus 3pt}

\cleardoublepage
\pagenumbering{arabic}

\setcounter{chapter}{0}
\chapter{Introduction}

\label{chap:introduction}

\section{Background}
In the broadest terms, the study of physics is the ultimate endeavour to make sense of the world we observe around us. 
From our observations, we build mathematical models that describe and predict all kinds of natural phenomena, ranging from Newton's apples to human psychology. 
However, we usually refer to physics by those more ``fundamental'' models. 
What is  fundamental is a subjective and evolving concept. 
For a theoretical physicist, this most likely means the theories describing the most fundamental forces and particles, by which we mean they cannot be further reduced to simpler constituents.
The best fundamental theories we currently have are
\vspace{-\parskip} 
\begin{itemize}
	\item the \emph{standard model} describing the electroweak and the strong forces,
	\item  and the theory of \emph{general relativity} describing gravitation. 
\end{itemize}
These two theories are not fully compatible nor complete\footnote{It is debatable if there can ever be a complete theory of physics. My view is that there would always be physics beyond our current understanding.}. The research in theoretical physics usually focuses on some specific issues concerning these two theories. 

In particular, the strong force described by quantum chromodynamics in the standard model is not yet fully understood. The main obstacle is that, unlike in quantum electrodynamics, the perturbative approach widely used for scattering experiments does not work in this regime where the interactions are strong. One of the basic assumptions for the Feynman diagrams is that the interaction is weak. There have been enormous efforts going into understanding the strong force. For example, string theory was originally invented as a theory of the strong force, which has evolved into a much more ambitious framework with deep insights in both physics and mathematics. Lattice gauge theory is another well-established probe into quantum chromodynamics utilising Monte Carlo simulations.

A different approach to deal with the non-perturbative nature of the standard model is obviously to understand non-perturbatively the underlying framework, \emph{quantum field theory}. 
It is easier said than done. The standard model is far too messy to deal with directly. 
So we adopt our best weapon, idealised toy models. This is ubiquitous across branches of science, where we tend to first study the simplist model that admits certain behaviours of interests. Our aim is to understand the general framework of quantum field theories by studying interesting toy models.
Therefore we are free to impose further symmetries and adjust model parameters, as long as the phenomena of interest are manifest. 
We sincerely hope that our spherical cows would taste the same as the normally shaped ones.

An important technique is to implement \index{supersymmetry} \emph{supersymmetry}, which is a proposed symmetry between the force carriers, bosons and the matter constituents, fermions. 
There was a time when we were hopeful that supersymmetry is a genuine symmetry of nature. Unfortunately, the recent results from the Large Hadron Collider have ruled out the most plausible supersymmetric extensions to the standard model. 
However, supersymmetry remains as a powerful tool to probe the structure of quantum field theories. 
For instance, many \index{infra-red duality} infra-red dualities ubiquitous in quantum field theories are first discovered in supersymmetric theories.
The Seiberg-Witten thoery~\cite{seiberg1994}, i.e., supersymmetric quantum chromodynamics retains the signature phenomenon of \index{confinement} confinement from non-supersymmetric quantum chromodynamics, while offering exactly computable observables. It also provides insight into topology via connections to Donaldson invariants~\cite{donaldson1996}. This is far from an isolated example contributing to mathematics. 

In addition to the motivation of better understanding the non-perturbative aspects of quantum field theories, the research in supersymmetry has been proven to deliver striking insights into many areas of pure mathematics, including algebraic geometry, topology, non-commutative algebras, and representation theory. In particular, the study two-dimensional supersymmetric theories has revolutionary contributions to symplectic and algebraic geometry via the development of homological \index{mirror symmetry} mirror symmetry, Gromov-Witten theory, and \index{quantum cohomology} quantum cohomology.

The research documented in this thesis specifically studies supersymmetric quantum field theories in three spacetime dimensions, where we are slightly distorted from the reality by tuning down the number of dimensions from four to three. 
The focus is on \index{gauge theory} gauge theories, which are a type of quantum field theory invariant under local gauge symmetries.
They are of significant interests in both physics and mathematics.
A three-dimensional gauge theory permits Chern-Simons terms which is used to describe the fractional Hall effect in condensed matter physics. 
Equipped with supersymmetry, it can be considered as a lift to three dimensions of two-dimensional supersymmetric gauge theories, which are closely related to quantum cohomology and mirror symmetry. 
Therefore it is expected to produce analogous mathematical results. In particular, there exists three-dimensional \index{mirror symmetry} mirror symmetry and \index{quantum K-theory} quantum K-theory, which can be considered as a lift respectively from homological mirror symmetry and quantum cohomology. 
More importantly, these three-dimensional analogues admit mathematically distinct properties.
Therefore the study of three-dimensional gauge theories are expected to lead to new mathematics.
Indeed the focus of this thesis is on those mathematical structures arising from three-dimensional supersymmetric gauge theories.

\section{Setup}
The primary class of theories considered in this thesis consists of $\mathcal{N}=2$ supersymmetric  Chern-Simons abelian gauge theories with chiral matter contents, on some three-dimensional base manifolds.
\vspace{-\parskip}
\begin{itemize}
	\item The gauge symmetry is taken to be $U(1)$. A discussion of higher rank groups $U(1)^K$ is left to Chapter~\ref{chap:U1K-gauge-theories}.
	\item The matter contents are $N > 0$ chiral superfields $\{\Phi_j \mid j = 1,\ldots, N\}$ charged under the gauge group with charges $Q_j \neq 0$. 
	\item The chiral multiplet $\Phi_j$ has R-charge $r_j$. 
	\item The theories contain various Chern-Simons terms mixing the gauge, flavour, and the R-symmetry.
	\item There exists a real Fayet-Iliopoulos parameter $\zeta$ associated to the topological symmetry $U(1)_\text{t}$.
\end{itemize}

\section{Results}

There are two main strands of original research conducted in this thesis, which are closely intertwined.
\begin{itemize}
	\item The first strand is on the geometry of the twisted indices of these supersymmetric gauge theories on $S^1 \times \Sigma$, where the circle $S^1$ is the compactified temporal dimension and the space is taken to be an arbitrary Riemann surface $\Sigma$ as shown below. This collaborative work has been published in~\cite{xu2022}.
	
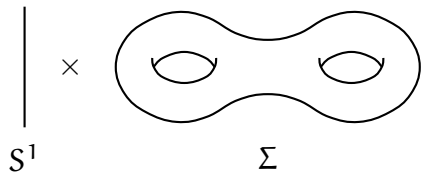
\smallskip
	\begin{center}

	\begin{tikzpicture}[scale=0.4]

	\draw [thick] (-8, -2)--(-8, 2);

	\draw [thick] plot [smooth cycle,tension=0.7] coordinates {(1,-1.1) (0,-0.9) (-1,-1.1) (-2,-1.65) (-3,-1.82) (-4,-1.5) (-4.7,-0.9) (-5,0) (-4.7,0.9) (-4,1.5) (-3,1.82) (-2,1.65) (-1,1.1) (0,0.9) (1,1.1) (2,1.65) (3,1.82) (4,1.5) (4.7,0.9) (5,0) (4.7,-0.9) (4,-1.5) (3,-1.82) (2,-1.65)};

	\draw [thick] plot [smooth,tension=0.7] coordinates {(1.7,0.3) (1.75,0) (2.05,-0.34) (2.75,-0.53) (3.45,-0.34) (3.75,0) (3.8,0.3)};
	\draw [thick] plot [smooth,tension=0.7] coordinates {(1.75,0) (2.05,0.3) (2.75,0.5) (3.45,0.3) (3.75,0)};

	\draw [thick] plot [smooth,tension=0.7] coordinates {(-1.7,0.3) (-1.75,0) (-2.05,-0.34) (-2.75,-0.53) (-3.45,-0.34) (-3.75,0) (-3.8,0.3)};
	\draw [thick] plot [smooth,tension=0.7] coordinates {(-1.75,0) (-2.05,0.3) (-2.75,0.5) (-3.45,0.3) (-3.75,0)};

	\node at (0,-3) {$\Sigma$};
	\node at (-8,-3) {$S^1$};
	\node at (-6.5,0) {$\times$};

	\end{tikzpicture}

	\end{center}
	We obtain an algebro-geometric interpretation of the twisted indices via the technique of \index{supersymmetric localisation} supersymmetric localisation~\cite{xu2022}, which is one of the most powerful tool offered by supersymmetry. It allows us to compute the path integrals exactly by ``localising'' the contributions to spaces of finite dimensions.
	Supersymmetry enables us to introduce exact deformations and scaling limits that lead to different mathematical models of the same partition function. 
	The topological twist is performed using the unbroken R-symmetry, which preserved an $\mathcal{N} = (0, 2)$ quantum mechanics on the circle $S^1$. 
	
	In general, using different classes of localisation schemes, the \index{twisted index} twisted index localises to different types of integrals~\cite{willet2017}. 
	In Coulomb branch localisation scheme, the path integral localises onto configurations where the vectormultiplet scalar is non-zero and the gauge group is broken into a maximum torus~\cite{kapustin2010,hama2011,hama2011a}. 
	This formulates the \index{twisted index} twisted index as Jeffrey-Kirwan contour integrals~\cite{benini2015,benini2017,closset2016}. 
	In Higgs branch localisation, the path integral localises onto configurations solving vortex equations. 
	This interprets the \index{twisted index} twisted index as integrals of characteristic classes over the moduli spaces of vortices~\cite{fujitsuka2014,benini2014}.

	We develop a Higgs branch localisation scheme, which leads to novel topological saddle points, in addition to the vortex configurations. The topological vacua are controlled by the Chern-Simons levels. Their existence is essential to preserve the index during wall-crossing.
	This is achieved by considering a different scaling limit in the path integral, with an exact deformation to the lagrangian depending on a one-dimensional Fayet-Iliopoulos parameter~\cite{bullimore2019}. 
	It allows us to unambiguously interpret the \index{twisted index} twisted index as the Witten index of the quantum mechanics on $S^1$, in each individual magnetic sector labelled by $\mathfrak{m}$, schematically given by~\cite{bullimore2019,bullimore2019a}
\begin{equation*}
	\mathcal{I} = \sum_{\mathfrak{m} \in \mathbb{Z}} q^\mathfrak{m} \int \hat{A}(\mathfrak{M}_\mathfrak{m}) \ch(\mathcal{V}_\mathfrak{m}) \,,
\end{equation*}
where $\mathfrak{M}_\mathfrak{m}$ denotes the moduli space parametrising saddle points of the localised path integral with magnetic flux $\mathfrak{m} \in \mathbb{Z}$, and $\mathcal{V}_\mathfrak{m}$ represents a complex of vector bundles arising from the massive fluctuations of chiral multiplets and Chern-Simons terms. 
For vortex saddles, the moduli space consists of symmetric products of the curve
\begin{equation*}
	\mathfrak{M}_\mathfrak{m} = \sum_i \mathrm{sym}^{d_i} \Sigma
\end{equation*}
for each possible non-vanishing chiral multiplet $\Phi_i$, where $\mathrm{sym}^{d_i} \Sigma$ are the loci of vortices.
For topological saddles, the moduli space is roughly a Picard variety parametrising holomorphic line bundles on $\Sigma$
\begin{equation*}
	\mathfrak{M}_\mathfrak{m} \sim \mathrm{Pic}^\mathfrak{m} (\Sigma) \simeq T^{2g}.
\end{equation*}
With a careful analysis of the index bundle and Chern-Simons terms, the characteristic classes are computed using the \index{Grothendiek-Riemann-Roch theorem} Grothendiek-Riemann-Roch theorem. 
This approach reproduces the results obtained using the Jeffrey-Kirwan residue prescription under the conventional scheme of Coulomb branch localisation.

	\item The second strand of research concerns the connection between these quantum field theories and quantum K-theory, to be published in~\cite{xu2022a}. 
	
	Quantum K-theory~\cite{givental2000,givental2001,lee2001,givental2015}\index{quantum K-theory} is a K-theoretical extension to quantum cohomology. It is also referred to as \index{qasi-map K-theory}quasi-map K-theory~\cite{Zhang:2020rou} in the mathematical literature. 
	Generally speaking, it studies the intersection theory of complex vector bundles over the spaces of holomorphic curves in K\"ahler manifolds. Many directions of quantum K-theory are under active research. 
	For example, recent works~\cite{braverman2009,maulik2012,okounkov2015} have developed deep connections to geometric representation theory and quantum integrable systems. 
	Furthermore, a striking correspondence between three-dimensional gauge theories and \index{quantum K-theory} quantum K-theory~\cite{jockers2018,jockers2019,jockers2019a,Ueda:2019qhg} has also been under active research.
	
	Quantum K-theory has an additional parameter called the ``level''~\cite{ruan2019,ruan2020}, compared to quantum cohomology. 
	We propose that this level can be interpreted as the Chern-Simons level in the class of supersymmetric gauge theories we have been studying. 
	While this correspondence has been studied before in ad hoc fashions, we aim to give a formal interpretation via our geometric construction of twisted indices.
	The Higgs branch localisation developed in the previous strand of research leads to novel topological saddle points, in addition to the vortex saddle points. 
	These topological saddle points exist when the effective Chern-Simons level in the asymptotic regions is non-zero. 
	This correspondence between these two distinct phenomena in physics and mathematics can be demonstrated directly from the semi-classical vacua, the Bethe ansatz equations, and the twisted indices. 
	Thus this phenomenon is a new physical interpretation to the \index{window phenomenon} window phenomenon in \index{quantum K-theory} quantum K-theory~\cite{jockers2019a}, which also opens up the study on how to define quantum K-theory outside the window by taking into account the topological saddles.
\end{itemize}

\section{Outline}

The bulk of this thesis is organised by the topology of the spacetime base manifold of the quantum field theories. 
\begin{itemize}
	\item The case of flat manifold $\mathbb{R}^3$ is discussed in Chapter~\ref{chap:flat-base-manifold}. We review the semi-classical vacuum equations and the type of their solutions, roughly following~\cite{aharony1997,intriligator2013}. The non-generic types of vacua, i.e., the Higgs branch and the topological branch vacua are studied in details. 
	Then the window phenomenon for Chern-Simons levels is introduces, which is the observation that the topological vacua can only co-exist with the Higgs vacua for the Chern-Simons level in a certain \index{critical window} critical window. We propose this as the physical interpretation of the critical window in \index{quantum K-theory} quantum K-theory with level structures. This discussion is based on the work to be published in~\cite{xu2022a}.
	\item We investigate the corresponding theories on $S^1 \times \mathbb{R}^2$ in Chapter~\ref{chap:twisted-chiral-ring}, where the circle $S^1$ of radius $\beta$ is the compactified temporal dimension and the space is taken to be the plane $\mathbb{R}^2$. 
	In this chapter we study the theory by viewing it as a lift from the \index{A-twist} A-twist of the two-dimensional $\mathcal{N}=(2,2)$ gauge  theory on $\mathbb{R}^2$.
	The behaviour of the Bethe ansatz equations are studied in the different limits of the radius $\beta$. 
	The solutions are shown to reproduce the behaviour of supersymmetric vacua in flat spacetime in the large radius limit. 
	In the small radius limit, the solutions corresponding to the Higgs vacua in flat spacetime are shown to be captured by a two-dimensional gauge theory, while the topological vacua are de-coupled into a collection of disjoint two-dimensional theories. 
	The Bethe equations define the \index{quantum K-theory} quantum K-theory ring, which reduces to the twisted chiral ring in the small radius limit if and only if the Chern-Simons level lies in a critical window.
	This chapter is based on~\cite{xu2022a}.  
	\item The more general case of a topological twist on $S^1 \times \Sigma$ is studied in Chapter~\ref{chap:twist-on-surface}, where the space is taken to be a generic Riemann surface $\Sigma$ of genus $g$. A topological twist~\cite{witten1991,Nekrasov:2014xaa} is performed to preserve supersymmetry on curved spaces, resulting in a $\mathcal{N}=(0,2)$ quantum mechanics on $S^1$. Instead of viewing the theory as a lift from two-dimensional theories in Chapter~\ref{chap:twisted-chiral-ring}, we study it from the perspective of this quantum mechanics. We schematically construct an algebro-geometric interpretation of the \index{twisted index} twisted index as the supersymmetric index of a quantum mechanics on $S^1$, which reproduces the Jeffrey-Kirwan contour integral formula. The full constructions are documented in Chapter~\ref{sec:vortex-vacuum} and Chapter~\ref{sec:topological-vacuum}. It also offers a geometric explanation of the \index{window phenomenon} window phenomenon via the construction in Chapter~\ref{sec:CS-from-det} of Chern-Simons contributions as determinant line bundles of auxiliary chiral multiplets. This chapter is based on the research from~\cite{xu2022,xu2022a}. 
	\item The full algebro-geometric construction of the twisted index for vortex saddles is discussed in Chapter~\ref{sec:vortex-vacuum}. This chapter is based on~\cite{xu2022}.
	\item The full algebro-geometric construction of the twisted index for topological saddles is discussed in Chapter~\ref{sec:topological-vacuum}. This chapter is based on~\cite{xu2022}.
	\item Chapter~\ref{sec:CS-from-det} gives an algebro-geometric construction for the Chern-Simons contributions in twisted indices. The Chern-Simons contributions are constructed from the determinant line bundles of auxiliary chiral multiplets as a result of integrating out. This is analogous to the level structure in quantum K-theory. This chapter is based on~\cite{xu2022a}.
	\item A brief exploration of higher rank abelian theories is conducted in Chapter~\ref{chap:U1K-gauge-theories}. Instead of trying to give a complete treatment, we only aim to set up the notations and discuss some expectations. These more general theories admit three-dimensional mirror symmetry. We investigate with some examples by computing their twisted indices. This chapter is based on~\cite{xu2022a}.
	\item The computational technique of multi-variate Jeffrey-Kirwan contour integrals is proposed in Appendix~\ref{chap:multivariate-JK} via a transformation formula. The validity is substantiated by correctly reproducing the twisted indices of a mirror pair of $\mathcal{N}=4$ abelian linear quiver gauge theories. This appendix is based on my unpublished work.
	\item Some foundational background materials for this thesis are reviewed in Appendix~\ref{chap:prelim-physics} and Appendix~\ref{chap:prelim-mathematics}, including gauge theories, bundles on Riemann surfaces, and abelian vortex equations.
\end{itemize}
\chapter{Supersymmetric Vacua}
\label{chap:flat-base-manifold}

The first half of this chapter reviews some general aspects of $\mathcal{N}=2$ supersymmetric gauge theories in flat space $\mathbb{R}^3$, closely following~\cite{aharony1997,intriligator2013}. In particular we discuss the classification of supersymmetric vacua of abelian Chern-Simons gauge theories in the presence of real masses and Fayet–Iliopoulos parameters. There exist two types of generic vacua: the Higgs branch and the topological branch. The discussion sets up the notation for this thesis, and provides a basis for discussing theories on more sophisticated manifolds in later chapters.

In later sections of this chapter, we introduce the \index{window phenomenon} window phenomenon for the Chern-Simons levels, where a specific window of Chern-Simons levels ensures the absence of topological vacua in a given chamber for the Fayet–Iliopoulos parameters. This observation suggests that it may be identified with the \index{window phenomenon} window phenomenon in the \index{quantum K-theory} quantum K-theory of the Higgs branch with level structures~\cite{xu2022a}. This preliminary identification is also evidenced from the perspective of twisted chiral rings in Chapter~\ref{chap:twisted-chiral-ring}, and finally justified with an algebro-geometric interpretation of twisted indices in Chapter~\ref{chap:twist-on-surface} and Chapter~\ref{sec:CS-from-det}.

Two examples are provided in the end to illustrate the vacuum structure and the window phenomenon .

\section{Supersymmetric Quantum Field Theory}
\label{sec:N=2-SUSY}

\subsection{Supersymmetry Algebra}
\label{sec:SUSY-algebra}

Assuming a lorentzian signature $\eta_{\mu \nu} = \diag(-1, +1, +1)$ for the metric, the $\mathcal{N}=2$ \index{supersymmetry} supersymmetry in three dimensions contains four \index{supercharge} supercharges~\cite{intriligator2013}, satisfying the following anti-commutation relations
\begin{subequations}
\begin{align}
	\left\{Q_\alpha, Q_\beta\right\} &= 0 \,, \quad \left\{\widebar{Q}_\alpha, \widebar{Q}_\beta\right\} = 0 \,, \\
	\left\{Q_\alpha, \widebar{Q}_\beta\right\} &= 2 \gamma^\mu_{\alpha \beta} P_\mu + 2 i \epsilon_{\alpha \beta} Z \,.
\end{align}
\end{subequations}
The complex supercharges $Q$ and $\widebar{Q}$ are labelled by the spinor indices $\alpha, \beta \in \{1, 2\}$. 
The gamma matrices $\{\gamma^\mu \mid \mu = 0, 1, 2\}$ satisfying
\begin{equation}
	\left( \gamma^\mu \right)_\alpha^{\ \ \rho} \left( \gamma^\nu \right)_\rho^{\ \ \beta} \
	= \eta^{\mu \nu} \delta_\alpha^{\ \ \beta}
	+ \epsilon^{\mu \nu \rho} \left( \gamma_\rho \right)_\alpha^{\ \ \beta}
\end{equation}
can be chosen as
\begin{equation}
	\begin{pmatrix}
		\gamma^1 \\
		\gamma^2 \\
		\gamma^3
	\end{pmatrix}
	= 
	\begin{pmatrix}
		- \mathbf{1} \\
		\sigma^1 \\
		\sigma^3
	\end{pmatrix}
	\,,
\end{equation}
where $\sigma$ denotes the Pauli matrices.
The symmetric term $\gamma^\mu P_\mu$ contains the momentum $P_\mu$, while the anti-symmetric term $\epsilon Z$ is from the real central charge $Z$. This algebra can be obtained by dimension reduction from the $\mathcal{N}=1$ supersymmetry~\cite{Wess:1992cp} in four dimensions, where the central charge $Z$ comes from the reduced $P_3$ momentum. 

In the superspace formulation, the supercharges are represented by differential operators $\mathcal{Q}_\alpha$ and $\widebar{\mathcal{Q}}_\alpha$ acting on a superfield $\mathcal{O}$ according to
\begin{equation}
	\delta_\epsilon \mathcal{O} = i \left[\epsilon Q - \widebar{\epsilon} \widebar{Q}, \mathcal{O}\right] 
	= \left(\epsilon \mathcal{Q} - \widebar{\epsilon} \widebar{\mathcal{Q}}\right) \mathcal{O} \,.
\end{equation}
The superspace derivatives $\DD_\alpha$ and $\widebar{\DD}_\alpha$ are defined to be anti-commuting with the differential operators $\mathcal{Q}_\alpha$ and $\widebar{\mathcal{Q}}_\alpha$. In the superspace coordinates $(x, \theta, \widebar{\theta})$, they can be explicitly written as
\begin{subequations}
\begin{align}
	\mathcal{Q}_\alpha &= \phantom{+} \frac{\partial}{\partial \theta^\alpha} - i \gamma^\mu_{\alpha \beta} \widebar{\theta}^\beta \partial_\mu \,,\\
	\widebar{\mathcal{Q}}_\alpha &= -\frac{\partial}{\partial \widebar{\theta}^\alpha} + i \theta^\beta \gamma^\mu_{\beta \alpha} \partial_\mu \,,
\end{align}
\end{subequations}
and 
\begin{subequations}
\begin{align}
	\DD_\alpha &= \phantom{+} \frac{\partial}{\partial \theta^\alpha} + i \gamma^\mu_{\alpha \beta} \widebar{\theta}^\beta \partial_\mu \,,\\
	\widebar{\DD}_\alpha &= -\frac{\partial}{\partial \widebar{\theta}^\alpha} - i \theta^\beta \gamma^\mu_{\beta \alpha} \partial_\mu \,.
\end{align}
\end{subequations}

A general superfield $F (x, \theta, \widebar{\theta})$ is a function of the superspace, which can be expanded in powers of $\theta$ and $\widebar{\theta}$ as
\begin{align}
	F (x, \theta, \widebar{\theta}) = f(x) &+ \theta \, \phi(x) + \widebar{\theta} \, \widebar{\chi}(x) \nonumber \\
	&
	+ \theta \theta \, m(x) + \widebar{\theta} \widebar{\theta} \, n(x) + \theta \gamma^\mu \widebar{\theta} \, v_\mu (x)  \nonumber \\
	&
	+ \theta \theta \widebar{\theta} \, \widebar{\lambda} (x) + \widebar{\theta} \widebar{\theta} \theta \, \psi (x)  \nonumber \\
	&
	+ \theta \theta \widebar{\theta} \widebar{\theta} D(x) \,.
\end{align}
Superfields form linear representations of the supersymmetry algebra, which are in general reducible. We may impose covariant constraints to eliminate some extra component fields.

The field contents of $\mathcal{N}=2$ gauge theories involve the following representations of the supersymmetry algebra:
\vspace{-\parskip}
\begin{itemize}
	\item A \index{superfield !chiral} chiral superfield $\Phi$ satisfies 
	\begin{equation}
	\widebar{\DD}_\alpha \Phi = 0 \,,
	\end{equation}		
	usually serving as matter contents. This constraint can be easily solved in terms of $y^\mu = x^\mu + i \theta \gamma^\mu \widebar{\theta}$ and $\theta$ satisfying 
	\begin{equation}
		\widebar{\DD}_\alpha y^\mu = \widebar{\DD}_\alpha \theta = 0 \,.
	\end{equation}
	In components, the chiral superfield $\Phi = (\phi, \psi, F)$ can be written as
	\begin{align}
	\label{eq:chiral-components}
		\Phi(x, \theta, \widebar{\theta}) &= \phi(y) + \sqrt{2} \theta \, \psi(y) + \theta \theta \, F(y) \nonumber \\
		&= \phi(x) + i \theta \gamma^\mu \widebar{\theta} \, \partial_\mu \phi(x) + \frac{1}{4} \theta \theta \widebar{\theta} \widebar{\theta} \, \partial^2 \phi(x) 
		\nonumber \\
		& \qquad
		+ \sqrt{2} \theta \, \psi(x) - \frac{i}{\sqrt{2}} \theta \theta \, \partial_\mu \psi(x) \gamma^\mu \, \widebar{\theta} + \theta \theta \, F(x) \,.
	\end{align}
	\item Similarly an \index{superfield !anti-chiral} anti-chiral superfield $\widebar{\Phi}$ satisfies 
	\begin{equation}
		\DD_\alpha \widebar{\Phi} = 0 \,.
	\end{equation}
	Its components can be obtained by directly taking the conjugation of a chiral superfield.
	\item A \index{superfield !vector} vector superfield $V$ obeys 
	\begin{equation}
		V = V^\dagger \,.
	\end{equation}		
	In the Wess-Zumino gauge~\cite{Wess:1992cp}, the vector multiplet $V = (\sigma, A_\mu, \lambda, \widebar{\lambda}, D)$ can be written as
	\begin{align}
	\label{eq:vector-components}
		V = - i \theta \widebar{\theta} \, \sigma - \theta \gamma^\mu \widebar{\theta} \, A_\mu + i \theta \theta \, \widebar{\lambda} - i \widebar{\theta} \widebar{\theta} \theta  \, \lambda + \frac{1}{2} \theta \theta \widebar{\theta} \widebar{\theta} \, D \,.
	\end{align}
	\item A \index{superfield !linear} linear superfield $\Sigma$ obeys 
	\begin{equation}
		\epsilon^{\alpha \beta} \DD_\alpha \DD_\beta \Sigma =  \epsilon^{\alpha \beta} \widebar{\DD}_\alpha \widebar{\DD}_\beta \Sigma = 0\,.
	\end{equation}
	For example, the gauge field strength is in the real linear multiplet
	\begin{align}
	\label{eq:linear-components}
		\Sigma :=& -\frac{i}{2}  \epsilon^{\alpha \beta} \widebar{\DD}_\alpha \DD_\beta V 
		\nonumber \\
		=& \sigma + \theta \, \widebar{\lambda} + \widebar{\theta} \, \lambda 
		\nonumber \\
		& \phantom{ \sigma } + \frac{1}{2} \theta \gamma^\mu \widebar{\theta} \, F^{\nu \rho} \epsilon_{\mu \nu \rho} + i \theta \widebar{\theta} \, D 
		\nonumber \\
		& \phantom{ \sigma } 
		+ \frac{i}{2} \widebar{\theta} \widebar{\theta} \theta \gamma^\mu \, \partial_\mu \lambda - \frac{i}{2} \theta \theta \widebar{\theta} \, \gamma^\mu \partial_\mu \widebar{\lambda} 
		\nonumber \\
		& \phantom{ \sigma } 
		+ \frac{1}{4} \theta \theta \widebar{\theta} \widebar{\theta} \, \partial^2 \sigma \,.
	\end{align}
	In general, a conserved global current $J$ satisfying $\DD \DD J = \widebar{\DD} \widebar{\DD} J = 0$ can be viewed as a component of linear multiplets. The gauge field strength~\eqref{eq:linear-components} contains such a component $J^\mu = \epsilon^{\mu \nu \rho} F_{\nu \rho}$, which generates the global $U(1)_J$ symmetry.
\end{itemize}

\subsection{Supersymmetric Lagrangian}
The general Wess-Zumino lagrangian involving chiral and anti-chiral superfields $\Phi, \widebar{\Phi}$ is
\begin{equation}
	\int \dd^4 \, \theta K(\Phi, \widebar{\Phi}) + \left( \int \dd^2 \theta \, W(\Phi) + \text{h.c.} \right)  \,,
\end{equation}
where $K(\Phi, \widebar{\Phi})$ is a general kinetic term, $W(\Phi)$ is the \index{superpotential} superpotential, and $\text{h.c.}$ denotes the hermitian conjugate. Note that the mass dimension of $\Phi$ in three dimensions is $\frac{1}{2}$, hence the classically marginal interaction term is $\Phi^4$. The theory with superpotential $W = \Phi^3$ flows to an interacting fixed point in the infra-red. In comparison, the Wess-Zumino theories in four dimensions always flow to non-interacting gaussian fixed points. 

For an abelian \index{gauge theory} gauge theory in three dimensions, the gauge vector superfield $V$ in~\eqref{eq:vector-components} contains an additional real scalar $\sigma$ valued in the adjoint representation of the gauge group, coming from the vector potential of the four-dimensional theory in the reduced direction. The kinetic term of the gauge vector field $V$ can be written as
\begin{equation}
\label{eq:kinetic-term-gauge-vector}
	- \frac{1}{g^2} \int \dd^2 \theta \, W^2_\alpha + \text{h.c.} \,,
\end{equation}
where $W_\alpha$ is the chiral field strength constructed via
\begin{equation}
	W_\alpha = - \frac{1}{4} (\widebar{\DD} \widebar{\DD}) \DD_\alpha V \,.
\end{equation}
The lowest component of the chiral field $W_\alpha$ is a gaugino. Alternatively, the gauge kinetic term can be written as
\begin{equation}
	- \frac{1}{e^2} \int \dd^4 \theta \, \Sigma^2
\end{equation}
in terms of a linear superfield $\Sigma =-\frac{i}{2} \epsilon^{\alpha \beta} \widebar{\DD}_\alpha \DD_\beta V$ in~\eqref{eq:linear-components}, whose lowest component is the vector multiplet scalar $\sigma$. The linear superfield $\Sigma$ is invariant under the gauge transformation
\begin{equation}
	V \mapsto V + i (\Lambda - \Lambda^\dagger) \,.
\end{equation}
In three dimensions, the vector can be dualised into a scalar, turning the linear multiplet into a chiral multiplet.

Each $U(1)$ factor in the gauge group has a \index{Fayet-Iliopoulos term} Fayet-Iliopoulos term in the form of
\begin{equation}
	- \frac{\zeta}{2 \pi} \int \dd^4 \theta \, V \,,
\end{equation}
where $\zeta$ is the \index{Fayet-Iliopoulos parameter} Fayet-Iliopoulos parameter.

There can be a supersymmetric \index{Chern-Simons term} Chern-Simons term of the general form
\begin{equation}
	- \frac{\kappa}{4 \pi} \int \dd^4 \theta \, \Sigma V
\end{equation}
at \index{Chern-Simons level} Chern-Simons level \index[sym]{$\kappa$} $\kappa$. Gauge invariance restricts~\cite{aharony1997} the Chern-Simons level to be integers, $\kappa \in \mathbb{Z}$. However, this term breaks parity at the classical level, which is often referred to as the \index{parity anomaly} parity anomaly. The parity anomaly gives an analogue of the 't Hooft anomaly matching conditions. In four dimensions, the 't Hooft anomalies associated with gauging global symmetries must match between the ultra-violet and infra-red theories. In three dimensions, the parity anomaly matching gives a weaker $\mathbb{Z}_2$ condition where whether the gauged global symmetry has a parity anomaly must match between the microscopic and low energy theories.

The matter contents are chiral superfields $\Phi_j$ in~\eqref{eq:chiral-components} giving a lagrangian
\begin{equation}
	\sum_j \int \dd^4 \theta \, \Phi_j^\dagger e^V \Phi_j \,.
\end{equation}
It contains a potential for the squarks \index[sym]{$\phi_j$} of the form
\begin{equation}
	\sum_j |\sigma \phi_j|^2 \,,
\end{equation}
which contributes to the \index{real mass} real mass for the matter fields.

The total lagarangian of the $U(1)_\kappa$ gauge theory with $N$ chiral superfields $\Phi_j$ of gauge charges $Q_j$ and real masses $m_j$ is
\begin{equation}
\label{eq:U(1)-lagrangian}
	\mathcal{L} = \int \dd^4 \theta \,
	\left( 
	\Phi_j^\dagger e^{Q_j V + i m_j \theta \widebar{\theta}} \Phi_j
	- \frac{1}{e^2} \Sigma^2
	-\frac{\kappa}{4 \pi} \Sigma V
	- \frac{\zeta}{2 \pi} V
	\right) \,.
\end{equation} 

\subsection{Supersymmetric Index}
\label{sec:flat-SUSY-Index}
In supersymmetric quantum mechanics, there is a natural ``invariant'' \index[sym]{$\Tr (-1)^F$}
\begin{equation}
\label{eq:susy-index-def}
	\Tr (-1)^F  := \Tr_\mathcal{H} (-1)^F e^{-\beta H}
\end{equation}
called the \index{supersymmetric index} supersymmetric index~\cite{witten1982} or \index{Witten index} Witten index. The base space is taken to be a circle of radius $\beta$, and $H = \{Q, \widebar{Q}\}$ is the hamiltonian operator. Crucially, the supersymmetric index is invariant under supersymmetric deformations to the lagrangian. The Hilbert space $\mathcal{H}$ is graded by the Fermion number operator $F$. If the spectrum is gapped, it only receives contributions from the supersymmetric vacua. In fact, it counts the difference of the numbers of bosonic and fermionic ground states. Geometrically the supersymmetric index is~\cite{hori2003} the \index{Euler characteristic} Euler characteristic of the $Q$-complex. 
\vspace{-\parskip}
\begin{itemize}
	\item For a sigma model of a quantum mechanics with a riemannian target manifold $M$, it is identified with the Euler characteristic of $M$,
	\begin{equation}
		\Tr (-1)^F = \sum_{f} (-1)^f \dim H^f(Q) = \chi (M)
	\end{equation}
	via \index{de Rham cohomology} de Rham cohomology. 
	\item For a sigma model to a Kh\"aler target manifold $M$ endowed with a holomorphic vector bundle $E$, it is identified~\cite{bullimore2019} with the holomorphic Euler characteristic 
	\begin{equation}
	\label{eq:susy-index-hol-euler-char}
		\Tr (-1)^F = \chi(M, K^{\sfrac{1}{2}} \otimes E) = \int_M \hat{A}(TM) \ch(E)
	\end{equation}
	via \index{Dolbeault cohomology} Dolbeault cohomology.
\end{itemize}

It plays a central role in studying the geometry of supersymmetric quantum field theories. For a three-dimensional theory, the supersymmetric index can be computed by splitting the base manifold into the product $S^1 \times T^2$ of a temporal circle and a spatial torus, and reducing to a quantum mechanics on $S^1$. This \index{torus index} torus index gives us the number of supersymmetric vacua~\cite{intriligator2013} weighted by their multiplicities. In addition to being invariant under $Q$-exact deformations, the index is also invariant under deformations to the mass and Fayet-Iliopoulos paraters.

\section{Abelian Gauge Theory}
\label{sec:U1-gauge-theories}

Consider an $\mathcal{N}=2$ supersymmetric $U(1)_\kappa$ Chern-Simons \index{gauge theory} gauge theory at level $\kappa$ \index[sym]{$\kappa$} \footnote{We focus on supersymmetric Chern-Simons theories of rank $1$, leaving a discussion of higher rank theories to Appendix~\ref{chap:U1K-gauge-theories}.} with $N > 0$ \index{chiral multiplet} chiral multiplets 
\index[sym]{$\Phi_j$} 
$\{\Phi_j\}_{j = 1}^N$ of \index{gauge charge} gauge charges \index[sym]{$Q_{j}$} $Q_{j} \neq 0$ and integer \index{R-charge} R-charge $r_j \in \mathbb{Z}$. The complex \index{superpotential} superpotential is set to vanish $\mathcal{W}=0$ so that we can maximise the global symmetry.

There is a real \index{Fayet-Iliopoulos parameter} Fayet-Iliopoulos parameter $\zeta$ \index[sym]{$\zeta$} associated to the global topological symmetry $T_t = U(1)$, which could be enhanced to a non-abelian symmetry in the infra-red. 

There is a global flavour symmetry with maximal torus
\begin{equation}
\label{eq:flavour-sym}
T_{\text{f}} \cong \Big[ \bigotimes_{j=1}^N U(1)_j  \Big] / U(1) \,,
\end{equation}
where $U(1)_j$ rotates $\Phi_j$ with charge $+1$ and the quotient is by the gauge group. 
Correspondingly, we introduce \index{real mass} real mass parameters $\{ m_j\}_{j =1}^N$ \index[sym]{$m_j$} associated to a flavour symmetry $T_\text{f}$ for each of the chiral multiplets. 
A linear combination of these mass parameters can be absorbed by shifting the real \index{vector multiplet scalar} vector multiplet scalar \index[sym]{$\sigma$} $\sigma$, 
\begin{equation}
\begin{aligned}
	\sigma &\mapsto \sigma + \delta \sigma \,, \\
	\zeta &\mapsto \zeta - \kappa \delta \sigma \,, \\
	m_j &\mapsto m_j -  Q_{j} \delta\sigma \,.
\end{aligned}
\label{eq:shift}
\end{equation}
This transformation leaves invariant the combination $\zeta + \kappa \sigma$ and the total effective mass \index{effective mass}
\begin{equation}
	M_j(\sigma) := Q_j\sigma +m_j
\end{equation}
of each chiral multiplet. 
There are therefore only $(N-1)$ independent real mass parameters associated to the flavour symmetry $T_\text{f} \cong U(1)^{N-1}$. 
This may form a maximal torus of a non-abelian flavour symmetry, e.g., $PSU(N)$ if $Q_j=1$.

It is convenient to fix the independent mass parameters $\{m_{a'}\}_{a' = 2}^N$ by writing \index[sym]{$m_{a'}$}
\begin{equation}
\label{eq:U(1)-independent-mass-parameters}
	m_j = \sum_{a'=2}^N q^{a'}{}_j m_{a'}
\end{equation}
in terms of the integer \index{flavour charge} flavour charge matrix $q^{a'}{}_j$. The flavour charge matrix is defined up to shifts $q^{a'}{}_j \to q^{a'}{}_j + f^{\alpha'}Q_j$, which may be absorbed by $\delta \sigma = f^{\alpha'} m_{\alpha'}$.

We can also combine the \index{gauge charge} gauge and \index{flavour charge} flavour charges into a single \index{extended charge matrix} extended charge matrix \index[sym]{$Q^i_{\ \ j}$} $Q^i_{\ \ j}$, where the top rows $i=a=1,\ldots,K$ encode the gauge charges and the other rows $i=a'=K+1,\ldots,N$ are the flavour charges. The extended charge matrix is particularly useful when generalising to theories of higher rank gauge groups in Chapter~\ref{chap:U1K-gauge-theories}. In the case of 
$U(1)$ theories where $K=1$, the matrix elements are
\begin{align}
\label{eq:U(1)-extended-charge-matrix}
	Q^1_{\ \ j} &:= Q_j \,, \nonumber \\
	Q^{a'}_{\ \ j} &:= q^{a'}_{\ \ j} \quad \text{for} \quad a'=2,\ldots,N\,.
\end{align}

We also introduce various mixed supersymmetric Chern-Simons terms between the gauge symmetry, the flavour symmetry, and the R-symmetry. 
We focus here on the dynamical Chern-Simons terms involving the gauge symmetry. 
In addition to the pure gauge Chern-Simons term at level $\kappa$, there are \index{Chern-Simons level !mixed} also the mixed gauge-flavour terms at levels \index[sym]{$\kappa_{a'}$} $\{ \kappa_{a'}\}_{a' = 2}^N$ and the mixed gauge-R term at level \index[sym]{$\kappa_R$} $\kappa_R$. 
In the presence of real masses, both the parameters $\kappa$ and $\kappa_{a'}$ play a role in the determination of supersymmetric vacua below and in Chapter~\ref{chap:twisted-chiral-ring}. 
The parameter $\kappa_R$ also becomes important in Chapter~\ref{chap:twist-on-surface}, due to the presence of an R-symmetry background.

\subsection{Semi-Classical Vacua}

We now consider the supersymmetric vacua on $\mathbb{R}^3$ as a function of the real masses $m_j$ and the Fayet–Iliopoulos parameter $\zeta$. 

Integrating out chiral multiplets in the presence of generic real masses generates additional contributions to the Chern-Simons term~\cite{Redlich:1983kn,Redlich:1983dv,Alvarez-Gaume:1983ihn} from one-loop diagrams. 
The geometric interpretation of this mechanism is discussed in Section~\ref{sec:CS-from-det}. The resulting \index{Chern-Simons level !mixed} \index{Chern-Simons level !effective} mixed effective Chern-Simons levels $\kappa^\eff$ are given by \index[sym]{$\kappa^\eff$}
\begin{subequations}
\label{eq:U(1)-effective-CS-levels}
\begin{align}
	\kappa^\eff(\sigma) & = \kappa + \frac{1}{2} \sum_{j=1}^N Q^2_j \sign(M_j(\sigma)) \,, \label{eq:U(1)-effective-CS-level-gauge}\\
	\kappa^\eff_{a'}(\sigma) & = \kappa_{a'} + \frac{1}{2} \sum_{j=1}^N Q_j Q_{a'j} \sign(M_j(\sigma))  \,, \\
	\kappa_R^\eff(\sigma) & = \kappa_R + \frac{1}{2}\sum_{j=1}^N Q_j(r_j-1) \text{sign}( M_j(\sigma) )  \,.
\end{align}
\end{subequations}
These are piecewise constant functions of the vector multiplet scalar $\sigma$ that jumps discontinuously at points $\sigma = - m_j / Q_j$ where the effective mass $M_j(\sigma) = 0$. 
They are required to be integer-valued to cancel potential parity anomalies, i.e.,
\begin{subequations}
\label{eq:U(1)-CS-anomaly-cancellation-cond} 
\begin{alignat}{3}
	&\kappa + \frac{1}{2} \sum_{j=1}^N Q_j^2 \quad &&\in \mathbb{Z} \,, \label{eq:U(1)-CS-quantisation-cond} \\
	&\kappa_{a'} + \frac{1}{2} \sum_{j=1}^N Q_j Q_{a'j} \quad &&\in \mathbb{Z} \,, 
	\label{eq:U(1)-CS-quantisation-cond-fl} \\
	&\kappa_{R} + \frac{1}{2} \sum_{j=1}^{N} Q_j (r_j - 1) \quad &&\in \mathbb{Z} \,.
	\label{eq:U(1)-CS-quantisation-cond-R} 
\end{alignat}
\end{subequations}

We define the \index{Chern-Simons level !asymptotic} asymptotic Chern-Simons levels by \index[sym]{$\kappa^\pm$}
\begin{subequations}
\label{eq:U(1)-asymp-CS-levels}
\begin{align}
	\kappa^{\pm} &:= \kappa^\eff(\sigma\rightarrow \pm\infty) = \kappa \pm \frac{1}{2} \sum_{j=1}^N |Q_{j}| Q_{j} \,, \label{eq:U(1)-asymp-CS-level-gauge}\\
	\kappa_{a'}^{\pm} &:= \kappa_{a'}^\eff(\sigma \rightarrow \pm \infty) = \kappa_{a'} \pm \frac{1}{2} \sum_{j=1}^N |Q_{j}| q_{a'j} \,, \\
	\kappa_R^{\pm} &:= \kappa_{R}^\eff(\sigma \rightarrow \pm \infty) = \kappa_{R} \pm \frac{1}{2} \sum_{j=1}^N |Q_{j}| (r_j-1) \,,
\end{align}
\end{subequations}
which control the \index{gauge charge} gauge charge, the \index{flavour charge} flavour charge and the \index{R-charge} R-charge of Bogomol'nyi–Prasad–Sommerfield \index{monopole} monopole operators of topological charges $\pm 1$ respectively.

Hence after integrating out auxiliary fields from the lagrangian~\eqref{eq:U(1)-lagrangian}, the semi-classical scalar potential is obtained as~\cite{dorey2000,intriligator2013}
\begin{equation}
\label{eq:U(1)-scalar-potential}
	U =  e^2 \left( \sum_{j=1}^{N} Q_j |\phi_j|^2  - F(\sigma) \right)^2 + \sum_{j=1}^N M_j(\sigma)^2 |\phi_j|^2 \,,
\end{equation}
where $\phi_j$ \index[sym]{$\phi_j$} is the \index{chiral multiplet scalar} chiral multiplet scalar. The effective parameter \index[sym]{$F(\sigma)$} $F(\sigma)$ can be written as
\begin{align}
	F(\sigma) 
	&=  \zeta^\eff(\sigma) + \kappa^\eff(\sigma) \sigma \nonumber \\
	&= \zeta + \kappa \sigma +\sum_{a'=2}^N \kappa_{a'} m_{a'} + \frac{1}{2} \sum_{j=1}^N Q_j |M_j| \,,
\label{eq:U(1)-F-a'}
\end{align}
where the \index{Fayet-Iliopoulos parameter !effective} effective Fayet-Iliopoulos parameter is
\begin{equation}
	\zeta^\eff(\sigma) 
	=  \zeta + \sum_{a'=2}^{N} \kappa_{a'}^{\eff}(\sigma)m_{a'} \,.
\end{equation}
It captures the combined effects of the effective Chern-Simons terms generated by integrating out chiral multiplets. The effective parameter $F(\sigma)$ is a continuous piecewise linear-function of $\sigma$ whose slope jumps discontinuously at the points $\sigma = - m_j / Q_j$. Both $\zeta^\eff$ and $\kappa^\eff$ are piece-wise constant functions, which respectively determine the intercept and slope of $F(\sigma)$. This qualitative difference enables us to tune $\zeta$ by hand as a background parameter, while $\kappa$ is regarded as a dynamical parameter.

It is sometimes more convenient to introduce \index{Chern-Simons level !mixed} mixed Chern-Simons levels $$\{\kappa_j, \kappa_{Rj} \mid j =1,\ldots,N\}$$ for each\index{chiral multiplet} chiral multiplet satisfying $\kappa_j + \frac{1}{2}Q_j \in \mathbb{Z}$, from which we can recover 
\begin{subequations}
\label{eq:U(1)-mixed-CS-levels-for-each-chiral}
\begin{align}
	&\kappa = \sum_{j=1}^N Q_{j} \kappa_j \,, \\ 
	&\kappa_{a'} = \sum_{j=1}^N Q_{a'j} \kappa_j \,,\\
	&\kappa_{Ra'} = \sum_{j=1}^N Q_{a'j} \kappa_{Rj} \,.
\end{align}
\end{subequations}
It allows us to write the \index{Chern-Simons level !effective} effective Chern-Simons levels $\kappa^{\eff}$ and $\kappa^{\eff}_{a'}$ in~\eqref{eq:U(1)-effective-CS-levels} simply as
\begin{equation}
	\kappa^\eff_j (\sigma) = \kappa_j + \frac{1}{2}	Q_j \sign(M_j(\sigma)) 
\end{equation}
with an anomaly cancellation condition
\begin{equation}
	\kappa_j + \frac{1}{2}	Q_j \quad \in \mathbb{Z} \,. 
\end{equation}
In this notation, the effective parameter \index[sym]{$F(\sigma)$} $F(\sigma)$ becomes
\begin{align}
	F(\sigma) 
	&=  \zeta^\eff(\sigma) + \kappa^\eff(\sigma) \sigma \nonumber \\
	&= \zeta + \kappa \sigma +\sum_{j=1}^N \kappa_{j} m_{j} + \frac{1}{2} \sum_{j=1}^N Q_j |M_j| \,,
\end{align}
where the effective Fayet-Iliopoulos parameter is
\begin{equation}
	\zeta^\eff(\sigma) 
	=  \zeta + \sum_{j=1}^{N} \kappa_{j}^{\eff}(\sigma)m_{a'} \,.
\end{equation}

The semi-classical supersymmetric vacua \index{vacuum !semi-classical} are constant solutions to the following set of\index{vortex equation} vortex equations
\begin{subequations}
\label{eq:U(1)-vacuum}
\begin{align}
	\sum_{j=1}^{N} Q_{j} |\phi_j|^2  &= F(\sigma) \,,
	\label{eq:U(1)-vacuum-D}	
	\\
	M_j(\sigma) \phi_j &= 0  \,,
\end{align}
\end{subequations}
where~\eqref{eq:U(1)-vacuum-D} is the \index{D-term equation}D-term equation.

\subsection{Classification of Vacua}
\label{sec:flat-vacuum-classes}
The solutions to these equations display an intricate dependence on the Fayet–Iliopoulos parameters $\zeta$ and mass parameters $m_j$. The type of solutions fall into the following trichotomy.

\subsubsection*{Higgs Branch}

Higgs branch vacua \index{vacuum !Higgs branch} are solutions where at least one chiral multiplet from $\{\phi_j\}_{j=1}^N$ is non-zero and the gauge symmetry is broken to a discrete subgroup. The vector multiplet scalar $\sigma$ is completely fixed by the D-term equations $M_j(\sigma) =0$ for all non-vanishing chiral multiplets $\phi_j$. 
	
A Higgs branch solution where $n$ chiral multiplets are non-vanishing  may only appear on a real co-dimension $(n-1)$ hyperplane in the space $\mathbb{R}^{N-1}$ of real mass parameters. Suppose the non-vanishing chiral multiplets are $\{\phi_{i_\alpha}\mid \alpha = 1,\ldots,n\}$, then the locus of the hyperplane is
\begin{equation}
m_{i_1}/Q_{i_1} = \cdots  = m_{i_n}/Q_{i_n} \, .
\end{equation}
The non-zero expectation values must further satisfy
		\begin{equation}
			\sum_{\alpha =1}^n Q_{i_\alpha} |\phi_{i_\alpha}|^2 = F(\sigma)\,,
		\end{equation}
forming an $(n-1)$-dimensional complex \index{toric variety} toric variety of Higgs branch solutions. 

If the charges of the chiral multiplets receiving expectation values obey 
$$\text{gcd}(Q_{i_1},\ldots,Q_{i_n}) >1 
\,,$$ 
there is an unbroken discrete gauge symmetry. Then this branch of the moduli space should be regarded as a \index{toric stack} toric stack.

When all the mass parameters vanish, $m_j = 0$, for a given choice of the Fayet–Iliopoulos parameter $\zeta \neq 0$, there is a maximal complex $N$-dimensional Higgs branch at $\sigma = 0$. 
We typically denote this $N$-dimensional \index{toric stack} toric stack by $X$. 
It can be understood explicitly as a weighted projective stack
\begin{equation}
	X \cong 
	\begin{cases}
	\mathbb{CP}(Q_1,\ldots,Q_N) \quad & \text{if} \quad \zeta > 0 \\
	\mathbb{CP}(-Q_1,\ldots,-Q_N) \quad & \text{if} \quad \zeta < 0 
	\end{cases}
	\, .
\end{equation}
When $\zeta >0$ it is empty if $Q_j <0$ for all $j = 1,\ldots,N$. Similarly when $\zeta <0$ it is empty if $Q_j > 0$ for all $j = 1,\ldots,N$.
This is the space whose \index{quantum K-theory} quantum K-theory with level structure we wish to study using supersymmetric gauge theory.

When turning on a non-vanishing mass parameter $m_i$, the remaining Higgs branch moduli space can be regarded as the fixed points of the action of the one-parameter subgroup of $T_\text{f}$ generated by the mass parameters $m_i$ on $X$. 
In the extreme case of generic mass parameters, there are $N$ isolated Higgs branch vacua $\{h_i \mid i=1,\ldots, N\}$ where $\phi_i \neq 0$, which are the fixed points of the $T_\text{f}$ action on $X$. 
Each isolated Higgs vacuum $h_i$ is fixed at $\sigma_i = - m_i / Q_i$ by the D-term equation, and the remaining vortex equation 
\begin{equation}
	 Q_{i} |\phi_i|^2  = F(\sigma_i)
\end{equation}
demands
\begin{equation}
	\sign Q_i = \sign F(\sigma_i) \,.
\end{equation}

If $|Q_i| >1$, there is an unbroken $\mathbb{Z}_{|Q_i|} \subset U(1)$ gauge symmetry in the vacuum $h_i$. 
If the mass parameters remain small compared to $\zeta$, the $N$ isolated vacua can be identified with the fixed loci as the classifying space $h_i := B\mathbb{Z}_{|Q_i|} \subset X$ of the $T_f$ action on $X$.

The contribution to the \index{supersymmetric index} torus supersymmetric index from this isolated vacuum is
\begin{equation}
	\Tr(-1)^F = Q_i^2 \,,
\end{equation}
which can be understood physically as coming from the $|Q_i|$ \index{Wilson line} Wilson lines screened by the chiral multiplet $\Phi_i$ wrapping around each of the two non-trivial cycles of the torus $T^2$.

\subsubsection*{Topological Branch}

Topological branch \index{vacuum !topological branch} are solutions where the $U(1)$ gauge symmetry is unbroken, $\phi_j = 0$ for all $j = 1,\ldots,N$. The vector multiplet scalar 
\begin{equation}
	\sigma = - \frac{\zeta^\eff}{\kappa^\eff} 
\end{equation}	
is a fixed solution to $F(\sigma) = 0$. 
For generic mass parameters, this can only occur if $\kappa^\eff \neq 0$ and
\begin{equation}
\label{eq:U(1)-topological-vacua-sign-alignment}
	\sign \zeta^\eff = \pm \sign \kappa^\eff
\end{equation}
for $\pm \sigma < 0$.

It is a low-energy effective $U(1)_{\kappa^\eff}$ theory without matter. The contribution to the \index{torus index} torus index is 
\begin{equation}
	\Tr(-1)^F = |\kappa^\eff| \,,
\end{equation}
where the \index{Wilson line} Wilson lines are screened by the Bogomol'nyi–Prasad–Sommerfield \index{monopole} monopoles~\cite{Manton:1981mp} of gauge charge $\kappa^\eff$. However, those Wilson lines wrapping on the two different non-trivial cycles of $T^2$ become correlated by quantum effects, thus no longer contribute independently to the index.

In the absence of mass parameters, the effective Chern-Simons level reduces to the asymptotic levels
\begin{align}
	\kappa^{\eff}(\sigma)  
	= 
	\begin{cases}
	\kappa^+ &  \quad \text{if} \quad \sigma > 0  \\
	\kappa^- & \quad \text{if} \quad \sigma < 0 
	\end{cases}  \,.
\end{align}
The potential solutions become
\begin{equation}
	\sigma = - \frac{\zeta}{ \kappa^\pm }
\end{equation}
in the regions $\pm \sigma >0$. The existence of a topological vacuum with $\pm \sigma >0$ therefore requires both $\kappa^\pm \neq 0$ and $\mp \zeta / \kappa^\pm >0$. If the topological vacuum exists, then the infra-red theory is $U(1)_{\kappa^\pm}$,  contributing $|\kappa^\pm|$ to the supersymmetric index.

Turning on mass parameters, the analysis of potential topological vacua is more intricate, which depends on the chamber in the parameter space for the vector multiplet scalar $\sigma$ separated by the walls $\sigma = - m_j / Q_j$. These walls collapse to the origin when the masses are turned off.

\subsubsection*{Coulomb Branch}

Coulomb branch vacua \index{vacuum !Coulomb branch} refers to the non-isolated solutions of the vector multiplet scalars $\sigma_a$ where $F(\sigma)=\kappa^\eff(\sigma)=0$ when all \index{chiral multiplet} chiral multiplet scalars $\phi_i$ vanish. 
It is a continuous parameter space of solutions where the gauge symmetry $U(1)$ is unbroken. 
This requires special tuning on the Fayet-Iliopoulos parameter $\zeta$.

In the absence of mass parameters, there is non-compact Coulomb branch parametrised by $\pm \sigma  \geq 0$ whenever $\kappa^\pm = 0$ at the Fayet-Iliopoulos parameter $\zeta = 0$. Generally in the presence of mass parameters, there exists a Coulomb branch whenever there is a chamber for the vector multiplet scalar $\sigma$ where $\kappa^{\eff}(\sigma) = 0$ and $F(\sigma)$ has zero slope. Tuning the Fayet-Iliopoulos parameter such that $\zeta^\eff = 0$ then ensures that $F(\sigma) = 0$ for all values of the vector multiplet scalar in this chamber. In this case, the Coulomb branch may be compact or non-compact depending on the chamber.

The focus in this thesis is on the vacua occurring with generic parameters where the Coulomb branch does not appear.

\section{Window Phenomenon}

Now consider the collection of supersymmetric gauge theories $\{\mathcal{T}_\kappa\}$ with fixed charges $\{Q_j\}_{j = 1}^N$ and varying supersymmetric \index{Chern-Simons level} Chern-Simons level $\kappa$. Let us first set the mass parameters to zero, and fix the sign of \index{Fayet-Iliopoulos parameter} Fayet–Iliopoulos parameter $\zeta$ such that the associated $N$-dimensional Higgs branch $X$ exists. 

We are interested in the set of supersymmetric Chern-Simons levels where there are no additional topological vacua. 
We define this set as the \index{critical window} critical window for the supersymmetric Chern-Simons levels. 
\begin{defn}
	The critical window in $\kappa$ consists of those theories $\mathcal{T}_\kappa$ that do not admit topological vacua in addition to the Higgs branch $X$.
\end{defn}
\vspace{-\parskip}
This window depends on the choice of chamber for the Fayet-Iliopoulos parameter $\zeta$ since it affects the existence of Higgs vacua. In the $U(1)$ case, the chambers are simply the two rays determined by $\sign \zeta$.

Loosely speaking, inside the critical window the gauge theory flows to a sigma model on $X$, and supersymmetric observables capture the geometry of $X$. In particular, if $X$ is a toric Deligne-Mumford stack, this will involve an unbroken discrete gauge symmetry.

Now we define the \index{Chern-Simons level !critical} critical Chern-Simons level $\kappa^\crit (\sigma)$ to be the bare Chern-Simons levels such that the effective Chern-Simons level in \eqref{eq:U(1)-effective-CS-levels} vanish, i.e., \index[sym]{$\kappa^\crit$}
\begin{equation}
\label{eq:U(1)-critical-CS-level}
	0 =  \kappa^\crit (\sigma) + \frac{1}{2} \sum_{j=1}^N Q_j Q_j \sign M_j(\sigma) \,.
\end{equation}
The critical level $\kappa^\crit (\sigma)$ depends on the sign of the effective masses $M_j(\sigma)$, thus has a piece-wise dependence on $\sigma$.

When the real masses are set to vanish, the critical Chern-Simons level reduces to
\begin{equation}
	\kappa^\crit (\sigma) = 
	\begin{cases}
		- \frac{1}{2} \sum_{j=1}^N |Q_{j}| Q_{j} \,,\quad &\text{if } \sigma > 0 \\
		+ \frac{1}{2} \sum_{j=1}^N |Q_{j}| Q_{j} \,,\quad &\text{if } \sigma < 0
	\end{cases} \,.
\end{equation}
Given the geometric regime $\zeta>0$ for the Higgs branch to exist, the condition~\eqref{eq:U(1)-topological-vacua-sign-alignment} for the existence of topological vacua becomes
\begin{equation}
	\begin{cases}
		\kappa < - \frac{1}{2} \sum_{j=1}^N |Q_{j}| Q_{j} \,,\quad &\text{if } \sigma > 0 \\
		\kappa > + \frac{1}{2} \sum_{j=1}^N |Q_{j}| Q_{j} \,,\quad &\text{if } \sigma < 0
	\end{cases} \,.
\end{equation}
It is convenient to define an \index[sym]{$\tilde{\kappa}^\crit$} asymptotic critical level
\begin{equation}
	\tilde{\kappa}^\crit := \kappa^\crit(\sigma \to -\infty) = \frac{1}{2} \sum_{j=1}^N |Q_{j}| Q_{j} \,.
\end{equation}
\vspace{-\parskip}
\begin{itemize}
	\item If $\tilde{\kappa}^\crit > 0$, then the critical window where Higgs branch exists without topological branch is simply the finite interval 
$$
	\left[ - \tilde{\kappa}^\crit,  \tilde{\kappa}^\crit \right]
$$ 
in the space of Chern-Simons parameters.
	\item If $\tilde{\kappa}^\crit < 0$, then the critical window is infinite containing two rays
$$
	\left[ - \infty,  \tilde{\kappa}^\crit \right] \cup \left[ -\tilde{\kappa}^\crit, \infty\right] \,.
$$ 
\end{itemize}
Inside the critical window the theory flows to a sigma model to $X = \mathbb{CP}(Q_1,\ldots,Q_N)$. A similar analysis applies for $\zeta < 0$ giving a sigma model to $X = \mathbb{CP}(-Q_1,\ldots,-Q_N)$ within a critical window complementary to the $\zeta>0$ case.

Quantum K-theory of toric varieties with level structures~\cite{ruan2019,ruan2020} also exhibits a window phenomenon where it is well-defined only if the level is within a critical window. We conjecture that this is the same as the \index{critical window} critical window of Chern-Simons level in three-dimensional supersymmetric gauge theory. This identification is investigated throughout this thesis from the perspectives of semi-classical vacua, twisted chiral rings, and twisted indices. Via this identification, the interpretation is that quantum K-theory is well behaved inside the window of levels where the supersymmetric vacua consist of only Higgs branch described by toric varieties. The addition of topological vacua outside the window is not taken into account in the current literature in \index{quantum K-theory} quantum K-theory, making the theory ill-defined.

In addition to the massless case, we are also interested in the case where real masses are turned on. It enables us to determine if topological vacua can co-exist with an individual disjoint component in the Higgs vacua. In this case, the space of $\sigma$ is divided into chambers by the hyperplanes at $M_j(\sigma) = 0$. Each chamber admits a different  critical level. Away from these critical levels, there may exist topological vacua in addition to Higgs vacua.
Although each individual chamber requires separate analysis to determine what values of $\kappa$ allow for the co-existence of Higgs and topological vacua.
We are primarily interested in turning on generic mass parameters that are small compared to $\zeta$ so the isolated Higgs vacua $\{h_1, \ldots,h_N\}$  can still be identified with the fixed loci on $X$.  
We have observed that for $U(1)$ theories these chamber-specific critical levels still fall into the same critical window of \index{Chern-Simons level} Chern-Simons levels as in the massless case. This is illustrated in Section~\ref{subsubsec:U(1)+2-vacuum-example} with an explicit example. 

The \index{window phenomenon} window phenomenon is already manifest by carefully studying the vacuum equations \eqref{eq:U(1)-vacuum}. Let us look at some examples in details to illustrate it.


\section{Examples}

For a set of generic parameters $\{\zeta, m_j\}$, we would like to understand what values of \index{Chern-Simons level} Chern-Simons levels $\kappa$ allow for the existence of topological vacua, given the existence of Higgs vacua.

\subsection[\texorpdfstring{$U(1)_\kappa$ with One Chiral Field}%
                        {U(1) with One Chiral Field}]%
        {$U(1)_\kappa$ with One Chiral Field}
\label{subsubsec:U(1)+1-vacuum-example}
Consider one \index{chiral multiplet} chiral multiplet $\Phi$ of charge $Q>0$. We can set the mass parameter $m=0$ by shifting the vector multiplet scalar $\delta \sigma = m/Q$, so there is no flavour symmetry. There is however a potential one-form symmetry of the form $\mathbb{Z}_{\text{gcd}(k,Q)}$.

The \index{Chern-Simons level !effective} effective Chern-Simons level is given by
\begin{equation}
	\kappa^\eff(\sigma) = \kappa + \frac{Q^2}{2} \sign \sigma =
	\begin{cases}
		\kappa + \frac{Q^2}{2} \,, \quad &\text{if } \sigma > 0 \\
		\kappa - \frac{Q^2}{2} \,, \quad &\text{if } \sigma < 0
	\end{cases} \,,
\end{equation}
and the effective parameter is 
\begin{equation}
F(\sigma) = \zeta + \kappa \sigma + \frac{Q^2}{2}|\sigma| \, .
\end{equation}
There are therefore two chambers for the vector multiplet scalar with critical levels
\begin{equation}
	\kappa^\crit = 
	\begin{cases}
		- \frac{Q^2}{2} \,, \quad &\text{if } \sigma > 0 \\
		+ \frac{Q^2}{2} \,, \quad &\text{if } \sigma < 0
	\end{cases} \,.
\end{equation}
The semi-classical vacua are solutions to
\begin{equation}
	Q|\phi|^2  = F(\sigma)\,, \qquad
	\sigma \phi = 0 \,.
\end{equation}
We consider here the geometric regime $\zeta>0$ where a Higgs vacuum exists.

First, there is the isolated \index{vacuum !Higgs branch} Higgs branch solution with $|\phi|^2 = \zeta / Q$ with an unbroken gauge symmetry $\mathbb{Z}_Q$. The associated $\mathbb{Z}_Q$ orbifold leads to a Higgs vacua multiplicity~\cite{intriligator2013} of $Q^2$. Therefore the Higgs branch contributes $Q^2$ to the \index{supersymmetric index} supersymmetric index. Formally this solution should be regarded as the stack $X = \left[\text{pt}/\mathbb{Z}_Q\right]$.

In comparison, the \index{vacuum !topological branch} topological branch contributes according to the effective Chern-Simons level. The existence of a topological vacuum depends on the Chern-Simons level in the following way:
\vspace{-\parskip}
\begin{itemize}
	\item If $\kappa < -\frac{Q^2}{2}$, there is a topological vacuum with
		\begin{equation}
			\sigma = - \frac{\zeta}{\kappa+\frac{Q^2}{2}} > 0  \,.
		\end{equation}
		This corresponds to a low-energy $U(1)_{\kappa+\frac{Q^2}{2}}$ gauge theory without matter. It contributes the topological multiplicity $\left|k+\sfrac{Q^2}{2}\right|$ to the supersymmetric index.
	\item If $\kappa > \frac{Q^2}{2}$, there is a topological vacuum with 
		\begin{equation}
			\sigma = - \frac{\zeta}{\kappa-\frac{Q^2}{2}} < 0 \,.
		\end{equation}
		Similarly this corresponds to a $U(1)_{\kappa-\frac{Q^2}{2}}$ gauge theory with no matter, and contributes $\left|k-\sfrac{Q^2}{2}\right|$ to the supersymmetric index.
	\item If $-\frac{Q^2}{2} \leq \kappa \leq \frac{Q^2}{2}$ there are no topological vacua.
\end{itemize}
The \index{critical window} critical window is therefore $-\frac{Q^2}{2} \leq \kappa \leq \frac{Q^2}{2}$. Combining contributions from both the Higgs and the topological vacua, the supersymmetric index is $Q^2$ inside the critical window and $|k|+\frac{Q^2}{2}$ outside,
\begin{equation}
	\Tr (-1)^F = \begin{cases}
	Q^2 \,, & \quad \text{if} \quad |\kappa| \leq \frac{Q^2}{2}  \\
	|\kappa|+\frac{Q^2}{2} \,, & \quad \text{if} \quad |\kappa| > \frac{Q^2}{2} 
\end{cases} \,.
\end{equation}

\begin{figure}[!h]
\centering
\caption{Lower Limit of $F(\sigma)$ for U(1) Gauge Theory with One Chiral Multiplet}
\label{fig:effective-FI-U(1)+1}

\begin{tikzpicture}[scale=0.7]
	\draw[gray] 
		(-4,0) -- (5,0)  node[pos=1, right] {$\sigma$};
	\draw[gray] 
		(0,-3.5) -- (0,5.5) node[pos=1, left] {$F(\sigma)$};;
    
	\node[gray] at (-2.5,0.7) {$\kappa^\crit=\frac{1}{2}$};
	\node[gray] at (2.5,-0.7) {$\kappa^\crit=-\frac{1}{2}$};
  
	\draw[red,thick] 
		(0,0) -- (2.5,5)  node[pos=1, right] {$\kappa = \frac{3}{2}$};
	\draw[red,thick] 
		(0,0) -- (-3,-3);  
	
	\draw[darkgreen,thick] 
		(0,0) -- (3,3)  node[pos=1, right] {$\kappa = \frac{1}{2}$};
	\draw[darkgreen,thick] 
		(0,0) -- (-3,0);  
	
	\draw[blue,thick] 
		(0,0) -- (4,0) node[pos=1, above] {$\kappa = -\frac{1}{2}$};
	\draw[blue,thick] 
		(0,0) -- (-3,3); 
	
	\draw[cyan,thick] 
		(0,0) -- (3,-3) node[pos=1, right] {$\kappa = -\frac{3}{2}$};
	\draw[cyan,thick] 
		(0,0) -- (-2.5,5); 
\end{tikzpicture}
\end{figure}
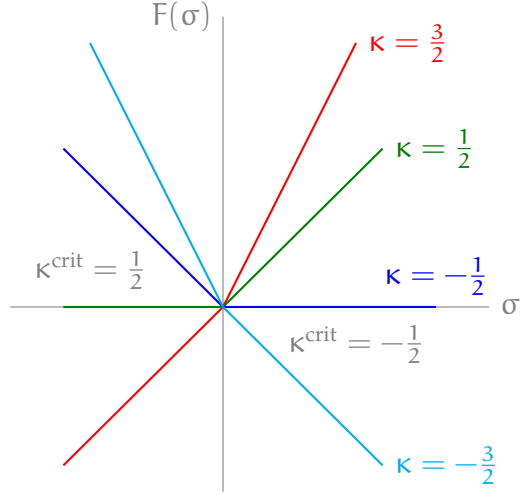

This can also be seen directly by plotting the limit of $F(\sigma)$, i.e.,
\begin{equation}
	F(\sigma) > \kappa \sigma + \frac{1}{2} |\sigma| = 
	\begin{cases}
		\left(\kappa + \frac{1}{2}\right) \sigma  \,, \quad &\text{if } \sigma \geq 0 \\
		\left(\kappa - \frac{1}{2}\right) \sigma  \,, \quad &\text{if } \sigma < 0
	\end{cases} \,,
\end{equation}
where $\zeta>0$ is assumed to ensure the existence of the Higgs branch. The plots for different $\kappa$ are shown in Figure~\ref{fig:effective-FI-U(1)+1} assuming $Q=1$. The effective parameter $F(\sigma)$ is obtained by a constant shift upwards from the corresponding lower limit. A Higgs vacuum exists at $\sigma = 0$ when $F(\sigma=0) > 0$. A topological vacuum appears at each isolated zeros of $F(\sigma)$. It is obvious that shifting the limits of $\kappa = -\frac{1}{2}, \frac{1}{2}$ does not produce any isolated zeros, while the shifts of $\kappa = -\frac{3}{2}, \frac{3}{2}$ do. When $|\kappa| = \frac{1}{2}$ there are no topological vacua, but instead a non-compact Coulomb branch opens up in the limit $\zeta \to 0^+$.

\subsection[\texorpdfstring{$U(1)_\kappa$ with Two Chiral Fields}%
                        {U(1) with Two Chiral Fields}]%
        {$U(1)_\kappa$ with Two Chiral Fields}
\label{subsubsec:U(1)+2-vacuum-example}

We focus here on the case of two \index{chiral multiplet} chiral multiplets with $Q_1 = Q_2 = 1$. Let us set the \index{flavour charge} flavour charges to $q_1 = 0$ and $q_2 = 1$. The full extended charge matrix is then
\begin{equation}
	Q^i_{\ \ j}=
	\begin{pmatrix}
		1 & 1 \\
		0 & 1
	\end{pmatrix} \,.
\end{equation}
Without loss of generality, one of the real masses is set to $m_1=0$ by shifting the vector multiplet scalar by $\delta \sigma = m_1$, and the other real mass is assumed to be $m_2 = -m < 0$. The other case can be analysed analogously, which has no structural difference. For convenience, we set the bare gauge-flavour Chern-Simons level to $\kappa_{2} = -\frac{1}{2}$, which contributes $\frac{1}{2} m$ in the scalar potential.

In the three chambers of $\sigma$, the \index{Chern-Simons level !effective} effective Chern-Simons level is then
\begin{align}
\label{eq:effective-CS-U(1)+2-example}
	\kappa^\eff &= \kappa + \frac{1}{2} \sign \sigma + \frac{1}{2} \sign (\sigma - m) \nonumber \\
	&=
	\begin{cases}
		\kappa + 1 \,, \quad &\text{if } \sigma > m \\
		\kappa \,, \quad &\text{if } 0 < \sigma < m \\
		\kappa - 1 \,, \quad &\text{if } \sigma < 0
	\end{cases} \,,
\end{align}
giving critical levels
\begin{equation}
	\kappa^\crit = 
	\begin{cases}
		-1 \,, \quad &\text{if } \sigma > m \\
		0 \,, \quad &\text{if } 0 < \sigma < m \\
		+1 \,, \quad &\text{if } \sigma < 0
	\end{cases} \,.
\end{equation}

The vacuum equations \eqref{eq:U(1)-vacuum} read as 
\begin{subequations}
\begin{align}
	|\phi_1|^2 + |\phi_2|^2 &= F(\sigma) = \zeta + \frac{1}{2} m + \kappa \sigma + \frac{1}{2} |\sigma| + \frac{1}{2} |\sigma-m| \,,\\
	\sigma \phi_1 &= 0 \,, \\
	(\sigma-m) \phi_2 &= 0 \,.
\end{align}
\end{subequations}

\subsubsection*{Higgs Branch}
A \index{vacuum !Higgs branch} Higgs branch can exist for $\phi_1 \neq 0$ at $\sigma = 0$, or $\phi_2 \neq 0$ at $\sigma = m$ according to the D-term equation~\eqref{eq:U(1)-vacuum-D}. The vacuum can then be found as the solutions to the vortex equation
\begin{equation}
	|\phi_1|^2 = \zeta + m \qquad \text{or} \qquad |\phi_2|^2 = \zeta + \left(\kappa + 1 \right) m \,,
\end{equation}
which forms a solution space $S^1$ provided the conditions
\begin{subequations}
\label{eq:higgs-cond-U(1)+2}
\begin{align}
	h_1: &\qquad \zeta > - m \,, \quad \\
	h_2: &\qquad \zeta > - \left(\kappa + 1\right) m\,.
\end{align}
\end{subequations}
Each of the Higgs vacua $\{h_1, h_2\}$ contribute $1$ to the supersymmetric index. The two disjoint Higgs vacua merge into a moduli space $\mathbb{CP}^1$ when the mass parameter $m \rightarrow 0$.

\subsubsection*{Topological Branch}
A \index{vacuum !topological branch} topological branch can be found by solving for isolated solutions to 
\begin{equation}
	F(\sigma) = \zeta + \frac{1}{2} m + \kappa \sigma + \frac{1}{2} |\sigma| + \frac{1}{2} |\sigma-m| = 0 \,.
\end{equation}
The solution depends on $\sign \sigma$ and $\sign (\sigma-m)$:
\vspace{-\parskip}
\begin{itemize}
	\item In the chamber $\sigma > m$, we have 
		\begin{equation}
			\sigma = - \frac{\zeta}{\kappa+1} \,.
		\end{equation}
		Requiring $\sigma > m$ gives
		\begin{equation}
		\label{eq:top-condition-sigma>m}
			\begin{cases}
				\zeta < - \left( \kappa + 1 \right) m \,, \quad &\text{if } \kappa > -1\\
				\zeta > - \left( \kappa + 1 \right) m \,, \quad &\text{if } \kappa < -1 
			\end{cases} \,.
		\end{equation}
		
	\item In the chamber $0 < \sigma < m$, we have 
		\begin{equation}
			\sigma = - \frac{\zeta + m}{\kappa} \,.
		\end{equation}
		Requiring $0 < \sigma < m$ gives
		\begin{equation}
		\label{eq:top-condition-0<sigma<m}
			\begin{cases}
				- (\kappa + 1) m < \zeta < - m \,, \quad &\text{if } \kappa > 0 \\
				- m < \zeta < - \left( \kappa + 1 \right) m \,, \quad &\text{if } \kappa < 0 
			\end{cases} \,.
		\end{equation}
		
		\item In the chamber $\sigma < 0$, we have 
		\begin{equation}
			\sigma = - \frac{\zeta + m}{\kappa-1} \,.
		\end{equation}
		Requiring $\sigma < 0$ gives
		\begin{equation}
		\label{eq:top-condition-sigma<0}
			\begin{cases}
				\zeta > - m \,, \quad &\text{if } \kappa > 1 \\
				\zeta < - m \,, \quad &\text{if } \kappa < 1 
			\end{cases} \,.
		\end{equation}
\end{itemize}

\subsubsection*{Supersymmetric Index}

The total contributions to the supersymmetric index can be obtained by carefully examining the conditions where a topological vacuum can co-exist with either of the Higgs vacua $\{h_1, h_2\}$. 
\begin{itemize}
	\item In the chamber $\sigma > m$, 
		combining the condition~\eqref{eq:top-condition-sigma>m} for a topological vacuum to exist with the Higgs existence condition \eqref{eq:higgs-cond-U(1)+2} leads to constraints on the parameters. 
		\begin{subequations}		
		For a topological vacuum to co-exist with the Higgs vacuum $h_1$, the constraints are
		\begin{equation}
			\hspace{-39pt}			
			h_1: \quad
			\begin{cases}
				-m < \zeta < -(\kappa+1) m \,, \quad &\text{if } \kappa > -1 \\
				-(\kappa+1) m < \zeta \,, \quad &\text{if } \kappa < -1 
			\end{cases} \,.
		\end{equation}
		For a topological vacuum to co-exist with the Higgs vacuum $h_2$, the constraints read
		\begin{equation}
			\hspace{42pt}
			h_2: \quad
			\begin{cases}
				-(\kappa+1) m < \zeta < -(\kappa+1) m \,, \quad &\text{if } \kappa > -1 \\
				-(\kappa+1) m < \zeta \,, \quad &\text{if } \kappa < -1 
			\end{cases} \,.
		\end{equation}
		\end{subequations}
		
		When $\kappa > -1$, the combined $h_1$ constraint $-1<\kappa<0$ gives no valid value of $\kappa$ satisfying the anomaly cancellation condition \eqref{eq:U(1)-CS-quantisation-cond}, and the combined $h_2$ constraint is clearly always false. So there is no topological vacuum in addition to either of the Higgs vacua $h_1$ and $h_2$ for $\kappa > -1$. But a topological vacuum can always co-exist with both of the Higgs vacua for $\kappa < -1$. Therefore the total contribution to supersymmetric index is respectively $2$ for $\kappa>-1$, and $|\kappa+1| + 2 = -\kappa + 1$ for $\kappa \leq -1$.
		
	\item In the chamber $0 < \sigma < m$, combining~\eqref{eq:top-condition-0<sigma<m} with the Higgs existence condition \eqref{eq:higgs-cond-U(1)+2} leads to the following constraints.
		\begin{subequations}		
		For a topological vacuum to co-exist with the Higgs vacuum $h_1$, the constraints are
		\begin{equation}
			\hspace{-39pt}			
			h_1: \quad
			\begin{cases}
				-m < \zeta < -m \,, \quad &\text{if } \kappa > 0 \\
				-m < \zeta < -(\kappa+1) m  \,, \quad &\text{if } \kappa < 0 
			\end{cases} \,.
		\end{equation}
		For a topological vacuum to co-exist with the Higgs vacuum $h_2$, the constraints read
		\begin{equation}
			\hspace{42pt}
			h_2: \quad
			\begin{cases}
				-(\kappa+1) m < \zeta < -m \,, \quad &\text{if } \kappa > 0 \\
				-(\kappa+1) m < \zeta < -(\kappa+1) m \,, \quad &\text{if } \kappa < 0 
			\end{cases} \,.
		\end{equation}
		\end{subequations}
		
		There is a topological vacuum accompanying a single Higgs vacuum $h_1$ if $\kappa < 0$, and a topological vacuum accompanying $h_2$ if $\kappa>0$. So there can always exist a topological vacuum in addition to either one of the Higgs vacua in this chamber. The contribution to the supersymmetric index is then $|\kappa| + 1$ as the effective level~\eqref{eq:effective-CS-U(1)+2-example} is just $\kappa$ in this chamber.
		
		\item In the chamber $\sigma < 0$, combining~\eqref{eq:top-condition-sigma<0} with the Higgs existence condition \eqref{eq:higgs-cond-U(1)+2} leads to the following constraints.
		\begin{subequations}		
		For a topological vacuum to co-exist with the Higgs vacuum $h_1$, the constraints are
		\begin{equation}
			\hspace{-39pt}			
			h_1: \quad
			\begin{cases}
				-m < \zeta \,, \quad &\text{if } \kappa > 1 \\
				-m < \zeta < -m  \,, \quad &\text{if } \kappa < 1 
			\end{cases} \,.
		\end{equation}
		For a topological vacuum to co-exist with the Higgs vacuum $h_2$, the constraints read
		\begin{equation}
			\hspace{42pt}
			h_2: \quad
			\begin{cases}
				-m < \zeta \,, \quad &\text{if } \kappa > 1 \\
				-(\kappa+1) m < \zeta < -m \,, \quad &\text{if } \kappa < 1 
			\end{cases} \,.
		\end{equation}
		\end{subequations}		
		
		So there is no topological vacuum for $\kappa < 1$ but exists a topological vacuum for $\kappa > 1$, in addition to either of the Higgs vacua. Hence the total contribution to supersymmetric index is respectively $2$ for $\kappa<1$, and $|\kappa-1| + 2 = \kappa + 1$ for $\kappa \geq 1$.
\end{itemize}

In summary the supersymmetric index is
\begin{equation}
	\Tr (-1)^F = \begin{cases}
	2 \,, & \quad \text{if} \quad |\kappa| \leq 1  \\
	|\kappa|+1 \,, & \quad \text{if} \quad |\kappa| > 1 
\end{cases} \,.
\end{equation}

When all three chambers are considered for the whole theory, there is a topological vacuum in addition to both of the Higgs vacua, if and only if the Chern-Simons levels are not in the \index{critical window} critical window, i.e.,
\begin{equation}
	k \not\in \{-1,0,1\} \,.
\end{equation}

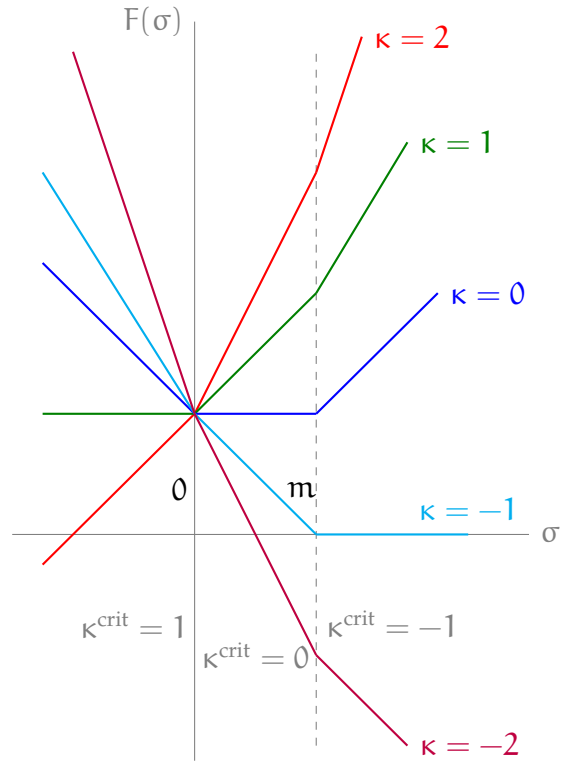
\begin{figure}[!h]

\centering

\caption{Plot of $F(\sigma)$ at $\zeta=0$ for U(1) Gauge Theory with Two Chiral Multiplet}
\label{fig:effective-FI-U(1)+2}

\begin{tikzpicture}[scale=0.4]
	\draw[gray] 
		(-6,-2) -- (11,-2)  node[pos=1, right] {$\sigma$};
	\draw[gray] 
		(0,-9.5) -- (0,15) node[pos=1, left] {$F(\sigma)$};
    
	\draw[gray,dashed]
		(4,-9) -- (4,14);

	\node[gray] at (-2,-5) {$\kappa^\crit=1$};
	\node[gray] at (2,-6) {$\kappa^\crit=0$};
	\node[gray] at (6.5,-5) {$\kappa^\crit=-1$};
  
  	\draw[red,thick] 
		(4,10) -- (5.5,14.5) node[pos=1, right] {$\kappa = 2$};
	\draw[red,thick] 
		(0,2) -- (4,10);
	\draw[red,thick] 
		(0,2) -- (-5,-3);  
	
	\draw[darkgreen,thick] 
		(4,6) -- (7,11)  node[pos=1, right] {$\kappa = 1$};
	\draw[darkgreen,thick] 
		(0,2) -- (4,6);
	\draw[darkgreen,thick] 
		(0,2) -- (-5,2);  
	
	\draw[blue,thick] 
		(4,2) -- (8,6) node[pos=1, right] {$\kappa = 0$};
	\draw[blue,thick] 
		(0,2) -- (4,2);
	\draw[blue,thick] 
		(0,2) -- (-5,7); 
	
	\draw[cyan,thick] 
		(4,-2) -- (9,-2) node[pos=1, above] {$\kappa = -1$};
	\draw[cyan,thick] 
		(0,2) -- (4,-2);
	\draw[cyan,thick] 
		(0,2) -- (-5,10); 
		
	\draw[purple,thick] 
		(4,-6) -- (7,-9) node[pos=1, right] {$\kappa = -2$};
	\draw[purple,thick] 
		(0,2) -- (4,-6);
	\draw[purple,thick] 
		(0,2) -- (-4,14); 
	
	\node[] at (-0.5,-0.5) {$0$};
	\node[] at (3.5,-0.5) {$m$};
\end{tikzpicture}

\end{figure}

This can be understood more intuitively by inspecting the plots of $F(\sigma)$ as shown in Figure~\ref{fig:effective-FI-U(1)+2}. Consider the case $\kappa = -1$ for example, the vacuum solutions depends the value of $\zeta$ as follows:
\vspace{-\parskip}
\begin{itemize}
	\item When $\zeta > 0$, the plot is shifted upwards such that it has no intersection with the $\sigma$-axis. Both Higgs vacua at $\sigma=0$ and $\sigma=m$ exist since $F(\sigma) > 0$ at these points. But there is no topological vacuum. 
	\item When $-m < \zeta < 0$, the plot is shifted downwards which has a single intersection with the $\sigma$-axis. There exist a single Higgs vacuum at $\sigma=0$, and a single topological vacuum at the intersection point $\sigma = \zeta + m$.
	\item When $\zeta \leq -m$, the plot is shifted downwards such that it intersects the $\sigma$-axis at $\sigma < 0$. There exist only a single topological vacuum at the intersection point $\sigma = \frac{\zeta}{2}+\frac{m}{2}$, with a multiplicity of two. The multiplicity becomes clear in Section~\ref{subsubsec:large-radius-N=2}.
\end{itemize}

At the non-generic value $\zeta = 0$, the horizontal part overlaps with the $\sigma$-axis. A Coulomb branch at $\sigma \geq m$ opens up, in addition to a signle Higgs vacuum at $\sigma=0$.

\subsection[\texorpdfstring{$U(1)_\kappa$ with $N$ Chiral Fields}%
                        {U(1) with N Chiral Fields}]%
        {$U(1)_\kappa$ with $N$ Chiral Fields}

Now consider $U(1)_\kappa$ with $N$ chiral multiplets of charge $Q_j=1$. There is a flavour symmetry $PSU(N)$ with maximal torus $T_f = U(1)^{N-1}$. One linear combination of the associated mass parameters $\{ m_j\}_{j=1}^N$ can be removed by the shift~\eqref{eq:shift}.
The semi-classical vacua are solutions to
\begin{subequations}
\begin{equation}
	\sum_j |\phi_j|^2  = F(\sigma)\,, \qquad
	(\sigma+m_j)  \phi_j = 0 \, ,
\end{equation}
where
\begin{equation}
F(\sigma) = \zeta + \kappa \sigma + \frac{1}{2} \sum_{j=1}^N | \sigma+m_j| \, .
\end{equation}
\end{subequations}
Here we have omitted the mixed gauge-flavour Chern-Simons levels $\kappa_{a'}$ from~\eqref{eq:U(1)-F-a'}. They do not play a role in the analysis below, assuming they are chosen appropriately so no parity anomalies appear.
The existence of a Higgs vacuum with $\phi_i \neq 0$ requires $F(-m_i) >0$ while topological vacua are solutions to $F(\sigma) = 0$.

Here we set the mass parameters to zero. Therefore the effective parameter reduces to
\begin{equation}
	F(\sigma) = \zeta + \kappa \sigma + \frac{N}{2} | \sigma| \, ,
\end{equation}
which is a piece-wise linear function whose slop jumps discontinuously from $\kappa - \frac{N}{2}$ to $\kappa+\frac{N}{2}$ at $\sigma = 0$. There are therefore two chambers $\sigma <0$ and $\sigma>0$ for the vector multiplet scalar.

Let us first consider the geometric regime $\zeta >0$. There is a Higgs branch $X \cong \mathbb{CP}^{N-1}$ at $\sigma = 0$ with K\"ahler parameter $\zeta >0$, which contributes $\chi(\mathbb{CP}^{N-1}) = N$ to the supersymmetric index. The existence of topological and Coulomb vacua depends on the Chern-Simons level:
\vspace{-\parskip}
\begin{itemize}
	\item If $\kappa < -\frac{N}{2}$, there is a topological vacuum with
		\begin{equation}
			\sigma = - \frac{\zeta}{\kappa+\frac{N}{2}} > 0  \,.
		\end{equation}
		Integrating out the massive chiral multiplet leaves a $U(1)_{\kappa+\frac{N}{2}}$ gauge theory and contributes $|\kappa+\frac{N}{2}|$ to the supersymmetric index.
	\item If $-\frac{N}{2} \leq \kappa \leq \frac{N}{2}$ there are no topological vacua. At the critical levels $\kappa = \pm 
\frac{N}{2}$, a non-compact Coulomb branch $\mp \sigma >0$ opens up as $\zeta \to 0^+$.
	\item If $\kappa > \frac{N}{2}$, there is a topological vacuum with 
		\begin{equation}
			\sigma = - \frac{\zeta}{\kappa-\frac{N}{2}} < 0 \,.
		\end{equation}
		Integrating out the massive chiral multiplet leaves a $U(1)_{\kappa-\frac{N}{2}}$ gauge theory and contributes $|\kappa-\frac{N}{2}|$ to the supersymmetric index.	
\end{itemize}
The critical window in this regime is therefore $-\frac{N}{2} \leq \kappa \leq \frac{N}{2}$. The supersymmetric index is $N$ inside the critical window, and $|\kappa|+\frac{N}{2}$ outside. 

In the opposite regime $\zeta <0$, the Higgs branch is empty and topological vacua exist for any level $\kappa$. The existence of topological vacua depends on the level as follows:
\vspace{-\parskip}
\begin{itemize}
\item If $\kappa \leq -\frac{N}{2}$, there is a single topological vacuum with $\sigma = - \frac{\zeta}{\kappa-\frac{N}{2}} < 0$, contributing $|\kappa| + \frac{N}{2}$ to the supersymmetric index. At $\kappa = - \frac{N}{2}$, a Coulomb branch $\sigma > 0$ opens up as $\zeta \to 0^-$.
\item If $-\frac{N}{2} < \kappa < \frac{N}{2}$, there are topological vacua both at $\sigma = - \frac{\zeta}{\kappa-\frac{N}{2}} < 0$ and $\sigma = - \frac{\zeta}{\kappa+\frac{N}{2}} > 0$, contributing $N$ to the supersymmetric index.
\item If $\kappa > \frac {N}{2}$, there is a single topological vacuum with $\sigma = - \frac{\zeta}{\kappa+\frac{N}{2}} > 0$, contributing $|\kappa| = \frac{N}{2}$ to the supersymmetric index. At $\kappa = \frac{N}{2}$, a Coulomb branch $\sigma <0$ opens up as $\zeta \to 0^-$.
\end{itemize}
Although there is no critical window in this regime, the contributions to the supersymmetric index agree with the computation at $\zeta >0$.
\chapter{Twisted Chiral Ring}
\label{chap:twisted-chiral-ring}

In this chapter we consider the supersymmetric ground states on $S^1 \times \mathbb{R}^2$, with finite radius $\beta$ of the circle $S^1$ in the presence of generic mass parameters.

When the chiral multiplets are all integrated out, the remaining vector multiplet scalar $\sigma$ becomes the lowest component of a two-dimensional twisted chiral multiplet in the \index{A-twist} A-twist of the $\mathcal{N}=(2,2)$ theory on $\mathbb{R}^2$. The two-dimensional twisted superpotential can be determined exactly since it is one-loop exact. The supersymmetric vacua of the effective theory are solutions to the ring relations of the so-called twisted chiral ring, which are identified with states of quantum integrable systems via the Bethe gauge correspondence~\cite{Nekrasov:2009ui,Nekrasov:2014xaa}.

We explore the \index{window phenomenon} window phenomenon from this perspective, and the examine the large radius limit $\beta \to \infty$, and the small radius limit $\beta \to 0$.
\begin{itemize}
	\item In the large radius limit we reproduce to the results from the flat space $\mathbb{R}^3$ discussed in Chapter~\ref{chap:flat-base-manifold} as expected. The solutions to the \index{Bethe ansatz} Bethe ansatz equation are in one-to-one correspondence with the supersymmetric vacua in flat space, where the Higgs and topological solutions admit qualitatively distinct behaviours.
	\item In the small radius limit the theory is reduced to two-dimensional theories. Since we are interested in the three-dimensional generalisation of the two-dimensional correspondence between sigma models and quantum cohomology, we expect this dimension reduction to produce new insights on the \index{window phenomenon} window phenomenon. In particular, the Higgs vacua are fully captured by a two-dimensional gauge theory, while the topological vacua become fully de-coupled theories in two dimensions. In analogy, the \index{quantum K-theory} quantum K-theory of a toric variety in three dimensions can only be interpreted as the lift from \index{quantum cohomology} quantum cohomology in two dimensions, if the level is within the \index{critical window} critical window. This is consistent with the \index{window phenomenon} window phenomenon in gauge theories where the three-dimensional theory is only fully captured by the two-dimensional theory, if the \index{Chern-Simons level} Chern-Simons level is within the critical window. The obstruction to the lifting is from the de-coupled two-dimensional theories corresponding to the topological vacua, which is not accounted for in the quantum K-theory of the Higgs vacua.
\end{itemize}

\section{Bethe Ansatz Equation}

Consider a general three-dimensional $\mathcal{N}=2$ supersymmetric $U(1)_\kappa$ theory with chiral multiplets $\{\Phi_i\}$ of charges $\{Q_i\}$ introduced in Chapter~\ref{chap:flat-base-manifold}, and put it on the base manifold $S^1 \times \mathbb{R}^2$. The \index{Fayet-Iliopoulos parameter} Fayet-Iliopoulos parameter and the \index{real mass} real mass parameters are naturally complexified by \index{Wilson line} Wilson lines associated with the global symmetries around the $S^1$ component. The corresponding \index{fugacity} fugacities can then be written as \index[sym]{$q$} \index[sym]{$y_j$}
\begin{subequations}
\begin{align}
	q &= e^{-2 \pi \beta (\zeta + i A_\text{t})} \,, \label{eq:topological-fugacity}\\
	y_j &= e^{-2 \pi \beta (m_j + i A_\text{f})} \label{eq:flavour-fugacity}\,,
\end{align}
\end{subequations}
where $A_\text{t}$ and $A_\text{f}$ are the constant background connections for the topological and flavour symmetry respectively. Similarly the \index{fugacity !gauge} gauge fugacity is \index[sym]{$x$}
\begin{equation}
\label{eq:gauge-fugacity}
	x = e^{-2 \pi \beta (\sigma + i A)}
\end{equation}
for the gauge connection along $S^1$.
The independent \index{fugacity !flavour} flavour fugacities \index[sym]{$y_{a'}$} $\{y_{a'}\}_{a' = 2}^N$ can be defined such that
\begin{equation}
	y_i \equiv \prod_{b'=2}^{N} y_{b'}^{q^{b'}_{i}}
\end{equation}
according to the definition~\eqref{eq:U(1)-independent-mass-parameters} of the independent mass parameters.

After integrating out the chiral multiplets, the supersymmetric vacua of the two-dimensional effective theory are determined by the \index{Bethe ansatz} Bethe ansatz equation~\cite{Nekrasov:2009ui, Nekrasov:2010ka}
\begin{equation}
	\exp(i \frac{\partial W}{\partial \sigma}) = 1 \,,
	\label{eq:U(1)-bethe}
\end{equation}
computed from the effective twisted superpotential $W(\sigma)$, which is one-loop exact. 

It can be interpreted as the ring relation for the \index{twisted chiral ring} twisted chiral ring.
From the perspective of the three-dimensional theory, the twisted chiral operators arise from Bogomol'nyi–Prasad–Sommerfield line operators wrapping the $S^1$ component of the base manifold. 
In particular, a supersymmetric Wilson loop of charge $q$ is represented by the monomial $x^q$. The Bethe ansatz equation then describes the ring structure inherited from the operator product expansion of the parallel line operators. The limit $\beta \to 0$ therefore reproduces the twisted chiral ring of the two-dimensional theory obtained by compactification on $S^1$. These line operators are local operators from the perspective of the two-dimensional theory on $\mathbb{R}^2$, which can be interpreted as elements of the twisted chiral ring. 
Hence the ring relation may be interpreted as Ward identities for the line operators.

Then for the effective two-dimensional $\mathcal{N}=(2,2)$ theory, the \index{Bethe ansatz} Bethe ansatz equation can be explicitly expressed in the following two equivalent forms
\begin{subequations}
\begin{align}
&\prod_{\{i\mid Q_i >0\}} \left( 1-x^{Q_i}\prod_{a'} y_{a'}^{q_i^{a'}} \right)^{Q_i} \nonumber \\
= & (-1)^{N_-} q x^{\kappa^+} \left( \prod_{a'}y_{a'}^{\kappa^+_{a'}}\right)  \prod_{\{i\mid Q_i <0\}} \left( 1-x^{-Q_i}\prod_{a'} y_{a'}^{-q_i^{a'}} \right)^{-Q_i}  \,, \label{eq:U(1)-general-BAE-+}\\
&\prod_{\{i\mid Q_i >0\}} \left( 1-x^{-Q_i}\prod_{a'} y_{a'}^{-q_i^{a'}} \right)^{Q_i} \nonumber \\
= & (-1)^{N_+} q x^{\kappa^-} \left( \prod_{a'}y_{a'}^{\kappa^-_{a'}} \right) \prod_{\{i\mid Q_i <0\}} \left( 1-x^{Q_i}\prod_{a'} y_{a'}^{q_i^{a'}} \right)^{-Q_i} \,, \label{eq:U(1)-general-BAE--}
\end{align}
\end{subequations}
where $N_{\pm} := \sum_{\{i \mid Q_i \gtrless 0\}} Q_i$ are the number of positively or negatively charged chiral multiplets if $|Q_i|=1$, and $\kappa^\pm$, $\kappa^\pm_{a'}$ are the \index{Chern-Simons level !asymptotic} asymptotic Chern-Simons levels~\eqref{eq:U(1)-asymp-CS-levels} acting as the gauge and flavour charges of the Bogomol'nyi-Prasad-Sommerfield monopole operators of topological charge $\pm1$. The two expressions correspond to arranging the formula such that left hand side is a polynomial in $x$ and $x^{-1}$ respectively, when the the monopole operators at $\sigma \rightarrow \pm \infty$ are positively charged, i.e., $\kappa^\pm > 0$. The anomaly cancellation conditions~\eqref{eq:U(1)-CS-anomaly-cancellation-cond} are crucial for these to be polynomial equations.

\section{Window Phenomenon}
The number of solutions to the \index{Bethe ansatz} Bethe ansatz equation depends on $\kappa_\pm$ as it alters the degree of the equations. Consider the first equation~\eqref{eq:U(1)-general-BAE-+}. The degree of $x$ on the left hand side is $\sum_{\{i \mid Q_i >0\}} Q_i^2$, while the degree on the right hand side is $\kappa^+ + \sum_{\{i \mid Q_i <0\}} Q_i^2$. The sign of the difference $$\kappa^+ + \sum_{\{i \mid Q_i <0\}} Q_i^2 - \sum_{\{i \mid Q_i >0\}} Q_i^2 = \kappa^-$$ determines which side of the equation dominates. Consider the case $\kappa^+ \geq 0$, i.e., $\kappa \geq -\frac{1}{2} \sum_{i=1}^N |Q_i| Q_i$. We expect the number of solutions to be
\begin{align*}
	\# &= 
	\begin{cases}
		\sum_{\{i \mid Q_i >0\}} Q_i^2 &\quad \text{if} \quad  \kappa^- < 0 \text{ and } \kappa^+ \geq 0\\ 
		\kappa^+ + \sum_{\{i \mid Q_i <0\}} Q_i^2 &\quad \text{if} \quad  \kappa^- \geq 0  \text{ and } \kappa^+ \geq 0	
	\end{cases} \\
	&= \begin{cases}
		\sum_{\{i \mid Q_i >0\}} Q_i^2 &\quad \text{if} \quad  -\frac{1}{2} \sum_{i=1}^N |Q_i| Q_i \leq \kappa < \frac{1}{2} \sum_{i=1}^N |Q_i| Q_i  \\ 
		\kappa + \frac{1}{2} \sum_{i=1}^N Q_i^2 &\quad \text{if} \quad  \kappa \geq \frac{1}{2} \left| \sum_{i=1}^N |Q_i| Q_i \right|
	\end{cases} \,.
\end{align*}
To count the number of solutions in the case $\kappa^+ < 0$, we need to rearrange the equation by moving the factor $x^{\kappa^+}$ to the left hand side. The same reasoning produces a complementary expression 
\begin{equation*}
	\# = \begin{cases}
		-k + \frac{1}{2} \sum_{i=1}^N Q_i^2 &\quad \text{if} \quad  \kappa < - \frac{1}{2} \left| \sum_{i=1}^N |Q_i| Q_i \right| \\ 
		\sum_{\{i \mid Q_i <0\}} Q_i^2 &\quad \text{if} \quad  \frac{1}{2} \sum_{i=1}^N |Q_i| Q_i 	\leq \kappa <  -\frac{1}{2} \sum_{i=1}^N |Q_i| Q_i
	\end{cases} \,.
\end{equation*}

Combining both gives the final formua for the number of solutions to the \index{Bethe ansatz} Bethe ansatz equation~\eqref{eq:U(1)-bethe} as
\begin{equation}
\label{eq:U(1)-num-of-solutions-to-BAEs}
	\# = 
	\begin{cases}
		\begin{cases}
		\sum_{\{i \mid Q_i > 0\}} Q_i^2 & \quad \text{if} \quad  |\kappa| \leq \tilde{\kappa}^\crit \\
		\sum_{\{i \mid Q_i < 0\}} Q_i^2 & \quad \text{if} \quad  |\kappa| \leq -\tilde{\kappa}^\crit \\
		\end{cases} &
		\\
		|\kappa| + \frac{1}{2} \sum_{i=1}^N Q_i^2 \hspace{8pt} \quad \text{if} \quad |\kappa| > |\tilde{\kappa}^\crit| &
	\end{cases} \,,
\end{equation}
where \index[sym]{$\tilde{\kappa}^\crit$} the asymptotic critical level
\begin{equation}
\label{eq:U(1)-BAE-critical-level}
	\tilde{\kappa}^\crit := \kappa^\crit (\sigma \to -\infty) = \frac{1}{2} \sum_{i=1}^N |Q_i| Q_i
\end{equation}
is defined as the \index{Chern-Simons level !critical} critical Chern-Simons level~\eqref{eq:U(1)-critical-CS-level} at $\sigma \to -\infty$, which roughly measures the difference between the number of positively charged chiral multiplets and the the number of negatively charged chiral multiplets. 
Note that only one of the first two lines can apply at a time, depending on the sign of $\tilde{\kappa}^\crit$.
The same expression emerges by considering the other equation~\eqref{eq:U(1)-general-BAE--}.

The \index{window phenomenon} window phenomenon can be seen here since the number of solutions has a qualitative change when the Chern-Simons level $\kappa$ goes from inside the interval
$$
	\left[-\left|\tilde{\kappa}^\crit\right|, \left|\tilde{\kappa}^\crit\right| \right]\,,
$$
suggesting the existence of different types of vacua.

\section{Large Radius Limit}
\label{sec:large-radius-limit}

In the limit $\beta \to \infty$, we expect the solutions to the \index{Bethe ansatz} Bethe ansatz equation reproduce the semi-classical vacua discussed in Chapter~\ref{chap:flat-base-manifold}. 
In a given chamber for the Fayet-Iliopoulos parameter, as $\beta \to \infty$ the solutions of the \index{Bethe ansatz} Bethe ansatz equation may tend towards \index{vacuum !Higgs branch} Higgs branch vacua at fixed $\sigma = -m_j / Q_j$, or \index{vacuum !topological branch} topological vacua at $\sigma = -\zeta^\eff / \kappa^\eff$. 
We refer to these as Higgs and topological solutions. 
We expect these two types solutions to behave like $\sigma \sim m$ and $\sigma \sim \zeta$ respectively.

\subsection[\texorpdfstring{$U(1)_\kappa$ with One Chiral Field}%
                        {$U(1)$ with One Chiral Field}]%
        {$U(1)_\kappa$ with One Chiral Field}
\label{subsubsec:large-radius-N=1}

Let us again consider $U(1)_\kappa$ with one \index{chiral multiplet} chiral multiplet of charge $Q>0$. The \index{twisted chiral ring} twisted chiral ring relations are equivalently
\begin{equation}
	(1-x^Q)^Q = q x^{\kappa+\frac{Q^2}{2}} \, .
\end{equation} 
The number of solutions depends on the relative degree of the polynomials on the left and right and therefore on the Chern-Simons level $\kappa$. The asymptotic critical level is $\tilde{\kappa}^\crit = \frac{1}{2} Q^2 > 0$. Therefore the first and third lines in~\eqref{eq:U(1)-num-of-solutions-to-BAEs} apply. The result is then
\begin{equation}
	\# = \begin{cases}
	Q^2 & \quad \text{if} \quad |\kappa| \leq \frac{Q^2}{2}  \\
	|\kappa|+\frac{Q^2}{2} & \quad \text{if} \quad |\kappa| > \frac{Q^2}{2} 
\end{cases}
\end{equation}
in agreement with the supersymmetric index computed in Section~\ref{subsubsec:U(1)+1-vacuum-example}.

In order to compare with the supersymmetric ground states on $\mathbb{R}^3$, we set $q = e^{-\beta \zeta}$ and $x = e^{-\beta \sigma}$. We then expand solutions for $\sigma$ in the limit $\beta \to \infty$ with $\zeta >0$, in order to directly compare to the analysis in Section~\ref{subsubsec:U(1)+1-vacuum-example}. Now let us assume $Q=1$ for simplicity and compare with the plots in Figure~\ref{fig:effective-FI-U(1)+1}.

Consider the classic case $\kappa = -\frac{1}{2}$. The solution to the \index{Bethe ansatz} Bethe ansatz equation $1-x = q$ is simply 
$$
	x = 1 - q \,,
$$
giving $\sigma = 0$ at the limit $\beta \rightarrow \infty$, which matches to the single Higgs solution in the $\zeta > 0$ chamber. 

In the case $\kappa = -\frac{3}{2}$, the equation $1-x = q x^{-1}$ has two solutions
$$
	x_\pm = \frac{1 \pm \sqrt{1 - 4 q}}{2} \,.
$$
In the limit $\beta \rightarrow \infty$, the two solutions correspond to 
$$
\sigma_+ = 0\,, \quad \sigma_- = \zeta\,,
$$ 
which are the Higgs solution and the topological solution respectively.

\subsection[\texorpdfstring{$U(1)_\kappa$ with Two Chiral Fields}%
                        {$U(1)$ with Two Chiral Fields}]%
        {$U(1)_\kappa$ with Two Chiral Fields}
\label{subsubsec:large-radius-N=2}

Consider the same example of two \index{chiral multiplet} chiral multiplets discussed in Section~\ref{subsubsec:U(1)+2-vacuum-example} with the extended charge matrix
\begin{equation}
	Q^i_{\ \ j}=
	\begin{pmatrix}
		1 & 1 \\
		0 & 1
	\end{pmatrix} \,.
\end{equation}
Denote the mixed gauge-flavour \index{Chern-Simons level} Chern-Simons level as $\kappa_{12}$.

The \index{Bethe ansatz} Bethe ansatz equation is then
\begin{equation}
	(1-x) (1-x y_2) = q x^{\kappa + 1} y_2^{\kappa_{12}+\sfrac{1}{2}} \,.
\end{equation}
The asymptotic critical level is $\tilde{\kappa}^\crit = 1 > 0$. Therefore the number of solutions is
\begin{equation}
	\# = 
	\begin{cases}
		2 &\quad \text{if} \quad  |\kappa| \leq 1 \\ 
		|\kappa| + 1  &\quad \text{if} \quad  |\kappa| >  1
	\end{cases} \,.
\end{equation}

To match the solutions with the choice of \index{real mass} real masses in Section~\ref{subsubsec:U(1)+2-vacuum-example}, we set $\kappa_{12}=-\frac{1}{2}$. For simplicity, we also drop the subscript on $y_2$ without causing ambiguities, giving the \index{Bethe ansatz} Bethe ansatz equation
\begin{equation}
	(1-x) (1-x y) = q x^{\kappa + 1} \,.
\end{equation}
It is instructive to explore its solutions.

Consider the simple case with $\kappa=-1$. The two solutions are
\begin{equation}
	x = \frac{1}{2 y}\left(1+y \pm \sqrt{1 + 4 q y - 2 y + y^2}\right) \,.
\end{equation}
By re-writing the \index{fugacity} fugacities in terms of the parameters again via
\begin{subequations}
\begin{align}
	x &= e^{-\beta \sigma} \,, \\
	y &= e^{-\beta m_2}=e^{\beta m} \,, \\
	q &= e^{-\beta \zeta} \,,
\end{align}
\end{subequations}
we can explore the flat space limits by sending the radius $\beta \rightarrow \infty$. The corresponding two solutions in terms of $\sigma$ in this limit are
\begin{equation}
	\sigma = 
	\begin{cases}
		0 \quad \text{and} \quad m \,, \quad &\text{if } \zeta > 0 \\
		0 \quad \text{and} \quad \zeta + m \,, \quad &\text{if } -m < \zeta < 0 \\
		\frac{\zeta}{2} + \frac{m}{2} \quad \text{and} \quad \frac{\zeta}{2} + \frac{m}{2} \,, \quad &\text{if }  \zeta \leq -m \\
	\end{cases} \,.
\end{equation}
This is in perfect agreement with the results from studying the vacuum equations in Section~\ref{subsubsec:U(1)+2-vacuum-example}. This case corresponds to the $\kappa=-1$ plot of $F(\sigma)$ at $\zeta=0$ in Figure~\ref{fig:effective-FI-U(1)+2}. 
\vspace{-\parskip}
\begin{itemize}
	\item When $\zeta > 0$, the plot is shifted upwards such that it has no intersection with the $\sigma$-axis. Both Higgs vacua at $\sigma=0$ and $\sigma=m$ exist since $F(\sigma) > 0$ at these points. But there is no topological vacuum. 
	\item When $-m < \zeta < 0$, the plot has a single intersection with the $\sigma$-axis. There exist a single Higgs vacuum at $\sigma=0$, and a single topological vacuum at the intersection point $\sigma = \zeta + m$.
	\item When $\zeta \leq -m$, the plot is shifted downwards such that it intersects the $\sigma$-axis at $\sigma < 0$. There exist only a single topological vacuum at the intersection point $\sigma = \frac{\zeta}{2}+\frac{m}{2}$, with a multiplicity of two.
\end{itemize}

There also exists a non-compact Coulomb branch solution at the non-generic value $\zeta = 0$ of the Fayet-Iliopoulos parameter, which exhibits more subtle behaviour from this perspective and requires further analysis.

The Higgs and topological solutions are qualitatively different. The locations of the Higgs solutions only depend the masses, while the topological solutions have dependence on the Fayet-Iliopoulos parameter $\zeta$. This distinction allows us to distinguish the two types of solutions without referencing to the original three-dimensional theories. In particular, all the topological solutions are separated if we turn $\zeta \to \infty$.

\section{Small Radius Limit}

We also consider the small radius limit $\beta \to 0$, as investigated in~\cite{aharony2018}, where it reduces to two-dimensional theories. 
In comparison with the large radius limit analysed in Section~\ref{sec:large-radius-limit}, our general expectation is the following:
\vspace{-\parskip}
\begin{itemize}
	\item The Higgs solutions at $\beta \to \infty$ lie in the regime $\beta \sigma \ll 1$ in the limit $\beta \to 0$, since the Higgs vacua appear at finite values $\sigma$.
	\item On the other hand, topological solutions at $\beta \to \infty$ lay outside the region $\beta \sigma \ll 1$ in the limit $\beta \to 0$, because $\sigma$ can be shifted to arbitarily large values in the topological vacua.
\end{itemize}
 
Within the geometric regime for the Fayet-Iliopoulos parameter $\zeta$ where Higgs vacua always exist, the Chern-Simons level determines the existence of topological vacua.
\vspace{-\parskip} 
\begin{itemize}
	\item Consequently, if the \index{Chern-Simons level} Chern-Simons level $\kappa$ lies within the critical window, the small radius limit can be fully captured by a corresponding two-dimensional $\mathcal{N} = (2,2)$ gauge theory in the regime $\beta \sigma \ll 1$. This property is independent of the specific values of the \index{Chern-Simons level} Chern-Simons level provided it remains in the critical window. 
	\item However, if $\kappa$ lies outside the window, there exist additional solutions outside the regime $\beta \sigma \ll 1$. In this case the small radius limit cannot be fully captured by a two-dimensional $\mathcal{N}=(2,2)$ gauge theory. The additional solutions are de-coupled two-dimensional theories which we do not fully understand.
\end{itemize}

For a $U(1)$ gauge theory with $N$ chiral multiplets of charges $Q_i > 0$ and masses $m_i$ in three dimensions. The twisted \index{superpotential} superpotential $W$ is
\begin{equation}
	W (\sigma; m_i, Q_i) = \sum_{i=1}^N W_\chi^\text{3d}(Q_i \sigma + m_i) + 2 \pi \beta \zeta \sigma + \pi \beta  \kappa  \sigma \left( \sigma + \frac{i}{\beta} \right) + 2 \pi \beta \tilde{\kappa}_\text{GF} \sigma \,,
\end{equation}
where $W_\chi^\text{3d}$ is the contribution of a single three-dimensional chiral multiplet
\begin{equation}
	W_\chi^\text{3d}(\sigma) = \frac{1}{2 \pi \beta}\operatorname{Li}_2(e^{-2 \pi \beta \sigma}) + \frac{\pi}{2} \beta \sigma^2 \,,
\end{equation}
and 
\begin{equation}
	\tilde{\kappa}_\text{GF} = \sum_{a'=2}^N \left(\kappa_{a'} + \frac{1}{2} \sum_{j=1}^N Q_{a'j} \right) m_{a'}
\end{equation}
is the contribution from the mixed gauge-flavour Chern-Simons terms. The \index{Bethe ansatz} Bethe ansatz equation is given by
\begin{equation}
\label{eq:U(1)-3d-BAE-charge-Q}
	\prod_{i=1}^N (1 - x^{Q_i} y_i)^{Q_i} = q x^{\kappa + \frac{1}{2} N} \prod_{a'=2}^N y_{a'}^{\kappa_{a'} + \frac{1}{2} \sum_{j=1}^N Q_{a'j}} 
\end{equation}
up to an overall sign. Its number of solutions is
\begin{equation}
	\# = \begin{cases}
		\sum_{i=1}^N Q_i^2 &\quad \text{if} \quad  |\kappa| \leq \frac{1}{2} \sum_{i=1}^N Q_i^2 \\
		|\kappa| + \frac{1}{2} \sum_{i=1}^N Q_i^2 &\quad \text{if} \quad  |\kappa| > \frac{1}{2} \sum_{i=1}^N Q_i^2 	\\ 
	\end{cases} \,.
\end{equation}

Consider taking the limit $\beta \rightarrow 0$, while fixing the combination
\begin{equation}
	t = 2\pi \beta \zeta + \pi i \left( k + \frac{N}{2} \right) + 2 \pi \beta \tilde{\kappa}_\text{GF} \,,
\end{equation}
which acts as the effective Fayet-Iliopoulos parameter for the two-dimensional theory. As discussed in Section~\ref{sec:large-radius-limit}, it controls the spacing between the Higgs solutions and the topological solutions. The $t \to \infty$ helps us to distinguish the topological solutions from the Higgs solutions.

\begin{itemize}

\item In the regime $\beta \sigma \ll 1$, after renormalisation the effective twisted superpotential tends to 
\begin{equation}
	W \to \sum_{i=1}^N W_\chi^\text{2d}(Q_i \sigma + m_i) + t \sigma \,,
\end{equation}
where $W_\chi^\text{2d}$ is the contribution of a single two-dimensional\index{chiral multiplet} chiral multiplet
\begin{equation}
	W_\chi^\text{2d}(\sigma) = \sigma (\log \sigma - 1) \,.
\end{equation}
This results in a $U(1)$ gauge theory with $N$\index{chiral multiplet} chiral multiplets in two dimensions. The \index{Bethe ansatz} Bethe ansatz equation for the two-dimensional theory is
\begin{equation}
	\prod_{i=1}^N (Q_i \sigma + m_i)^{Q_i} = e^{-t} \,.
\end{equation}
which has $\sum_{i=1}^N Q_i$ solutions for generic parameters since it is a polynomial of degree $\sum_{i=1}^N Q_i$. 

However this only captures the real solutions but $\sigma$ is a complexified by the \index{Wilson line} Wilson lines. For each chiral multiplet $\Phi_i$, the periodicity is $Q_i \sigma \sim Q_i \sigma + 2 \pi i$. Hence each real solution of $\sigma$ in two dimensional corresponds to $Q_i$ solutions of $x = e^\sigma$ in three dimensions. Physically, the $Q^2$ dependence in the number of vacua in three dimensions comes from the $|Q|$ topological Wilson lines wrapping on the two non-trivial cycles of the spatial torus $T^2$. When it is reduced to two dimensions, one of the cycles is contracted. The theory is decomposed~\cite{Robbins:2020msp} into $|Q|$ distinct topological sectors. 

When the decomposition is taken into account, the effective two-dimensional theory really captures $\sum_{i=1}^N Q_i^2$ solutions, which corresponds to all of the Higgs vacua. But this regime $\beta \sigma \ll 1$ does not fully capture all the solutions from the three-dimensional theory when $\kappa > \frac{1}{2} \sum_{i=1}^N Q_i^2$, where topological solutions appear. 

\item Outside the region $\beta \sigma \ll 1$, we would like to find finite solutions in terms of $\Sigma := 2 \pi \beta \sigma$. The effective twisted superpotential becomes 
\begin{equation}
	W \rightarrow \frac{1}{2 \pi \beta} \left( \sum_{i=1}^N \operatorname{Li}_2(e^{-Q_i \Sigma}) + \frac{1}{2} \left(\kappa+\frac{N}{2}\right) \Sigma^2 + t \Sigma \right) + \mathcal{O}(1) \,,
\end{equation}
which looks like a $U(1)$ theory of $N$ massless\index{chiral multiplet} chiral multiplets. The \index{Bethe ansatz} Bethe ansatz equation becomes
\begin{equation}
	\prod_{i=1}^N (1-X^{Q_i})^{Q_i} = e^{-t} X^{\kappa + \sfrac{N}{2}} \,,
\end{equation}
where $X := e^{-\Sigma}$. The distinct solutions decouple~\cite{aharony2018} into separate theories as the values of the twisted superpotential in different vacua differ by a diverging factor $\frac{1}{\beta}$.

Naively, the number of solutions to this two-dimensional \index{Bethe ansatz} Bethe ansatz equation is either $\sum_{i=1}^N Q_i^2$ or $\left( |\kappa| + \frac{1}{2} \sum_{i=1}^N Q_i^2 \right)$ just as in the original three-dimensional equation~\eqref{eq:U(1)-3d-BAE-charge-Q}. 

However, some of the solutions reside in the other regime $\Sigma \sim \beta \sigma \ll 1$, which need to be discarded to avoid double counting. The new solutions can be separated by considering $t \to \infty$. In particular, these are the $\sum_{i=1}^N Q_i^2$ solutions ``close'' to $X=1$, which resides in the other regime $\beta \sigma \ll 1$ and needs to be discarded. Hence there are only $|\kappa| - \frac{1}{2} \sum_{i=1}^N Q_i^2$ solutions residing outside of the regime $\beta \sigma \ll 1$, which correspond to the topological vacua. This can be demonstrated explicitly with examples.

\end{itemize}

\subsection[\texorpdfstring{$U(1)_\kappa$ with One Chiral Field}%
                        {$U(1)$ with One Chiral Field}]%
        {$U(1)_\kappa$ with One Chiral Field}

Consider again one \index{chiral multiplet} chiral multiplet with $Q=1$. The twisted superpotential $W$ is
\begin{equation}
	W (\sigma) = W_\chi^\text{3d}(\sigma ) + 2 \pi \beta \zeta \sigma + \pi \beta  \kappa  \sigma \left( \sigma + \frac{i}{\beta} \right) \,.
\end{equation}
The \index{Bethe ansatz} Bethe ansatz equation reads
\begin{equation}
	1 - x = q x^{\kappa + \sfrac{1}{2}} \,,
\end{equation}
with the number of solutions given by
\begin{equation}
	\# = \begin{cases}
		1 & \quad \text{if} \quad |\kappa| \leq \frac{1}{2}  \\
		|\kappa|+\frac{1}{2} & \quad \text{if} \quad |\kappa| > \frac{1}{2} 
	\end{cases} \,.
\end{equation}

Consider taking the limit $\beta \rightarrow 0$, while fixing the combination
$$
t = 2\pi \beta \zeta + \pi i \left( k + \frac{1}{2} \right)  \,.
$$

\begin{itemize}
\item In the regime $\beta \sigma \ll 1$, the effective twisted superpotential tends to 
\begin{equation}
	W \rightarrow W_\chi^\text{2d}(\sigma) + t \sigma \,,
\end{equation}
giving an effective Bethe equation
\begin{equation}
	1 = e^t \sigma
\end{equation}
with one solution at $\sigma = e^{-t}$. This captures the single Higgs vacuum of the three-dimensional theory. 

This can be explicitly seen by comparing to the large radius limit in Section~\ref{subsubsec:large-radius-N=1}, where $\sigma = e^{-t}$ corresponds to the Higgs solution.

\item Outside of the regime $\beta \sigma \ll 1$, we define $\Sigma := 2 \pi \beta \sigma$ being finite. The twisted superpotential becomes
\begin{equation}
	W \rightarrow \frac{1}{2 \pi \beta} \left(  \operatorname{Li}_2(e^{-\Sigma}) + \frac{1}{2} \left(\kappa+\frac{1}{2}\right) \Sigma^2 + t \Sigma \right) + \mathcal{O}(1) \,,
\end{equation}
giving the \index{Bethe ansatz} Bethe ansatz equation
\begin{equation}
	1-X = e^{-t} X^{\kappa + \sfrac{1}{2}} \,,
\end{equation}
where $X = e^{-\Sigma}$.

At $\kappa = -\frac{1}{2}$, there is a single solution $X = 1 - e^{-t}$ matching with the large radius solution $x = 1 - q$ from the Higgs vacuum. 
When taking the limit $t \to \infty$, it tends to $X = 1$, corresponding to $\Sigma = 0$. Therefore it is not a new solution.

At $\kappa = -\frac{3}{2}$, there are two solutions $ X_\pm = \frac{1}{2} \left( 1 \pm  \sqrt{1 - 4 e^{-t}} \right)$ matching with the large radius solutions $x_\pm = \frac{1}{2} \left( 1 \pm  \sqrt{1 - 4 q} \right)$ from the Higgs and topological vacuum respectively. 
In the $t \to \infty$ limit, the solutions tend to $X_+ = 1$ and $X_- = 0$, corresponding to $\Sigma_+ = 0$ and $\Sigma_- = \infty$.
Therefore the only new solution in this regime is 
$$
	X_- = \frac{1}{2} \left( 1 -  \sqrt{1 - 4 e^{-t}} \right) 
$$
corresponding to the topological vacuum.

\end{itemize}

\chapter{Twisted Index}
\label{chap:twist-on-surface}

Consider the same class of theories on the product manifold  $S^1 \times \Sigma$, where the flat $\mathbb{R}^2$ in Chapter~\ref{chap:twisted-chiral-ring} is replaced with a closed Riemann surface of genus $g$. 
These theories fall into the general class introduced in~\cite{Closset:2017zgf}.
To preserve supersymmetry on a curved space, a topological twist is performed by mixing the $U(1)$ Lorentz rotation on the plane with the unbroken R-symmetry. This is lift from the topological \index{A-twist} A-twist~\cite{witten1991} in two dimensions. 
The metric tensor on the plane becomes exact, and therefore closed observables are independent of the metric. 
This enables us to put a sub-sector of the theory onto arbitary curved space $\Sigma$. We are interested in the twisted indices discussed in~\cite{Nekrasov:2014xaa,benini2015,closset2016,benini2017}.

The topological twist preserves an $\mathcal{N}=(0,2)$ quantum mechanics on $S^1$ with a pair of of \index{supercharge} supercharges $Q$ and $\widebar{Q}$.
We would like to give an algebro-geometric interpretation of  the \index{twisted index} twisted index via quantum mechanics. 
This approach is different from the perspective of the topological \index{A-model} A-model. We are reducing the theory to a quantum mechanics on the temporal circle, rather than a two-dimensional theory on the spatial component.

This chapter is based on my research conducted in~\cite{xu2022,xu2022a}, with the following outline.
\begin{itemize}
	\item Section~\ref{sec:twisted-index} sets up the notation, and constructs two different forms of the \index{twisted index} twisted index from the perspectives of quantum mechanics and the two-dimensional A-model respectively.
	\item We review the contour integral formula~\cite{bullimore2019} in Section~\ref{sec:contour-integral-formula} which we wish to reproduce from the geometric construction.
	\item A schematic construction for the algebro-geometric interpretation is given in Section~\ref{sec:geometric-interpretation}. The detailed computations for the vortex and topological saddles are respectively left to Chapter~\ref{sec:vortex-vacuum} and Chapter~\ref{sec:topological-vacuum}. The full construction of Chern-Simons contributions are discussed in Chapter~\ref{sec:CS-from-det}.
	\item The \index{window phenomenon} window phenomenon and the connection to \index{quantum K-theory} quantum K-theory are then discussed in Section~\ref{sec:index-window-phenomenon} in the context of twisted indices.
	\item Finally two simple examples are provided in Section~\ref{sec:twisted-index-examples}.
\end{itemize}

\section{Topological Twist}
\label{sec:topological-twist}

In the absence of central charges, the three-dimensional $\mathcal{N}=2$ supersymmetry introduced in Section~\ref{sec:SUSY-algebra} reduces to 
\begin{equation}
	\left\{Q_\alpha, \widebar{Q}_\beta\right\} = 2 \gamma^\mu_{\alpha \beta} P_\mu  = P_{\alpha \beta}\,.
\end{equation}
We implement a \index{topological twist} topological twist equivalent to the topological A-twist on $\Sigma$. The Lorentz group on the plane is mixed with an unbroken $U(1)_R \subset SU(2)_R$ symmetry. The twist assigns a new spin
\begin{equation}
	L' = L + \frac{1}{2} R
\end{equation}
to a field of charge $L$ under rotations on the plane. It preserves the supercharges that commute with the new Lorentz group, resulting a supersymmetry algebra
\begin{equation}
	\left\{Q, \widebar{Q}\right\} = P_0 \,
\label{eq:twisted-SUSY-algebra}
\end{equation}
where $P_0$ generates translations on $S^1$.

After the twisted is performed, it is more convenient to adopt the twisted field notation~\cite{closset2016,closset2015,bullimore2019}.


The vector multiplet is denoted as
\begin{equation}
	V = (\sigma, A_\mu, \lambda, \widebar{\lambda}, \Lambda_1, \widebar{\Lambda}_{\bar{1}}, D) \,,
\end{equation}
where $\sigma$ is the real scalar, $A_\mu$ is the gauge connection, $D$ is the auxiliary field, and $\lambda$, $\widebar{\lambda}$, $\Lambda$, $\widebar{\Lambda}$ are gauginos. The abelian Yang-Mills lagrangian is
\begin{align}
\label{eq:twisted-YM-lagrangian}
	\mathcal{L}_\text{YM} =& \frac{1}{2} F_{01} F_{0\bar{1}} 
	+ \frac{1}{2}(-2i F_{1\bar{1}})^2 
	+ \frac{1}{2} D^2 
	+ \frac{1}{2} \left|\DD_\mu \sigma\right|^2
	\nonumber \\
	& \phantom{\frac{1}{2} F_{01} F_{0\bar{1}}}
	- i \widebar{\lambda} D_0 \lambda
	- i \widebar{\Lambda}_{\bar{1}} D_0 \Lambda_1
	+ 2 i \widebar{\Lambda}_{\bar{1}} D_1 \lambda
	- 2 i \Lambda_1 D_{\bar{1}} \widebar{\lambda} 
	\,,
\end{align}
which is exact with respect to $Q$ and $\widebar{Q}$ in~\eqref{eq:twisted-SUSY-algebra}. The Chern-Simons term is given by
\begin{equation}
\label{eq:twisted-CS-lagrangian}
	\mathcal{L}_\text{CS} = \frac{\kappa}{4 \pi} 
	\left[
	i \epsilon^{\mu \nu \rho} \left( A_\mu \partial_\mu A_\rho - \frac{2i}{3} A_\mu A_\nu A_\rho \right)
	- 2 D \sigma + 2i \widebar{\lambda} \lambda + 2i \widebar{\Lambda}_{\bar{1}} \Lambda_1
	\right] \,,
\end{equation}
which is not exact.

The twisted chiral multiplet is denoted as
\begin{equation}
	\Phi = (\phi, \psi, \eta, F) \,.
\end{equation}
Its lagrangian is given by
\begin{align}
\label{eq:twisted-chiral-lagrangian}
	\mathcal{L}_\Phi =& \phi^\dagger \left( -D_0^2 - 4 D_1 D_{\bar{1}} + \sigma^2 + i D - 2 i F_{1\bar{1}} \right) \phi
	- F^\dagger F
	\nonumber \\
	& \phantom{\phi^\dagger (}
	- \frac{i}{2} \widebar{\psi} (D_0 + \sigma) \psi
	- 2i \widebar{\eta} (D_0 - \sigma) \eta
	\nonumber \\
	& \phantom{\phi^\dagger (}
	+ 2i \widebar{\psi} D_1 \eta
	- 2i \widebar{\eta} D_{\bar{1}} \psi
	\nonumber \\
	& \phantom{\phi^\dagger (}
	- i \widebar{\psi} \widebar{\lambda} \phi
	+ i \phi^\dagger \lambda \psi
	\nonumber \\
	& \phantom{\phi^\dagger (}
	- 2 i \phi^\dagger \Lambda_1 \eta
	+ 2 i \widebar{\eta} \widebar{\Lambda}_{\bar{1}} \phi \,,
\end{align}
which is exact under $Q$ and $\widebar{Q}$.

In addition, a background vector multiplet $V_\text{f}$ for a maximal torus of a flavour symmetry $T_\text{f} \subset G_\text{f}$ contains a real scalar component $m_\text{f}$ valued in the Cartan sub-algebra of the flavour group, and a background connection $A_\text{f}$. For example we can turn on a real mass of the background vector multiplet for the $U(1)_\text{t}$ topological symmetry, which is the Fayet-Iliopoulos parameter $\zeta$. Then the Fayet-Iliopoulos term is 
\begin{equation}
\label{eq:twisted-FI-lagrangian}
	\mathcal{L}_\text{FI} = - \frac{i \zeta}{2 \pi} D \,.
\end{equation}

\section{Supersymmetric Index}
\label{sec:twisted-index}
In the operator formalism, the \index{twisted index} twisted index counts the supersymmetric ground states $\mathcal{H}$ annihilated by the supercharges $Q$ and $\widebar{Q}$. The space $\mathcal{H}$ of supersymmetric ground states forms a representation of the global symmetry $T_\text{t} \times T_\text{f}$ where $T_\text{t}$ is topological and $T_\text{f}$ is flavour. The \index{fugacity} fugacities can be defined via
\begin{subequations}
\begin{align}
	q &= e^{-2 \pi \beta (\zeta + i A_\text{t})} \,, \\
	y_j &= e^{-2 \pi \beta (m_j + i A_\text{f})} \,, \\
	x &= e^{-2 \pi \beta (\sigma + i A)} \,,
\end{align}
\end{subequations}
the same way as in Chapter~\ref{chap:twisted-chiral-ring}. 

The \index{twisted index} twisted index is then in the form
\begin{equation}
\label{eq:U(1)-index-def}
	\mathcal{I} = \Tr (-1)^F q^{J} \prod_{i=1}^N y_i^{J_i} \,,
\end{equation}
where $J$ is the Cartan generator of the topological symmetry $T_\text{t}=U(1)_\text{t}$, and $J_i$ is the generator of $U(1)_i$ in $T_\text{f}$. The trace is performed over the Hilbert space $\mathcal{H}$ which is assumed to be locally finitely graded so that the coefficient of a monomial in $q$ and $y_i$ is a finite integer.

From the perspective of the \index{A-twist} A-twist in two dimensions, the \index{twisted index} twisted index for gauge group $G = U(1)$ can be computed via the Bethe formula~\cite{benini2017,Nekrasov:2014xaa}
\begin{equation} 
\label{eq:U(1)-index-bethe-formula}
	\mathcal{I} = \sum_{x = x_i} Z(x; \mathfrak{m}) \eval_{\mathfrak{m}=0} H(x)^{g-1} \,,
\end{equation}
over solutions $x_i$ of the Bethe ansatz equation~\eqref{eq:U(1)-bethe}, where $Z(x; \mathfrak{m})$ denotes the classical and one-loop contribution in \index{supersymmetric localisation} supersymmetric localisation and 
 \begin{equation}
	H = \frac{\partial^2 \log Z}{\partial \log x \partial \mathfrak{m}} \,.
\end{equation}
is the hessian determinant independent of the flux $\mathfrak{m}$. The \index{twisted index} twisted index is a meromorphic function of the \index{fugacity} fugacities $q$ and $y_i$. When $g=1$, it reduces to the \index{torus index} torus index discussed in Section~\ref{sec:flat-SUSY-Index}.

Note that in this notation, the \index{Bethe ansatz} Bethe ansatz equation~\eqref{eq:U(1)-bethe} can be expressed in terms of the classical and one-loop determinant $Z$ instead of the twisted superpotential $W$,
\begin{equation}
\label{eq:U(1)-bethe-Z}
	\exp(i \frac{\partial \log Z}{\partial \mathfrak{m}}) = 1  \,.
\end{equation}

For a general $U(1)$ gauge theory of the type considered in Chapter~\ref{chap:flat-base-manifold}, the classical and one-loop determinant takes the following form
\begin{align}
\label{eq:U(1)-one-loop-det}
	Z = & q^{\mathfrak{m}}  
	x^{\kappa \mathfrak{m}}
	\left( \prod_{b'=2}^N y_{b'}^{\kappa_{b'} \mathfrak{m}} \right)
	x^{\kappa_{R} (g-1)}
	\left( \prod_{b'=2}^N y_{b'}^{\kappa_{Rb'} (g-1)} \right)	
	\quad \times \nonumber \\
	& \hspace{18ex} \left[ \prod_{i=1}^N \left( \frac{ x^{\sfrac{Q^1_{\ \ i}}{2}} \prod_{b'=2}^{N} y_{b'}^{ \sfrac{Q^{b'}_{\ \ i} }{2} } }{ 1 -  x^{Q^1_{\ \ i}} \prod_{b'=2}^{N} y_{b'}^{Q^{b'}_{\ \ i}} } \right)^{\left(  Q^1_{\ \ i} \mathfrak{m} \right) + (r_i - 1)(g - 1)} \right]\,. 
\end{align}
Here the notation of extended charge matrices~\eqref{eq:U(1)-extended-charge-matrix} has been used, where $Q^1_{\ \ i} := Q_i$ are the \index{gauge charge} gauge charges and $\{ Q^{b'}_{\ \ i} \}_{b'=2}^N$ are the \index{flavour charge} flavour charges. The independent \index{fugacity !flavour} flavour fugacities $\{y_{b'}\}_{b' = 2}^N$ are defined such that
\begin{equation}
	y_i \equiv \prod_{b'=2}^{N} y_{b'}^{Q^{b'}_{\ \ i}} \,.
\end{equation} 
The corresponding hessian is given by
\begin{equation}
H(x) 
 = \kappa + \sum_{i=1}^N  (Q^1_{\ \ i})^2 \left(\frac{1}{2} + \frac{x^{Q^1_{\ \ i}} \prod_{b'=2}^{N} y_{b'}^{Q^{b'}_{\ \ i}}}{1 - x^{Q^1_{\ \ i}} \prod_{b'=2}^{N} y_{b'}^{Q^{b'}_{\ \ i}} }\right)\, .
\end{equation}

It is convenient to use the \index{Chern-Simons level !mixed} mixed Chern-Simons levels~\eqref{eq:U(1)-mixed-CS-levels-for-each-chiral} 
$$\{\kappa_j, \kappa_{Rj} \mid j =1,\ldots,N\}$$ 
for each chiral multiplet satisfying $\kappa_j + \frac{1}{2}Q_j \in \mathbb{Z}$.
It enables us to write the classical and one-loop determinant exclusively in terms of the flavour fugacities \index[sym]{$y_j$} $y_j$ for each chiral multiplet as
\begin{align}
\label{eq:U(1)-one-loop-det-yi}
	Z = & q^{\mathfrak{m}}  
	x^{\kappa \mathfrak{m}}
	\left( \prod_{j=1}^N y_{j}^{\kappa_{j} \mathfrak{m}} \right)
	x^{\kappa_{R} (g-1)}
	\left( \prod_{j=1}^N y_{j}^{\kappa_{Rj} (g-1)} \right)
	\quad \times \nonumber \\
	& \hspace{28ex} \left[ \prod_{i=1}^N \left( \frac{ x^{ \sfrac{Q_i}{2}} y_i^{\sfrac{1}{2}} }{ 1 -  x^{Q_i} y_i } \right)^{\left(  Q_i \mathfrak{m} \right) + (r_i - 1)(g - 1)} \right]\,.
\end{align}
Then the hessian simply reads
\begin{equation}
\label{eq:U(1)-hessian-yi}
	H(x) 
	= \kappa + \sum_{i=1}^N  (Q_i)^2 \left(\frac{1}{2} + \frac{x^{Q_i} y_i }{1 - x^{Q_i} y_i }\right)\, .
\end{equation}

In the case of $g = 1$, the classical and one-loop determinant is trivial, 
$$Z(x; \mathfrak{m}) \eval_{\mathfrak{m}=0} = 1 \,.$$
Hence the \index{twisted index} twisted index~\eqref{eq:U(1)-index-bethe-formula} indeed returns the number of solutions of the Bethe ansatz equation, in agreement with the expected \index{supersymmetric index}\index{torus index}torus supersymmetric index discussed in Chapter~\ref{chap:flat-base-manifold}.

\section{Contour Integral Formula}
\label{sec:contour-integral-formula}

The contour integral formula of the \index{twisted index} twisted index can be derived using the Coulomb branch \index{supersymmetric localisation} supersymmetric localisation. In this section, we briefly review the derivation of the contour integral formula~\cite{bullimore2019}, which is based on a modification to the localisation scheme in~\cite{benini2015,closset2016,benini2017}. Schematically, the modified Coulomb localisation scheme considers the lagrangian
\begin{equation}
\label{eq:Coulomb-localisation-lagrangian}
	\mathcal{L} = \frac{1}{t^2} \left( \frac{1}{e^2} \mathcal{L}_\text{YM} + \mathcal{L}_\tau \right)  + \frac{1}{g^2} \mathcal{L}_\Phi + \mathcal{L}_\text{CS} + \mathcal{L}_\text{FI} \,,
\end{equation}
with the exact deformation
\begin{equation}
\label{eq:1d-FI-lagrangian}
	\mathcal{L}_\tau =\frac{i \tau}{2} \left(Q + \widebar{Q}\right) \left(\lambda + \bar{\lambda}\right) = -i \tau D_{\text{1d}} \,,
\end{equation}
where
\begin{equation}
	D_{\text{1d}} = D - 2 F_{1 \bar{1}}
\end{equation}
is an auxiliary field in the one-dimensional vector multiplet.
The modification is to introduce this one-dimensional \index{Fayet-Iliopoulos parameter} Fayet-Iliopoulos parameter $\tau$ to ensure a meaningful result for each individual flux $\mathfrak{m} \in \mathbb{Z}$. This feature is necessary if we want to unambiguously interpret the coefficient of $q^\mathfrak{m}$ as counting the supersymmetric ground states with $U(1)_t$ charge $\mathfrak{m}$, as in the hamiltonian definition~\eqref{eq:U(1)-index-def}.

In the limit $t^2 \rightarrow 0$ and $e^2$ finite, the path integral localises to solutions of the following equations
\begin{subequations}
\begin{align}
	* F_A  + i D &= 0 \,, \\
	\bar{\partial}_A \phi_i &=0\,, \\
	\dd_A \sigma &=0\,,  \\
	F_{01} = F_{0\bar{1}} &= 0 \,, \\
	\left(m_i + Q_i \sigma\right) \phi_i &= 0 \,.
\end{align}
\end{subequations}
After integrating out the fermionic zero modes, the localised path integral leads to \index{contour integral formula} the contour integral formula~\cite{benini2017}
\begin{equation}
\label{eq:U(1)-contour-integral-formula}
	\mathcal{I} = \sum_{\mathfrak{m} \in \mathbb{Z}}  \frac{1}{2 \pi i} \oint_\text{JK} \frac{\dd x}{x} \, H^g Z(x; \mathfrak{m}) \,.
\end{equation}
The contour is explicitly given by the \index{Jefrrey-Kirwan residue} Jefrrey-Kirwan residues
\begin{equation}
\label{eq:U(1)-JK-prescription}
	\frac{1}{2 \pi i} \oint_\text{JK} \frac{\dd x}{x} = \sum_{x_*} \JK{x_*}(Q_*, \eta) \frac{\dd x}{x} \,,
\end{equation}
with
\begin{equation}
\label{eq:U(1)-JK-res}
	\JK{0}(Q, \eta) \frac{\dd x}{x} := \Theta (Q \eta ) \, \sign(Q) \,,
\end{equation}
where $\eta \neq 0$ is an auxiliary real parameter, and $Q_*$ denotes the \index{Jeffrey-Kirwan charge} Jeffrey-Kirwan charge of the pole at $x_*$. 
\begin{itemize}
	\item For the interior poles at $x^{Q_i} y_i = 1$ coming from the\index{chiral multiplet} chiral multiplets $\Phi_i$, their Jeffrey-Kirwan charges are taken to be the corresponding \index{gauge charge} gauge charges $Q_i$. 
	\item For the boundary poles at $x = 0$ and $x = \infty$ associated with the monopole operators, the Jeffrey-Kirwan charges are taken to be 
		\begin{subequations}
		\label{eq:JK-charges-boundary}
		\begin{align}
			Q_{+} &= 
			\begin{dcases}
				- \kappa^+ \quad &\text{if} \quad \kappa^+  \neq 0 \\
				\mathfrak{m}- \tau' \quad &\text{otherwise} 
			\end{dcases}\,, \\
			Q_{-} &= 
			\begin{dcases}
				+ \kappa^-\quad &\text{if} \quad \kappa^- \neq 0 \\
			\mathfrak{m}- \tau' \quad &\text{otherwise} 
			\end{dcases}\,,
		\end{align}
		\end{subequations}
		where the one-dimensional Fayet-Iliopoulous parameter $\tau$ is re-scaled as
		\begin{equation}
		\label{eq:rescaled-tau}
			\tau':=\frac{e^2 \mathrm{vol}(\Sigma)}{2 \pi} \tau\,.
		\end{equation}		
		
\end{itemize}

The contribution from each magnetic flux $\mathfrak{m}\in \mathbb{Z}$ is separately independent of the auxiliary parameter $\eta$ provided $\tau' \neq \mathfrak{m}$. However, the \index{twisted index} twisted index might jump across the wall $\tau ' = \mathfrak{m}$ according to
\begin{align}
\label{eq:wc}
	&\mathcal{I}(\tau' = \mathfrak{m}+\epsilon) - \mathcal{I}(\tau' = \mathfrak{m} - \epsilon) \nonumber \\
	=& q^\mathfrak{m}\left[ \, \delta_{\kappa^+,0} \; \underset{x=0}{\mathrm{Res}}+ \delta_{\kappa^-,0} \; \underset{x=\infty}{\mathrm{Res}} \,\right]\frac{\dd x}{x} \, H(x)^g Z(x; \mathfrak{m}) \, .
\end{align}
where $\epsilon  \to 0^+$. Therefore we require $\tau' \notin \mathbb{Z}$, which ensures that the Jeffrey-Kirwan charges are always non-vanishing and the contribution to the \index{twisted index} twisted index from each flux $\mathfrak{m} \in \mathbb{Z}$ is well-defined.

Finally, the contour prescription used in~\cite{benini2015,closset2016,benini2017} is recovered by sending $\tau' \to +\infty$ with $\eta >0$ or $\tau' \to - \infty$ with $\eta <0$. That this is independent of the auxiliary parameter $\eta$ is equivalent to the statement that sum of~\eqref{eq:wc} over $\mathfrak{m} \in \mathbb{Z}$ is proportional to a formal delta function at $q = 1$.

In general, a pole from a \index{chiral multiplet} chiral multiplet in the localisation formula corresponds to a solution of the \index{Bethe ansatz} Bethe ansatz equation~\eqref{eq:U(1)-bethe-Z}. However, each individual residue does not reproduce the corresponding summand in the Bethe formula~\eqref{eq:U(1)-index-bethe-formula}. This can be understood as the results of taking different limits in the ``de-compactification'' procedure. The Bethe ansatz equation evaluates at the limit $\beta \rightarrow 0$, while the localisation formula was derived by taking $\beta \rightarrow \infty$. Nevertheless, both formulae agree only after summing up all the contributions from each chiral multiplets.

\section{Geometric Interpretation}
\label{sec:geometric-interpretation}

The definition \eqref{eq:U(1)-index-def} of the \index{twisted index} twisted index can be interpreted as the \index{supersymmetric index} supersymmetric index~\eqref{eq:susy-index-def} of the supersymemtric quantum mechanics obtained by the twist. The supersymmetric index is identified with the holomorphic \index{Euler characteristic} Euler characteristic~\eqref{eq:susy-index-hol-euler-char}. Therefore the geometric interpretation of the \index{twisted index} twisted index is expected to be in the following form~\cite{xu2022}
\begin{equation*}
	\mathcal{I} = \sum_{\mathfrak{m}} q^{\mathfrak{m}} \int \hat{A}(\mathfrak{M}_{\mathfrak{m}}) \ch (\mathcal{V}_{\mathfrak{m}}) \,,
\end{equation*}
where $\mathfrak{m}$ labels the magnetic sectors, $\mathfrak{M}_{\mathfrak{m}}$ denotes the moduli space parametrising the saddle points of the localised path integral, and $\mathcal{V}_{\mathfrak{m}}$ is roughly a complex of vector bundles encoding the massive fluctuations of the \index{chiral multiplet} chiral multiplets and the \index{Chern-Simons term} Chern-Simons terms. 

For such an interpretation to be meaningful, it is necessary for the contribution to the \index{twisted index} twisted index from each individual flux $\mathfrak{m} \in \mathbb{Z}$ to be unambiguous. This necessitates the introduction of the one-dimensional Fayet-Iliopoulos parameter $\tau$ in~\eqref{eq:1d-FI-lagrangian}. The  wall-crossing phenomena in $\tau$ are then reflected in jumps in the structure of the moduli spaces $\mathfrak{M}_{\mathfrak{m}}$ and the complexes $\mathcal{V}_\mathfrak{m}$.

This general expectation was verified in previous work~\cite{bullimore2019a,bullimore2019} for a class of theories with $\mathcal{N}=4$ supersymmetry, where for generic $\tau' \neq \mathfrak{m}$ the moduli spaces $\mathfrak{M}_\mathfrak{m}$ exclusively parametrise vortex-like configurations on $\Sigma$ where the gauge group is either completely broken or broken to a discrete subgroup. Here we extend the geometric interpretation to theories with topological saddle points, where there is an unbroken continuous gauge symmetry and the moduli spaces $\mathfrak{M}_{\mathfrak{m}}$ must be described as quotient stacks. 

There is an important distinction between saddle points where the unbroken gauge symmetry is the whole $G = U(1)$ or a discrete subgroup. The latter involves a relatively mild extension of~\cite{bullimore2019} to deal with moduli spaces with orbifold singularities. We only focus on theories without orbifold singularities with the constraint $|Q_i| = 1$. We therefore consider theories with topological saddle points where $G = U(1)$ is fully unbroken with all matter fields set to vanish. The moduli space $\mathfrak{M}_{\mathfrak{m}}$ has a component that is roughly the Picard variety $\pic{\mathfrak{m}}{\Sigma} \cong T^{2g}$ parametrising degree $\mathfrak{m}$ holomorphic line bundles on $\Sigma$.

\subsection{Higgs Branch Localisation}
\label{sec:higgs-branch-localisation}

For the class of theories of our interest, this geometric interpretation can be realised using the Higgs branch localisation scheme~\cite{xu2022} such that both topological and vortex saddle points appear. It is similar to the Higgs branch localisation schemes implemented for two-dimensional $\mathcal{N}=(2,2)$ theories in~\cite{benini2015a,doround2013,closset2015}, and for three-dimensional $\mathcal{N}=2$ theories in~\cite{fujitsuka2014,benini2014}. 
In this localisation scheme we take the lagrangian~\eqref{eq:Coulomb-localisation-lagrangian} with an additional exact term 
\begin{equation}
	\mathcal{L}_\text{H} = -\frac{i}{2} \left(Q + \widebar{Q}\right) \left(\lambda + \bar{\lambda}\right)  \mu(\phi) \,,
\end{equation}
where $\mu(\phi) = \sum_{j =1}^N  Q_j |\phi_j|^2$ is the moment map for the gauge action.
After setting $g=t$, the resulting lagrangian 
\begin{equation}
	\mathcal{L} = \frac{1}{t^2} \left( \frac{1}{e^2} \mathcal{L}_\text{YM} + \mathcal{L}_\Phi  + \mathcal{L}_\tau + \mathcal{L}_\text{H} \right) + \mathcal{L}_\text{CS} + \mathcal{L}_\text{FI}
\end{equation}
is taken to the limit $t^2 \rightarrow 0$ while $e^2$ is kept finite. The supersymmetric saddle points are then solutions to the following set of generalised vortex equations,
\begin{subequations}
\label{eq:higgs-localisation-vacuum}
\begin{align}
	\frac{1}{e^2} * F_A + \sum_{j =1}^N  Q_j |\phi_j|^2 - t^2\sigma \frac{\kappa^\eff(\sigma) }{2\pi} - \tau &= 0 \,, \label{eq:higgs-vortex-eqaution} \\
	\bar{\partial}_A \phi_i &=0\,, \\
	\dd_A \sigma &=0\,, \\
	\left(m_i + Q_i \sigma\right) \phi_i &= 0  \label{eq:higgs-m-phi}
\end{align}
\end{subequations}
for all $i  = 1,\ldots, N$, where $F_A$ is the curvature of the gauge connection $A$, and $*$ is the Hodge star operator on $\Sigma$. In these equations, $\phi_j$ should be understood as a section of $K_\Sigma^{\sfrac{r_j}{2}} \otimes L^{Q_j}$, where $K_\Sigma$ is the canonical bundle on $\Sigma$ and $L$ is the holomorphic gauge bundle on $\Sigma$.

Note that the dependence on the three-dimensional \index{Fayet-Iliopoulos parameter} Fayet-Iliopoulos parameter $\zeta$ has dropped out but the equations depend critically on the one-dimensional Fayet-Iliopoulos parameter $\tau$. The term proportional to the \index{Chern-Simons level !effective} effective Chern-Simons level $\kappa^\eff$ is kept to capture potential saddle points where $|\sigma| \to \infty$ with $\sigma_0 : = t^2\sigma$ finite.

The solutions to equations \eqref{eq:higgs-localisation-vacuum} fall into topologically distinct sectors labelled by the flux
\begin{equation}
\mathfrak{m} := \frac{1}{2\pi}\int_\Sigma F_A \in \mathbb{Z} \, .
\end{equation}
A constraint on the existence of saddle points with a given flux $\mathfrak{m}$ is found by integrating the \index{D-term equation} D-term equation~\eqref{eq:higgs-vortex-eqaution} over the Riemann surface $\Sigma$ to give
\begin{equation}
\left(\tau' - \mathfrak{m}\right) +\frac{e^2\vol(\Sigma)}{4\pi^2}  t^2 \sigma \kappa^{\eff}(\sigma) = \sum_{j=1}^N  Q_j \, \norm{\phi_j}^2 \,,
\label{eq:integrated-vortex-equation}
\end{equation}
where $\tau'$ is rescaled according to~\eqref{eq:rescaled-tau}, and
\begin{equation}
\norm{\phi_j}^2 := \frac{e^2}{2\pi}\int_\Sigma \bar\phi_j \wedge * \, \phi_j
\end{equation}
is a positive definite inner product on sections of $K^{r_j}_\Sigma \otimes L^{Q_j}$.

\subsection{Classification of Saddles}
\label{sec:twisted-vacuum-classes}

Assuming the one-dimensional Fayet-Iliopoulos parameter is generic, i.e., $\tau' \neq \mathfrak{m}$, there are two classes of solutions in each magnetic flux sector $\mathfrak{m} \in \mathbb{Z}$. They are analogous to the Higgs vacua and topological vacua for generic mass parameters in flat spacetime, which are discussed in Section~\ref{sec:flat-vacuum-classes}.
\begin{itemize}
	\item {\bf Vortex saddles} \index{saddle point !vortex} are solutions where $\sigma$ remains finite in the limit $t^2 \to 0$ and the term proportional the effective Chern-Simon level $k^{\eff}$ in equation \eqref{eq:higgs-vortex-eqaution} can be ignored. The saddle equations~\eqref{eq:higgs-localisation-vacuum} then reduces to the abelian \index{vortex equation} vortex equations
\begin{equation}
\frac{1}{e^2} * F_A + \sum_{j =1}^N  Q_j |\phi_j|^2  = \tau \, , \quad \bar\partial_A \phi_i = 0 \, , \quad (m_i + Q_i \sigma) \phi_i = 0 \, .
\label{eq:vortex-eq-1}
\end{equation}
For generic \index{real mass} real mass parameters $m_i$, the space of solutions decomposes as a disjoint union of components where a single $\phi_i$ is non-vanishing and $\sigma = -m_i / Q_i$. From the constraint \eqref{eq:integrated-vortex-equation}, a component of the moduli space where $\phi_i$ is non-vanishing exists if and only if
\begin{equation}
\label{eq:vortex-saddles-sign-alignment}
	\sign(\tau'-\mathfrak{m}) = \sign(Q_i) \, .
\end{equation}

	\item {\bf Topological saddles} \index{saddle point !topological} are solutions where $|\sigma| \to \infty$ in the limit $t^2 \to 0$, such that the combination $\sigma_0:=t^2 \sigma$ remains finite and has a unique non-vanishing solution. This requires $\phi_j = 0$ for all $j=1,\ldots,N$ and therefore the constraint \eqref{eq:integrated-vortex-equation} becomes 
	\begin{equation}
		\tau' - \mathfrak{m} = - \frac{e^2\vol(\Sigma)}{4\pi^2} \sigma_0 \kappa^\pm  \,.
	\label{eq:vortex-topological}
	\end{equation}		

If a solution exists, then it is in the form
\begin{equation}
	\sigma_0 \sim - \frac{\tau' - \zeta}{\kappa^\pm} \,,
\end{equation}
ignoring the positive multiplicative factor.
\begin{subequations}
When $\sigma \to + \infty$, a solution at
\begin{equation}
	\sigma_0 \sim - \frac{\tau' - \zeta}{\kappa^+} > 0 \,.
\end{equation}
exists if and only if $\kappa^+ \neq 0$ and $\sign(\tau'-\mathfrak{m}) = - \sign \kappa^+$.
Similarly when $\sigma \to - \infty$, a solution at
\begin{equation}
	\sigma_0 \sim - \frac{\tau' - \zeta}{\kappa^-} < 0 \,.
\end{equation}
\end{subequations}
exists if and only if $\kappa^- \neq 0$ and $\sign(\tau'-\mathfrak{m}) =  \sign \kappa^-$.

In summary, a unique solution with $\pm\sigma_0 >0$ exists, if $\kappa^\pm \neq 0$ and 
\begin{equation}
\label{eq:topological-saddles-sign-alignment}
	\sign(\tau'-\mathfrak{m}) = \sign(Q_\pm) \, ,
\end{equation}	
where $Q_\pm$ are the Jeffrey-Kirwan charges~\eqref{eq:JK-charges-boundary} at the boundary poles, acting as the gauge charges of the monopole operators. In analogy to the condition~\eqref{eq:vortex-saddles-sign-alignment} on vortex saddles, this is consistent with the interpretation of attributing the topological vacua to monopole operators.

\end{itemize}

In addition, if $\kappa^\pm = 0$ then a non-compact Coulomb branch parametrised by $\pm\sigma_0 >0$ appears at $\tau'=\mathfrak{m}$, which is responsible for the wall-crossing phenomena in equation~\eqref{eq:wc}. These three classes are analogous to the trichotomy of flat space supersymmetric vacua~\cite{intriligator2013} discussed in Section~\ref{sec:flat-vacuum-classes}.

If we align the auxiliary parameter $\eta$ in the Jeffrey-Kirwan residue~\eqref{eq:U(1)-JK-res} as
\begin{equation}
\sign(\tau'-\mathfrak{m}) = \sign(\eta) \, ,
\end{equation}
then the components of the moduli space of saddles with flux $\mathfrak{m}$ are in one-to-one correspondence with the poles selected by the contour prescription in Section~\ref{sec:contour-integral-formula}. There is a component of the vortex moduli space with $\phi_i \neq 0$ when the pole at $x^{Q_i} y_i = 1$ is selected. Similarly, there is a topological saddle point with $\pm\sigma_0 >0$ whenever the residue at $x^{\pm1} \to 0$ is selected. 

\subsection{Geometric Integral Formula}
\label{sec:geomtric-integral-formula}

The moduli space \index[sym]{$\widebar{\mathfrak{M}}$}$\widebar{\mathfrak{M}}$ of saddle points \index{moduli space !of saddles} of the path integral splits into disjoint unions of topologically distinct components
$$
	\widebar{\mathfrak{M}} = \bigsqcup_\mathfrak{m \in \mathbb{Z}}  \widebar{\mathfrak{M}}_\mathfrak{m}
$$
labelled by the flux $\mathfrak{m}$. 
When all the mass parameters $\{m_j = 0\}_{j = 1}^N$ are set to vanish, the twisted index localises to the following schematic form
\begin{equation}
\label{eq:U(1)-integral-full-moduli}
	\sum_{\mathfrak{m} \in \mathbb{Z}} q^\mathfrak{m} \int  \hat{A}(\widebar{\mathfrak{M}}_\mathfrak{m}) \ch\left( \mathcal{L}_\mathfrak{m} \right) \,,
\end{equation}
where $\widebar{\mathfrak{M}}_\mathfrak{m}$ parametrises both the vortex and topological saddles, and $\mathcal{L}_\mathfrak{m}$ encodes the Chern-Simons terms.  

As we turn on non-vanishing mass parameters $\{m_j \neq 0\}_{j = 1}^N$, the remaining moduli space $\mathfrak{M}_{\mathfrak{m}}$ is the fixed points of the of the flavour $T_\text{f}$ action on $\widebar{\mathfrak{M}}_\mathfrak{m}$. Based on the structure of supersymmetric quantum mechanics, the \index{supersymmetric localisation} supersymmetric localisation leads to the following expression~\cite{bullimore2019,xu2022} for the \index{twisted index} twisted index
\begin{align}
\label{eq:U(1)-integral-of-characteristic-classes}
	\mathcal{I} &=  \sum_{\mathfrak{m} \in \mathbb{Z}} 
	q^{\mathfrak{m}} 
	\int \hat{A}(\mathfrak{M}_{\mathfrak{m}}) 
	\frac{\ch\big( \bigotimes_{\alpha} \mathcal{L}_{\mathfrak{m}, \alpha} \big)}{\ch \big(\hat{\wedge}^\bullet \mathcal{E}_{\mathfrak{m}} \big)} 
	 \nonumber \\
	&= \sum_{\mathfrak{m} \in \mathbb{Z}} 
	q^{\mathfrak{m}} 
	\int \hat{A}(\mathfrak{M}_{\mathfrak{m}}) 
	\frac{\hat{A}(\mathcal{E}_{\mathfrak{m}})}{\mathrm{e}(\mathcal{E}_{\mathfrak{m}})} 
	\ch\left( \bigotimes_{\alpha} \mathcal{L}_{\mathfrak{m}, \alpha} \right) \,,
\end{align}
where the \index{index bundle} index bundle \index[sym]{$\mathcal{E}_{\mathfrak{m}}$} $\mathcal{E}_{\mathfrak{m}}$ is formally a perfect complex of sheaves encoding the massive fluctuations of the\index{chiral multiplet} chiral multiplets, and $\mathcal{L}_{\mathfrak{m}, \alpha}$ are holomorphic line bundles arising from various \index{Chern-Simons term} Chern-Simons terms. This is analogous to the supersymmetric index~\eqref{eq:susy-index-hol-euler-char}. The notation $\hat{\wedge}$ denotes the normalised exterior algebra~\cite{okounkov2015}
\begin{align*}
	\hat{\wedge}^\bullet V &:= (\det V)^{\sfrac{-1}{2}} \otimes \wedge^\bullet V  \\
	&= (\det V)^{\sfrac{-1}{2}} \bigotimes_{i \geq 0} (-1)^i \wedge^i V \,.
\end{align*}
The integral should be understood equivariantly with respect to the flavour symmetry $T_{\text{f}}$ in~\eqref{eq:flavour-sym}, leading to the dependence on the \index{real mass} real mass parameters. 
The Dirac genus $\hat{A}(\mathfrak{M}_{\mathfrak{m}})$ comes from the tangent directions of the $T_{\text{f}}$ action on $\widebar{\mathfrak{M}}_\mathfrak{m}$, while the Chern character $\ch \big(\hat{\wedge}^\bullet \mathcal{E}_{\mathfrak{m}} \big)$ from the index bundle $\mathcal{E}_{\mathfrak{m}}$ appears as the contributions of the normal directions of the $T_{\text{f}}$ action on $\widebar{\mathfrak{M}}_\mathfrak{m}$.

The integrand consists of some characteristic classes, namely the Dirac genus $\hat{A}(\mathcal{F})$, the Euler class $\mathrm{e}(\mathcal{F})$, and the total Chern character $\ch(\mathcal{F})$ of some fibre bundle $\mathcal{F} \xrightarrow{\pi} \mathcal{M}$. 
In terms of the Chern roots $\alpha_i$ defined by the diagonalised curvature $\mathcal{A}$ of the bundle $E$
\begin{equation}
	\mathcal{A} := g^{-1} \left(\frac{i \mathcal{F}}{2 \pi}\right) g = \mathrm{diag}(\alpha_1,\ldots,\alpha_N) \,,
\end{equation}
the relevant characteristic classes\index[sym]{$\ch$}\index[sym]{$\mathrm{e}$}\index[sym]{$\hat{A}$}\index[sym]{$\mathrm{c}_1$}\index[sym]{$\td$} can be computed~\cite{nakahara2003} as
\begin{subequations}
\label{eq:characteristic-classes}
\begin{align}
	\ch(\mathcal{F}) &= \sum_{j=1}^N  e^{\alpha_j} \,,\\
	\mathrm{e}(\mathcal{F}) &= \prod_{j=1}^N  \alpha_j \,,\\
	\hat{A}(\mathcal{F}) &= \prod_{j=1}^N  \frac{\alpha_j}{2\sinh(\sfrac{\alpha_j}{2})} \,.
\end{align}
In addition, the first Chern class and the Todd class 
\begin{align}
	\mathrm{c}_1(\mathcal{F}) &= \sum_{j=1}^N \alpha_j \,, \\
	\td(\mathcal{F}) &= \prod_{j=1}^N \frac{\alpha_j}{1 - e^{-\alpha_j}} 
\end{align}
are needed for later computations.
\end{subequations}

It is expected to reproduce the contour integral formula~\eqref{eq:U(1)-contour-integral-formula} of twisted indices with the appropriate geometric objects. The contributions from the moduli spaces and the gauge bundles are explicitly constructed in Chapter~\ref{sec:vortex-vacuum} and Chapter~\ref{sec:topological-vacuum}, while the mechanism of generating the Chern-Simons contributions is further discussed in Chapter~\ref{sec:CS-from-det}.

\subsection{Schematic Construction}

Before diving into the details in Chapter~\ref{sec:vortex-vacuum} and Chapter~\ref{sec:topological-vacuum}, let us give a brief schematic for the specific geometric construction. 
Consider a single component $\mathfrak{M}_{\mathfrak{m}}$ of the moduli space. For vortex saddles, each component moduli space $\mathfrak{M}_{\mathfrak{m}}$ is a symmetric product of the curve $\Sigma$ from the non-vanishing \index{chiral multiplet} chiral multiplet. For topological saddles, $\mathfrak{M}_{\mathfrak{m}}$ are Picard stacks. The massive fluctuations of chiral multiplets $\Phi_j$ generate an \index{index bundle} index bundle $\mathcal{E}_j^\bullet$ , which is a perfect complex of sheaves as the derived push-forward
\begin{equation}
	\mathcal{E}_j^\bullet := R^\bullet \pi_* (\mathcal{L}^{Q_j} \otimes \mathcal{K}^{\sfrac{r_j}{2}})\,,
\label{eq:U(1)-complex-of-sheaves}
\end{equation}
where $\mathcal{L}$, $\mathcal{K}$ are respectively the universal line bundle and the canonical bundle on the product space $\mathfrak{M}_{\mathfrak{m}} \times \Sigma$, and $Q_j$, $r_j$ are the gauge and the \index{R-charge} R-charge of $\Phi_j$. The class $\ch(\mathcal{E}_j^\bullet) = \ch(\mathcal{E}_j^0) - \ch(\mathcal{E}_j^1)$ makes sense in equivariant K-theory and the complex behaves like a vector bundle of rank $d_j - g + 1$ for the purpose of such computations.

\begin{itemize}

\item For vortex saddles, the full construction can be found in Chapter~~\ref{sec:vortex-vacuum}. 
The moduli space is a symmetric product
\begin{equation}
	\mathfrak{M}^\text{vor}_{\mathfrak{m}} =  \mathrm{Sym}^{d_i} \Sigma \equiv \Sigma_{d_i} \,,
\end{equation}
coming from the single non-vanishing\index{chiral multiplet} chiral multiplet $\Phi_i$.
Following computations in Section~\ref{sec:vortex-tangent-directions}, the tangent direction contributes a class
\begin{align}
&\hat{A}(\Sigma_{d_i})
	=  \left(\frac{\eta e^{\sfrac{-\eta}{2}}}{1-e^{-\eta}}\right)^{d_i-g+1}    
	\exp[Q_i^2 \theta\left(\frac{1}{2}-\frac{1}{\eta}+\frac{e^{-\eta}}{1 - e^{-\eta}}\right)] \nonumber \,.
\end{align}
The index bundle contributions are discussed in~\ref{sec:vortex-index-bundle}. An application of the \index{Grothendiek-Riemann-Roch theorem} Grothendiek-Riemann-Roch theorem~\eqref{eq:G-R-R} gives
\begin{equation*}
	\ch(\mathcal{E}^\bullet_j)  
	= e^{Q_j \eta}\left[(d_j -2g +1) + \sum_{a=1}^g e^{-Q_j^2\theta_a} \theta\right] \,.
\end{equation*}
for the massive fluctuations of remaining vanishing\index{chiral multiplet} chiral multiplets $\{\Phi_j\}_{j \neq i}$.
Therefore the Chern roots can be read off as
\begin{equation*}
	\big( \underbrace{Q_j \eta,\ldots,Q_j \eta}_{\mathclap{d_j - 2g + 1}},Q_j\eta-Q_j^2\theta_1,\ldots,Q_j\eta-Q_j^2\theta_g \big) \,.
\end{equation*}
It can be completed to $T_\text{f}$-equivariant forms by adding the \index{real mass} real mass $\left(m_j - Q_j m_i\right)$ as the equivariant parameter, promoting to a $T_\text{f}$-equivariant Chern character
\begin{equation*}
\ch(\mathcal{E}^\bullet_j) = z_j e^{Q_j \eta}\left((d_j -g +1) - Q_j^2 \theta\right) \, ,
\end{equation*}
where $z_j := y_j / y_i^{Q_j}$.
The contribution of these fluctuations to the twisted index can be evaluated as
\begin{equation*}
	\frac{\hat{A}(\mathcal{E})}{\mathrm{e}(\mathcal{E} )}  
=  \prod_{j\neq i} \, \left(\frac{(e^{-Q_j\eta}z_j)^{\sfrac{1}{2}}}{1-e^{-Q_j\eta}z_j}\right)^{d_j-g+1} \exp[Q_j^2 \theta \left(\frac{1}{2} + \frac{e^{-Q_j\eta}z_j}{1-e^{-Q_j\eta}z_j}\right)] \,.
\end{equation*}
After integrating over the symmetric product $\Sigma_{d_i}$, this gives the contribution to the twisted index from the vortex saddle with $\phi \neq 0$.

\item For the topological saddles, the full construction can be found in Chapter~\ref{sec:topological-vacuum}.
The moduli stack $\mathfrak{M}^\text{top}_{\mathfrak{m}}$ can be decomposed into
\begin{equation}
	\mathfrak{M}^\text{top}_{\mathfrak{m}} = \mathrm{Pic}^\mathfrak{m} (\Sigma) \times [\text{pt} / \mathbb{C}^*] \,,
\end{equation}
if we choose a base point on $\Sigma$. Then the characteristic classes in the integral~\eqref{eq:U(1)-integral-of-characteristic-classes} can be understood as $\mathbb{C}^*$-equivariant classes on the moduli space 
\begin{equation}
\mathcal{M}_\mathfrak{m} = \mathrm{Pic}^\mathfrak{m} \simeq T^{2 g} \nonumber \,. 
\end{equation}
The integral splits into two parts:
\vspace{-\parskip} 
\begin{itemize}
	\item An integral of the equivariant classes over the moduli space $\mathcal{M}_\mathfrak{m}$,
	\item A contour integral
		\begin{equation}
			\frac{1}{2 \pi i} \oint \frac{\dd x}{x} \,,	\nonumber
		\end{equation}
		where $x$ is the Chern character of the trivial $\mathbb{C}^*$-equivariant holomorphic vector bundles with weight $+1$.
\end{itemize}

For the contribution from the tangent directions, we have $\hat{A}(\mathcal{M}_\mathfrak{m}) = 1$ because the tangent bundle is flat. Following computations in Section~\ref{sec:topological-index-bundle}, the Chern character of the index bundle can be computed via the Grothendieck-Riemann-Roch theorem~\cite{grothendieck1971} as
\begin{align}
	\ch(\mathcal{E}_j^\bullet) = (d_j-2g+1) + \sum_{a=1}^g e^{-Q_j^2\theta_a} \nonumber \,,
\end{align}
from which the Chern roots manifest as
\begin{equation}
	\big(\underbrace{0,\ldots,0}_{\mathclap{d_j - 2g + 1}},-Q_j^2\theta_1,\ldots,-Q_j^2\theta_g \big)  \nonumber \,.
\end{equation}
The degree of the complex is denoted as $d_j := \deg(\mathcal{E}_j^\bullet)$. Completing to an $\mathbb{C}^*$-equivariant form~\cite{libine2010} adds the \index{real mass} real mass
$
	Q_j \sigma + m_j
$
of the fluctuation of $\Phi_j$ to all the Chern roots, leading to an overall multiplicative factor of 
$
	x^{Q_j} y_j
$
on the Chern character. 
The factor generated from the chiral multiplet $\Phi_j$ in~\eqref{eq:U(1)-integral-of-characteristic-classes} can be evaluated to be 
\begin{align}
	\frac{\hat{A}( \mathcal{E}_j^\bullet )}{\mathrm{e}( \mathcal{E}_j^\bullet )} &= 
	\exp[ \left( \frac{1}{2} + \frac{  x^{Q_j} y_j }{ 1 -  x^{Q_j} y_j } \right)  (Q_j)^2 \theta ]   \left( \frac{  x^{\sfrac{Q_j}{2}} y_j^{\sfrac{1}{2}} }{ 1 -  x^{Q_j} y_j } \right)^{\left(  Q_j \mathfrak{m} \right) + (r_j - 1)(g - 1)}  \nonumber \,.
\end{align}
After integrating over the Picard variety $\mathrm{Pic}^\mathfrak{m}(\Sigma)$, this gives the contribution to the twisted index from the topological saddle.

\end{itemize}

Lastly the line bundles $\mathcal{L}_\alpha$ in~\eqref{eq:U(1)-integral-of-characteristic-classes} can be interpreted as the determinant line bundles obtained by integrating out additional auxiliary\index{chiral multiplet} chiral multiplets. This is consistent with the physical phenomenon that integrating out heavy fermions in three-dimensional theories induces effective Chern-Simons terms as a low-energy effect~\cite{Redlich:1983kn,Redlich:1983dv,Alvarez-Gaume:1983ihn}. It is these determinant line bundles that give the level structure in the corresponding \index{quantum K-theory} quantum K-theory.

\section{Window Phenomenon}
\label{sec:index-window-phenomenon}

Schematically, the \index{quantum K-theory} quantum K-theory of a \index{toric stack} toric stack $X$ concerns with integrals in the form of the holomorphic \index{Euler characteristic} Euler characteristic
\begin{equation}
	\sum_{\mathfrak{m}} q^\mathfrak{m}
	\int \hat{A}(\widetilde{\mathfrak{M}}_\mathfrak{m}) \ch\left( \mathcal{L}_\mathfrak{m} \right) \,,
\end{equation}
where  \index[sym]{$\widetilde{\mathfrak{M}}$} $\widetilde{\mathfrak{M}}_\mathfrak{m}$ is the moduli space of stable quasi-maps \index{moduli space !of quasi-maps} of degree $\mathfrak{m}$ from the curve $\Sigma$ into $X$, and $\mathcal{L}_\mathfrak{m}$ is some determinant line bundles giving arise to the level structure.

In three-dimensional gauge theory, the twisted index localises to an integral~\eqref{eq:U(1)-integral-full-moduli} of the same form over the full moduli space $\widebar{\mathfrak{M}}_\mathfrak{m}$ of saddle points when all mass parameters are vanishing. The connection between these saddles and quantum K-theory is the following.
\begin{itemize}
	\item For \index{saddle point !vortex} vortex saddles, the moduli space $\widebar{\mathfrak{M}}_\mathfrak{m}$ parametrises  generalised vortices~\cite{Bullimore:2016hdc}, which are identified algebraically via the \index{Hitchin-Kobayashi correspondence} Hitchin-Kobayashi correspondence~\cite{Jaffe:1980mj,Garcia-Prada:1993usn,Alvarez-Consul:2001mqd} as stable quasi-maps into the \index{vacuum !Higgs branch} Higgs branch $X$. The \index{D-term equation} D-term equation~\eqref{eq:higgs-vortex-eqaution} serves as the stability condition. The loci of quasi-maps contain a finite number of ``unstable'' points at which the vortices are located. 
	
	If no topological saddles exist, the moduli space $\widetilde{\mathfrak{M}}_\mathfrak{m}$ coincides with the moduli space $\widebar{\mathfrak{M}}_\mathfrak{m}$ of vortices, and the twisted index corresponds to integrals of the quantum K-theory of $X$, which has been well studied in the literature as a lift from \index{quantum cohomology} quantum cohomology. Specifically the twisted index is the trivial correlator $\langle 1 \rangle_\Sigma$ in K-theoretic languages~\cite{givental2015,jockers2018,jockers2019,jockers2019a,Ueda:2019qhg}.

	The Chern-Simons contribution $\mathcal{L}_\mathfrak{m}$ is constructed from determinant line bundles~\eqref{eq:U(1)-det-bundle-chiral} of auxiliary chiral multiplets, which are analogous to the level structure in quantum K-theory~\cite{ruan2019}. This gives a natural ground for our conjecture that the Chern-Simons level in gauge theories is equivalent to the quantum K-theory level. 

	The \index{window phenomenon} window phenomenon refers to the fact that quantum K-theory is ill-defined outside the \index{critical window} critical window where topological saddles do appear.
	\item For \index{saddle point !topological} topological saddles, the moduli space $\widebar{\mathfrak{M}}_\mathfrak{m}$ of saddles parametrises a more general type of maps into a quotient stack, where the stability condition~\eqref{eq:higgs-vortex-eqaution} does not apply. The loci of the maps contain an infinite number of ``unstable'' points. The precise mathematical description for topological saddles is currently missing in the literature of quantum K-theory. As topological saddles have not been taken into account in quantum K-theory, we expect this is the key missing ingredient to properly define quantum K-theory outside the \index{critical window} critical window.

\end{itemize}

The \index{window phenomenon} window phenomenon is  also manifest from the \index{contour integral formula} contour integral formula~\eqref{eq:U(1)-contour-integral-formula} of twisted indices. The Chern-Simons level determines the power of the exponential parameter $x$ appearing in the contour integral, which are responsible for the boundary poles at $x = 0$ and $x = \infty$. These boundary poles are contributions from the topological saddle points. Therefore, the Chern-Simons level controls the existence of topological vacua. 

Consider the $g=1$ case where the \index{twisted index} twisted index is expected to reproduce the sueprsymmetric index. The power of $x$ is simply
\begin{equation}
	\left( \kappa + \frac{1}{2} \sum_{j=1}^N Q_j^2 \right) \mathfrak{m}
\end{equation}
as the hessian factor drops out.
The pole at $x=0$ only appears if
\begin{equation}
	\kappa < - \frac{1}{2} \sum_{j=1}^N Q_j^2 \,,
\end{equation}
which coincides with the critical Chern-Simons levels from~\eqref{eq:U(1)-BAE-critical-level} and~\eqref{eq:U(1)-critical-CS-level}, up to a sign. The same story holds for the pole at $x=\infty$.

In summary, we have demonstrated that the \index{window phenomenon} window phenemenon of Chern-Simons levels in three-dimensional gauge theories coincide with the \index{window phenomenon} window phenomenon in \index{quantum K-theory} quantum K-theory. The observations from the semi-classical vacua and twisted chiral rings are both strong evidence for this connection. Finally the construction in Chapter~\ref{sec:CS-from-det} of Chern-Simons contributions to the twisted indices gives a natural interpretation from a geometric point of view. We expect the topological vacua are the key ingredients to define \index{quantum K-theory} quantum K-theory properly at levels outside of the \index{critical window} critical window. Furthermore, three-dimensional \index{mirror symmetry} mirror symmetry should provide insights into \index{quantum K-theory} quantum K-theories of dual spaces.

\section{Examples}
\label{sec:twisted-index-examples}

We consider a $U(1)$ Chern-Simons theory at level $\kappa \in \frac{1}{2}+\mathbb{Z}_{\geq 0}$ with one\index{chiral multiplet} chiral multiplet $\Phi$ of \index{R-charge} R-charge $r=1$ and charge $Q=+1$. The flavour symmetry $T_\text{f}$ is trivial and there are no \index{real mass} real mass parameters. The \index{Chern-Simons level !effective} effective Chern-Simons level~\eqref{eq:U(1)-effective-CS-level-gauge} is
\begin{equation}
	\kappa^\eff(\sigma) = \kappa + \frac{1}{2} \, \sign( \sigma) 
\end{equation}
and so the asymptotic level~\eqref{eq:U(1)-asymp-CS-level-gauge} is
\begin{equation}
	\kappa^\pm = \kappa \pm\frac{1}{2} \, .
\end{equation}
The cases $\kappa=\frac{1}{2}$ and $\kappa > \frac{1}{2}$ are quite different. The former has a neutral monopole operator and is mirror to a free chiral multiplet. This difference is reflected in the structure of the saddle points in our computation of the \index{twisted index} twisted index and therefore we treat the two cases separately. We also restrict attention to the twisted index with $g>0$.

\subsection[\texorpdfstring{$U(1)_{\frac{1}{2}}$ with One Chiral Field}%
                        {$U(1)$ at Level 1/2 with One Chiral Field}]%
        {$U(1)_{\frac{1}{2}}$ with One Chiral Field}
\label{subsec:Example-U(1)-1/2-1-chiral}

First consider $\kappa = \frac{1}{2}$. In this case $\kappa^\eff(\sigma) = \frac{1}{2}(1+\sign( \sigma) )$, and therefore $\kappa^+=1$ and $\kappa^- = 0$. There is a neutral \index{monopole} monopole operator and the theory is mirror to a free \index{chiral multiplet} chiral multiplet, together with specific background mixed Chern-Simons couplings.

The contour integral~\eqref{eq:U(1)-contour-integral-formula} for the \index{twisted index} twisted index is
\begin{equation}
	\mathcal{I} = \sum_{\mathfrak{m}\in \mathbb{Z}}  \frac{(-q)^\mathfrak{m}}{2 \pi i} \oint_\mathcal{C} \frac{\dd x}{x} \, \frac{x^{\mathfrak{m}}}{(1-x)^{\mathfrak{m}+g}}\,.
\label{eq:U(1)-k-1/2-index}
\end{equation}
where we have shifted $q\to-q$. In the presence of a one-dimensional Fayet-Iliopoulos parameter $\tau$, the contour is a Jeffrey-Kirwan residue prescription with charges
\begin{equation}
Q_{+} = -1 \, , \quad Q_1 = 1 \, ,\quad Q_- = \mathfrak{m}-\tau' \, .
\label{eq:example-charges-1}
\end{equation}
Note that the charge $Q_-$ associated to the residue at $x=\infty$ now depends on the one-dimensional Fayet-Iliopoulos parameter $\tau$ according to equation~\eqref{eq:JK-charges-boundary} since $\kappa^-=0$.

For $g>0$ the residue at $x = \infty$ vanishes. So there is no wall-crossing phenomena. The \index{twisted index} twisted index is given by computing the residue at $x=1$ with $\eta>0$, or equivalently minus the residue at $x=0$ with $\eta<0$, giving the result
\begin{equation}
	\mathcal{I} = (-1)^g q^{1-g}(1-q)^{g-1}\, .
\end{equation}
While the twisted index is non-zero only for fluxes $1-g\leq \mathfrak{m}\leq 0$, there are in fact~\cite{bullimore2019a} supersymmetric ground states for all $\mathfrak{m} \geq 1-g$. 

We now reproduce this result by evaluating the contributions from vortex and topological saddle points. The existence of vortex and topological saddle points is constrained by equation \eqref{eq:integrated-vortex-equation}, which becomes
\begin{equation}
	\left(\tau' - \mathfrak{m}\right) +\frac{e^2\vol(\Sigma)}{4\pi^2}\sigma_0 \kappa^{\eff}(\sigma) =  \norm{\phi}^2 \,,
\end{equation}
together with the equation $\sigma \phi = 0$. The existence of solutions depends on the sign of $\tau'-\mathfrak{m}$.
\vspace{-\parskip}
\begin{itemize}
\item When $\tau'-\mathfrak{m}>0$, there are vortex saddle points with $\sigma_0 = 0$. The moduli space of vortex solutions with flux $\mathfrak{m}$ is the symmetric product $\mathfrak{M}_\mathfrak{m} = \Sigma_d$ where $d = \mathfrak{m}+g-1$. Following the computations in Chapter~\ref{sec:vortex-vacuum}, the contribution to the twisted index is
\begin{align}
	\int_{\Sigma_d} \hat{\text{A}}(T\Sigma_d) \, \ch(\mathcal{L}^{\sfrac{1}{2}})
	& = \int_{\Sigma_d} \, \left(\frac{\eta e^{-\eta}}{1-e^{-\eta}}\right)^{\mathfrak{m}}    \exp[\theta\left(1-\frac{1}{\eta}+\frac{e^{-\eta}}{1 - e^{-\eta}}\right)] 
	\nonumber \\
	& = \frac{1}{2\pi i} \int_{x = 1} \frac{\dd x}{x} \, \frac{x^\mathfrak{m}}{(1-x)^{\mathfrak{m}+g}}\, .
\end{align}
\item When $\tau'-\mathfrak{m}<0$, there are topological saddle points with
\begin{equation}
	\phi=0\, , \quad \sigma_0 = - \frac{4\pi^2}{e^2\text{vol}(\Sigma)}(\tau' - \mathfrak{m}) > 0 \, .
\end{equation}
The moduli space of topological solutions with flux $\mathfrak{m}$ is the Picard variety $\mathfrak{M}_\mathfrak{m} = \text{Pic}^{\mathfrak{m}}\Sigma$. Following the computations in Chapter~\ref{sec:topological-vacuum}, the contribution to the twisted index is
\begin{align}
	\int_{\Sigma_d} \hat{\text{A}}(T\Sigma_d) \, \ch(\mathcal{L}^{\sfrac{1}{2}}) 
	& = \frac{1}{2\pi i} \int_{x=0} \frac{\dd x}{x} \int_{\text{Pic}^\mathfrak{m}\Sigma} \, \left(\frac{x}{1-x}\right)^{\mathfrak{m}}    \exp[\left(\frac{1}{1 - x}\right)\theta ]
	\nonumber \\
	& = \frac{1}{2\pi i} \int_{x =0} \frac{\dd x}{x} \, \frac{x^\mathfrak{m}}{(1-x)^{\mathfrak{m}+g}} \, ,
\end{align}
where the residue at $x=0$ is taken since $\sigma_0>0$.
\end{itemize}
A Coulomb branch of solutions with $\sigma_0 <0$ opens at $\tau'-\mathfrak{m}=0$ so there is the potential for wall-crossing. However, the vanishing of the residue at $x = \infty$ means that the twisted index is independent of $\tau$. This reproduces the Jeffrey-Kirwan residue prescription with charges~\eqref{eq:example-charges-1} and $\sign(\eta) = \sign( \tau'-\mathfrak{m})$. The result is independent of $\eta$ for each flux $\mathfrak{m}$ by construction.

\subsection[\texorpdfstring{$U(1)_\kappa$ with One Chiral Field}%
                        {$U(1)$ at Level κ with One Chiral Field}]%
        {$U(1)_\kappa$ with One Chiral Field}

Now consider one \index{chiral multiplet} chiral multiplet with $\kappa > \frac{1}{2}$ such that $\kappa^\pm = \kappa \pm \frac{1}{2} > 0$. There are no gauge neutral \index{monopole} monopole operators and the structure of the \index{twisted index} twisted index differs considerably.

The contour integral~\eqref{eq:U(1)-contour-integral-formula} for the \index{twisted index} twisted index is now
\begin{equation}
	\mathcal{I} = \sum_{\mathfrak{m}\in \mathbb{Z}}  \frac{(-q)^\mathfrak{m}}{2 \pi i} \oint_\mathcal{C} \frac{\dd x}{x} \, x^{\kappa \mathfrak{m}} \, \left( \frac{x^{\sfrac{1}{2}}}{1-x}\right)^\mathfrak{m} \left( \kappa +\frac{1}{2}\frac{1+x}{1-x}\right)^g\,,
\label{eq:U(1)-k-index}
\end{equation}
where the contour is a Jeffrey-Kirwan residue prescription with charges
\begin{equation}
Q_+ = -\kappa-\frac{1}{2}<0\, ,  \quad Q_1 = 1\, , \quad Q_- = \kappa - \frac{1}{2} > 0 \, .
\label{eq:example-charges-2}
\end{equation}
The index is now manifestly independent of $\tau$ and there is no wall-crossing. We therefore enumerate the residues at $x=1$ and $x=\infty$ with $\eta >0$, or equivalently minus the residues at $x=0$ with $\eta <0$. 

In this case it is illuminating to spell out the contributions from individual residues. For example, with $g=2$ and $k = \frac{3}{2}$ we find
\begin{subequations}
\begin{align}
	-\mathcal{I}_0 & =\frac{1}{q} - 4 \, , \\
	\mathcal{I}_1 & =\frac{1}{q} - 3 - q - 2q^2 -5q^3-14q^4- \cdots  = \frac{1-4 q+\sqrt{1-4 q}}{2 q} \, , \\
	\mathcal{I}_\infty & = -1 + q + 2q^2 +5q^3+14q^4+ \cdots =  \frac{1-4 q-\sqrt{1-4 q}}{2 q} \, .
\end{align}
\end{subequations}
Notice that the contributions $\mathcal{I}_1$ and $\mathcal{I}_\infty$ are not rational function of $q$ and so cannot individually reproduce a reasonable index. In fact they do not count honest supersymmetric ground states but only perturbative ground states. These are subject to instanton corrections that remove pairs of perturbative ground states corresponding to cancelations in the sum $\mathcal{I}_1 +\mathcal{I}_\infty = - \mathcal{I}_0$.

We can reproduce these contributions from an analysis of vortex and topological saddle points. 
The saddle points are again constrained by 
\begin{equation}
	\left(\tau' - \mathfrak{m}\right) +\frac{e^2\vol(\Sigma)}{4\pi^2}\sigma_0 \kappa^{\eff}(\sigma) =  \norm{\phi}^2 \,,
\end{equation}
and depend on the sign of $\tau'-\mathfrak{m}$.
\vspace{-\parskip}
\begin{itemize}
	\item When $\tau'-\mathfrak{m}> 0$, there are both vortex saddle points and topological saddle points with $\sigma_0 < 0$. The contributions from these saddle points reproduce the residues at $x = 1$ and $x=\infty$ respectively.
	\item When $\tau'-\mathfrak{m}< 0$, there are topological saddle points with $\sigma_0 >0$, whose contribution reproduces the residue at $x = 0$.
\end{itemize}
There is no Coulomb branch at $\tau'-\mathfrak{m} = 0$ and the twisted index is independent of $\tau$. This reproduces precisely the Jeffrey-Kirwan residue prescription with charges~\eqref{eq:example-charges-2} and $\sign(\eta) = \sign(\tau' - \mathfrak{m})$. The result is independent of $\eta$ for each flux $\mathfrak{m}$ by construction.

\chapter{Vortex Saddles}
\label{sec:vortex-vacuum}

This chapter contains the full algebro-geometric construction of the twisted index for \index{saddle point !vortex} vortex saddle points, which is introduced in Chapter~\ref{chap:twist-on-surface}. The schematics can be found in Section~\ref{sec:geometric-interpretation}. We first analyse the structure of the moduli space of the vortex saddles. Then the individual contributions from the tangent directions, the normal directions, and the Chern-Simons terms are constructed. Finally the total contribution to the twisted index is evaluated, which is in agreement with the contour integral formula.

\section{Moduli Space}
\label{sec:vortex-moduli}

The \index{moduli space !of saddles} moduli space of vortex saddle points consists of solutions to 
\begin{equation}
\begin{gathered}
\frac{1}{e^2} * F_A + \sum_{j =1}^N  Q_j |\phi_j|^2  = \tau \, , \\ 
\bar\partial_A \phi_j = 0 \, ,\quad (m_j + Q_j \sigma) \phi_j = 0 \, ,
\end{gathered}
\label{eq:vortex-eq-2}
\end{equation}
for all\index{chiral multiplet} chiral multiplets $j \in \{ 1,\ldots, N \}$, modulo gauge transformations. The moduli space is a disjoint union of topologically distinct components $\mathfrak{M}_\mathfrak{m}$ labelled by the magnetic flux $\mathfrak{m}\in \mathbb{Z}$. The entire moduli space is realised as an infinite-dimensional K\"ahler quotient. Under our assumption $|Q_j|=1$ each component $\mathfrak{M}_\mathfrak{m}$ is a finite-dimensional smooth K\"ahler manifold. Unless stated otherwise, we have kept $Q_j$ explicit throughout to better understand each contributions. In particular, the dependence on $Q_j^2$ would have been easily lost during long computations if it is set to $Q_j^2=1$ beforehand.
 
For generic mass parameters $m_i$, the moduli space further decomposes as a disjoint union of components
\begin{equation}
	\mathfrak{M}_\mathfrak{m} = \bigotimes_i \mathcal{M}_{\mathfrak{m},i}
\end{equation}
where a single chiral multiplet $\phi_i$ is non-vanishing and $\sigma = - m_i / Q_i$. Each component parametrises solutions to the abelian vortex equations
\begin{equation}
\frac{1}{e^2} * F_A + Q_i |\phi_i|^2  = \tau \, ,  \quad \bar\partial_A \phi_i = 0  \, ,
\label{eq:vortex-eq-part}
\end{equation}
where $\phi_i$ transforms as a section of $K_\Sigma^{\sfrac{r_j}{2}} \otimes L^{Q_i}$ on $\Sigma$. A small modification of the standard analysis applies and each component is either a symmetric product of the curve $\Sigma$ or empty,
\begin{align}
Q_i = +1: \qquad \mathcal{M}_{\mathfrak{m},i} = \begin{cases}
\Sigma_{d_i} & \text{if}  \quad \tau' > \mathfrak{m} \\
\emptyset & \text{if}  \quad \tau' < \mathfrak{m}
\end{cases} \\
Q_i = -1: \qquad \mathcal{M}_{\mathfrak{m},i} = \begin{cases}
\emptyset & \text{if}  \quad \tau' > \mathfrak{m} \\
\Sigma_{d_i} & \text{if}  \quad \tau' < \mathfrak{m}
\end{cases}
\label{eq:reduced-vortex}
\end{align}
where
\begin{equation}
d_i : = Q_i \mathfrak{m} + r_i(g-1)
\end{equation}
is the degree of $K_\Sigma^{\sfrac{r_j}{2}} \otimes L^{Q_i}$ and $\Sigma_d:=\text{Sym}^{d}\Sigma$ with the understanding that this is empty for $d<0$. The symmetric product $\Sigma_{d_i}$ parametrises the positions of the vortices. The assumption $|Q_i| = 1$ is important to get a symmetric product, otherwise the moduli space has orbifold singularities where a discrete gauge subgroup is unbroken.

Note that if the auxiliary parameter $\eta$ is aligned with $\tau' - \mathfrak{m}$, i.e.,
\begin{equation*}
\text{sign}(\tau' - \mathfrak{m}) = \text{sign}(\eta) \, ,
\end{equation*}
then the component $\mathcal{M}_{\mathfrak{m},i}$ of the moduli space is non-empty whenever the Jeffrey-Kirwan residue prescription selects the pole at $x^{Q_i}y_i = 1$ from the chiral multiplet $\Phi_i$. 
The task in the remainder of this chapter is to reproduce the residue at this pole from supersymmetric localisation.

It is useful to use an algebraic description of moduli spaces of abelian vortices in terms of holomorphic pairs. Let us assume $\sign(\tau'-\mathfrak{m}) = \sign(Q_i)$ so that the vortex moduli space $\mathcal{M}_{\mathfrak{m},i}$ is non-empty. Then the \index{Hitchin-Kobayashi correspondence} Hitchin-Kobayashi correspondence~\cite{Jaffe:1980mj,Garcia-Prada:1993usn,Alvarez-Consul:2001mqd} says that there is an algebraic description parametrising pairs $(L,\phi_i)$ where $L$ is a holomorphic line bundle of degree $\mathfrak{m}$ and $\phi_i$ is a non-zero section of $K_\Sigma^{\sfrac{r_j}{2}} \otimes L^{Q_i}$. The symmetric product $\Sigma_{d_i}$ in equation~\eqref{eq:reduced-vortex} parametrises the zeros of the section $\phi_i$.

For simplicity and to avoid a profusion of signs at intermediate steps in the calculation, we assume 
\begin{equation}
	Q_i = +1
\label{eq:assumption-Q+1}
\end{equation}
for the non-vanishing\index{chiral multiplet} chiral multiplet $\Phi_i$, along with $\tau' > \mathfrak{m}$. A similar computation applies to $Q_i = -1$ and the final result is presented in a uniform way for both cases.

\section{Contributions to Index}
\label{sec:vortex-contributions}

The contribution to the twisted index from a component $\mathcal{M}_{\mathfrak{m},i}$ of the vortex moduli space is
\begin{align}
	\mathcal{I}_{\mathfrak{m},i} 
	& = \int \hat{A} ( \mathcal{M}_{\mathfrak{m},i} )   
	\frac{\hat{A}(\mathcal{E}  )}{\mathrm{e}( \mathcal{E} )} 
	\ch\Big( \bigotimes_{\alpha} \mathcal{L}_
\alpha \Big) 
	\label{eq:vortex-contribution-integral} 
\end{align}
where $\mathcal{E} $ is a perfect complex of sheaves encoding the massive fluctuations of the remaining\index{chiral multiplet} chiral multiplets $\Phi_{j \neq i}$ of vanishing expectation values around the vortex configurations, and $\mathcal{L}_
\alpha$ are holomorphic line bundles arising from the  mixed Chern-Simons terms. Note that we have omitted the labels $\mathfrak{m},i$ from the bundles $\mathcal{E}$ and $\mathcal{L}_\alpha$ for brevity.

This integral should be understood equivariantly with respect to the flavour symmetry $T_\text{f}$ with parameters $y_i$. It can be evaluated using intersection theory on symmetric products, and converted into a contour integral following~\cite{macdonald1962,arbarello1985}. This extends a computation performed in~\cite{bullimore2019} to a wider class of theories.

\subsection{Tangent Direction}
\label{sec:vortex-tangent-directions}

We first consider the contribution from the directions tangent to $\mathcal{M}_{\mathfrak{m},i}$, which is the symmetric product $\Sigma_{d_i}$ when $\tau'>\mathfrak{m}$ and otherwise empty.

Let us briefly summarise some notations for the intersection theory on a symmetric product. There are standard generators 
\begin{subequations}
\begin{align}
 	\zeta_a, \bar{\zeta}_a & \quad \in H^1(\Sigma_d,\mathbb{Z}) \simeq H^1(\Sigma,\mathbb{Z}) \qquad \forall  a \in \{1,\ldots,g\}\,, \\
	\eta & \quad \in H^2(\Sigma_d,\mathbb{Z}) \simeq H^2(\Sigma,\mathbb{Z})
\end{align}
\end{subequations}
arising from cohomology classes~\cite{macdonald1962,arbarello1985} on $\Sigma$. We then define the \index{Dolbeault cohomology} Dolbeault cohomology classes 
\begin{equation}
	\theta_a := \zeta_a \wedge \bar\zeta_a \quad \in H^{1,1}(\Sigma_d,\mathbb{Z}) \simeq H^{1,1}(\Sigma,\mathbb{Z})\,,
\end{equation}
and for convenience their sum
\begin{equation}
	\theta := \sum_{a=1}^g \theta_a \,.
\end{equation}

The Chern character of the tangent bundle~\cite{macdonald1962} is given by
\begin{align}
	\ch( T\Sigma_{d_i}) 
	& = (g-1) + [(d_i-2g+1)-\theta] e^\eta \nonumber \\
	& = (g-1) + (d_i-2g+1)e^{\eta} + \sum_{a=1}^ge^{\eta- Q_i^2\theta_a} \,,
\end{align}
from which the curvature of the symmetric product $\Sigma_{d_i}$ can be read off as
\begin{equation}
	\mathcal{A}_{\Sigma_{d_i}} = \diag\big[\underbrace{\eta,\ldots,\eta}_{\mathclap{d_i-2g+1}},
	\left(\eta-Q_i^2\theta_1\right),\ldots,\left(\eta-Q_i^2\theta_g\right),
	\underbrace{0,\ldots,0}_{\mathclap{g-1}}\big] \,.
\label{eq:chern-roots-of-sym}
\end{equation}

From here we can evaluate the $\hat A$-genus 
\begin{align}
	\hat{A}(\Sigma_{d_i}) 
	=& \prod_{n=1}^{d_i} \frac{\alpha_n}{2\sinh(\sfrac{\alpha_n}{2})} 
	= \prod_{n=1}^{d_i} e^{-\sfrac{\alpha_n}{2}}\frac{\alpha_n}{1 - e^{-\alpha_n}} \nonumber\\
	=&  e^{-\frac{1}{2} \mathrm{c}_1\left(\Sigma_{d_i}\right)} 
	\td(\Sigma_{d_i}) 
\end{align}
as follows.

The factor involving the first Chern class is simply
\begingroup
\allowdisplaybreaks[0]
\begin{align}
\label{eq:vortex-A-roof-factor-1}
	e^{-\frac{1}{2}\mathrm{c}_1\left(\Sigma_{d_i}\right)} 
	&= \exp(-\frac{d_i-g+1}{2}\eta + Q_i^2 \frac{\theta}{2}) \nonumber \\
	&= \left( e^{-\eta} \right)^{\frac{1}{2} (d_i - g + 1)}  \exp(\frac{1}{2} Q_i^2 \theta)\,.
\end{align}
\endgroup

The Todd class factor can be split into
\begin{equation}
	\td(\Sigma_{d_i}) = \prod_{n=1}^{d_i} \frac{\alpha_n}{1 - e^{-\alpha_n}} 
	= 
	\left(\frac{\eta}{1-e^{-\eta}}\right)^{d_i-2g+1} 
	\prod_{a=1}^g \frac{\eta- Q_i^2 \theta_a}{1 - e^{-\eta+ Q_i^2 \theta_a}} \,,
\end{equation}
where the $g$-fold product can be manipulated as
\begin{align}
	\prod_{a=1}^g \frac{\eta- Q_i^2 \theta_a}{1 - e^{-\eta+ Q_i^2 \theta_a}} &= 
	\prod_{a=1}^g \frac{\eta- Q_i^2 \theta_a}{1 - e^{-\eta}(1+ Q_i^2 \theta_a)} \nonumber \\
	&= \prod_{a=1}^g \frac{\eta- Q_i^2 \theta_a}{(1 - e^{-\eta})-e^{-\eta}  Q_i^2 \theta_a} \nonumber\\
	&= \prod_{a=1}^g \frac{\eta- Q_i^2 \theta_a}{1 - e^{-\eta}}\left(1-\frac{e^{-\eta} }{1 - e^{-\eta}}Q_i^2 \theta_a\right)^{-1} \nonumber\\
	&= \prod_{a=1}^g \frac{\eta}{1 - e^{-\eta}}\left(1-\frac{ Q_i^2 \theta_a}{\eta}+\frac{e^{-\eta} }{1 - e^{-\eta}}Q_i^2 \theta_a\right) \nonumber \\
	&= \left(\frac{\eta}{1 - e^{-\eta}} \right)^g
	\prod_{a=1}^g    \exp[Q_i^2 \theta_a\left(-\frac{1}{\eta}+\frac{e^{-\eta}}{1 - e^{-\eta}}\right)] 
\end{align}
by repeatedly utilising the nilpotency $\theta_a^2 = 0$ for any $a =1,\ldots,g$.
Hence the Todd class can be written as
\begin{align}
\label{eq:vortex-A-roof-factor-2}
	\td(\Sigma_{d_i}) 
	&= \left(\frac{\eta}{1-e^{-\eta}}\right)^{d_i-g+1} 
	\prod_{a=1}^g    \exp[Q_i^2 \theta_a\left(-\frac{1}{\eta}+\frac{e^{-\eta}}{1 - e^{-\eta}}\right)] \nonumber \\
	&= \left(\frac{\eta}{1-e^{-\eta}}\right)^{d_i-g+1} 
	\exp[Q_i^2 \theta \left(-\frac{1}{\eta}+\frac{e^{-\eta}}{1 - e^{-\eta}}\right)] \,.
\end{align}

Combining~\eqref{eq:vortex-A-roof-factor-1} and~\eqref{eq:vortex-A-roof-factor-2} gives the desired form of the $\hat{A}$-genus as
\begin{equation}
\label{eq:U(1)-vortex-A-roof-sym}
	\hat{A}(\Sigma_{d_i}) =
	\left(\frac{\eta e^{\sfrac{-\eta}{2}}}{1-e^{-\eta}}\right)^{d_i-g+1}    
	\exp[Q_i^2 \theta\left(\frac{1}{2}-\frac{1}{\eta}+\frac{e^{-\eta}}{1 - e^{-\eta}}\right)] \,.
\end{equation}

\subsection{Index Bundle}
\label{sec:vortex-index-bundle}

We now consider the \index{index bundle} index bundle $\mathcal{E}_{j \neq i}$, encoding fluctuations of each of the remaining massive \index{chiral multiplet} chiral multiplets $\Phi_{j \neq i}$ around configurations in $\mathcal{M}_{\mathfrak{m},i}$. These are coming from the normal directions of the flavour $T_\text{f}$ action on the full moduli space $\widebar{\mathfrak{M}}_\mathfrak{m}$. 

At a point $(L,\phi_i)$ on the moduli space $\mathcal{M}_{\mathfrak{m},i}$, each \index{chiral multiplet} chiral multiplet $\Phi_{j \neq i}$ generates one-dimensional $\mathcal{N}=(0,2)$ chiral multiplet and Fermi multiplet fluctuations. Due to the supersymmetry they are valued in the following cohomologies.
\vspace{-\parskip}
\begin{itemize}
	\item Chiral multiplet fluctuations are valued in
	\begin{subequations}
	\begin{equation}
		E_j^0 := H^0\left(\Sigma,  L^{Q_j} \otimes K_\Sigma^{\sfrac{r_j}{2}}\right) \,.
	\end{equation}		
	\item Fermi multiplet fluctuations are valued in 
	\begin{equation}
		E_j^1 : = H^1\left(\Sigma,L^{Q_j} \otimes K_\Sigma^{\sfrac{r_j}{2}}\right) \,.
	\end{equation}
	\end{subequations}
\end{itemize}
The line bundle $L$ is the \index{gauge bundle} gauge bundle with the gauge charge $Q_j$ inherited from the three-dimensional chiral multiplet $\Phi_j$. The canonical bundle $K_\Sigma$ comes from the topological twist mixing in the R-symmetry.
When we move around in the moduli space $\mathcal{M}_{\mathfrak{m},i}$, the dimensions of these vector spaces may jump. But the difference of their dimensions is constant by the \index{Riemann-Roch theorem} Riemann-Roch theorem
\begin{align}
\label{eq:U(1)-vortex-dim-E}
	\dim E_j^0 - \dim E_j^1 
	& = h^0 \left( L^{Q_j}\otimes K_{\Sigma}^{\sfrac{r_j}{2}} \right) - h^1 \left( L^{Q_j}\otimes K_{\Sigma}^{\sfrac{r_j}{2}} \right)  \nonumber \\
	& \equiv d_j -g +1 \nonumber \\
	& = Q_j \mathfrak{m}+ (g-1)(r_j-1) \, , 
\end{align}
where $d_j$ is the degree of the line bundle
\begin{align}
	d_j 
	&= \deg\left(L^{Q_j}\otimes K_{\Sigma}^{\sfrac{r_j}{2}}\right) \nonumber \\
	&= Q_j \deg(L)  + \frac{r_j}{2} \deg(K_{\Sigma}) \nonumber \\
	&= Q_j \mathfrak{m}  + \frac{r_j}{2} (2g-2) \,.
\end{align}
We can therefore formally regard the difference of these vector spaces as the fibre of a holomorphic vector bundle on the moduli space $\mathcal{M}_{\mathfrak{m},i}$ of rank $d_j-g+1$. At least this  defines a reasonable K-theory class for use in the computation of the twisted index. 

The above schematics needs to be defined more precisely for computations, which requires the \index{universal construction} universal construction. The massive fluctuations should be described by a complex of sheaves. We recall the construction of the universal bundle on a symmetric product. We consider the pair of projection maps
\begin{equation}
\begin{tikzcd}[column sep=small]
	& {\Sigma_{d_i} \times \Sigma} \arrow[dl,"\pi", swap]\arrow[dr,"p"] & \\ {\Sigma_{d_i}} && {\Sigma}
\end{tikzcd} \,.
\label{eq:sym-projections}
\end{equation}
There is a unique universal line bundle $\mathcal{L}$ on $\Sigma_{d_i} \times \Sigma$ with the property that its restriction to a point $q = (L,\phi_i)$ on the symmetric product $\Sigma_{d_i}$ is the holomorphic line bundle $L$ on $\Sigma$, i.e., $\mathcal{L} \eval_q = L$. We also define $\mathcal{K} := p^*K_\Sigma$ to be the pull-back of the canonical bundle on the curve. With this in hand, the fluctuations of $\Phi_j$ transform in a perfect complex of sheaves on $\Sigma_{d_i}$ defined by the derived push-forward 
\begin{equation}
	\mathcal{E}^\bullet_j := R^\bullet \pi_*(\mathcal{L}^{Q_j} \otimes \mathcal{K}^{\sfrac{r_j}{2}}) \, .
\label{eq:U(1)-vortex-chiral-derived-push-forward}
\end{equation}
The stalks of $\mathcal{E}^\bullet_j$ over a point $(L,\phi_i)$ on the symmetric product are the vector spaces $E_j^\bullet$.

We can extract the Chern roots of $\mathcal{E}^\bullet_j$ following standard computations~\cite{arbarello1985}. Here we abuse the notations and identify the cohomology classes $\eta$ and $\theta$ with their pull-back by $\pi$. 
The first Chern class of $\mathcal{L}$ is then
\begin{equation}
	\mathrm{c}_1(\mathcal{L}) = \mathfrak{m}\eta_{\scriptscriptstyle \Sigma} +  \gamma + \eta \,,
\end{equation}
where $\eta_{\scriptscriptstyle \Sigma}$ is the pull-back of the class of a point on $\Sigma$, and $\gamma$ is the pull-back of the class of the diagonal $\Sigma \otimes \Sigma$ satisfying
\begin{align*}
	\gamma^2 &= -2 \eta_{\scriptscriptstyle \Sigma} \theta \,,\\
	\gamma^3 &= \eta_{\scriptscriptstyle \Sigma} \gamma = 0 \,.
\end{align*}
The class $\gamma$ does not play a role in what follows.

Following a small modification to the standard computation~\cite{arbarello1985}, the Chern character $\ch\left(\mathcal{L}^{Q_j}\right)$ reads
\begin{align}
	\ch\left(\mathcal{L}^{Q_j}\right)  &= e^{Q_j \mathrm{c}_1(\mathcal{L})} 
	= \sum_{n=0}^\infty \frac{(Q_j)^n (\mathfrak{m}\eta_{\scriptscriptstyle \Sigma} + \gamma + \eta )^n}{n!} \nonumber \\
	&= 1 + Q_j \mathfrak{m}\eta_{\scriptscriptstyle \Sigma} + Q_j \gamma + Q_j \eta + \frac{Q_j^2}{2} (-2 \eta_{\scriptscriptstyle \Sigma} \theta) \nonumber \\
	&= 1 + Q_j \mathfrak{m}\eta_{\scriptscriptstyle \Sigma} + Q_j \gamma + Q_j \eta - Q_j^2 \eta_{\scriptscriptstyle \Sigma} \theta \nonumber \\
	&= e^{Q_j \eta} \left(1 + Q_j \mathfrak{m} \,\eta_{\scriptscriptstyle \Sigma} + Q_j \gamma - Q_j^2 \eta_{\scriptscriptstyle \Sigma} \theta \right) \,,
\end{align}
where $\theta$ denotes the pull-back of the class in $H^{1,1}(\Sigma_{d_i},\mathbb{Z})$.

To compute $\ch\left(\mathcal{K}^{\sfrac{r_j}{2}}\right)$, we want to know the first Chern class of the canonical bundle $K_\Sigma$ on $\Sigma$. By inverting the definition 
\begin{equation}
	\deg\left(K_\Sigma\right)=2g-2 \equiv \int_\Sigma \mathrm{c}_1(k) \,,
\end{equation}
the first Chern class $\mathrm{c}_1(K_\Sigma)$ can be written as
\begin{equation}
	\mathrm{c}_1(K_\Sigma) = (2g-2) \omega_\Sigma \,,
\label{eq:c1-canonical-bundle}
\end{equation}
where $\omega_\Sigma$ is the volume form on $\Sigma$.
Then the Chern character of the pull-back is
\begin{align}
	\ch\left(\mathcal{K}^{\sfrac{r_j}{2}}\right) &= e^{\frac{r_j}{2} \mathrm{c}_1(p^*(K_\Sigma))} \nonumber \\
	&= e^{\frac{r_j}{2} (2g-2) \eta_{\scriptscriptstyle \Sigma}} \nonumber \\
	&= 1 + r_j(g-1) \eta_{\scriptscriptstyle \Sigma} \,.
\label{eq:ch-kappa-r-over-2}
\end{align}

The Chern character $\ch(\mathcal{E}^\bullet_j)$ is related to $\ch(\mathcal{L}^{Q_j} \otimes \mathcal{K}^{\sfrac{r_j}{2}})$ via the \index{Grothendiek-Riemann-Roch theorem} Grothendiek-Riemann-Roch theorem~\cite{grothendieck1971,arbarello1985}
\begin{equation}
	\ch(R^\bullet \pi_* \mathcal{F}) \td(Y) = \pi_*(\ch(\mathcal{F})  \td(X)) \,,
\label{eq:G-R-R}
\end{equation}
where the appropriate identifications are
\begin{subequations}
\begin{align}
	\mathcal{F} &= \mathcal{L}^{Q_j} \otimes \mathcal{K}^{\sfrac{r_j}{2}}\,, \\
	X &= \Sigma \times \Sigma_{d_i}\,, \\
	Y &= \Sigma_{d_i}\,.
\end{align}
\end{subequations}
The proper morphism $\pi$ in \eqref{eq:G-R-R} is  identified with the projection $\pi:\Sigma \times \Sigma_{d_i} \rightarrow \Sigma_{d_i}$ in \eqref{eq:sym-projections}. The application of the Grothendiek-Riemann-Roch theorem thus gives
\begin{align}
	\ch(\mathcal{E}^\bullet_j) \td(\Sigma_{d_i}) 
	&= \pi_* \left[ \ch\left(\mathcal{L}^{Q_j} \otimes \mathcal{K}^{\sfrac{r_j}{2}} \right)  \td\left(\Sigma \times \Sigma_{d_i} \right) \right] \nonumber \\
	&= \pi_* \left[ \ch\left( \mathcal{L}^{Q_j} \right) \ch\left( \mathcal{K}^{\sfrac{r_j}{2}} \right) \td(\Sigma) \right]  \td(\Sigma_{d_i}) \,,
\end{align}
where $td(\Sigma_{d_i})$ cancels on both sides, and
\begin{equation}
	td(\Sigma) = 1 + \frac{1}{2} \mathrm{c}_1(\Sigma) = 1 - (g-1)\eta_{\scriptscriptstyle \Sigma \,.}
\end{equation}
Therefore the Chern character $\ch(\mathcal{E}^\bullet_j)$ is computed as follows
\begin{align}
	\ch\left(\mathcal{E}^\bullet_j\right) 
	&= \pi_* \left[ \ch\left( \mathcal{L}^{Q_j} \right) \ch\left( \mathcal{K}^{\sfrac{r_j}{2}} \right) (1-(g-1)\eta_{\scriptscriptstyle \Sigma} \right] \nonumber \\
	&= \pi_*\left[e^{Q_j \eta}\left( 
	1+(g-1)(r_j-1)\eta_{\scriptscriptstyle \Sigma} - Q_j \mathfrak{m}\eta_{\scriptscriptstyle \Sigma} + Q_j \gamma - Q_j^2 \eta_{\scriptscriptstyle \Sigma} \theta
	\right) \right] \nonumber \\
	&= e^{Q_j \eta} \left( Q_j \mathfrak{m}+ (g-1)(r_j-1) - Q_j^2 \theta \right) \nonumber \\
	&= e^{Q_j \eta}\left((d_j -g +1) - Q_j^2 \theta\right) \nonumber \\
	&= e^{Q_j \eta}\bigg[(d_j -2g +1) + \sum_{a=1}^g e^{-Q_j^2\theta_a} \theta\bigg]
	\,.
\label{eq:U(1)-vortex-ch-E}
\end{align}
The push-forward $\pi_*$ effectively acts as an integration, picking out the coefficients of $\eta_{\scriptscriptstyle \Sigma}$.

Combining the result of $\ch(\mathcal{E}^\bullet_j)$ in~\eqref{eq:U(1)-vortex-ch-E} and $$\dim(\mathcal{E}^\bullet_j) = d_j -g +1$$ from~\eqref{eq:U(1)-vortex-dim-E},
the curvature of the complex $\mathcal{E}^\bullet_i$ can therefore be effectively written as
\begin{equation}
	\mathcal{A}_{\mathcal{E}^\bullet_j} = \diag\big( \underbrace{Q_j \eta,\ldots,Q_j \eta}_{\mathclap{d_j - 2g + 1}},Q_j\eta-Q_j^2\theta_1,\ldots,Q_j\eta-Q_j^2\theta_g \big) \,.
\label{eq:U(1)-vortex-chiral-roots}
\end{equation} 
It is easy to check that this expression is consistent with the above computations.

We need to take into account the equivariancy as well. On vortex saddle points parametrised by the moduli space $\mathcal{M}_{\mathfrak{m},i} = \Sigma_{d_i}$, the real vectormultiplet scalar is fixed to $\sigma = - m_i$, where the \index{real mass} real mass of the $\Phi_j$ fluctuations is
\begin{equation}
	m_j - Q_j m_i \,.
\end{equation}
This promotes to a $T_\text{f}$-equivariant Chern character
\begin{equation}
	\ch(\mathcal{E}^\bullet_j) = z_j e^{Q_j \eta}\left((d_j -g +1) - Q_j^2 \theta\right) \, ,
\end{equation}
where we have identified 
\begin{equation}
	y_l = e^{-m_l}
\end{equation}
up to a scaling for all $l=1,\ldots,N$, and defined
\begin{equation}
z_j := y_j / y_i^{Q_j} \,.
\end{equation}
Effectively all the Chern roots~\eqref{eq:U(1)-vortex-chiral-roots} gain an addition term $(m_j - Q_j m_i)$ from the equivariancy, becoming
\begin{align}
	\big(
	&\underbrace{Q_j \eta + m_j - Q_j m_i, \ldots, Q_j \eta + m_j - Q_j m_i}_{\mathclap{d_j - 2g + 1}}, 
	\nonumber \\
	& \hspace{10ex} Q_j\eta-Q_j^2\theta_1 + m_j - Q_j m_i,\ldots,Q_j\eta-Q_j^2\theta_g + m_j - Q_j m_i \big) \,.
\label{eq:U(1)-vortex-chiral-equiv-roots}
\end{align}

These equivariant Chern roots have a similar structure to those in \eqref{eq:chern-roots-of-sym}, so the contribution $\frac{\hat{A}(\mathcal{E}^\bullet_j)}{\mathrm{e}(\mathcal{E}^\bullet_j)}$ to the twisted index from an individual chiral multiplet $\Phi_j$ can be evaluated in a similar way via
\begin{align}
	\frac{\hat{A}(\mathcal{E}^\bullet_j)}{\mathrm{e}(\mathcal{E}^\bullet_j)}
	&= \prod_{n=1}^{d_j - g + 1}  \frac{1}{2 \sinh(\sfrac{\alpha_n}{2})}
	= \prod_{n=1}^{d_j - g + 1} \frac{e^{-\sfrac{\alpha_n}{2}}}{1-e^{-\alpha_n}} \,.
\end{align}
The numerator and the denominator can be evaluated separately. The numerator exponentiates the first Chern class as
\begin{align}
	e^{-\frac{1}{2} \mathrm{c}_1 \left( \mathcal{E}^\bullet_j \right) ) }
	&= \prod_{n=1}^{d_j - g + 1} e^{-\sfrac{\alpha_n}{2}}
	\nonumber \\
	&= \exp(\frac{1}{2} (d_j - g + 1)(-Q_j \eta - m_j + Q_j m_i)  + \frac{1}{2} Q_j^2 \theta)  
	\nonumber \\	
	&=\left( e^{-Q_j \eta} z_j \right)^{\frac{1}{2} (d_i - g + 1)}  \exp(\frac{1}{2} Q_j^2 \theta)	\,.
\end{align}
The denominator reads
\begin{align}
	\prod_{n=1}^{d_j - g + 1} \frac{1}{1-e^{-\alpha_n}}
	&= 
	\left( \prod_{n=1}^{d_j - 2g + 1} \frac{1}{1 - e^{-Q_j \eta} z_j} \right)
	\prod_{a=1}^{g} \frac{1}{1 - e^{-Q_j \eta} z_j \exp(Q_j^2 \theta_a) }  
	\nonumber \\
	&=
	\left( \prod_{n=1}^{d_j - 2g + 1} \frac{1}{1 - e^{-Q_j \eta} z_j} \right)
	\prod_{a=1}^{g} \frac{1}{1 - e^{-Q_j \eta} z_j \left( 1 + Q_j^2 \theta_a \right) }  
	\nonumber \\
	&=
	\left( \prod_{n=1}^{d_j - g + 1} \frac{1}{1 - e^{-Q_j \eta} z_j} \right)
	\prod_{a=1}^{g}  \frac{1}{1 - \frac{e^{-Q_j \eta} z_j}{1-e^{-Q_j \eta} z_j}Q_j^2 \theta_a}  
	\nonumber \\
	&=
	\left( \frac{1}{1 - e^{-Q_j \eta} z_j} \right)^{d_j - g + 1}
	\prod_{a=1}^{g}  \left( 1 + \frac{e^{-Q_j \eta} z_j}{1-e^{-Q_j \eta} z_j}Q_j^2 \theta_a \right)  
	\nonumber \\
	&=
	\left( \frac{1}{1 - e^{-Q_j \eta} z_j} \right)^{d_j - g + 1}
	\exp(\frac{e^{-Q_j \eta} z_j}{1-e^{-Q_j \eta} z_j}Q_j^2 \theta) \,,
\end{align}
where $\theta_a^2 = 0$ has been used repeatedly.
Finally multiplying them together gives the full expression
\begin{align}
	\frac{\hat{A}(\mathcal{E}^\bullet_j)}{\mathrm{e}(\mathcal{E}^\bullet_j)}
	&= \left( \frac{ e^{-Q_j \eta / 2} z_j^{\sfrac{1}{2}} }{1 - e^{-Q_j \eta} z_j} \right)^{d_j - g + 1}
	\exp[Q_j^2 \theta \left( \frac{1}{2} + \frac{e^{-Q_j \eta} z_j}{1-e^{-Q_j \eta} z_j} \right)]
\end{align}
for the contribution from a single chiral multiplet $\Phi_j$.

In total, the fluctuations from all the massive\index{chiral multiplet} chiral multiplets $\{\Phi_j\}_{j \neq i}$ are encoded in the product
\begin{equation}
	\mathcal{E} = \bigotimes_{j\neq i} \mathcal{E}^\bullet_j \, .
\end{equation}
Therefore the final result for their contributions to the index is
\begin{align}
	\frac{\hat{A}(\mathcal{E})}{\mathrm{e}(\mathcal{E} )}  
=  \prod_{j\neq i} \, \left(\frac{(e^{-Q_j\eta}z_j)^{\sfrac{1}{2}}}{1-e^{-Q_j\eta}z_j}\right)^{d_j-g+1} \exp[Q_j^2 \theta \left(\frac{1}{2} + \frac{e^{-Q_j\eta}z_j}{1-e^{-Q_j\eta}z_j}\right)] \,.
\label{eq:U(1)-vortex-A-roof-over-e}
\end{align}

\subsection{Chern-Simons Term}

The supersymmetric Chern-Simons terms generate holomorphic line bundles on the moduli space $\mathfrak{M}_{\mathfrak{m},i} \cong \Sigma_{d_i}$ according to the general mechanism in~\cite{collie2008}. A careful translation into the algebraic framework of this paper leads to the conclusion that the \index{Chern-Simons level} Chern-Simons levels $\kappa$, $\kappa_{a'}$, $\kappa_R$, and $\kappa_{Ra'}$ generate holomorphic line bundles 
\begin{equation}
	\bigotimes_{\alpha} \mathcal{L}_
\alpha 
	=\left( \mathcal{L}^\kappa \otimes \mathcal{L}_{R}^{\kappa_R} \right)
	\bigotimes_{a'=2}^N 
	\left( \mathcal{L}_{a'}^{\kappa_{a'}} \otimes \mathcal{L}_{Ra'}^{\kappa_{Ra'}} \right)
	\,.
\end{equation}
Here a full construction of these bundles are not required since we only need the their first Chern classes for our computations. We come back to their construction later in Chapter~\ref{sec:CS-from-det}.

The pure gauge and R-symmetry bundles $\left( \mathcal{L}^\kappa \otimes \mathcal{L}_{R}^{\kappa_R} \right)$ have
\begin{align}
	\mathrm{c}_1(\mathcal{L}) & = \theta-\mathfrak{m} \eta \, , \\
	\mathrm{c}_1(\mathcal{L}_R) & = -(g-1)\eta \, .
\end{align}
The contribution to the integrand of equation~\eqref{eq:vortex-contribution-integral} is therefore
\begin{equation}
	\ch(\mathcal{L}^\kappa \otimes \mathcal{L}_R^{\kappa_R}) = e^{\kappa (\theta - \mathfrak{m}\eta)} e^{-\kappa_R(g-1)\eta} \, .
\label{eq:U(1)-vortex-CS}
\end{equation}
This result passes a consistency check. It is compatible with the contribution~\eqref{eq:U(1)-vortex-A-roof-over-e} from massive fluctuations of chiral multiplets and the fact that integrating out a massive chiral multiplet of charge $Q_j$ and \index{R-charge} R-charge $r_j$ with \index{real mass} real mass $m_j \to \pm\infty$ shifts
\begin{align}
	\kappa & \to \kappa \pm \frac{Q_j^2}{2} \, ,\\
	\kappa_{R} & \to \kappa_{R} \pm \frac{Q_j}{2}(r_j-1) \, .
\label{eq:shifts}
\end{align}

The mixed symmetry bundles $\left( \mathcal{L}_{a'}^\kappa \otimes \mathcal{L}_{Ra'}^{\kappa_{Ra'}} \right)$ have Chern characters 
\begin{subequations}
\begin{align}
	\ch(\mathcal{L}_{a'}^{\kappa_{a'}}) &= y_{a'}^{\kappa_{a'}\mathfrak{m}} \,, \\
	\ch(\mathcal{L}_{Ra'}^{\kappa_{Ra'}}) & = y_{a'}^{\kappa_{Ra'}(g-1)} \, .
\end{align}
\end{subequations}

Here we have mostly relied on consistency arguments. In Chapter~\ref{sec:CS-from-det}, we show that these line bundles can be precisely built from the determinant line bundles of auxiliary \index{chiral multiplet} chiral multiplets.

\section{Evaluation of Supersymmetric Index}

Collecting the contribution~\eqref{eq:U(1)-vortex-A-roof-sym} from the tangent directions to the moduli space, the fluctuations~\eqref{eq:U(1)-vortex-A-roof-over-e} of massive chiral multiplets, and the supersymmetric Chern-Simons terms~\eqref{eq:U(1)-vortex-CS}, the contribution~\eqref{eq:vortex-contribution-integral} to the twisted index from the component $\mathcal{M}_{\mathfrak{m},i}$ of the vortex moduli space is 
\begin{align}
	\mathcal{I}_{\mathfrak{m},i} 
	=\int_{\Sigma_{d_i}} \,&  e^{k (\theta - \mathfrak{m}\eta)} e^{-k_R(g-1)\eta}   
	\left(\frac{\eta e^{\sfrac{-\eta}{2}}}{1-e^{-\eta}}\right)^{d_i-g+1}    
	\exp[\theta\left(\frac{1}{2}-\frac{1}{\eta}+\frac{e^{-\eta}}{1 - e^{-\eta}}\right)]    
	\quad \times 
	\nonumber \\
	& \hspace{3ex} \prod_{j\neq i}^N \left(\frac{(e^{-Q_j\eta}z_j)^{\sfrac{1}{2}}}{1-e^{-Q_j\eta}z_j}\right)^{d_j-g+1}   
	\exp[ Q_j^2 \theta \left(\frac{1}{2} + \frac{e^{-Q_j\eta}z_j}{1-e^{-Q_j\eta}z_j}\right)]    
\label{eq:vortex-integral-before-conversion}
\end{align}
if $\tau' > \mathfrak{m}$, and vanishes otherwise.

The final step is to convert the integration over the symmetric product into a contour integral using the Don Zagier formula~\cite{thaddeus1992},
\begin{equation}
\label{eq:don-zagier}
	\int_{\Sigma_d} A(\eta) e^{\theta B(\eta)} = \oint_{u=0} \frac{\dd u}{u} \, \frac{A(u) \, [1+u \, B(u)]^g}{u^d} \,.
\end{equation}
The integral in equation~\eqref{eq:vortex-integral-before-conversion} has precisely this form with
\begin{align}
	A(\eta) &= e^{-k\mathfrak{m}\eta} e^{-k_R(g-1)\eta} \left(\frac{\eta e^{-\eta/2}}{1-e^{-\eta}}\right)^{d_i-g+1} \prod_{j \neq i}   \left[\frac{(e^{-Q_j\eta}z_j)^{\sfrac{1}{2}}}{1-e^{-Q_j\eta}z_j}\right]^{d_j-g+1}  \,,\\
	B(\eta) &= k+ \left(\frac{1}{2}-\frac{1}{\eta}+\frac{e^{-\eta}}{1 - e^{-\eta}}\right) + \sum_{j\neq i}  Q_j^2\left(\frac{1}{2} + \frac{e^{-Q_j\eta}z_j}{1-e^{-Q_j\eta}z_j}\right)  \, , 
\end{align}
and therefore we find
\begin{align}
	\mathcal{I}_{\mathfrak{m},i} & =\oint_{u=0} \dd u \,  
	e^{-k\mathfrak{m}u} e^{-k_R(g-1)u}  
	\left(\frac{e^{\sfrac{-u}{2}}}{1-e^{-u}}\right)^{d_i-g+1}   
	\prod_{j\neq i}^N  \left[\frac{(e^{-Q_ju}z_j)^{\sfrac{1}{2}}}{1-e^{-Q_ju}z_j}\right]^{d_j-g+1}   
	\quad \times 
	\nonumber \\
	& \hspace{22ex} \left[k+ \frac{1}{2}\left(\frac{1+e^{-u}}{1 - e^{-u}}\right) + \sum_{j\neq i}^N  \frac{Q_j^2}{2}\left(\frac{1 + e^{-Q_ju}z_j}{1-e^{-Q_ju}z_j}\right) \right]^g \nonumber \\
	& =\oint_{x=y_i^{-1}} \frac{\dd x}{x} \, 
	x^{k\mathfrak{m}+k_R(g-1)} 
	\left(\frac{(x y_i)^{\sfrac{1}{2}}}{1-x y_i}\right)^{d_i-g+1}   
	\prod_{j\neq i}^N    \left[\frac{(x^{Q_j}y_j)^{\sfrac{1}{2}}}{1-x^{Q_j}y_j}\right]^{d_j-g+1}    
	\quad \times 
	\nonumber \\
	&  \hspace{22ex} \left[k+\frac{1}{2}\left(\frac{1+x y_i}{1 - x y_i}\right) + \sum_{j\neq i}^N  \frac{Q_j^2}{2}\left(\frac{1+x^{Q_j}y_j}{1-x^{Q_j}y_j}\right) \right]^g \,, 
\label{eq:vortex-contribution-for-1}
\end{align}
where the substitution $e^{-u} = x y_i$ has been made in the second line. Note that we have assumed $Q_i = +1$ from~\eqref{eq:assumption-Q+1} for our computations. A similar calculation can be performed in the case $Q_i = -1$.

The final result, under the assumption that $|Q_i|=1$ is that the contribution to the twisted index from vortex saddle points parametrised by $\mathcal{M}_{\mathfrak{m},i}$ is
\begin{align}
	\mathcal{I}_{\mathfrak{m},i}=\oint_{x=y_i^{\sfrac{-1}{Q_i}}} 
	& \frac{\dd x}{x} \, 
	x^{k\mathfrak{m}+k_R(g-1)} 
	\quad \times
	\nonumber \\  
	& \hspace{1ex} \prod_{j=1}^N \left[\frac{(x^{Q_i}y_j)^{\sfrac{1}{2}}}{1-x^{Q_j}y_j}\right]^{d_j-g+1}       
	\left[k+\sum_{j=1}^N Q_j^2 \left(\frac{1}{2} + \frac{x^{Q_j}y_j}{1-x^{Q_j}y_j}\right) \right]^g 
\label{eq:vortex-contribution}
\end{align}
if $\sign(\tau'-\mathfrak{m}) = \sign(Q_i)$, and zero otherwise. This  exactly reproduces the contribution to the twisted index from the pole at $x^{Q_i}y_i = 1$ when the auxiliary parameter $\eta$ is chosen such that $\sign(\eta)  = \sign(\tau' - \mathfrak{m})$.


\chapter{Topological Saddles}
\label{sec:topological-vacuum}

This chapter contains the full algebro-geometric construction of the twisted index for \index{saddle point !topological} topological saddle points, which is introduced in Chapter~\ref{chap:twist-on-surface}. The schematics can be found in Section~\ref{sec:geometric-interpretation}.  We first analyse the structure of the moduli space of the topological saddles. Then the individual contributions from the tangent directions, the normal directions, and the Chern-Simons terms are constructed. Finally the total contribution to the twisted index is evaluated, which is in agreement with the contour integral formula.

\section{Moduli Space}

Topological saddle points are configurations where $\phi_j = 0$ for all $j=1,\ldots,N$ and there is a unique finite expectation value for $\sigma_0 : = t^2\sigma$ that solves the equation
\begin{equation}
		\tau' - \mathfrak{m} = - \frac{e^2\vol(\Sigma)}{4\pi^2} \sigma_0 \kappa^\pm  \,
	\end{equation}
in the region $\pm \sigma_0 >0$. Topological saddle points exist provided $\kappa^\pm \neq 0$ and $\sign(\tau'-\mathfrak{m}) = \sign(Q_\pm)$. If we choose the auxiliary parameter such that $\sign(\eta) = \sign(\tau'-\mathfrak{m})$, there are topological saddle points with $\pm \sigma_0 > 0$ whenever the Jeffrey-Kirwan residue prescription selects the poles at $x^{\pm1} \to 0$. The task in this chapter is to reproduce the residues at these poles.

The only massless bosonic fluctuations around a topological saddle are those of the gauge connection $A$. Topological saddle points with flux $\mathfrak{m}\in\mathbb{Z}$ are therefore  parametrised by connections $A$ on a principle $U(1)$ bundle satisfying
\begin{equation}
	* F_A = \frac{2\pi }{\vol(\Sigma)} \mathfrak{m} \,,
\label{eq:top-sol}
\end{equation}
modulo gauge transformations on $\Sigma$. As for vortex saddle points, the contribution to the twisted index is expected to be the \index{supersymmetric index} supersymmetric index of a supersymmetric quantum mechanics whose target is the \index{moduli space !of saddles} moduli space of solutions to these equations. However, gauge transformations act trivially on $F_A$ and $\sigma_0$, so the $U(1)$ gauge symmetry is unbroken and the  quantum mechanics is gauged.

To describe the supersymmetric quantum mechanics concretely, we use the algebraic description of solutions to $\eqref{eq:top-sol}$ as holomorphic line bundle $L$ of degree $\mathrm{c}_1(L) = \mathfrak{m}$. We then expect a supersymmetric sigma model to the Picard variety $\pic{\mathfrak{m}}{\Sigma}$, parametrised by the complex structure $\bar\partial_A$ which transforms as a\index{chiral multiplet} chiral multiplet under $\mathcal{N}=(0,2)$ supersymmetry. 

However, any holomorphic line bundle has a $\mathbb{C}^*$ worth of automorphisms, corresponding to unbroken complexified gauge transformations. It is therefore more appropriate to describe the supersymmetric quantum mechanics as a sigma model to the Picard stack,
\begin{equation}
\mathfrak{M}_{\mathfrak{m}} = \mathfrak{Pic}^{\mathfrak{m}}(\Sigma) \, .
\end{equation}
We can make this more concrete at the cost of introducing an auxiliary base point $p \in \Sigma$. Decomposing complex gauge transformations into those trivial at $p$ and constant gauge transformations, we have
\begin{equation}
	\mathfrak{M}_\mathfrak{m}  = \mathcal{M}_\mathfrak{m} \times [\text{pt} / \mathbb{C}^*] \, ,
\end{equation}
where the moduli space is a Picard variety isomorphic to the complex torus
\begin{equation}
\mathcal{M}_\mathfrak{m} = \pic{\mathfrak{m}}{\Sigma} \cong T^{2g}\, .
\end{equation}
In this way, the supersymmetric quantum mechanics is a hybrid of a non-linear sigma model with target space $T^{2g}$ and a $U(1)$ gauge theory.

The supersymmetric quantum mechanics is not, however, a product due to the massive fluctuations of the chiral multiplets $\Phi_j$. They transform in a perfect complex on $\mathfrak{M}_\mathfrak{m}$ generated by fluctuations annihilated by $\bar\partial_A$. Choosing an auxiliary base point as above, this becomes a $\mathbb{C}^*$-equivariant complex on $\mathcal{M}_\mathfrak{m}$. So the fluctuations roughly transform as sections of a holomorphic vector bundle on the target space $T^{2g}$ of the sigma model and are also charged under the unbroken $U(1)$ gauge symmetry.

\section{Contributions to Index}
\label{sec:top-contributions}

The contributions to the twisted index from topological saddle points can be expressed in the same form as vortex saddle points~\eqref{eq:top-contribution-integral}, given by
\begin{align}
	\int \hat{A} ( \mathfrak{M}_\mathfrak{m} ) 
	\frac{\hat{A}(\mathcal{E}  )}{\mathrm{e}( \mathcal{E} )}   
	\ch\Big( \bigotimes_{\alpha} \mathcal{L}_
\alpha \Big) \, ,
 \label{eq:top-contribution-integral}
\end{align}
where $\mathcal{E}$ is a perfect complex arising from fluctuations of the massive\index{chiral multiplet} chiral multiplets $\{\Phi_j\}_{j=1}^N$, and $\mathcal{L}_
\alpha$ are holomorphic line bundles arising from the mixed \index{Chern-Simons term} Chern-Simons terms. We have again omitted the topological label $\mathfrak{m}$ on the bundles $\mathcal{E}$ and $\mathcal{L}_\alpha$ for brevity. Note that however they are not labelled by a chiral index $i$, unlike the vortex case.

To make this more precise, we choose an auxiliary base point on $\Sigma$ and decompose the moduli stack $\mathfrak{M}_\mathfrak{m} = \mathcal{M}_\mathfrak{m}\times [\text{pt} / \mathbb{C}^*]$. The characteristic classes in equation~\eqref{eq:top-contribution-integral} are then to be understood as $\mathbb{C}^*$-equivariant classes on $\mathcal{M}_\mathfrak{m}$. The integral over the moduli stack decomposes into two parts:
\vspace{-\parskip}
\begin{itemize}
	\item A regular integral over the moduli space $\mathcal{M}_\mathfrak{m} = \pic{\mathfrak{m}}{\Sigma}$. This is the usual contribution from an $\mathcal{N}=(0,2)$ supersymmetric non-linear sigma model.
	\item A contour integral 
	$$
	\frac{1}{2\pi i} \oint_{\mathcal{C}} \frac{\text{d}x}{x} \, ,
	$$
	where $x$ is the Chern character of the trivial $\mathbb{C}^*$-equivariant holomorphic vector bundle with weight $+1$. This is the contribution due to the unbroken $U(1)$ gauge symmetry.
\end{itemize}
The purpose of the contour integral over ${\mathcal{C}}$ is of course to project onto gauge invariant contributions. This is not meaningful as it stands because the integrals of $\mathbb{C}^*$-equivariant classes in equation~\eqref{eq:top-contribution-integral} over the moduli space $\mathcal{M}_\mathfrak{m}$ produce rational functions of $x$. It is therefore necessary to specify whether the integrand should be expanded inside or outside the unit circle, which correspond to the residues at $x = 0$ or $x = \infty$ respectively. 

Our prescription is guided by physical intuition. First, note that the path integral construction identifies $x = e^{-2\pi \beta (\sigma+iA)}$ where $\sigma$ is the real vectormultiplet scalar and $A$ is a constant gauge connection around the circle. Topological saddle points with $\sigma_0>0$ are therefore associated with the region $x \to 0$, while those with $\sigma_0 <0$ are associated with $x\to \infty$. The natural expectation for the contour $\mathcal{C}$ is therefore 
\begin{subequations}
\label{eq:top-res}
\begin{align}
& \sigma_0 > 0: \quad \frac{1}{2\pi i}\int_{x = 0} \frac{\text{d}x}{x} \,, \\
& \sigma_0 < 0: \quad \frac{1}{2\pi i}\int_{x=\infty}
\frac{\text{d}x}{x} \, .
\end{align}
\end{subequations}

This gains further support from the hamiltonian interpretation of the twisted index as counting supersymmetric ground states. The supersymmetric ground states depend on the sign of the \index{real mass} real mass of fluctuations, which is dominated by $\sigma$ as $|\sigma| \to \infty$. For example, the ground state wavefunctions of a one-dimensional chiral multiplet of charge $+1$ are
\begin{subequations}
\begin{alignat}{2}
	& \sigma > 0 : \quad \phi^n e^{-\sigma |\phi|^2}\, , &&\quad n \geq 0 \,, \\ 
	& \sigma < 0 : \quad \widebar\phi^n e^{-\sigma |\phi|^2} \widebar\psi \, ,&&\quad n \geq 0
\end{alignat}
\end{subequations}
with contributions to the index
\begin{subequations}
\begin{alignat}{2}
 	\sigma > 0 & : \quad 1 + x + x^2 +\cdots  &&= \frac{1}{1-x} \,, \\
 	\sigma < 0 & : \quad -x^{-1}-x^{-2} + \cdots &&= \frac{1}{1-x} \, .
\end{alignat}
\end{subequations}
So projecting onto uncharged states at the level of the index is equivalent to
\begin{subequations}
\begin{alignat}{2}
	\sigma>0 & : \quad \frac{1}{2\pi i}\int_{x = 0} \frac{\text{d}x}{x} \frac{1}{1-x}  &&= 1 \,, \\
	\sigma< 0 & : \quad  \frac{1}{2\pi i}\int_{x=\infty} \frac{\text{d}x}{x} \frac{1}{1-x} &&= 0 \,,
\label{eq:res-example}
\end{alignat}
\end{subequations}
which select the coefficient of $x^0$ in the expansions around $x = 0$ and $x =\infty$ respectively. The general prescription~\eqref{eq:top-res} is basically a broad generalisation of this observation.

In summary, we have two contributions from potential topological vacua with $\sigma_0 >0$ and $\sigma_0<0$ are given by the following integrals
\begin{alignat}{2}
	I_0  &= \frac{1}{2\pi i}
	\int_{x = 0} \frac{\text{d}x}{x}  \,
	&&\int \hat{A} ( \mathcal{M}_\mathfrak{m} ) 
	\frac{\hat{A}(\mathcal{E}  )}{\mathrm{e}( \mathcal{E} )}   
	\ch(\mathcal{L}^k\otimes \mathcal{L}^{k_R}_R)  
	\, ,\\
	I_\infty & = \frac{1}{2\pi i}
	\int_{x = \infty} \frac{\text{d}x}{x} \, 
	&&\int \hat{A} ( \mathcal{M}_\mathfrak{m} ) 
	\frac{\hat{A}(\mathcal{E}  )}{\mathrm{e}( \mathcal{E} )}   
	\ch(\mathcal{L}^k\otimes \mathcal{L}^{k_R}_R)  \, ,
\end{alignat}
where we interpret $\mathcal{E}$, $\mathcal{L}$, and $\mathcal{L}_R$ as $\mathbb{C}^*$-equivariant objects on the moduli space $\mathcal{M}_\mathfrak{m} \cong T^{2g}$. In the next section we evaluate these explicitly and show that they reproduce the appropriate contributions to the twisted index according to the contour prescription~\eqref{eq:U(1)-JK-prescription}.

\subsection{Tangent Direction}

Let us first summarise some notation for intersection theory on the Picard variety $\mathcal{M}_\mathfrak{m} \cong T^{2g}$. The \index{Dolbeault cohomology} Dolbeault cohomology ring is generated by classes 
\begin{subequations}
\begin{align}
 	\zeta_a \quad
 	&\in H^{1,0}(T^{2g},\mathbb{Z}) \qquad \forall  a \in \{1,\ldots,g\}\,, \\
 	\bar{\zeta}_a \quad
 	&\in H^{0,1}(T^{2g},\mathbb{Z}) \qquad \forall  a \in \{1,\ldots,g\}\,.
\end{align}
\end{subequations}
We define 
\begin{equation}
	\theta_a := \zeta_a \wedge \bar\zeta_a \quad \in H^{1,1}(T^{2g},\mathbb{Z})\,,
\end{equation}
and 
\begin{equation}
	\theta := \sum_{a=1}^g \theta_a
\end{equation}
with normalisation
\begin{equation}
\int_{T^{2g}}\frac{\theta^g}{g!} = 1 \, .
\end{equation}

The tangent bundle is flat, and therefore the contribution from the tangent directions is simply
\begin{equation}
	\hat A(\mathcal{M}_\mathfrak{m})=1 \,.
\end{equation}

\subsection{Index Bundle}
\label{sec:topological-index-bundle}

We now consider the \index{index bundle} index bundle $\mathcal{E}_{j}$ encoding the massive fluctuations of each of the  \index{chiral multiplet} chiral multiplets $\Phi_{j}$ around configurations in $\mathcal{M}_{\mathfrak{m}}$. These are coming from the normal directions of the flavour $T_\text{f}$ action on the full moduli space $\widebar{\mathfrak{M}}_\mathfrak{m}$. 

At a point on the moduli space corresponding to a holomorphic line bundle $L$, each chiral multiplet generates chiral and Fermi multiplet fluctuations solving 
\begin{align}
\bar\partial_A \phi_j = 0  \, , \quad
\bar\partial_A \eta_j = 0 \, ,
\end{align}
where $\phi_j$ and $\eta_j$ transform as zero-form and one-form sections of $L^{Q_j} \otimes K_\Sigma^{\sfrac{r_j}{2}}$ respectively. The fluctuations of $\Phi_j$ around this point therefore generate the following vector spaces.
\vspace{-\parskip}
\begin{itemize}
	\item Chiral multiplet fluctuations are valued in
	\begin{equation}
		E_j^0 := H^0\left(\Sigma,  L^{Q_j} \otimes K_\Sigma^{\sfrac{r_j}{2}}\right) \,.
	\end{equation}		
	\item Fermi multiplet fluctuations are valued in 
	\begin{equation}
		E_j^1 : = H^1\left(\Sigma,L^{Q_j} \otimes K_\Sigma^{\sfrac{r_j}{2}}\right) \,.
	\end{equation}
\end{itemize}
The line bundle $L$ is the \index{gauge bundle} gauge bundle with the gauge charge $Q_j$ inherited from the three-dimensional chiral multiplet $\Phi_j$. The canonical bundle $K_\Sigma$ comes from the topological twist mixing in the R-symmetry.
As $L$ varies in $\pic{\mathfrak{m}}{\Sigma}$ the dimensions of these vector spaces may jump. But by the \index{Riemann-Roch theorem} Riemann-Roch theorem the difference is a constant equal to
\begin{align}
\label{eq:U(1)-top-dim-E}
	\dim E_j^0 - \dim E_j^1 
	& = h^0 \left( L^{Q_j}\otimes K_{\Sigma}^{\sfrac{r_j}{2}} \right) - h^1 \left( L^{Q_j}\otimes K_{\Sigma}^{\sfrac{r_j}{2}} \right)  \nonumber \\
	& \equiv d_j -g +1 \nonumber \\
	& = Q_j \mathfrak{m}+ (g-1)(r_j-1) \, , 
\end{align}
where $d_j$ is the degree of the line bundle
\begin{align}
	d_j 
	&= \deg\left(L^{Q_j}\otimes K_{\Sigma}^{\sfrac{r_j}{2}}\right) \nonumber \\
	&= Q_j \deg(L)  + \frac{r_j}{2} \deg(K_{\Sigma}) \nonumber \\
	&= Q_j \mathfrak{m}  + \frac{r_j}{2} (2g-2) \,.
\end{align}
This means the difference behave like the fibre of a holomorphic vector bundle on $\text{Pic}^{\mathfrak{m}}(\Sigma)$ for the purpose of K-theoretic computations involved in the twisted index. 

More precisely we need to describe the massive fluctuations as a complex of sheaves. It is again useful to consider the \index{universal construction} universal construction. This is canonical for the moduli stack $\mathfrak{M}_\mathfrak{m}$. But for concreteness we pick a base point $b \in \Sigma$ and pass to the moduli space $\mathcal{M}_\mathfrak{m} = \pic{\mathfrak{m}}{\Sigma}$. There is a corresponding diagram
\begin{equation}
\begin{tikzcd}[column sep=small]
	& { \pic{\mathfrak{m}}{\Sigma} \times \Sigma  } \arrow[dl,"\pi", swap]\arrow[dr,"p"] & \\ {\pic{\mathfrak{m}}{\Sigma}} && {\Sigma}
\end{tikzcd} 
\label{eq:pic-projections}
\end{equation}
A universal line bundle $\mathcal{L}$ is defined such that on restriction to a point $q \in \pic{\mathfrak{m}}{\Sigma}$ corresponding to a holomorphic line bundle $L$ on $\Sigma$, i.e., $\mathcal{L}\eval_q \simeq L$. The universal line bundle is not unique due to the $\mathbb{C}^*$ automorphisms. There is the possibility to transform $\mathcal{L} \to \mathcal{L} \otimes \pi^*\mathcal{N}$. However, this can be fixed by demanding $\mathcal{L}$ is trivial on restriction to the base point $b \in \Sigma$. Note that there is a unique universal line bundle on the moduli stack $\mathfrak{M}_\mathfrak{m} \times \Sigma$ without such a choice. We also define the pull-back of the canonical bundle $\mathcal{K} = p^*K_\Sigma$.

The massive fluctuations of the\index{chiral multiplet} chiral multiplet $\Phi_i$ generate a perfect complex $\mathcal{E}^\bullet_i$ of sheaves defined by the derived push-forward 
\begin{equation}
	 \mathcal{E}^\bullet_j   := R^\bullet \pi_*\left(\mathcal{L}^{Q_j} \otimes \mathcal{K}^{\sfrac{r_j}{2}}\right) \,.
\label{eq:def-V_i-topological}
\end{equation}
The stalks of $\mathcal{E}^\bullet_j$ at $L \in \pic{\mathfrak{m}}{\Sigma}$ are the vector spaces $E_j^\bullet$ considered above. The class $ \ch\left(\mathcal{E}^\bullet_j\right)= \ch(\mathcal{E}^0_j) - \ch(\mathcal{E}^1_j)$ makes sense in equivariant K-theory and the complex behaves like a vector bundle of rank $d_j-g+1$ for the purpose of such computations.

To compute the contribution to the twisted index, we begin by computing the Chern character of $\mathcal{E}^\bullet_j$. This is a small modification of a standard argument presented in~\cite{arbarello1985}. In what follows, we again abuse notation and identify the class $\theta$ with its pull-back via $\pi$. Similarly $\eta_{\scriptscriptstyle \Sigma}$ denotes the class of a point on $\Sigma$ and its pull-back via $p$. The following computations are similar to those in Section~\ref{sec:vortex-index-bundle}.

First, the Chern class of the universal line bundle is
\begin{equation}
\mathrm{c}_1(\mathcal{L}) = \mathfrak{m} \eta_{\scriptscriptstyle \Sigma} + \gamma \, ,
\end{equation}
where the class $\gamma$ satisfies 
\begin{align*}
	\gamma^2 &= -2 \eta_{\scriptscriptstyle \Sigma} \theta \,,\\
	\gamma^3 &= \eta_{\scriptscriptstyle \Sigma} \gamma = 0 \,.
\end{align*}
The class $\gamma$ does not play a role in what follows.

We therefore find
\begin{align}
	\ch\left(\mathcal{L}^{Q_i}\right)  &= e^{Q_i \mathrm{c}_1(\mathcal{L})} \nonumber\\
	&= 1 + Q_i \mathfrak{m}\eta_{\scriptscriptstyle \Sigma} + Q_i \gamma + \frac{Q_i^2}{2} \gamma^2 \nonumber \\
	&= 1 + Q_i \mathfrak{m}\eta_{\scriptscriptstyle \Sigma} + Q_i \gamma - Q_i^2 \eta_{\scriptscriptstyle \Sigma} \theta \,.
\label{eq:ch-L-Q}
\end{align}
Similarly, using
\begin{equation*}
	\mathrm{c}_1(\mathcal{K}) = (2g-2) \eta_{\scriptscriptstyle \Sigma}
\end{equation*}
from~\eqref{eq:c1-canonical-bundle}, we find
\begin{align}
	\ch\left(\mathcal{K}^{\sfrac{r_j}{2}}\right) 
	&= e^{\frac{r_j}{2} \mathrm{c}_1(\mathcal{K})} \nonumber \\
	&= e^{r_j (g-1) \eta_{\scriptscriptstyle \Sigma}} \nonumber \\
	&= 1 + r_j(g-1) \eta_{\scriptscriptstyle \Sigma} \,.
\label{eq:ch-kappa-r-over-2-vortex}
\end{align}
Combining these results
\begin{equation}
	\ch\left(\mathcal{L}^{Q_j} \otimes \mathcal{K}^{\sfrac{r_j}{2}}\right) 
	= 1 + d_j\eta_{\scriptscriptstyle \Sigma} + Q_j \gamma - Q_j^2 \eta_{\scriptscriptstyle \Sigma} \theta \,.
\end{equation}

Using the \index{Grothendiek-Riemann-Roch theorem} Grothendiek-Riemann-Roch theorem~\eqref{eq:G-R-R}, we can now compute the Chern character of $\mathcal{E}^\bullet_j$ as
\begin{align}
	\ch\left(\mathcal{E}^\bullet_j\right) 
	& = \pi_* \left[ \ch(\mathcal{L}^{Q_j} \otimes \mathcal{K}^{\sfrac{r_j}{2}} ) \, \td (\text{Pic}^\mathfrak{m}\Sigma \times \Sigma) \right] \nonumber \\
	&= \pi_*\left[  \,   \ch\left(\mathcal{L}^{Q_i} \otimes \mathcal{K}^{\sfrac{r_i}{2}}\right)(1-(g-1)\eta_{\scriptscriptstyle \Sigma}) \right] \nonumber\\
	&= \pi_*\left[1+(d_j-g+1)\eta_{\scriptscriptstyle \Sigma} + Q_i \gamma - Q_i^2 \eta_{\scriptscriptstyle \Sigma} \theta\right] \nonumber \\
	&= (d_j-g+1) - Q_j^2 \theta \nonumber \\
	& = (d_j-2g+1) + \sum_{a=1}^g e^{-Q_j^2\theta_a} \, ,
\label{eq:U(1)-top-ch-E}
\end{align}
where in the final line we have expressed the result in such a way that the Chern roots are manifest.

Combining the result of $\ch(\mathcal{E}^\bullet_j)$ in~\eqref{eq:U(1)-top-ch-E} and $$\dim(\mathcal{E}^\bullet_j) = d_j -g +1$$ from~\eqref{eq:U(1)-top-dim-E},
the curvature of the complex $\mathcal{E}^\bullet_i$ can therefore be effectively written as
\begin{equation}
	\mathcal{A}_{\mathcal{E}^\bullet_j} = \diag\big( \underbrace{0,\ldots,0}_{\mathclap{d_j - 2g + 1}},-Q_j^2\theta_1,\ldots,-Q_j^2\theta_g \big) \,.
\label{eq:U(1)-top-chiral-roots}
\end{equation}

We need to take into account the $\mathbb{C}^*$-equivariancy. On topological saddle points parametrised by the moduli space $\mathcal{M}_{\mathfrak{m}} = \pic{\mathfrak{m}}{\Sigma}$,  the \index{real mass} real mass of the $\Phi_j$ fluctuations is
\begin{equation}
	m_j + Q_j \sigma \,.
\end{equation}
This promotes to a $\mathbb{C}^*$-equivariant Chern character
\begin{equation}
	\ch\left(\mathcal{E}^\bullet_j\right) = x^{Q_j}y_j \bigg[(d_j-2g+1) + \sum_{a=1}^g e^{-Q_j^2\theta_a} \bigg] \, , 
\end{equation}
where we have identified 
\begin{subequations}
\begin{align}
	x &= e^{-\sigma} \,, \\
	y_j &= e^{-m_j}
\end{align}
\end{subequations}
up to a scaling.
Effectively all the Chern roots~\eqref{eq:U(1)-top-chiral-roots} gain an addition term $(m_j + Q_j \sigma )$ from the equivariancy, becoming
\begin{align}
\label{eq:U(1)-top-equiv-roots}
	\big(
	&\underbrace{m_j + Q_j \sigma , \ldots, m_j + Q_j \sigma }_{\mathclap{d_j - 2g + 1}}, 
	\nonumber \\
	& \hspace{20ex} -Q_j^2\theta_1 + m_j + Q_j \sigma ,\ldots, -Q_j^2\theta_g + m_j + Q_j \sigma  \big) \,.
\end{align}

The contribution to the twisted index is now given by the equivariant $\hat A$-genus of the complex $\mathcal{E}_j^\bullet$. This is straightforward to compute from the equivariant Chern roots~\eqref{eq:U(1)-top-equiv-roots} by a now familiar set of manipulations.

\begin{align}
	\frac{\hat{A}(\mathcal{E}^\bullet_j)}{\mathrm{e}(\mathcal{E}^\bullet_j)}
	&= \prod_{n=1}^{d_j - g + 1}  \frac{1}{2 \sinh(\sfrac{\alpha_n}{2})}
	= \prod_{n=1}^{d_j - g + 1} \frac{e^{-\sfrac{\alpha_n}{2}}}{1-e^{-\alpha_n}} \,.
\end{align}
The numerator and the denominator can be evaluated separately. The numerator exponentiates the first Chern class as
\begin{align}
	e^{-\frac{1}{2} \mathrm{c}_1 \left( \mathcal{E}^\bullet_j \right) ) }
	&= \prod_{n=1}^{d_j - g + 1} e^{-\sfrac{\alpha_n}{2}}
	\nonumber \\
	&= \exp(\frac{1}{2} (d_j - g + 1)( - m_j - Q_j \sigma)  + \frac{1}{2} Q_j^2 \theta)  
	\nonumber \\	
	&=\left( x^{Q_j} y_j \right)^{\frac{1}{2} (d_i - g + 1)}  \exp(\frac{1}{2} Q_j^2 \theta)	\,.
\end{align}
The denominator reads
\begin{align}
	\prod_{n=1}^{d_j - g + 1} \frac{1}{1-e^{-\alpha_n}}
	&= 
	\left( \prod_{n=1}^{d_j - 2g + 1} \frac{1}{1 - x^{Q_j} y_j} \right)
	\prod_{a=1}^{g} \frac{1}{1 - x^{Q_j} y_j \exp(Q_j^2 \theta_a) }  
	\nonumber \\
	&=
	\left( \prod_{n=1}^{d_j - 2g + 1} \frac{1}{1 - x^{Q_j} y_j} \right)
	\prod_{a=1}^{g} \frac{1}{1 - x^{Q_j} y_j \left( 1 + Q_j^2 \theta_a \right) }  
	\nonumber \\
	&=
	\left( \prod_{n=1}^{d_j - g + 1} \frac{1}{1 - x^{Q_j} y_j} \right)
	\prod_{a=1}^{g}  \frac{1}{1 - \frac{x^{Q_j} y_j}{1-x^{Q_j} y_j}Q_j^2 \theta_a}  
	\nonumber \\
	&=
	\left( \frac{1}{1 - x^{Q_j} y_j} \right)^{d_j - g + 1}
	\prod_{a=1}^{g}  \left( 1 + \frac{x^{Q_j} y_j}{1-x^{Q_j} y_j}Q_j^2 \theta_a \right)  
	\nonumber \\
	&=
	\left( \frac{1}{1 - x^{Q_j} y_j} \right)^{d_j - g + 1}
	\exp(\frac{x^{Q_j} y_j}{1-x^{Q_j} y_j}Q_j^2 \theta) \,,
\end{align}
where $\theta_a^2 = 0$ has been used repeatedly.
Finally multiplying them together gives the full expression
\begin{align}
	\frac{\hat{A}(\mathcal{E}^\bullet_j)}{\mathrm{e}(\mathcal{E}^\bullet_j)}
	&= \left( \frac{ x^{\sfrac{-Q_j}{2}} y_j^{\sfrac{1}{2}} }{1 - x^{Q_j} y_j} \right)^{d_j - g + 1}
	\exp[Q_j^2 \theta \left( \frac{1}{2} + \frac{x^{Q_j} y_j}{1-x^{Q_j} y_j} \right)]
\end{align}
for the contribution from a single chiral multiplet $\Phi_j$.

In total, the fluctuations from all the massive\index{chiral multiplet} chiral multiplets $\{\Phi_j\}_{j=1}^N$ are encoded in the product
\begin{equation}
	\mathcal{E} = \bigotimes_{j=1}^N \mathcal{E}^\bullet_j \, .
\end{equation}
Therefore the final result for their contributions to the index is
\begin{align}
	\frac{\hat{A}(\mathcal{E})}{\mathrm{e}(\mathcal{E} )}  =  
	\prod_{j=1}^N 
	\left( \frac{ x^{\sfrac{-Q_j}{2}} y_j^{\sfrac{1}{2}} }{1 - x^{Q_j} y_j} \right)^{d_j - g + 1}
	\exp[Q_j^2 \theta \left( \frac{1}{2} + \frac{x^{Q_j} y_j}{1-x^{Q_j} y_j} \right)] \,.
\label{eq:U(1)-top-A-roof-over-e}
\end{align}

\subsection{Chern-Simons Term}

The supersymmetric Chern-Simons again induce holomorphic line bundles over the moduli space $\mathcal{M}_\mathfrak{m} \cong T^{2g}$. In the algebraic framework the \index{Chern-Simons level} Chern-Simons levels $\kappa$, $\kappa_{a'}$, $\kappa_R$, and $\kappa_{Ra'}$ induce holomorphic line bundles 
\begin{equation}
	\bigotimes_{\alpha} \mathcal{L}_
\alpha 
	=\left( \mathcal{L}^\kappa \otimes \mathcal{L}_{R}^{\kappa_R} \right)
	\bigotimes_{a'=2}^N 
	\left( \mathcal{L}_{a'}^{\kappa_{a'}} \otimes \mathcal{L}_{Ra'}^{\kappa_{Ra'}} \right)
	\,.
\end{equation}
Here a full construction of these bundles are not required since we only need the their first Chern classes for our computations. We come back to their construction later in Chapter~\ref{sec:CS-from-det}.

The pure gauge and R-symmetry bundles $\left( \mathcal{L}^\kappa \otimes \mathcal{L}_{R}^{\kappa_R} \right)$ have
\begin{subequations}
\begin{align}
	\mathrm{c}_1(\mathcal{L}) & = \theta \, ,\\
	\mathrm{c}_1(\mathcal{L}_R) & = 0 \, ,
\end{align}
\end{subequations}
and transform equivariantly with weights $\mathfrak{m}$ and $(g-1)$ respectively. The equivariant Chern characters are therefore
\begin{subequations}
\begin{align}
	\ch(\mathcal{L}^\kappa) & = (x^\mathfrak{m} e^\theta)^\kappa = x^{\kappa \mathfrak{m}}e^{\kappa \theta} \,, \\
	\ch(\mathcal{L}_R^{\kappa_R}) & = x^{\kappa_R(g-1)} \, .
\end{align}
\end{subequations}
This is compatible with the contribution~\eqref{eq:U(1)-top-A-roof-over-e} from fluctuations of $\Phi_j$ and the fact that integrating out a massive\index{chiral multiplet} chiral multiplet of charge $Q_j$ and \index{R-charge} R-charge $R_j$ shifts the supersymmetric \index{Chern-Simons level} Chern-Simons levels as in equation~\eqref{eq:shifts}.

The mixed symmetry bundles $\left( \mathcal{L}_{a'}^\kappa \otimes \mathcal{L}_{Ra'}^{\kappa_{Ra'}} \right)$ have Chern characters 
\begin{subequations}
\begin{align}
	\ch(\mathcal{L}_{a'}^{\kappa_{a'}}) &= y_{a'}^{\kappa_{a'}\mathfrak{m}} \,, \\
	\ch(\mathcal{L}_{Ra'}^{\kappa_{Ra'}}) & = y_{a'}^{\kappa_{Ra'}(g-1)} \, .
\end{align}
\end{subequations}

Here we have mostly relied on consistency arguments. In Chapter~\ref{sec:CS-from-det}, we show that these line bundles can be precisely built from the determinant line bundles of auxiliary\index{chiral multiplet} chiral multiplets.

\section{Evaluation of Supersymmetric Index}

Combining all these contributions, the contribution to the integrand from the integration over the moduli space $\mathcal{M}_\mathfrak{m} = \pic{\mathfrak{m}}{\Sigma} \cong T^{2g}$ is
\begin{align}
	\, & \int_{\pic{\mathfrak{m}}{\Sigma}} x^{k\mathfrak{m}}e^{k\theta} x^{k_R(g-1)} 
	\prod_{j=1}^N \left( \frac{(x^{Q_j}y_j)^{^{\sfrac{1}{2}}}}{1-x^{Q_j}y_j}  \right)^{d_j-g+1}
	\exp \left( \left( \frac{1}{2}+\frac{x^{Q_j}y_j}{1-x^{Q_j}y_j} \right)Q_j^2 \theta   \right) 
	\nonumber \\
	=& x^{k\mathfrak{m}+k_R(g-1)} \prod_{j=1}^N \left( \frac{(x^{Q_j}y_j)^{^{\sfrac{1}{2}}}}{1-x^{Q_j}y_j}  \right)^{d_j-g+1} \int_{\pic{\mathfrak{m}}{\Sigma}}  
	\exp \left(k+\sum_{j=1}^N Q_j^2\left( \frac{1}{2}+\frac{x^{Q_j}y_j}{1-x^{Q_j}y_j} \right)    \right)\theta 
	\nonumber \\
	= & x^{k\mathfrak{m}+k_R(g-1)} \prod_{j=1}^N \left( \frac{(x^{Q_j}y_j)^{^{\sfrac{1}{2}}}}{1-x^{Q_j}y_j}  \right)^{d_j-g+1}  
	\left[k+\sum_{j=1}^N Q_j^2\left( \frac{1}{2}+\frac{x^{Q_j}y_j}{1-x^{Q_j}y_j} \right)    \right]^g \, .
\end{align}
Note that the integral $\int_{\pic{\mathfrak{m}}{\Sigma}}$ effectively picks out the coefficients of the volume form $\frac{\theta^g}{g!}$. The contributions from topological saddle points therefore exactly reproduce the potential residues at $x = 0$ and $x = \infty$ in the Jeffrey-Kirwan residue prescription with $\eta$ aligned with $\tau'-\mathfrak{m}$.

\chapter{Chern-Simons Term from Determinant Line Bundle}
\label{sec:CS-from-det}

It is a general phenomenon for three-dimensional theories that when heavy fermions are integrated out, they induce an effective \index{Chern-Simons term} Chern-Simons term as an low-energy effect~\cite{Redlich:1983kn,Redlich:1983dv,Alvarez-Gaume:1983ihn}. 
\vspace{-\parskip}
\begin{itemize}
	\item In this chapter we first develops the physical intuition of this mechanics from the abelian Higgs model in Section~\ref{sec:abelian-Higgs-model}.
	\item Then we propose an algebro-geometric construction of the Chern-Simons contributions to the twisted index from determinant line bundles of auxiliary chiral multiplets in Section~\ref{sec:det-line-bundle}.
	\item Finally we give the full construction of the Chern-Simons contributions for both vortex saddles and topological saddles in Section~\ref{sec:CS-term-line-bundle-example}, completing the discussion from Chapter~\ref{chap:twist-on-surface}.
\end{itemize}

\section{Abelian Higgs Model}
\label{sec:abelian-Higgs-model}
It can be understood at the level of lagrangian of vortices~\cite{collie2008} in the abelian Higgs model
\begin{equation*}
	\mathcal{L} = \frac{1}{2} g_{ab(X)} \dot{X}^a \dot{X}^b \,.
\end{equation*}
When considering a bosonic lagrangian without Chern-Simons interactions, adding suitable fermion content and integrating out its zero modes produces an effective \index{Chern-Simons term} Chern-Simons term 
\begin{equation}
\label{eq:CS-lagrangian}
	\mathcal{L}^{\eff} = - \kappa \mathcal{A}_a(X) \dot{X}^a \,.
\end{equation}

This has been shown explicitly for a $U(N)$ Yang-Mills Chern-Simons theory coupled to a real adjoint scalar $\sigma$ and scalars $\{q_i\}_{i = 1}^{n_f}$ in the fundamental representation of the gauge group~\cite{collie2008}. The \index{Chern-Simons term} Chern-Simons term in the lagrangian can be reproduced by integrating out\index{chiral multiplet} chiral multiplets $\{\Phi_i\}_{i = 1}^N$ in the anti-fundamental representation of the $U(N)$ gauge group. The resulting effective Chern-Simons lagrangian from a single chiral multiplet $\Phi$ is given by
\begin{equation}
\label{eq:CS-lagrangian-from-single-chiral}
	\mathcal{L}^{\eff} = - \frac{1}{2} \sign(m) \Tr(\omega_a) \dot{X}^a \,,
\end{equation}
in the limit the mass $m \rightarrow \infty$. The trace $\Tr(\omega_a)$ is over the the connection $\omega$ on the index bundle, which is the bundle over the vortex moduli space defined by the space of zero mode solutions of the Dirac equation
\begin{equation*}
	\left(i \slashed{D} - (\sigma + m)\right) \Psi = 0 \,,
\end{equation*}
where $\Psi$ is a Dirac fermion in the\index{chiral multiplet} chiral multiplet $\Phi$.

This formula~\eqref{eq:CS-lagrangian-from-single-chiral} is obtained by integrating out these zero modes. The dynamics of these zero modes is described in terms of the grassmannian coordiantes $\xi^l$ of the fibre of the index bundle by the kinetic term
\begin{equation}
	\bar{\xi}^l (i \mathrm{D}_t - m) \xi^l \,,
\end{equation}
where the covariant derivative is defined by
\begin{equation}
	\mathrm{D}_t \xi^l = \partial_t \xi^l + i (\omega_a)^l_n \dot{X}^a \xi^n \,.
\end{equation}
Integrating out the fermion $\xi$ in the path integral leads to the normalised determinant
\begin{equation}
	\det(\frac{i \mathrm{D}_t - m}{i \partial_t - m}) = \det(\frac{-\partial_{\tau} - i \omega_a \partial_{\tau} X^a - m}{-\partial_{\tau} - m})\,,
\end{equation}
where $\tau = i t$ is the compact euclidean time with periodicity $\tau \in [0,\beta)$. 

The solution $\chi$ of eigenvalue $\lambda$ to the equation
\begin{equation}
	 (-\partial_{\tau} - i \omega_a \partial_{\tau} X^a - m) \chi = \lambda \chi 
\end{equation}
is given by 
\begin{equation}
	\chi = e^{-(m+\lambda)\tau} V(\tau) \,,
\end{equation}
where $V(\tau)$ is the time-ordered product
\begin{equation}
	V(\tau) = \mathrm{T} \exp(-i \int_0^{\tau} \dd \tau' \, \omega_a \partial_{\tau'} X^a) \,.
\end{equation}
Denoting the eigenvalues of $V(\beta)$ as $e^{v_l}$ and imposing the periodicity condition $\chi(0) = \chi(\beta)$ gives
\begin{equation}
	\lambda_l = \frac{v_l + 2\pi i n}{\beta} - m\,, \quad n \in \mathbb{Z}\,.
\end{equation}
Now the determinant is obtained as
\begin{align}
	&\det(\frac{i \mathrm{D}_t - m}{i \partial_t - m}) \nonumber \\
	= &\prod_l \prod_n \frac{ \sfrac{2\pi i n}{\beta} + \sfrac{v_l}{\beta} - m}{2\pi i n / \beta - m} \nonumber \\
	= &\prod_l \left(1 - \frac{v_l}{\beta m}\right) \frac{\sinh( \sfrac{\beta m}{2} - \sfrac{v_l}{2})}{\sinh( \sfrac{\beta m}{2} )} \nonumber \\
	\rightarrow &\exp(-\frac{1}{2} \sign(m) \sum_l v_l)
\end{align}
as $\beta \rightarrow \infty$. This contribution to the path integral corresponds exactly to the effective lagrangian~\eqref{eq:CS-lagrangian-from-single-chiral}. The zero modes have induced an effective magnetic field $\mathcal{F}=\dd \mathcal{A}$, where $\mathcal{A} = \Tr(\omega)$ is the Chern-Simons one-form. If we integrate out $N = 2 \kappa$ chiral multiplets, then the \index{Chern-Simons term} Chern-Simons term~\eqref{eq:CS-lagrangian} is recovered.

In general, we would like to identify \index{Chern-Simons term} Chern-Simons terms as effects of some ``determinant'' of the bundles encoding the\index{chiral multiplet} chiral multiplets.

\section{Determinant Line Bundle}
\label{sec:det-line-bundle}

In quantum K-theory, the level structure is defined by \index{determinant line bundle} determinant line bundles~\cite{ruan2019}. Let $\chi$ be a Deligne-Mumford stack. The determinant line bundle for a locally free, finitely generated $\mathcal{O}_\chi$ module $E$ is defined as
\begin{equation}
\label{eq:det-of-vector-bundle}
	\det(E) := \wedge^{\mathrm{rank}(E)} E \,.
\end{equation} 
Similarly the determinant line bundle at level $k$ is defined to be
\begin{equation}
	D_k(E) := (\det E)^{-k} \,.
\end{equation}

Let $\mathcal{E}^\bullet$ be a complex of coherent sheaves on $\chi$, which has a bounded complex of locally free, finitely generated $\mathcal{O}_\chi$ modules $E^\bullet$ and a quasi-isomorphism
\begin{equation}
	E^\bullet \rightarrow \mathcal{E}^\bullet \,.
\end{equation}
Then the determinant of $\mathcal{E}^\bullet$ is defined as
\begin{equation}
	\det(\mathcal{E}^\bullet) := \bigotimes_n \det(E^n)^{(-1)^n} \,.
\end{equation}

To define the level structure in a quantum K-theory, let $\mathfrak{M}$ be the algebraic stack of pre-stable nodal curves, and $\mathfrak{Bun}_G$ be the relative moduli stack 
$$
	\mathfrak{Bun}_G \xrightarrow{\phi} \mathfrak{M}
	$$ 
of principal $G$-bundles on the fibres of the universal curve $\mathfrak{C} \rightarrow \mathfrak{M}$. Given a finite-dimensional representation $R$ of $G$, the level-$k$ determinant line bundle over the $\epsilon$-stable quasi-map space $\mathcal{Q}$ is defined as 
\begin{equation}
	\mathcal{D}^k := \left( \det( R \pi_* (\mathcal{L} \times_G R)) \right)^{-k} \,,
\end{equation}
where $\mathcal{L} \rightarrow \mathcal{C}$ is the universal principal bundle given by the pull-back of the universal principal $G$-bundle $\tilde{\mathfrak{P}} \rightarrow \mathfrak{C}_{\mathfrak{Bun}}$. Here $\mathfrak{C}_{\mathfrak{Bun}} \xrightarrow{\tilde{\pi}} \mathfrak{Bun}_G $ is the universal curve as the pull-back of $\mathfrak{C}$ along $\phi$, and $\mathcal{C} \xrightarrow{\pi} \mathcal{Q}$ is the universal curve on the quasi-map space.

\section{Chern Simons Term}
\label{sec:CS-term-line-bundle-example}

Consider an additional auxiliary\index{chiral multiplet} chiral multiplet $\Phi_j$. It can be interpreted to generate \index{Chern-Simons term} Chern-Simons terms via its determinant line bundle. The determinant line bundle of $\mathcal{E}_j^\bullet$ is 
\begin{equation}
\label{eq:U(1)-det-bundle-chiral}
	\det(\mathcal{E}_j^\bullet) = \det(\mathcal{E}_j^0 - \mathcal{E}_j^1) = \wedge^{d_j - g + 1} (\mathcal{E}_j^0 - \mathcal{E}_j^1) \,.
\end{equation}
Note that the first Chern class of this line bundle is the same as the one of $\mathcal{E}_j^\bullet$, 
\begin{equation}
	\mathrm{c}_1\left(\det(\mathcal{E}_j^\bullet)\right) = \mathrm{c}_1\left(\wedge^{d_j - g + 1} \mathcal{E}_j^\bullet\right) = \mathrm{c}_1\left(\mathcal{E}_j^\bullet\right) \,.
\end{equation}

\subsection{Vortex Saddles}

For vortex saddles, the first Chern class is 
\begin{equation}
	\mathrm{c}_1\left(\det(\mathcal{E}_j^\bullet)\right) = Q_j \eta (d_j - g + 1) -  Q_j^2 \theta 
\end{equation}
from the Chern roots~\eqref{eq:U(1)-vortex-chiral-roots}.
Since it is a line bundle, the Chern character of the determinant bundle at level $k$ is given by
\begin{equation}
	\ch(D_k(\mathcal{E}_j^\bullet)) 
	= \exp[\mathrm{c}_1\left((\det \mathcal{E}_j^\bullet)^{-k}\right)]
	= e^{- k Q_j \eta (d_j - g + 1)} e^{k Q_j^2 \theta} \,.
\end{equation}
Completing to equivariant forms~\eqref{eq:U(1)-vortex-chiral-equiv-roots} gives
\begin{align}
	\ch(D_k(\mathcal{E}_j^\bullet)) 
	&= 
	e^{k Q_j^2 \theta}
	e^{-k (d_j-g+1) ( Q_j \eta - Q_j m_i + m_j)} \nonumber \\
	&=   
	e^{k Q_j^2 \theta} 
	e^{- k Q_j^2 \mathfrak{m} \eta } y_i^{-k Q_j^2 \mathfrak{m}}
	y_j^{k Q_j \mathfrak{m}}
	\nonumber \quad \times \\
	& \hspace{10ex} e^{- k Q_j (r_j - 1) (g-1) \eta } y_i^{-k Q_j (r_j - 1)(g-1)} 
	y_j^{k (r_j - 1) (g-1)}	
	\,.
\end{align}
With the Don Zagier formula~\eqref{eq:don-zagier} and the appropriate substitution from~\eqref{eq:vortex-contribution-for-1}, the $\eta$ factors can be effectively re-written via $e^{-\eta} \mapsto x y_i$, producing
\begin{align}
	\ch(D_k(\mathcal{E}_j^\bullet)) 
	=   
	e^{k Q_j^2 \theta} 
	x^{ k Q_j^2 \mathfrak{m} } 
	y_j^{k Q_j \mathfrak{m}}
	x^{ k Q_j (r_j - 1) (g-1)  } 
	y_j^{k (r_j - 1) (g-1)}	
	\,.
\label{eq:U(1)-vortex-CS-det}
\end{align}

Note that it already reproduces the relevant factors
$$   
	x^{\kappa \mathfrak{m}}
	\left( \prod_{j=1}^N y_{j}^{\kappa_{j} \mathfrak{m}} \right)
	x^{\kappa_{R} (g-1)}
	\left( \prod_{j=1}^N y_{j}^{\kappa_{Rj} (g-1)} \right)
$$
from the one-loop determinant~\eqref{eq:U(1)-one-loop-det-yi}, if we identify the charges of this auxiliary chiral multiplet $\Phi_j$ with the \index{Chern-Simons term} Chern-Simons terms as follows
\begin{subequations}
\label{eq:U(1)-CS-det-identification}
\begin{align}
	k Q_j^2 &\mapsto \kappa \,,\\
	k Q_j &\mapsto \kappa_{j} \,,\\
	k Q_j (r_j-1) &\mapsto \kappa_{R} \,,\\
	k (r_j-1) &\mapsto \kappa_{Rj} \,.
\end{align}
\end{subequations}
After integration over the moduli space, the $\theta$ factor $e^{k  Q_j^2 \theta}$ contributes to the \index{Chern-Simons level} Chern-Simons level in the hessian~\eqref{eq:U(1)-hessian-yi}.

It is clear that introducing multiple auxiliary\index{chiral multiplet} chiral multiplets of various charges is sufficient to generate arbitrary \index{Chern-Simons level} Chern-Simons levels in the above factors.

\subsection{Topological Saddles}

In case of topological saddles, the first Chern class is a straightforward evaluation
\begin{equation}
	\mathrm{c}_1\left(\det(\mathcal{E}_j^\bullet)\right) = -  Q_j^2 \theta 
\end{equation}
from the Chern roots~\eqref{eq:U(1)-top-chiral-roots}.
Since it is a line bundle, the Chern character of the determinant bundle at level $k$ is given by
\begin{equation}
	\ch(D_k(\mathcal{E}_j^\bullet)) 
	= \exp[\mathrm{c}_1\left((\det \mathcal{E}_j^\bullet)^{-k}\right)]
	= e^{k Q_j^2 \theta} \,.
\end{equation}

Completing to equivariant forms~\eqref{eq:U(1)-top-equiv-roots} gives
\begin{align}
	\ch(D_k(\mathcal{E}_j^\bullet)) 
	&= 
	e^{k  Q_j^2 \theta} 
	e^{-k (d_j-g+1) ( Q_j \sigma + m_j)} \nonumber \\
	&=   
	e^{k Q_j^2 \theta} 
	x^{k Q_j^2 \mathfrak{m} } 
	y_j^{k Q_j \mathfrak{m}}
	x^{k Q_j (r_j - 1) (g-1) }   
	y_j^{k (r_j - 1) (g-1)}	
	\,,
\end{align}
which is the same expression~\eqref{eq:U(1)-vortex-CS-det} for the vortex case.

It already reproduces the relevant factors
$$   
	x^{\kappa \mathfrak{m}}
	\left( \prod_{j=1}^N y_{j}^{\kappa_{j} \mathfrak{m}} \right)
	x^{\kappa_{R} (g-1)}
	\left( \prod_{j=1}^N y_{j}^{\kappa_{Rj} (g-1)} \right)
$$
from the one-loop determinant~\eqref{eq:U(1)-one-loop-det-yi}, if we identify the charges of this auxiliary chiral multiplet $\Phi_j$ with the \index{Chern-Simons term} Chern-Simons terms according to~\eqref{eq:U(1)-CS-det-identification}.
After integration over the moduli space, the $\theta$ factor $e^{k  Q_j^2 \theta}$ contributes to the \index{Chern-Simons level} Chern-Simons level in the hessian~\eqref{eq:U(1)-hessian-yi}.

Thus we conclude that the Chern-Simons contributions can be obtained by integrating out additional \emph{auxiliary}\index{chiral multiplet} chiral multiplets.
\chapter{Higher Rank Abelian Theory}
\label{chap:U1K-gauge-theories}

In this chapter we briefly discuss a generalisation of the $\mathcal{N} = 2$ supersymmetric $U(1)$ gauge theories discussed in Chapter~\ref{chap:flat-base-manifold} to a higher rank abelian group $U(1)^K$. It is expected to inherit most of the structures from the rank one case. 
However, there is a potential mixing between topological and vortex vacua, which leads to more intricate \index{window phenomenon} window phenomenon and geometric interpretation.

Instead of trying to give a complete treatment, we aim to set up the notations and explore heuristically, following the general outline set up in the previous chapters.

Furthermore, these more general theories admit mirror symmetry, which we investigate with some examples in Section~\ref{sec:mirror-examples}.

\section{Semi-Classical Vacua}
Consider $\mathcal{N} = 2$ supersymmetric $U(1)^K$ gauge theories, with $N$\index{chiral multiplet} chiral multiplets $\{\Phi_j\}_{j = 1}^N$ for $ N \geq K$. The $j$-th chiral field $\Phi_j$ is assigned with \index{gauge charge} gauge charge $\{Q^a_{\ \ j}\}_{a=1}^K$ in the $a$-th $U(1)$ group, and \index{R-charge} R-charge $r_j$. Again, we set the complex \index{superpotential} superpotential to vanish, $W = 0$. 

We introduce various mixed \index{Chern-Simons term}Chern-Simons terms at the following \index[sym]{$\kappa$}\index[sym]{$\kappa_{a'}$}\index[sym]{$\kappa_R$}\index[sym]{$\kappa_{Ra'}$}levels: 
\vspace{-\parskip}
\begin{itemize}
	\item $\kappa_{ab}$ denote gauge levels,
	\item $\kappa_{ab'}$ denote gauge-flavour levels,
	\item $\kappa_{Ra}$ denote gauge-R levels,
	\item $\kappa_{Ra'}$ denote flavour-R levels.
\end{itemize}
The chiral multiplet index has been split into $i = (a, a')$, where $a' \in \{K+1,\ldots,N\}$ labels the $(N-K)$ independent \index{fugacity !flavour} flavour fugacities. The gauge and flavour levels can be combined into a symmetric matrix $\kappa_{ij}$ and a vector $\kappa_{Ri}$. Cancellation of parity anomalies requires
\begin{subequations}
\begin{alignat}{3}
	&\kappa_{ij} + \frac{1}{2} \sum_{k=1}^{N} Q^i_{\ \ k} Q^j_{\ \ k} \quad &&\in \mathbb{Z} \,, \\
	&\kappa_{Ri} + \frac{1}{2} \sum_{k=1}^{N} Q^i_{\ \ k} (r_k - 1) \quad &&\in \mathbb{Z} \,.
\end{alignat}
\end{subequations}
To have non-singular moduli spaces of vacua, the entries $Q^i_{\ \ j}$ of the extended $N \times N$ charge matrix are required to be have unit modulus, 
\begin{equation}
	\left| Q^i_{\ \ j} \right| = 1 \,.
\end{equation}

The \index{real mass} real mass \index[sym]{$m_j$} $m_j$ are introduced for the global flavour symmetry $G_\text{f}$. There are also three-dimensional \index{Fayet-Iliopoulos parameter} Fayet-Iliopoulos parameters \index[sym]{$\zeta$} $\zeta_a$ associated with the global topological symmetry $T_\text{t} = \bigotimes_{b=1}^K U(1)_b$. The global symmetry $T_\text{t} \times G_\text{f}$ for generic Fayet-Iliopoulos parameters $\zeta_a$ and masses $m_j$ contains a maximal torus 
\begin{equation}
	T_\text{f} = \left(\bigotimes_{j=1}^N U(1)_j \right) / U(1)^K \,.
\end{equation}
Since $K$ linear combinations of the $U(1)_j$ generators are gauged, $K$ linear combinations of $\zeta_a$ and $m_j$ can be absorbed into shifts of the vector multiplet scalar \index[sym]{$\sigma$} $\sigma_a$ following
\begin{subequations}
\begin{align}
	\sigma_a &\mapsto \sigma_a + \delta \sigma_a \,, \\
	\zeta_a &\mapsto \zeta_a - \sum_b \kappa_{ab} \delta \sigma_b \,, \\
	m_j &\mapsto m_j - \sum_a Q^a_{\ \ j} \delta \sigma_a \,.
\end{align}
\end{subequations}

After integrating out auxiliary fields, the classical scalar potential is obtained~\cite{dorey2000} as
\begin{equation}
	U = \sum_{a=1}^{K} e_a^2 \left( \sum_{j=1}^{N} Q^a_{\ \ j} |\phi_j|^2 - \sum_{b=1}^K \kappa_{ab}\sigma_b - \zeta_a \right)^2 + \sum_{j=1}^N M_j^2 (\sigma) |\phi_j|^2
	\,,
\end{equation}
where $\phi_j$ and $\sigma_a$ are the scalars in the\index{chiral multiplet scalar} chiral multiplet and \index{vector multiplet scalar} vector multiplet respectively, and $\zeta_a$ is the three-dimensional Fayet-Iliopoulos parameter.

The \index{effective mass} effective mass of $\phi_j$ is given by
\begin{equation}
\label{eq:effective-mass}
	M_j(\sigma) = \sum_{a=1}^K Q^a_{\ \ j} \sigma_a + m_j \,.
\end{equation}

The dynamically generated Chern-Simons terms give corrections to the mixed gauge-gauge, gauge-flavour, gauge-R, and flavour-R mixed \index{Chern-Simons level} Chern-Simons levels respectively as
\begin{subequations}
\begin{align}
\label{eq:effective-gauge-gauge-CS-level}
	\kappa_{ab}^\eff &= \kappa_{ab} + \frac{1}{2} \sum_{j=1}^N Q_{aj} Q_{bj} \sign M_j(\sigma) \,,\\
\label{eq:effective-gauge-flavour-CS-level}
	\kappa_{ab'}^\eff &= \kappa_{ab'} + \frac{1}{2} \sum_{j=1}^N Q_{aj} Q_{b'j} \sign M_j(\sigma) \,,\\
\label{eq:effective-gauge-R-CS-level}
	\kappa_{Rb}^\eff &= \kappa_{Rb} + \frac{1}{2} \sum_{j=1}^N Q_{bj} (r_j - 1) \sign M_j(\sigma) \,,\\
\label{eq:effective-flavour-R-CS-level}
	\kappa_{Rb'}^\eff &= \kappa_{Rb'} + \frac{1}{2} \sum_{j=1}^N Q_{b'j} (r_j - 1) \sign M_j(\sigma) \,.
\end{align}
\end{subequations}

We define the \index{Chern-Simons level !asymptotic} asymptotic Chern-Simons levels as
\begin{subequations}
\begin{align}
	\kappa_{ab}^{\pm} &:= \kappa_{ab}^\eff(\sigma_a \rightarrow \pm\infty) = \kappa_{ab} \pm \frac{1}{2} \sum_{j=1}^N |Q_{aj}| Q_{bj} \,, \\
	\kappa_{ab'}^{\pm} &:= \kappa_{ab'}^\eff(\sigma_a \rightarrow \pm \infty) = \kappa_{ab'} \pm \frac{1}{2} \sum_{j=1}^N |Q_{aj}| Q_{b'j} \,, \\
	\kappa_{Ra}^{\pm} &:= \kappa_{Ra}^\eff(\sigma_a \rightarrow \pm\infty) = \kappa_{Ra} \pm \frac{1}{2} \sum_{j=1}^N |Q_{aj}| (r_j-1) \,.
\end{align}
\end{subequations}

The effect of the dynamical generation of gauge-flavour \index{Chern-Simons term} Chern-Simons terms can be interpreted as the renormalisation of \index{Fayet-Iliopoulos parameter} Fayet-Iliopoulos parameter. The resulting \index{Fayet-Iliopoulos parameter !effective} effective Fayet-Iliopoulos parameter is
\begin{equation}
	\zeta_a^\eff = \zeta_a + \sum_{b'=K+1}^N \kappa_{ab'}^\eff m_{b'} \,,
\end{equation}
where \index[sym]{$m_{a'}$} $m_{a'}$ are the independent mass parameters satisfying 
\begin{equation}
	m_j = \sum_{a'=K+1}^N Q^{a'}_{\ \ j} m_{a'} \,.
\end{equation}

The combined effects of the dynamical generation of gauge-gauge and gauge-flavour Chern-Simons terms are captured by the effective parameter \index[sym]{$F_a(\sigma)$}
\begin{align}
\label{eq:U(1)K-effective-parameter}
	F_a(\sigma)&= \sum_{b=1}^K \kappa_{ab}^\eff \sigma_b + \zeta_a +  \sum_{b'=K+1}^N \kappa_{ab'}^\eff m_{b'} \nonumber \\
	&= \zeta_a + \sum_{b=1}^K \kappa_{ab} \sigma_b +\sum_{b'=K+1}^N \kappa_{ab'} m_{b'} + \frac{1}{2} \sum_{j=1}^N Q_{aj} |M_j(\sigma)|\,.
\end{align}

Hence the semi-classical scalar potential is obtained as
\begin{equation}
\label{eq:scalar-potential}
	U = \sum_{a=1}^{K} e_a^2 \left( \sum_{j=1}^{N} Q^a_{\ \ j} |\phi_j|^2  - F_a(\sigma)\right)^2 + \sum_{j=1}^N M_j^2 (\sigma) |\phi_j|^2 \,.
\end{equation}

As discussed in Chapter~\ref{chap:twist-on-surface}, the theory can be twisted onto a target space $\Sigma_g \times S^1$, where $\Sigma_g$ is a Riemann surface of genus $g$, and $S^1$ is a circle of radius $\beta$. The twisted is performed using the unbroken R-symmetry, which preserves an $\mathcal{N}=(0,2)$ quantum mechanics on $S^1$ with a pair of \index{supercharge} supercharges $Q$ and $\widebar{Q}$. 

Exponentiating $F_a(\sigma)$ gives a factor of the form 
\begin{equation}
	q_a \prod_{b=1}^K x_b^{\kappa_{ab}} \prod_{b'=K+1}^N y_{b'}^{\kappa_{a b'}} \,,
\end{equation}
where the \index{fugacity} fugacities are defined \index[sym]{$q$}\index[sym]{$x$}\index[sym]{$y_{a'}$}as 
\begin{subequations}
\begin{align}
	q_a &:= e^{-\beta \zeta_a} \,, \\
	x_b &:= e^{-\beta \sigma_b} \,, \\
	y_{b'} &:= e^{-\beta m_{b'}} \,.
\end{align}
\end{subequations}
This results in factors involving the mixed Chern-Simons terms in the one-loop determinant
\begin{align}
\label{eq:one-loop-det}
	\hspace{-1ex} 
	Z = & \left[ \prod_{a=1}^K q_a^{\mathfrak{m}_a}  
	\left( \prod_{b=1}^K x_b^{\kappa_{ab} \mathfrak{m}_a} \right)
	\left( \prod_{b'=K+1}^N y_{b'}^{\kappa_{a b'} \mathfrak{m}_a} \right)
	\right]
	\quad \times \nonumber \\
	& \hspace{5ex} \left( \prod_{b=1}^K x_b^{\kappa_{Rb} (g-1)} \right)
	\left( \prod_{b'=K+1}^N y_{b'}^{\kappa_{Rb'} (g-1)} \right)	
	\quad \times \nonumber \\
	& \hspace{5ex} \left[ \prod_{i=1}^N \left( \frac{ \prod_{b=1}^K x_b^{\sfrac{ Q^b_{\ \ i} }{ 2 }} \prod_{b'=K+1}^{N} y_{b'}^{ \sfrac{Q^{b'}_{\ \ i} }{ 2 } } }{ 1 - \prod_{b=1}^K x_b^{Q^b_{\ \ i}} \prod_{b'=K+1}^{N} y_{b'}^{Q^{b'}_{\ \ i}} } \right)^{\left( \sum_{c=1}^K Q^c_{\ \ i} \mathfrak{m}_c \right) + (r_i - 1)(g - 1)} \right]\,,
\end{align}
for the twisted theory on $S^1 \times \Sigma$.

The semi-classical vacua of the scalar potential \eqref{eq:scalar-potential} are the solutions to to the following set of\index{vortex equation} vortex equations
\begin{subequations}
\label{eq:vacuum}
\begin{alignat}{3}
	\sum_{j=1}^{N} Q_{aj} |\phi_j|^2  &= F_a(\sigma)\qquad &&\forall a\,, \label{eq:vacuum-D} \\
	M_j(\sigma) \phi_j &= 0 \qquad &&\forall j \label{eq:vacuum-effective-mass}\,, 
\end{alignat}
\end{subequations}
where~\eqref{eq:vacuum-D} is the \index{D-term equation}D-term equation.

Analogous to the $U(1)$ vacua discussed in Section~\ref{sec:flat-vacuum-classes}, the vacuum solutions can be classified into the following classes:
\vspace{-\parskip}
\begin{itemize}
	\item Higgs branch \index{vacuum !Higgs branch} vacua occur when at least $K$ chiral multiplet scalars $\{\phi_i \mid i \in I\}$ are non-vanishing. Their effective masses $M_i(\sigma)$ must vanish due to \eqref{eq:vacuum-D}, fixing the values of all vector multiplet scalars $\sigma_a$. The right hand side of the vortex equation \eqref{eq:vacuum-D} is now fixed
		\begin{equation}
			\sum_{i \in I} Q_{ai} |\phi_i|^2 = \zeta_a + \sum_{b=1}^K +  \kappa_{ab} \sigma_b + \frac{1}{2} \sum_{j \not\in I} Q_{aj} |M_j(\sigma)| \,,
		\end{equation}
		 for given gauge charges $Q_{ai}$ and real masses $m_i$. For generic mass parameters $m_i$, there are exactly $K$ non-vanishing\index{chiral multiplet} chiral multiplet scalars.
	\item Topological branch \index{vacuum !topological branch} vacua are discrete solutions of the scalars $\sigma_a$ to 
		$$F_a(\sigma) = \zeta_a^\eff + \sum_{b=1}^K \kappa_{ab}^\eff(\sigma) \sigma_b = 0$$ 
		at low energies where all chiral multiplet scalars $\phi_j=0$. This can only occur if $\kappa_{ab}^\eff(\sigma) \neq 0$ for generic mass parameters $m_j$.
	\item Coulomb branch \index{vacuum !Coulomb branch} refers to the non-isolated solutions of the vector multiplet scalars $\sigma_a$ where $F_a(\sigma)=\kappa_{ab}^\eff (\sigma)=0$ when all chiral multiplet scalars $\phi_i$ vanish.
	\item A mixing of Higgs and Coulomb branches can occur when the number $|I|$ of vanishing chiral multiplet scalars is $0 < |I| < K$, where the vacuum consists of both non-isolated values of the chiral multiplet scalars $\phi_i$ and the vector multiplet scalars $\sigma_a$. 
\end{itemize}

The vacuum structure is considerably more intricate compared to the the $U(1)$ case discussed in Section~\ref{sec:flat-vacuum-classes}. The mixed types are of particular difficulty for analogous analysis.

\section{Window Phenomenon}
\label{sec:higher-rank-window-phenomenon}

There also exists the \index{window phenomenon} window phenomenon on the space of \index{Chern-Simons level} Chern-Simons levels, where a Higgs branch can exist alone within a certain region but must accompany a topological branch outside the region.

Similar to \eqref{eq:U(1)-critical-CS-level}, the \index{Chern-Simons level !critical} critical Chern-Simons levels $\kappa^\crit_{ab}$ are defined to be the bare \index{Chern-Simons level} Chern-Simons levels such that the \index{Chern-Simons level !effective} effective Chern-Simons levels in \eqref{eq:effective-gauge-gauge-CS-level} vanish, 
\begin{equation}
	0 =  \kappa^\crit_{ab} + \frac{1}{2} \sum_{j=1}^N Q_{aj} Q_{bj} \sign M_j(\sigma) \,,
\label{eq:U(1)-critical-levels}
\end{equation}
where $\sign M_j(\sigma)$ depends on all $\sigma_a$ from each $U(1)$ component.
The space of $\sigma_a$ is divided into chambers by $M_j(\sigma)=0$. Each chamber admits a different set of critical levels. Away from these critical levels, there may exist topological vacua in addition to Higgs vacua. There may not exist a finite window for any choice of Fayet-Iliopoulos parameters

\section{Bethe Ansatz Equations}
The \index{Bethe ansatz} Bethe ansatz equations for $U(1)^K$ theories are given by 
\begin{equation}
	\exp(i \frac{\partial \log Z}{\partial \mathfrak{m}_a}) = 1 \quad \forall a \,.
\end{equation}
For the theories of our interests, substituting in the classical and one-loop determinant \eqref{eq:one-loop-det} gives 
\begin{align}
	&\prod_{\{j \mid Q^a_{\ \ j} > 0\}} \left( 1 - \prod_{b=1}^K x_b^{Q^b_{\ \ j}}\prod_{b'=K+1}^N y_{b'}^{Q^{b'}_{\ \ j}} \right)^{Q^a_{\ \ j}} 
	\nonumber \\
	=& (-1)^{N_-} q_a 
	\left( \prod_b x_b^{\kappa_{ab}^+} \right) 
	\left( \prod_{b'} y_{b'}^{\kappa_{ab'}^+} \right) 
	\quad \times \nonumber \\	
	& \hspace{25ex} 
	\prod_{\{j \mid Q^a_{\ \ j} < 0\}} \left( 1 - \prod_{b=1}^K x_b^{-Q^b_{\ \ j}}\prod_{b'=K+1}^N y_{b'}^{-Q^{b'}_{\ \ j}} \right)^{-Q^a_{\ \ j}} 
\end{align}
for each $a = 1,\ldots,K$. The constant exponent $N^\pm$ is defined as
$$
	N^\pm = \sum_{\{j \mid Q^a_{\ \ j} \gtrless 0\}} Q^a_{\ \ j} \,,
$$
which roughly counts the number of positively or negatively charged chiral multiplets.
The splitting of the index set into positive and negative charges is to make the equations close to polynomials of $x_a$. Equivalently, the Bethe ansatz equations can also be written as
\begin{align}
	&\prod_{\{j \mid Q^a_{\ \ j} > 0\}} \left( 1 - \prod_{b=1}^K x_b^{-Q^b_{\ \ j}}\prod_{b'=K+1}^N y_{b'}^{-Q^{b'}_{\ \ j}} \right)^{Q^a_{\ \ j}} 
	\nonumber \\
	=& (-1)^{Q^a_{\ \ j} N_+} q_a 
	\left( \prod_b x_b^{\kappa_{ab}^-} \right) 
	\left( \prod_{b'} y_{b'}^{\kappa_{ab'}^-} \right) 
	\quad \times \nonumber\\
	& \hspace{27ex} \prod_{\{j \mid Q^a_{\ \ j} < 0\}} \left( 1 - \prod_{b=1}^K x_b^{Q^b_{\ \ j}}\prod_{b'=K+1}^N y_{b'}^{Q^{b'}_{\ \ j}} \right)^{-Q^a_{\ \ j}} \,.
\end{align}

\section{Twisted Index}

Consider the twisted index on $S^1 \times \Sigma$, where $\Sigma$ is a closed orientable Riemann surface of genus $g$. In the operator formalism, the twisted index counts the supersymmetric ground states $\mathcal{H}$ annihilated by the \index{supercharge} supercharges $Q$ and $\widebar{Q}$. The space $\mathcal{H}$ of supersymmetric ground states forms a representation of the global symmetry $T_\text{t} \times G_\text{f}$. The twisted index is then in the form
\begin{equation}
\label{eq:index-def}
	\mathcal{I} = \Tr (-1)^F \prod_{a=1}^K q_a^{J_a} \prod_{i=1}^N y_i^{J_i} \,,
\end{equation}
where $J_a$ is the Cartan generator of the $a$-th $U(1)$ factor in the topological symmetry $T_\text{t}$, and $J_i$ is the generator of $U(1)_i$ in $G_\text{f}$. The Hilbert space $\mathcal{H}$ is assumed to be locally finitely graded.

This definition \eqref{eq:index-def} of the twisted index can be interpreted as the \index{supersymmetric index} supersymmetric index of the supersymmetric quantum mechanics obtained by the twist. The geometric construction can be expected to be in the following form~\cite{xu2022}
\begin{equation}
	\mathcal{I} = \sum_{\mathfrak{m}} q^{\mathfrak{m}} \int \hat{A}(\mathfrak{M}_{\mathfrak{m}}) \ch (\mathcal{E}_{\mathfrak{m}}) \,,
\end{equation}
where $\mathfrak{m}$ labels the magnetic sectors, $\mathfrak{M}_{\mathfrak{m}}$ denotes the moduli space parametrising the saddle points of the localised path integral, and $\mathcal{E}_{\mathfrak{m}}$ is a perfect complex of sheaves encoding the massive fluctuations of the\index{chiral multiplet} chiral multiplets.

It can be shown~\cite{xu2022} that with appropriate localisation schemes, this geometric construction reproduces the Jeffrey-Kirwan contour integral formula of twisted indices~\cite{benini2015}:
\begin{equation}
	\mathcal{I} = \sum_{\mathfrak{m}_1,\ldots,\mathfrak{m}_K} \left( \frac{1}{2 \pi i} \right)^K \oint_\text{JK} \frac{\dd x_1}{x_1} \ldots \frac{\dd x_K}{x_K} \, \det(H_{ab})^g Z(x_1,\ldots, x_K; \mathfrak{m}_1, \ldots, \mathfrak{m}_K) \,,
\end{equation}
where $Z$ is the classical and one-loop determinant~\eqref{eq:one-loop-det} and the hessian factors are
\begin{equation}
	H_{ab} = \frac{\partial^2 \log Z}{\partial \log x_a \partial \mathfrak{m}_b} \,.
\end{equation}

\subsection{Geometric Interpretation}
Generalising the work~\cite{xu2022} on the $U(1)$ case, the \index{moduli space !of saddles} moduli space of saddles splits into disjoint unions of topologically distinct components $\mathfrak{M}_{\mathfrak{m}}$, where for a $U(1)^K$ theory $$\mathfrak{m}=(\mathfrak{m}_1,\ldots,\mathfrak{m}_K)$$ labels the distinct magnetic sectors. For \index{saddle point !vortex} vortex saddles, each component moduli space $\mathfrak{M}_{\mathfrak{m}}$ is $K$ copies of symmetric products of the curve $\Sigma$ from the $K$ non-vanishing \index{chiral multiplet} chiral multiplets. For \index{saddle point !topological} topological saddles, $\mathfrak{M}_{\mathfrak{m}}$ are Picard stacks.

The contribution to the \index{twisted index} twisted index can then be written in the form
\begin{equation}
\label{eq:integral-of-characteristic-classes}
	\mathcal{I} = \sum_{\mathfrak{m} \in \mathbb{Z}^K} 
	q_1^{\mathfrak{m}_1} q_2^{\mathfrak{m}_2} \ldots q_K^{\mathfrak{m}_K} 
	\int \hat{A}(\mathfrak{M}_{\mathfrak{m}}) 
	\frac{\hat{A}(\mathcal{E}_{\mathfrak{m}})}{e(\mathcal{E}_{\mathfrak{m}})} 
	\ch\Big( \bigotimes_{\alpha} \mathcal{L}_\alpha \Big) \,,
\end{equation}
where $\mathcal{L}_\alpha$ are holomorphic line bundles arising from Chern-Simons terms. We have shown that they can be interpreted as the determinant line bundles of additional auxiliary\index{chiral multiplet} chiral multiplets. This is consistent with the physical phenomenon that integrating out heavy fermions in three-dimensional theories induces effective Chern-Simons terms as a low-energy effect~\cite{Redlich:1983kn,Redlich:1983dv,Alvarez-Gaume:1983ihn}. It is these determinant line bundles that give the level structure in the corresponding quantum K-theory.

\section{Mirror Symmetry}
\label{sec:mirror-examples}

In this section we explore two examples $U(1)^K$ gauge theories and their mirror pairs, under the three-dimensional \index{mirror symmetry} mirror symmetry~\cite{dorey2000}. Their twisted indices are verified to agree under the mirror, with the appropriate choices of Chern-Simons levels.

\subsection[\texorpdfstring{$U(1)^2$ with Three Chiral Fields}%
                        {U(1)² with Three Chiral Fields}]%
        {$U(1)^2$ with Three Chiral Fields}

Consider a $U(1)^2$ gauge theory with three chiral multiplets. Let the \index{gauge charge} gauge charge matrix $Q^a_{\ \ j}$ be
\begin{equation}
	Q^a_{\ \ j} = 
	\begin{pmatrix}
		1 & -1 & 0 \\
		0 & 1 & -1
	\end{pmatrix}
	\,.
\end{equation}
Without loss of generality, the real masses are set to 
\begin{subequations}
\label{eq:CP2-masses}
\begin{align}
	m_1 &= -m \,,\\
	m_2 &= 0 \,, \\
	m_3 &= -m 
\end{align}
\end{subequations}
with the associated \index{flavour charge} flavour charges $(1,0,1)$ for a single independent mass parameter $-m$.
The \index{extended charge matrix} extended charge matrix is then
\begin{equation}
	Q^i_{\ \ j} = 
	\begin{pmatrix}
		1 & -1 & 0 \\
		0 & 1 & -1 \\
		1 & 0 & 1
	\end{pmatrix}
	\,,
\label{eq:charge-C1-A}
\end{equation}
where the top two rows are the \index{gauge charge} gauge charges, and the third row is the \index{flavour charge} flavour charges. 

The effective masses of $\phi_j$ are then
\begin{subequations}
\begin{align}
	M_1 &= \sigma_1 - m \,, \\
	M_2 &= -\sigma_1 + \sigma_2 \,, \\
	M_3 &= -\sigma_2 - m \,.
\end{align}
\end{subequations}
The effective gauge Chern-Simons levels~\eqref{eq:effective-gauge-gauge-CS-level} and effective gauge-flavour Chern-Simons levels~\eqref{eq:effective-gauge-flavour-CS-level} are respectively given by
\begin{equation}
\label{eq:CP2-effective-CS-levels}
	\kappa_{ab}^\eff =
	\begin{pmatrix}
		\kappa_{11} + \frac{1}{2} \sign M_1 + \frac{1}{2} \sign M_2 & \kappa_{12} - \frac{1}{2} \sign M_2\\
		\kappa_{21} - \frac{1}{2} \sign M_2 & \kappa_{22} + \frac{1}{2} \sign M_2 + \frac{1}{2} \sign M_3
	\end{pmatrix}
\end{equation}
and
\begin{equation}
	\kappa_{ab'}^\eff =
	\begin{pmatrix}
		\kappa_{13}^\eff = \kappa_{13} + \frac{1}{2} \sign M_1 \\
		\kappa_{23}^\eff = \kappa_{23} - \frac{1}{2} \sign M_3
	\end{pmatrix} \,.
\end{equation}

The semi-classical vacuum equations~\eqref{eq:vacuum} become
\begin{subequations}
\begin{align}
	|\phi_1|^2 - |\phi_2|^2  &= F_1(\sigma) \,,\\
	|\phi_2|^2 - |\phi_3|^2  &= F_2(\sigma) \,,\\
	(\sigma_1 - m) \phi_1 &= 0 \,,\\
	(-\sigma_1 + \sigma_2) \phi_2 &= 0 \,,\\
	(-\sigma_2 - m) \phi_3 &= 0 \,,
\end{align}
\end{subequations}
where the effective parameters~\eqref{eq:U(1)K-effective-parameter} are given by
\begin{subequations}
\begin{align}
	F_1(\sigma) &= \zeta_1 
	+ \kappa_{11} \sigma_1 + \kappa_{12} \sigma_2
	- \kappa_{13} m
	+ \frac{1}{2} |\sigma_1 - m| - \frac{1}{2} |-\sigma_1 + \sigma_2|
	\,,\\
	F_2(\sigma) &= \zeta_2 
	+ \kappa_{21} \sigma_1 + \kappa_{22} \sigma_2
	- \kappa_{23} m
	+ \frac{1}{2} |-\sigma_1 + \sigma_2| - \frac{1}{2} |-\sigma_2 - m| \,.
\end{align}
\end{subequations}

The one-loop determinant \eqref{eq:one-loop-det} for this theory is
\begin{alignat}{3}
\label{eq:CP2-one-loop-det}
	Z
	=& \left( q_1^{\mathfrak{m}_1} q_2^{\mathfrak{m}_2} \right)		
	\left( x_1^{\kappa_{11}\mathfrak{m}_1 + \kappa_{21}\mathfrak{m}_2} 
	x_2^{\kappa_{12}\mathfrak{m}_1 + \kappa_{22}\mathfrak{m}_2}
	y_3^{\kappa_{13}\mathfrak{m}_1 + \kappa_{23}\mathfrak{m}_2} \right)
	\quad \times \nonumber \\ 
	& \hspace{22ex} \left( x_1^{\kappa_{R1} (g-1)} 
	x_2^{\kappa_{R2} (g-1)}
	y_3^{\kappa_{R3} (g-1)} \right) 
	\quad \times \nonumber \\
	& \hspace{26ex} \left( \frac{ x_1^{\sfrac{1}{2}} }{ 1 - x_1 } \right)^{\mathfrak{m}_1 + (r_1 - 1)(g - 1)}
	&&\quad \times \nonumber \\ 	
	& \hspace{26ex} \left( \frac{ x_1^{\sfrac{-1}{2}} x_2^{\sfrac{1}{2}} }{ 1 - x_1^{-1} x_2 } \right)^{-\mathfrak{m}_1 + \mathfrak{m}_2 + (r_2 - 1)(g - 1)}   
	&&\quad \times \nonumber \\ 	
	& \hspace{26ex} \left( \frac{ x_2^{\sfrac{-1}{2}} y_{3}^{\sfrac{1}{2}} }{ 1 - x_2^{-1} y_3 } \right)^{-\mathfrak{m}_2 + (r_3 - 1)(g - 1)} \,.
\end{alignat}

\subsubsection{Higgs Branch}
The Higgs branch is obtained at $\sigma_1 = \sigma_2 = m$ such that the effective masses all vanish. Hence the vacuum is the solutions to the remaining $D$-term equations
\begin{subequations}
\begin{align}
	|\phi_1|^2 - |\phi_2|^2 &= \zeta_1 
	+ (\kappa_{11} + \kappa_{12} - \kappa_{13} ) m
	\,,\\
	|\phi_2|^2 - |\phi_3|^2 &= \zeta_2 
	+ (\kappa_{21}  + \kappa_{22} - \kappa_{23}) m
	\,.
\end{align}
\end{subequations}

Via \index{symplectic quotient} symplectic quotient, the full moduli space $\mathcal{M}^\text{H}$ at $m=0$ can be constructed as 
\begin{equation}
	\frac{\mu_1^{-1}(\zeta_1) \cap \mu_2^{-1}(\zeta_2) }{ U(1)^2 } \,,
\end{equation}
where the corresponding moment maps $\mu_1$ and $\mu_2$ for the two $U(1)$ components of the gauge group $U(1)^2$ action are 
\begin{subequations}
\begin{align}
	\mu_1 := |\phi_1|^2 - |\phi_2|^2 &= \zeta_1 
	\,,\\
	\mu_2 := |\phi_2|^2 - |\phi_3|^2 &= \zeta_2 
	\,.
\end{align}
The moment map for the flavour symmetry is 
\begin{equation}
	\mu_3 := |\phi_1|^2 + |\phi_3|^2 \,.
\end{equation}
\end{subequations}
They generate flows on the $\mathbb{C}^3$ space of the chiral multiplets via the Poisson bracket
\begin{equation}
	\partial_i \phi_j = \{\mu_i, \phi_j\}_\omega = \omega \left( X_{\mu_i}, X_{\phi_j} \right) \,,
\end{equation}
where $\omega$ is the standard symplectic form
\begin{equation}
	\omega = i \left( \dd \phi_1 \wedge \dd \bar{\phi}_1 + \dd \phi_2 \wedge \dd \bar{\phi}_2 + \dd \phi_3 \wedge \dd \bar{\phi}_3\right) \,,
\end{equation}
and the hamiltonian vector fields $X$ are in the form
\begin{equation}
	X_H = \sum_{i=1}^3 
	\left( \frac{\partial H}{\partial \bar{\phi}_i} \frac{\partial}{\partial \phi_i} 
	- \frac{\partial H}{\partial \phi_i} \frac{\partial}{\partial \bar{\phi}_i} \right)\,.
\end{equation}
The induced maps are the gauge and flavour actions
\begin{subequations}
\begin{align}
	e^{\alpha \mu_1}:& \quad (\phi_1, \phi_2, \phi_3) \mapsto (e^{i \alpha} \phi_1, e^{-i \alpha} \phi_2, \phi_3) \,,\\
	e^{\beta \mu_2}:& \quad (\phi_1, \phi_2, \phi_3) \mapsto (\phi_1, e^{i \beta} \phi_2, e^{-i \beta} \phi_3) \,,\\
	e^{\gamma \mu_3}:& \quad (\phi_1, \phi_2, \phi_3) \mapsto (e^{i \gamma} \phi_1, \phi_2, e^{i \gamma} \phi_3) \,.
\end{align}
\end{subequations}
The flavour symmetry is degenerate at $\phi_1 = \phi_3 = 0$, which corresponds to the point $\mu_3 = 0$.

We can express the coordinates $\phi_j$ in terms of the moment maps as
\begin{subequations}
\begin{align}
	|\phi_1|^2 = \frac{1}{2} (\mu_3 + \zeta_1 + \zeta_2) \,, \\
	|\phi_2|^2 = \frac{1}{2} (\mu_3 - \zeta_1 + \zeta_2) \,, \\
	|\phi_3|^2 = \frac{1}{2} (\mu_3 - \zeta_1 - \zeta_2) \,.
\end{align}
\end{subequations}
This requires the flavour moment map to be
\begin{equation}
	\mu_3 \geq \max \left\{(\zeta_1 + \zeta_2), (-\zeta_1 + \zeta_2), (-\zeta_1 - \zeta_2) \right\} \,.
\end{equation}

When the Fayet-Iliopoulos parameters $\zeta_1$, $\zeta_2$ lie in the K\"ahler cone~\cite{dorey2000} of the moduli space $\mathcal{M}^\text{H}$, it can be constructed as a toric variety
\begin{equation}
	\mathcal{M}^\text{H} = \frac{\mathbb{C}^3 - F_\Delta}{G \times \Gamma} \,,
\end{equation}
where $F_\Delta$ is a subset of $\mathbb{C}^3$. It depends on the data encoded in the toric fan $\Delta$ which is constructed from linear relations among the gauge charges.
The group $G$ is the complexified gauge actions
\begin{subequations}
\begin{align}
	\mathbb{C}^*_1: & \quad (\phi_1, \phi_2, \phi_3) \mapsto (\lambda_1 \phi_1, \lambda_1^{-1} \phi_2, \phi_3) \,,\\
	\mathbb{C}^*_2:& \quad (\phi_1, \phi_2, \phi_3) \mapsto (\phi_1, \lambda_2 \phi_2, \lambda_2^{-1} \phi_3) \,.
\end{align}
\end{subequations}
The gauge charge vectors have a single linear relation 
\begin{equation}
	\begin{pmatrix}
		1\\
		0
	\end{pmatrix}
	+
	\begin{pmatrix}
		-1\\
		1
	\end{pmatrix}
	+
	\begin{pmatrix}
		0\\
		-1
	\end{pmatrix}
	= 0
\end{equation}
between them, giving three one-dimensional vectors in $\mathbb{Z}$
\begin{equation}
	v_1 = v_2 = v_3 = 1 \,,
\end{equation}
which is the gauge charges of its dual theory under mirror symmetry. 
This can be found easily by regarding the gauge charges as two vectors $\left\{ (1, -1, 0), (0, 1, -1) \right\}$ in $\mathbb{Z}^3$, and computing their orthogonal vector to be $(1,1,1)$. 
The vectors $\{v_j\}$ generate a one-dimensional cone belonging to the fan $\Delta$. The set $F_\Delta$ is found by associating $\phi_j$ with $v_j$, and taking all the loci $\phi_i = \cdots = \phi_j = 0$ whenever $\{v_i, \ldots, v_j\}$ do not span a cone in the fan $\Delta$. In this case the set is empty 
\begin{equation}
	F_\Delta = \emptyset \,.
\end{equation}
The generating vectors $\{v_1, v_2, v_3\}$ induce a map $\psi: \mathbb{C}^3 \to \mathbb{C}$ given by
\begin{equation}
	\psi: \quad (\phi_1, \phi_2, \phi_3) \mapsto \phi_1 \phi_2  \phi_3 \,,
\end{equation}
whose kernel $\ker \psi$ is the complexified gauge symmetry 
\begin{equation}
	G = \ker \psi = (\mathbb{C}^*)^2 \,.
\end{equation}
The group $\Gamma$ is discrete given by
\begin{equation}
	\Gamma = \frac{\mathbb{Z}}{\operatorname{span}_\mathbb{Z} \{v_1, v_2, v_3\}} \cong \mathbb{Z}_2 \,,
\end{equation}
which gives rise to a $\mathbb{Z}_2$ orbifold singularity.
The moduli space is then
\begin{equation}
	\mathcal{M}^\text{H} = \frac{\mathbb{C}^3}{(\mathbb{C}^*)^2 \times \mathbb{Z}_2} \,.
\end{equation}

\subsubsection{Coulomb Branch}
The Coulomb branch is the non-isolated solutions to
\begin{equation}
	F_1(\sigma) = F_2(\sigma) = 0
\end{equation}
when all chiral multiplets vanish, which requires the effective Chern-Simons levels~\eqref{eq:CP2-effective-CS-levels} to vanish.

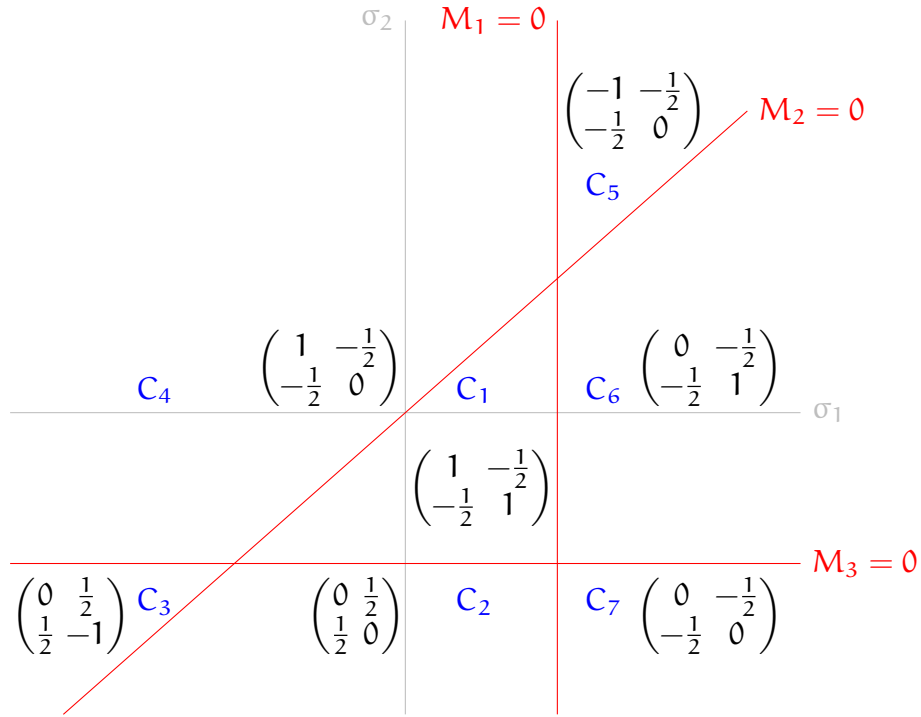
\begin{figure}[!h]
\centering
\caption{Critical Chern-Simons Levels on Toric Diagram}
\label{fig:critical-CS-levels-on-toric-diagram}

\begin{tikzpicture}[scale=1]
 
  \def\xa{-4}
  \def\xb{ 4}
  \def\ya{-4}
  \def\yb{ 4}
  \def\tmax{2.1}
  \def\a{1.3}
  \def\b{1}
  \def\c{{sqrt(\a^2+\b^2)}}
  \def\N{100} 
 
  \draw[mygrey] 
    (\xa*\a,0) -- (\xb*\a,0)  node[pos=1, right] {$\sigma_1$};
  \draw[mygrey] 
    (0,-4) -- (0,\yb*\a) node[pos=1, left] {$\sigma_2$};
    
  \draw[red]
    (\xa*\a,-2) -- (\xb*\a,-2)  node[pos=1, right] {$M_3=0$};
  \draw[red]
    (2,-4) -- (2,\yb*\a) node[pos=1, left] {$M_1=0$};
  \draw[red]
    (-4.5,-4) -- (4.5,4) node[pos=1, right] {$M_2=0$};

  \node[] at (1,-1) {$\begin{pmatrix} 1 & -\frac{1}{2} \\ -\frac{1}{2} & 1 \end{pmatrix}$};
  \node[] at (-0.65,-2.7) {$\begin{pmatrix} 0 & \frac{1}{2} \\ \frac{1}{2} & 0 \end{pmatrix}$};
  \node[] at (-4.4,-2.7) {$\begin{pmatrix} 0 & \frac{1}{2} \\ \frac{1}{2} & -1 \end{pmatrix}$};
  \node[] at (-1,0.6) {$\begin{pmatrix} 1 & -\frac{1}{2} \\ -\frac{1}{2} & 0 \end{pmatrix}$};
  \node[] at (3,4) {$\begin{pmatrix} -1 & -\frac{1}{2} \\ -\frac{1}{2} & 0 \end{pmatrix}$};
  \node[] at (4,0.6) {$\begin{pmatrix} 0 & -\frac{1}{2} \\ -\frac{1}{2} & 1 \end{pmatrix}$};
  \node[] at (4,-2.7) {$\begin{pmatrix} 0 & -\frac{1}{2} \\ -\frac{1}{2} & 0 \end{pmatrix}$};
  
  \node[blue] at (0.9,0.3) {$C_1$};
  \node[blue] at (0.9,-2.5) {$C_2$};
  \node[blue] at (-3.3,-2.5) {$C_3$};
  \node[blue] at (-3.3,0.3) {$C_4$};
  \node[blue] at (2.6,3) {$C_5$};
  \node[blue] at (2.6,0.3) {$C_6$};
  \node[blue] at (2.6,-2.5) {$C_7$};
\end{tikzpicture}
\end{figure}

The hypersurfaces 
\begin{equation}
\label{eq:theory-A-hypersurfaces}
	D_j := \left\{ (\sigma_1, \sigma_2) \in \mathbb{R}^2 \mid M_j(\sigma_1, \sigma_2) = 0 \right\} \qquad \text{for} \quad j = 1, 2, 3
\end{equation}
split the space of $\sigma_1$ and $\sigma_2$ into seven chambers.
Away from the hypersurface $M_i = 0$, the chiral multiplet $\phi_i$ must vanish in the vacua according to~\eqref{eq:vacuum-effective-mass}.
The critical Chern-Simons levels~\eqref{eq:U(1)-critical-levels} for the effective levels to vanish depend on the chamber on the $\sigma_1$--$\sigma_2$ ``toric'' diagram, as shown in Figure~\ref{fig:critical-CS-levels-on-toric-diagram} assuming $m>0$.

\begin{itemize}
\item In the finite triangular chamber $C_1$ bounded by
\begin{subequations}
\begin{align}
	\sign M_1 = -1 \quad &\Rightarrow \quad \sigma_1 < m \,, \\
	\sign M_2 = -1 \quad &\Rightarrow \quad \sigma_2 < \sigma_1 \,, \\
	\sign M_3 = -1 \quad &\Rightarrow \quad \sigma_2 > -m \,,
\end{align}
\end{subequations}
the critical Chern-Simons levels are
\begin{subequations}
\begin{equation}
\label{eq:CP2-CS-levels}
	\kappa^\crit_{ab} =
	\begin{pmatrix}
		1 & -\frac{1}{2} \\
		-\frac{1}{2} & 1
	\end{pmatrix}
	\,.
\end{equation}
\item In the infinite chamber $C_2$ below $C_1$, the critical Chern-Simons levels are
\begin{equation}
\label{eq:C2-CS-levels}
	\kappa^\crit_{ab} =
	\begin{pmatrix}
		0 & \frac{1}{2} \\
		\frac{1}{2} & 0
	\end{pmatrix}
	\,.
\end{equation}
\item In the infinite chamber $C_3$, the critical Chern-Simons levels are
\begin{equation}
\label{eq:C3-CS-levels}
	\kappa^\crit_{ab} =
	\begin{pmatrix}
		0 & \frac{1}{2} \\
		\frac{1}{2} & -1
	\end{pmatrix}
	\,.
\end{equation}
\item In the infinite chamber $C_4$, the critical Chern-Simons levels are
\begin{equation}
\label{eq:C4-CS-levels}
	\kappa^\crit_{ab} =
	\begin{pmatrix}
		1 & -\frac{1}{2} \\
		-\frac{1}{2} & 0
	\end{pmatrix}
	\,.
\end{equation}
\item In the infinite chamber $C_5$, the critical Chern-Simons levels are
\begin{equation}
\label{eq:C5-CS-levels}
	\kappa^\crit_{ab} =
	\begin{pmatrix}
		-1 & -\frac{1}{2} \\
		-\frac{1}{2} & 0
	\end{pmatrix}
	\,.
\end{equation}
\item In the infinite chamber $C_6$, the critical Chern-Simons levels are
\begin{equation}
\label{eq:C6-CS-levels}
	\kappa^\crit_{ab} =
	\begin{pmatrix}
		0 & -\frac{1}{2} \\
		-\frac{1}{2} & 1
	\end{pmatrix}
	\,.
\end{equation}
\item In the infinite chamber $C_7$, the critical Chern-Simons levels are
\begin{equation}
\label{eq:C7-CS-levels}
	\kappa^\crit_{ab} =
	\begin{pmatrix}
		0 & -\frac{1}{2} \\
		-\frac{1}{2} & 0
	\end{pmatrix}
	\,.
\end{equation}
\end{subequations}
\end{itemize}

A Coulomb branch vacua may open up when the effective Chern-Simons levels are vanishing. 
For example, in the chamber $C_1$ this requires the bare Chern-Simons levels to be set to the critical levels~\eqref{eq:CP2-CS-levels}. In addition the Fayet-Iliopoulos parameters $\zeta_a$ can be chosen appropriately such that $\zeta_a^\eff$ vanishes as well. Then there is a Coulomb branch with a toric moduli space $\mathbb{CP}^2$.
We refer to this theory at the critical levels~\eqref{eq:CP2-CS-levels} as~\cite{dorey2000} the theory $\mathcal{A}$. 

In other words, the theory $\mathcal{A}$ admits a Coulomb branch 
\begin{equation}
	\mathcal{M}^\text{C}_\mathcal{A} = \mathbb{CP}^2
\end{equation}
if and only if the vector multiplet scalars $\sigma_1$ and $\sigma_2$ are restricted to the triangular polytope $\nabla$ bounded by the hypersurfaces $\{D_j\}_{j=1}^3$ in Figure~\ref{fig:critical-CS-levels-on-toric-diagram}. The projective toric variety $\mathbb{CP}^2$ can be interpreted~\cite{Leung:1997tw,hori2003} as a $T^2$ fibration over the polytope $\nabla$.
It is isomorphic to the Higgs branch of its mirror symmetric theory.

The Coulomb branches in other chambers can be accessed by shifting the bare \index{Chern-Simons level} Chern-Simons levels accordingly.

\subsubsection{Mirror Symmetry}
Given a $U(1)^K$ theory with $N$ \index{chiral multiplet} chiral multiplets of charges $Q$, there exists a $U(1)^{N-K}$ theory under mirror symmetry~\cite{Batyrev:1993oya} with $N$ chiral multiplets of charges $\tilde{Q}$ satisfying
\begin{equation}
	\sum_{i=1}^N Q_{ai} \tilde{Q}_{bi} = 0
\label{eq:mirror-charges}
\end{equation}
for all $a=1,\ldots,K$ and $b=K+1,\ldots,N$. With this choice of labels for $\tilde{Q}$, the bottom $(N-K)$ rows are the \index{gauge charge} gauge charges while the top $K$ rows are \index{flavour charge} flavour charges in the mirror theory. Under mirror symmetry, the moduli spaces of Coulomb branch and Higgs branch are exchanged. To directly compare with $Q$ of the original theory, we  swap the rows of the charge matrix $\tilde{Q}$ to put the \index{gauge charge} gauge charges on top as $\{\tilde{Q}_{bi}\}_{b=1}^K$.

Now let us consider the mirror theory $\mathcal{B}$ to the theory $\mathcal{A}$ with charges~\eqref{eq:charge-C1-A}, mass parameters~\eqref{eq:CP2-masses}, and bare Chern-Simons levels~\eqref{eq:CP2-CS-levels}.
It is a supersymmetric quantum electrodynamics with charges
\begin{equation}
	\tilde{Q}^i_{\ \ j} = 
	\begin{pmatrix}
		1 & 1 & 1 \\
		1 & -1 & 0 \\
		0 & 1 & -1
	\end{pmatrix}
	\,,
\end{equation}
where the first row is the \index{gauge charge} gauge charges. Without loss of generality, we can set the one of the masses to vanish $\tilde{m}_2 = 0$.

The effective masses of the chiral multiplets $\tilde{\phi}_j$ are then
\begin{subequations}
\begin{align}
	\tilde{M}_1 &= \tilde{\sigma} + \tilde{m}_1 \,, \\
	\tilde{M}_2 &= \tilde{\sigma} \,, \\
	\tilde{M}_3 &= \tilde{\sigma} + \tilde{m}_3 \,.
\end{align}
\end{subequations}
The effective gauge Chern-Simons levels~\eqref{eq:effective-gauge-gauge-CS-level} and effective gauge-flavour Chern-Simons levels~\eqref{eq:effective-gauge-flavour-CS-level} are respectively given by
\begin{equation}
	\tilde{\kappa}_{11}^\eff =
	\tilde{\kappa}_{11} + \frac{1}{2} \sign \tilde{M}_1 + \frac{1}{2} \sign \tilde{M}_2 + \frac{1}{2} \sign \tilde{M}_3 
\end{equation}
and
\begin{equation}
	\tilde{\kappa}_{ab'}^\eff =
	\begin{pmatrix}
		\tilde{\kappa}_{12}^\eff = \tilde{\kappa}_{12} + \frac{1}{2} \sign \tilde{M}_1 - \frac{1}{2} \sign \tilde{M}_2 \\
		\tilde{\kappa}_{13}^\eff = \tilde{\kappa}_{13} + \frac{1}{2} \sign \tilde{M}_2 - \frac{1}{2} \sign \tilde{M}_3
	\end{pmatrix} \,.
\end{equation}

The semi-classical vacuum equations~\eqref{eq:vacuum} become
\begin{subequations}
\begin{align}
	|\tilde{\phi}_1|^2 + |\tilde{\phi}_2|^2 + |\tilde{\phi}_3|^2  &= F(\tilde{\sigma}) \,,\\
	(\tilde{\sigma} + \tilde{m}_1) \tilde{\phi}_1 &= 0 \,,\\
	\tilde{\sigma} \tilde{\phi}_2 &= 0 \,,\\
	(\tilde{\sigma} + \tilde{m}_2) \tilde{\phi}_3 &= 0 \,,
\end{align}
\end{subequations}
where the effective parameter~\eqref{eq:U(1)K-effective-parameter} is given by
\begin{equation}
	F(\tilde{\sigma}) = \tilde{\zeta}
	+ \tilde{\kappa}_{11} \tilde{\sigma} 
	- \tilde{\kappa}_{13} \tilde{m}_3
	+ \frac{1}{2} |\tilde{\sigma} + \tilde{m}_1| + \frac{1}{2} |\tilde{\sigma}| + \frac{1}{2} |\tilde{\sigma} + \tilde{m}_3|
	\,.
\end{equation}

The one-loop determinant \eqref{eq:one-loop-det} for this theory is
\begin{alignat}{3}
\label{eq:theory-B-one-loop-det}
	Z
	=& \tilde{q}^{\tilde{\mathfrak{m}}}	
	\left( \tilde{x}^{\tilde{\kappa}_{11} \tilde{\mathfrak{m}}} 
	\tilde{y}_2^{\tilde{\kappa}_{12}\tilde{\mathfrak{m}}}
	\tilde{y}_3^{\tilde{\kappa}_{13}\tilde{\mathfrak{m}}} 
	\right)
	\left( \tilde{x}^{\tilde{\kappa}_{R1} (g-1)} 
	\tilde{y}_2^{\tilde{\kappa}_{R2} (g-1)}
	\tilde{y}_3^{\tilde{\kappa}_{R3} (g-1)} \right) 
	\quad \times \nonumber \\
	& \hspace{26ex} \left( \frac{ \tilde{x}^{\sfrac{1}{2}} \tilde{y}_2^{\sfrac{1}{2}} }{ 1 - \tilde{x} \tilde{y}_2 } \right)^{\tilde{\mathfrak{m}} + (\tilde{r}_1 - 1)(g - 1)}
	&&\quad \times \nonumber \\ 	
	& \hspace{26ex} \left( \frac{ \tilde{x}^{\sfrac{1}{2}} \tilde{y}_2^{\sfrac{-1}{2}} \tilde{y}_3^{\sfrac{1}{2}} }{ 1 - \tilde{x} \tilde{y}_2^{-1} \tilde{y}_3 } \right)^{\tilde{\mathfrak{m}} + (\tilde{r}_2 - 1)(g - 1)}   
	&&\quad \times \nonumber \\ 	
	& \hspace{26ex} \left( \frac{ \tilde{x}^{\sfrac{1}{2}} \tilde{y}_{3}^{\sfrac{-1}{2}} }{ 1 - \tilde{x} \tilde{y}_3^{-1} } \right)^{\tilde{\mathfrak{m}} + (\tilde{r}_3 - 1)(g - 1)} \,.
\end{alignat}

With masses all vanishing, a Higgs branch opens up at $\tilde{\sigma} = 0$ with the vortex equation
\begin{equation}
	|\tilde{\phi}_1|^2 + |\tilde{\phi}_2|^2 + |\tilde{\phi}_3|^2 = \tilde{\zeta}
\end{equation}
This gives a moduli space 
\begin{equation}
	\mathcal{M}^{H}_\mathcal{B} = \mathbb{CP}^2
\end{equation}
for the Higgs branch after quotienting out the $U(1)$ gauge transformations, provided $\tilde{\zeta} > 0$. This is the same space as the Coulomb branch $\mathcal{M}^{C}_\mathcal{A}$ of theory $\mathcal{A}$.

\begin{figure}[!h]
\centering
\caption{Toric Fan and Polytope of $\mathbb{CP}^2$}
\label{fig:toric-diagram-B}

\begin{tikzpicture}[scale=0.8]
 
  \draw[mygrey] 
    (-7,0) -- (7,0) ;
  \draw[mygrey] 
    (0,-3) -- (0,7);
    
  \draw[red]
    (-6,-2) -- (2,-2);
  \node[red] at (-2,-2.4) {$\tilde{D}_3$};  
  \draw[red]
    (2,-2) -- (2,6) node[pos=0.5, right] {$\tilde{D}_1$};
  \draw[red]
    (-6,-2) -- (2,6) node[pos=0.628, left] {$\tilde{D}_2$};
    
  \draw[blue,thick]
    (0,0) -- (2,0) ;
  \node[blue] at (1,0.4) {$\tilde{v}_1$};
  \draw[blue,thick]
    (0,-2) -- (0,0) node[pos=0.3, right] {$\tilde{v}_3$};
  \draw[blue,thick]
    (-2,2) -- (0,0) node[pos=0.5, left] {$\tilde{v}_2$};
    
   \node[gray] at (2.7,0.4) {$(1,0)$};
   \node[gray] at (0.9,-2.4) {$(0,-1)$};
   \node[gray] at (-3,2.2) {$(-1,-1)$};
\end{tikzpicture}
\end{figure}
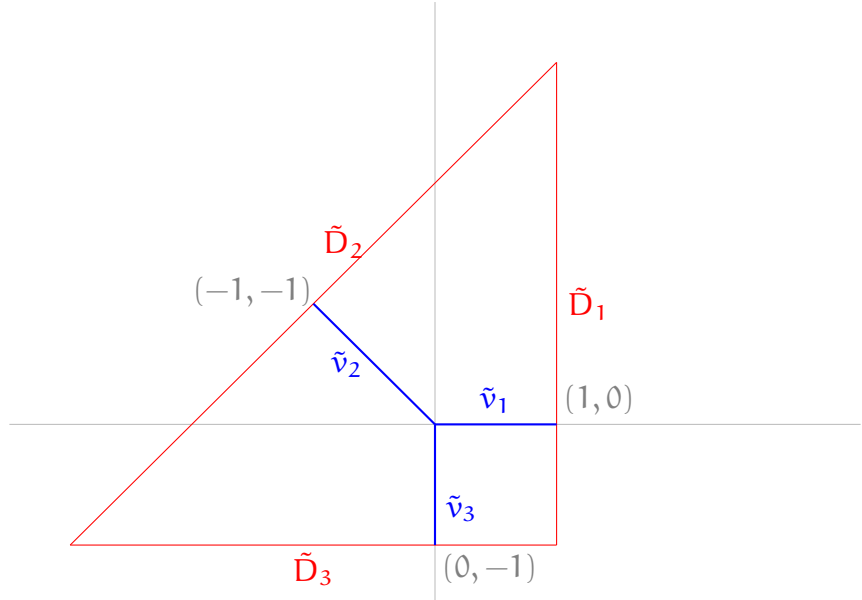

In terms of symplectic geometry, the moment maps are
\begin{subequations}
\begin{align}
	\tilde{\mu}_1 &= |\tilde{\phi}_1|^2 + |\tilde{\phi}_2|^2 + |\tilde{\phi}_3|^2 = \tilde{\zeta} \,, \\
	\tilde{\mu}_2 &= |\tilde{\phi}_1|^2 - |\tilde{\phi}_2|^2 \,, \\
	\tilde{\mu}_3 &= |\tilde{\phi}_2|^2 - |\tilde{\phi}_3|^2 \,.
\end{align}
\end{subequations}
The corresponding flows are
\begin{subequations}
\begin{align}
	e^{\alpha \tilde{\mu}_1}:& \quad (\tilde{\phi}_1, \tilde{\phi}_2, \tilde{\phi}_3) \mapsto (e^{i \alpha} \tilde{\phi}_1, e^{i \alpha} \tilde{\phi}_2, e^{i \alpha} \tilde{\phi}_3) \label{eq:theory-B-gauge-flow}\,,\\
	e^{\beta \tilde{\mu}_2}:& \quad (\tilde{\phi}_1, \tilde{\phi}_2, \tilde{\phi}_3) \mapsto (e^{i \beta}\tilde{\phi}_1, e^{-i \beta} \tilde{\phi}_2,  \tilde{\phi}_3) \,,\\
	e^{\gamma \tilde{\mu}_3}:& \quad (\tilde{\phi}_1, \tilde{\phi}_2, \tilde{\phi}_3) \mapsto ( \tilde{\phi}_1, e^{i \gamma} \tilde{\phi}_2, e^{- i \gamma} \tilde{\phi}_3) \,.
\end{align}
\end{subequations}
The moduli space can be realised as a symplectic quotient
\begin{equation}
	\mathcal{M}^{H}_\mathcal{B} = \tilde{\mu}_1^{-1}(\zeta) / U(1) \,,
\end{equation}
where $U(1)$ is the gauge action~\eqref{eq:theory-B-gauge-flow}.
The flavour flows become degenerate at 
\begin{subequations}
\begin{align}
	e^{\beta \tilde{\mu}_2}:& \quad \tilde{\phi}_1 = \tilde{\phi}_2 = 0 \,,\\
	e^{\gamma \tilde{\mu}_3}:& \quad \tilde{\phi}_2 = \tilde{\phi}_3 = 0 \,, \\
	e^{\alpha (\tilde{\mu}_1 + \tilde{\mu}_2)}:& \quad \tilde{\phi}_1 = \tilde{\phi}_3 = 0 \,,
\end{align}
\end{subequations}
giving the respective degeneration loci on the space of $\tilde{\mu}_2$ and $\tilde{\mu}_3$ as 
\begin{subequations}
\begin{align}
	(0, -1):& \quad \tilde{\mu}_2 = 0 \,, \tilde{\mu}_3 \leq 0 \,,\\
	(1, 0):& \quad \tilde{\mu}_2 \leq 0 \,, \tilde{\mu}_3 = 0 \,, \\
	(-1, 1):& \quad \tilde{\mu}_2 = - \tilde{\mu}_3 \leq 0 \,.
\end{align}
\end{subequations}
The generating vectors $\{(0, -1), (1, 0), (-1, 1)\}$ are shown in Figure~\ref{fig:toric-diagram-B}.

When $\tilde{\zeta} > 0$, the moduli space $\mathcal{M}^{H}_\mathcal{B}$ is toric. These degeneration loci correspond to the one-dimensional cones in the toric fan $\tilde{\Delta}$.
The moduli space can be constructed from the fan $\tilde{\Delta}$ as
\begin{equation}
	\mathcal{M}^{H}_\mathcal{B} = \frac{\mathbb{C}^3 - F_{\tilde{\Delta}}}{\tilde{G} \times \tilde{\Gamma}} \,.
\end{equation}
The one-dimensional cones in the toric fan $\tilde{\Delta}$ are generated by three vectors $\{\tilde{v}_1, \tilde{v}_2, \tilde{v}_3\}$ which are identified with the charge vectors of the theory $\mathcal{A}$ via the relation~\eqref{eq:mirror-charges} between mirror charges,
\begin{subequations}
\begin{align}
	\tilde{v}_1 &= (1,0) \,,\\
	\tilde{v}_2 &= (-1,1) \,,\\
	\tilde{v}_3 &= (0, -1) \,.
\end{align}
\end{subequations}
Each pair of the vectors generate a two-dimensional cone in $\tilde{\Delta}$. But there is no three-dimensional cone generated by all three of $\{\tilde{v}_1, \tilde{v}_2, \tilde{v}_3\}$. Hence the subset $F_{\tilde{\Delta}}$ is given by
\begin{equation}
	F_{\tilde{\Delta}} = \left\{ \tilde{\phi}_1 = \tilde{\phi}_2 = \tilde{\phi}_3 = 0 \right\} \,.
\end{equation}
The generating vectors $\{\tilde{v}_1, \tilde{v}_2, \tilde{v}_3\}$ induce a map $\tilde{\psi}: \mathbb{C}^3 \to \mathbb{C}^2$ given by
\begin{equation}
	\tilde{\psi}: \quad (\tilde{\phi}_1, \tilde{\phi}_2, \tilde{\phi}_3) \mapsto (\tilde{\phi}_1 \tilde{\phi}_2^{-1}, \tilde{\phi}_2 \tilde{\phi}_3^{-1}) \,,
\end{equation}
the kernel $\ker \psi$ of which is the complexified gauge symmetry 
\begin{equation}
	G = \ker \tilde{\psi} = \mathbb{C}^* \,.
\end{equation}
acting via
\begin{equation}
	G: \quad (\tilde{\phi}_1, \tilde{\phi}_2, \tilde{\phi}_3) \mapsto (\lambda \tilde{\phi}_1, \lambda \tilde{\phi}_2, \lambda \tilde{\phi}_3) \,.
\end{equation}
The group $\Gamma$ is discrete given by
\begin{equation}
	\Gamma = \frac{\mathbb{Z}^2}{\operatorname{span}_\mathbb{Z} \{v_1, v_2, v_3\}} \cong 1 \,.
\end{equation}
The moduli space is then
\begin{equation}
	\mathcal{M}^\text{H}_\mathcal{B} = \frac{\mathbb{C}^3 - \{0\}}{\mathbb{C}^*} \cong \mathbb{CP}^2 \,.
\end{equation}

This toric space is projective and therefore can also be encoded as a polytope. The polytope $\tilde{\nabla}$ is the region bounded by hypersurfaces $\{\tilde{D}_1, \tilde{D}_2, \tilde{D}_3\}$ orthogonal to the generating vectors, as shown in Figure~\ref{fig:toric-diagram-B}. They may be obtained as the appropriate bounds on the flavour moment maps $\tilde{\mu}_1$ and $\tilde{\mu}_2$ given by
\begin{subequations}
\begin{align}
	\tilde{\mu}_2 \leq \zeta \,, \\
	\tilde{\mu}_2 - \tilde{\mu}_3 +  2\zeta \geq 0 \,, \\
	\tilde{\mu}_3 + \zeta \geq 0 \,.
\end{align}
\end{subequations}
On the hypersurfaces $\{\tilde{D}_1, \tilde{D}_2, \tilde{D}_3\}$, some of the chiral multiplets are forces to vanish according to
\begin{subequations}
\begin{align}
	\tilde{D}_1: \quad \tilde{\phi}_2 = \tilde{\phi}_3 = 0  \,, \\
	\tilde{D}_2: \quad \tilde{\phi}_1 = \tilde{\phi}_3 = 0 \,, \\
	\tilde{D}_3: \quad \tilde{\phi}_1 = \tilde{\phi}_2 = 0 \,.
\end{align}
\end{subequations}
The hypersurfaces $\{\tilde{D}_1, \tilde{D}_2, \tilde{D}_3\}$ can be identified with the hypersurfaces $\{D_1, D_2, D_3\}$ in~\eqref{eq:theory-A-hypersurfaces} for the mirror theory with appropriate parameter maps. Then the polytope $\tilde{\nabla}$ is identified with the region $C_1$ where the Coulomb branch of theory $\mathcal{A}$ is isomorphic
\begin{equation}
	\mathcal{M}^\text{C}_\mathcal{A} = \mathcal{M}^\text{H}_\mathcal{B} = \mathbb{CP}^2 \,.
\end{equation}

\subsubsection{Bethe Ansatz Equations}

Taking the one-loop determinant \eqref{eq:CP2-one-loop-det} for the theory $\mathcal{A}$, the exponent 
\begin{equation}
	i B_a = \frac{\partial \log Z}{\partial \mathfrak{m}_a}
\end{equation}
of the Bethe ansatz equations are
\begin{subequations}
\begin{align}
	i B_1 &= \log q_1 + \kappa_{11} \log x_1 + \kappa_{12} \log x_2 + \kappa_{13} \log y_3  
	\quad + \nonumber \\
	& \hspace{31ex} 
	\log \left( \frac{ x_1^{\sfrac{1}{2}} }{ 1 - x_1 } \right) 
	- \log \left( \frac{ x_1^{\sfrac{-1}{2}} x_2^{\sfrac{1}{2}} }{ 1 - x_1^{-1} x_2 } \right) 
	\,, \\
	i B_2 &= \log q_2 + \kappa_{21} \log x_1 + \kappa_{22} \log x_2 + \kappa_{23} \log y_3 
	\quad + \nonumber \\
	& \hspace{28ex} \log \left( \frac{ x_1^{\sfrac{-1}{2}} x_2^{\sfrac{1}{2}} }{ 1 - x_1^{-1} x_2 } \right) - \log \left( \frac{ x_2^{\sfrac{-1}{2}} y_{3}^{\sfrac{1}{2}} }{ 1 - x_2^{-1} y_3 } \right) \,.
\end{align}
\end{subequations}
The Bethe ansatz equations $e^{i B_a} = 1$ are
\begin{subequations}
\begin{align}
	1 = e^{i B_1} &=  q_1 
	x_1^{\kappa_{11}} 
	x_2^{\kappa_{12}}
	y_3^{\kappa_{13}}
	\left( \frac{ x_1^{\sfrac{1}{2}} }{ 1 - x_1 } \right) 
	\left( \frac{ x_1^{\sfrac{-1}{2}} x_2^{\sfrac{1}{2}} }{ 1 - x_1^{-1} x_2 } \right)^{-1} 
	\,, \\
	1 = e^{i B_2} &= q_2 
	x_1^{\kappa_{12}} 
	x_2^{\kappa_{22}}
	y_3^{\kappa_{23}}
	\left( \frac{ x_1^{\sfrac{-1}{2}} x_2^{\sfrac{1}{2}} }{ 1 - x_1^{-1} x_2 } \right) 
	\left( \frac{ x_2^{\sfrac{-1}{2}} y_{3}^{\sfrac{1}{2}} }{ 1 - x_2^{-1} y_3 } \right)^{-1} \,.
\end{align}
\end{subequations}
Rearranging and relabelling $y_3 \mapsto y$ gives
\begin{subequations}
\label{eq:CP2-BAE}
\begin{align}
	q_1 &=  x_1^{-\kappa_{11} - 1} x_2^{-\kappa_{12} + \sfrac{1}{2}} (1 - x_1) (1 - x_1^{-1} x_2)^{-1} y^{-\kappa_{13}}\,, \\
	q_2 &=  x_1^{-\kappa_{21} + \sfrac{1}{2}} x_2^{-\kappa_{22} - 1} (1 - x_1^{-1} x_2) (1 - x_2^{-1} y)^{-1} y^{-\kappa_{23} + \sfrac{1}{2}} \,.
\end{align}
\end{subequations}

We can send $x_1 \mapsto x_1^{-1}$, $x_2 \mapsto x_2^{-1}$, and $y \mapsto y^{-1}$ to examine solutions at infinity. The effect is flipping the signs on \index{Chern-Simons level} Chern-Simons levels, up to an overall factor of $(-1)^{K}$. The results for this case $K=2$ are
\begin{subequations}
\begin{align}
	q_1 &=  x_1^{\kappa_{11} - 1} x_2^{\kappa_{21} + \sfrac{1}{2}} (1 - x_1) (1 - x_1^{-1} x_2)^{-1} y^{\kappa_{13}} \,, \\
	q_2 &= x_1^{\kappa_{12} + \sfrac{1}{2}} x_2^{\kappa_{22} - 1} (1 - x_1^{-1} x_2) (1 - x_2^{-1} y)^{-1} y^{\kappa_{23} + \sfrac{1}{2}} \,,
\end{align}
\end{subequations}
which are equivalent to \eqref{eq:CP2-BAE}. Substituting in the critical Chern-Simons levels from \eqref{eq:CP2-CS-levels} produces
\begin{subequations}
\begin{align}
	q_1 &= (1 - x_1) (1 - x_1^{-1} x_2)^{-1} y^{\kappa_{13}} \,, \\
	q_2 &= (1 - x_1^{-1} x_2) (1 - x_2^{-1} y)^{-1} y^{\kappa_{23} + \sfrac{1}{2}} \,.
\end{align}
\end{subequations}

To eliminate factors of $y$, we can set the mixed gauge-flavour \index{Chern-Simons level} Chern-Simons levels to be critical at $k_{13}=0$ and $k_{23}=-\frac{1}{2}$, i.e.,
\begin{equation}
\label{eq:CP2-CS-levels-full}
	\kappa_{ij} =
	\begin{pmatrix}
		1 & -\frac{1}{2} & 0 \\
		-\frac{1}{2} & 1 & -\frac{1}{2} \\
		0 & -\frac{1}{2} & 1 \,.
	\end{pmatrix}
	\,.
\end{equation}
Then the Bethe ansatz equations become
\begin{subequations}
\begin{align}
	q_1 &= (1 - x_1) (1 - x_1^{-1} x_2)^{-1}  \,, \\
	q_2 &= (1 - x_1^{-1} x_2) (1 - x_2^{-1} y)^{-1}  \,.
\end{align}
\end{subequations}

\subsubsection{Twisted Index}
For the theory $\mathcal{A}$, the hessian factors
\begin{equation}
	H_a = \frac{\partial^2 \log Z}{\partial \log x_a \partial \mathfrak{m}_a} = \frac{\partial i B_a}{\partial \log x_a}
\end{equation}
are computed to be
\begin{subequations}
\begin{align}
	H_1 &= \kappa_{11} + 1 + \frac{x_1}{1-x_1} + \frac{x_1^{-1} x_2}{1 - x_1^{-1} x_2} \,, \\
	H_2 &= \kappa_{22} + 1 + \frac{x_1^{-1} x_2}{1 - x_1^{-1} x_2} + \frac{x_2^{-1} y}{1 - x_2^{-1} y} \,.
\end{align}
\end{subequations}

The contour integral formula for the twisted index is then
\begin{equation}
	\mathcal{I} = \sum_{\mathfrak{m}_1} \sum_{\mathfrak{m}_2} \left( \frac{1}{2 \pi i} \right)^2 \oint_\text{JK} \frac{\dd x_1}{x_1} \frac{\dd x_2}{x_2} \, H_1^g H_2^g Z(x_1,x_2; \mathfrak{m}_1, \mathfrak{m}_2) \,,
\end{equation}
where the integral is evaluated with the Jeffrey-Kirwan prescription.

Consider the case $g=0$ where the hessian factor can be ignored. Take the gauge Chern-Simons levels to be the critical levels in \eqref{eq:CP2-CS-levels}. Set \index{R-charge} R-charges to $r_i = 0$ to compare with the mirror theory of R-charges $\tilde{r}_i=1$. The one-loop determinant \eqref{eq:CP2-one-loop-det} becomes
\begin{align}
	Z
	=& \left( q_1^{\mathfrak{m}_1} q_2^{\mathfrak{m}_2} \right)		
	\left( x_1^{\mathfrak{m}_1 - \frac{1}{2}\mathfrak{m}_2} 
	x_2^{-\frac{1}{2}\mathfrak{m}_1 + \mathfrak{m}_2}
	y_3^{\kappa_{13}\mathfrak{m}_1 + \kappa_{23}\mathfrak{m}_2} \right) 
	\left( x_1^{-\kappa_{R1}} 
	x_2^{-\kappa_{R2}}
	y_3^{-\kappa_{R3}} \right) 
	\quad \times \nonumber \\
	& \hspace{7ex} \left( \frac{ x_1^{\sfrac{1}{2}} }{ 1 - x_1 } \right)^{\mathfrak{m}_1 + 1}
	\left( \frac{ x_1^{\sfrac{-1}{2}} x_2^{\sfrac{1}{2}} }{ 1 - x_1^{-1} x_2 } \right)^{-\mathfrak{m}_1 + \mathfrak{m}_2 + 1}   
	\left( \frac{ x_2^{\sfrac{-1}{2}} y_{3}^{\sfrac{1}{2}} }{ 1 - x_2^{-1} y_3 } \right)^{-\mathfrak{m}_2 + 1} \,.
\end{align}

The contour integral is then
\begin{align}
	\mathcal{I} = \sum_{\mathfrak{m}_1, \mathfrak{m}_2} 
	& q_1^{\mathfrak{m}_1} q_2^{\mathfrak{m}_2} 
	y_3^{\kappa_{13} \mathfrak{m}_1 + (\kappa_{23} - \sfrac{1}{2}) \mathfrak{m}_2 + \sfrac{1}{2} - \kappa_{R3}} \left( \frac{1}{2 \pi i} \right)^2
	\quad \times \nonumber \\  
	& \hspace{4ex} \oint_\text{JK} \frac{\dd x_1}{x_1} \frac{\dd x_2}{x_2} \, 
	x_1^{2 \mathfrak{m}_1 - \mathfrak{m}_2  - \kappa_{R1}} x_2^{- \mathfrak{m}_1 + 2 \mathfrak{m}_2  - \kappa_{R2}}
	\quad \times \nonumber \\
	& \hspace{11ex} \left(\frac{1}{1-x_1}\right)^{\mathfrak{m}_1 + 1} 
	\left(\frac{1}{1-\frac{x_2}{x_1}}\right)^{-\mathfrak{m}_1 + \mathfrak{m}_2 + 1} 
	\left(\frac{1}{1-\frac{y_3}{x_2}}\right)^{-\mathfrak{m}_2 + 1} \,.
\end{align}

The Jeffrey-Kirwan charges $(Q_1,Q_2)$ for the three denominator factors are respectively
\begin{align*}
	Q^{x_1=1} &= (1,0) \,, \\
	Q^{x_1=x_2} &= (-1,1) \,, \\
	Q^{x_2=y_3} &= (0,-1) \,,
\end{align*}
which are responsible for the the interior poles at $(1,1)$, $(1,y_3)$, and $(y_3,y_3)$.
Taking $\eta=(1,1)$ selects only the residue at $(x_1,x_2)=(1,1)$. 

However, for $\mathcal{N}=2$ theory there are potentially additional residues from the topological vacua at $(0,0)$, $(0,y_3)$, $(0,\infty)$, $(1,0)$, $(1,y_3)$, $(1,\infty)$, $(\infty,0)$, $(\infty,y_3)$, and $(\infty,\infty)$. The Jeffrey-Kirwan charges for the $x_a$ factor can be assigned analogously to the $U(1)$ theory~\cite{xu2022} as follows
\begin{subequations}
\begin{align}
	Q^{x_a=0}_b = - \delta_{ab} \kappa_{aa}^\eff(\sigma_a \mapsto \infty) \,, \\
	Q^{x_a=\infty}_b = \delta_{ab} \kappa_{aa}^\eff(\sigma_a \mapsto -\infty) \,.
\end{align}
\end{subequations}
The resulting normalised charge vectors for this $U(1)^2$ theory $\mathcal{A}$re 
\begin{align*}
	Q^{x_1=0} &= (-1,0) \,,\\
	Q^{x_1=\infty} &= (1,0) \,,\\
	Q^{x_2=0} &= (0,-1) \,,\\
	Q^{x_2=\infty} &= (0,1) \,.
\end{align*}
Taking into account both interior and boundary poles, $\eta=(1,1)$ selects residues at $(1,1)$, $(1,\infty)$, and $(\infty,\infty)$.

The residues can be computed following the procedure in Appendix~\ref{chap:multivariate-JK}. The Jeffrey-Kirwan charge vectors are ordered anti-clockwise. For simplicity, these mixed \index{Chern-Simons level} Chern-Simons levels $\kappa_{R1}$ and $\kappa_{R2}$ are set to zero, which only shifts the residues at boundaries into different magnetic sectors. The index is then
\begin{equation}
\label{eq:theory-A-index-g0-r0}
	\frac{y_3^{\sfrac{1}{2}-\kappa_{R3}}}{1-y_3} \,,
\end{equation}
which is simply the zeroth sector contribution from the pole at $(1,1)$.

In the mirror theory $\mathcal{B}$, the classical and one-loop determinant~\eqref{eq:theory-B-one-loop-det} at $r_i=1$ is 
\begin{align}
	Z 
	=& \tilde{q}^\mathfrak{\tilde{m}} \tilde{x}^{\tilde{\kappa}_{11} \mathfrak{\tilde{m}}} \tilde{y}_2^{\tilde{\kappa}_{12} \mathfrak{\tilde{m}}} \tilde{y}_3^{\tilde{\kappa}_{13} \mathfrak{\tilde{m}}}
	\left( \tilde{x}_1^{-\tilde{\kappa}_{R1}} 
	\tilde{y}_2^{-\tilde{\kappa}_{R2}}
	\tilde{y}_3^{-\tilde{\kappa}_{R3}} \right) 
	\quad \times \nonumber \\
	& \hspace{20ex} \left( \frac{ \tilde{x}^{\sfrac{1}{2}} \tilde{y}_2^{\sfrac{1}{2}} }{ 1 - \tilde{x} \tilde{y}_2 } \right)^\mathfrak{\tilde{m}}
	\left( \frac{ \tilde{x}^{\sfrac{1}{2}} \tilde{y}_2^{\sfrac{-1}{2}} \tilde{y}_3^{\sfrac{1}{2}} }{ 1 - \tilde{x} \tilde{y}_2^{-1} \tilde{y}_3 } \right)^\mathfrak{\tilde{m}}
	\left( \frac{ \tilde{x}^{\sfrac{1}{2}} \tilde{y}_{3}^{\sfrac{-1}{2}} }{ 1 - \tilde{x} \tilde{y}_3^{-1} } \right)^\mathfrak{\tilde{m}} \,.
\end{align}
To compare with the mirror, set the gauge-flavour  Chern-Simons levels to be critical at $\tilde{\kappa}_{11} = -\frac{3}{2}$, $\tilde{\kappa}_{12} = -1$, and $\tilde{\kappa}_{13} = -\frac{1}{2}$. For simplicity, set $\tilde{\kappa}_{R1}=0$. The index at $g=0$ and  $r_i=1$ is then
\begin{equation}
	-\frac{\tilde{q} \tilde{y}_2^{-\kappa_{R2}} \tilde{y}_3^{-\kappa_{R3}} }{1 - \tilde{q}} \,.
\end{equation}
With the mirror map $\tilde{q} \rightarrow 1/y_3$, it becomes
\begin{equation}
\label{eq:theory-B-index-g0-r1}
	\frac{\tilde{y}_2^{-\kappa_{R2}} \tilde{y}_3^{-\kappa_{R3}} }{1 - y_3} \,,
\end{equation}
which matches \eqref{eq:theory-A-index-g0-r0} by setting
\begin{align*}
	\kappa_{R3}=\frac{1}{2} \,,\\
	\tilde{\kappa}_{R2}=0 \,,\\
	\tilde{\kappa}_{R3}=0 \,.\\
\end{align*}

\subsection{$U(1)$ with Two Chiral Fields}
Now consider simpler $U(1)$ mirror theories with two chiral multiplets. Take the theory $\mathcal{A}$ with charge matrix
\begin{equation}
	Q^i_{\ \ j} = 
	\begin{pmatrix}
		1 & -1 \\
		0 & 1
	\end{pmatrix} \,,
\end{equation}
masses $m_1=m_2=-m \leq 0$, and gauge Chern-Simons level $\kappa_{11}=1$ critical at $-m \leq \sigma \leq m$. The Coulomb branch opens at $\sigma \leq -m$ for $\zeta = 0$. The mirror theory $\mathcal{B}$ has charge matrix
\begin{equation}
	\tilde{Q}^i_{\ \ j} = 
	\begin{pmatrix}
		1 & 1 \\
		0 & 1
	\end{pmatrix} \,,
\end{equation}
masses $\tilde{m}_1=\tilde{m}_2=-\tilde{m} \leq 0$, and gauge Chern-Simons level $\tilde{\kappa}_{11}=-1$ critical at $\tilde{\sigma} \geq \tilde{m}$. The Higgs branch opens at $\tilde{\sigma}=\tilde{m}$.

The one-loop determinants are respectively
\begin{align}
	Z^A =& q^\mathfrak{m} x^\mathfrak{m} y_2^{\kappa_{12}\mathfrak{m}} x^{\kappa_{R1} (g-1)} y_2^{\kappa_{R2} (g-1)} 
	\quad \times \nonumber \\
	& \hspace{14ex} \left( \frac{x^\frac{1}{2}}{1-x} \right)^{\mathfrak{m}+(r_1-1)(g-1)}
	\left( \frac{ x^{-\frac{1}{2}} y_2^{\frac{1}{2}}  }{1 - x^{-1} y_2 }\right)^{\mathfrak{m}+(r_2-1)(g-1)} \,,
\end{align}
and
\begin{align}
	Z^B =& \tilde{q}^\mathfrak{\tilde{m}} \tilde{x}^{-\mathfrak{\tilde{m}}} \tilde{y}_2^{\tilde{\kappa}_{12}\mathfrak{\tilde{m}}} \tilde{x}^{\tilde{\kappa}_{R1} (g-1)} \tilde{y}_2^{\tilde{\kappa}_{R2} (g-1)} 
	\quad \times \nonumber \\
	& \hspace{16ex}	\left( \frac{\tilde{x}^\frac{1}{2}}{1-\tilde{x}} \right)^{\mathfrak{\tilde{m}}+(\tilde{r}_1-1)(g-1)}
	\left( \frac{ \tilde{x}^{\frac{1}{2}}  \tilde{y}_2^{\frac{1}{2}} }{1 - \tilde{x} \tilde{y}_2} \right)^{\mathfrak{\tilde{m}}+(\tilde{r}_2-1)(g-1)} \,.
\end{align}

For simplicity, $\kappa_{R1}=\tilde{\kappa}_{R1}=0$ is taken, which only shifts the boundary residues into different magnetic sectors. Taking $\eta=1$ for both theories selects the poles at $x=1$, $x=\infty$, $\tilde{x}=1$, and $\tilde{x}=\tilde{y}_2^{-1}$. The twisted indices at $g=0$ are then computed to be
\begin{equation}
	\mathcal{I}^A_{r_i=0} = -\frac{y^{\sfrac{1}{2}+\kappa_{R2}}}{1 - y} \,,
\end{equation}
and
\begin{align}
	\mathcal{I}^B_{r_i=1} &= -\frac{\tilde{q} \tilde{y}^{\sfrac{1}{2}+ \tilde{\kappa}_{12} + \tilde{\kappa}_{R2}}}{1 - \tilde{q} \tilde{y}^{\frac{1}{2} \tilde{\kappa}_{12}}} \\
	&\mapsto -\frac{y q^{\sfrac{1}{2}+ \tilde{\kappa}_{12} + \tilde{\kappa}_{R2}}}{1 - y q^{\frac{1}{2} \tilde{\kappa}_{12}}} \,,
\end{align}
where $\mathcal{I}^B_{r_i=1}$ is mapped under $\tilde{q} \mapsto y$ and $\tilde{y} \mapsto q$ in the last line. They agree with each other when $\kappa_{R2}=\frac{1}{2}$, $\tilde{\kappa}_{12}=-\frac{1}{2}$, and $\tilde{\kappa}_{R2}=0$.

\appendix
\chapter{Mutivariate Jeffrey-Kirwan Integral}
\label{chap:multivariate-JK}

In this Chapter we develop the computational techniques for multivariate Jeffrey-Kirwan residues.
\vspace{-\parskip}
\begin{itemize}
	\item The basics of multivariate residues is reviewed in Section~\ref{sec:multivariate-residue}. 
	\item We then propose a procedure to compute the multivariate Jeffrey-Kirwan residues in Section~\ref{sec:multivariate-JK-residue}.
	\item The conjecture is verified for twisted indices of abelian quiver gauge theories in Section~\ref{sec:example-abelian-quiver}.
\end{itemize}

\section{Multivariate Residue}
\label{sec:multivariate-residue}

Consider a meromorphic $n$-form
\begin{equation}
	\omega = \frac{h(z) \, \dd z_1 \wedge \ldots \wedge \dd z_n}{f_1(z) \ldots f_n(z)} \,,
\end{equation}
where $h(z): \mathbb{C}^n \rightarrow \mathbb{C}$ and $f(z)=(f_1(z), \ldots, f_n(z)): \mathbb{C}^n \rightarrow \mathbb{C}^n$ are holomophic functions.

\begin{defn}
	A pole of the meromorphic $n$-form $\omega$ is~\cite{larsen2019} a point $p \in \mathbb{C}^n$ where $f$ has an isolated zero, i.e.,
	\begin{equation}
		f(p) = 0 \,,
	\end{equation}
	and 
	\begin{equation}
		f^{-1}(0) \cap U = {p} \,,
	\end{equation}
	for a sufficiently small neighbourhood $U$ of $p$. 
\end{defn}

\begin{defn}
	The residue at a pole $p$ is defined as an integral over a product of $n$ circles, i.e., an $n$-torus,
	\begin{equation}
		\Res_p (\omega) := \frac{1}{(2 \pi i)^n} \oint_{\Gamma_\epsilon} \frac{h(z) \, \dd z_1 \wedge \ldots \wedge \dd z_n}{f_1(z) \ldots f_n(z)} \,,
	\end{equation}
	where $\Gamma_\epsilon := \{z \in \mathbb{C}^n \mid |f_i(z)|=\epsilon_i\}$ is the pre-image of an $n$-torus under $f$, and the integration cycle is oriented such that
	\begin{equation}
		\dd (\arg f_1) \wedge \ldots \wedge \dd (\arg f_n) \geq 0.
	\end{equation}
\end{defn}

It can be generalised to the case where there are different number of denominator factors $f(z)=(f_1(z), \ldots, f_m(z))$ than the number $n$ of variables. For $m<n$, the relevant construction is called a residual form. For $m>n$, the denominator factors need to be grouped into exactly $n$ partitions. The residue also depends the partitioning, in addition to the pole. Hence, unlike univariate residues, a generic multivariate residue is not solely determined by the pole.

When the jacobian determinant at a pole $p$
\begin{equation}
	J(p) := \det_{i,j} \left(\frac{\partial f_i}{\partial z_j} \right) \eval_{z=p}
\end{equation}
is non-vanishing, the residue is said to be non-degenerate. Non-degenerate residues can be directly evaluated by a coordinate transformation~\cite{arkani-hamed2009} $u = f(z)$ as
\begin{equation}
	\Res_p (\omega) = \frac{1}{(2 \pi i)^n} \oint_{|u_i| \leq \epsilon_i} \frac{h(f^{-1} (u)) \, \dd u_1 \wedge \ldots \wedge \dd u_n}{J(p) u_1(z) \ldots u_n(z)} = \frac{h(p)}{J(p)} \,.
\end{equation}
However, this formula immediately breaks down for higher order poles as they are degenerate.

To evaluate a generic residue, we need to utilise the transformation law~\cite[p.657-658]{griffiths1978}, which is general property of residues.

\begin{thm}[Transformation Law]
\label{thm:transformation-law}
Let $I = \langle f_1(z), \ldots, f_n(z) \rangle$ be the ideal generated by a finite set of holomorphic functions $f_i$ such that the solution to $$f_1(z)=\ldots=f_n(z)=0$$ is a finite set of points $\{p,\ldots, q\}$, i.e., zero-dimensional. Suppose the zero-dimensional ideal $J = \langle g_1(z), \ldots, g_n(z) \rangle$ is a subspace $J \subseteq I$. Then $J$ is related to $I$ by a holomorphic matrix $A$ such that
\begin{equation}
	g_i(z) = \sum_j a_{ij} f_j(z) \,.
\end{equation}
Then the residue at $p$ satisfies
\begin{equation}
\label{eq:transformation-formula}
	\Res_p \left( \frac{h(z) \, \dd z_1 \wedge \ldots \wedge \dd z_n}{f_1(z) \ldots f_n(z)} \right)
	= \Res_p \left( \frac{\det(A)(z) h(z)  \, \dd z_1 \wedge \ldots \wedge \dd z_n}{g_1(z) \ldots g_n(z)} \right) \,.
\end{equation}
\end{thm}

The transformation formula~\eqref{eq:transformation-formula} can simplify the computation of a multivariate residue to the product of univariate residues, by choosing all of $g_1,\ldots,g_n$ to be univariate. A set of these univariate polynomials can be obtained from the Gr\"obner bases of $\{f_1(z), \ldots, f_n(z)\}$ with different lexicographic monomial orders. 

A univariate polynomial $g_i(z_i)$ in $z_i$ is taken to be the first element of the Gr\"obner basis generated with the order $z_{i+1} \succ z_{i+2} \succ \cdots \succ z_n \succ z_1 \succ \cdots \succ z_{i}$. By computing all cyclic permutations of this ordering, we obtain a set of $n$ univariate polynomials $\{g_1(z_1),\ldots,g_n(z_n)\}$. 

The transformation matrix $A$ can be obtained using the algorithm implemented in~\cite{lichtblau2014}. Shown in Algorithm~\ref{listing:mathematica} is the Mathematica code for two ideal generators in two variables. The matrix $A$ is assembled row by row by taking the first row of \texttt{tT} with the corresponding lexicographic ordering. An improved version of this method is used by the function \texttt{MultivariateResidue}~\cite{larsen2019}.

\renewcommand{\lstlistingname}{Algorithm}
\lstset{ 
    language=Mathematica,
    basicstyle=\small\sffamily,
    numbers=left,
    numberstyle=\tiny,
    frame=tb,
    tabsize=4,
    columns=fixed,
    showstringspaces=false,
    showtabs=false,
    keepspaces,
    commentstyle=\color{gray},
}

\begin{lstlisting}[frame=single, caption={Mathematica Code for Computing Transformation Matrix}, label=listing:mathematica, numbers=left]
moduleGroebnerBasis[polys_, vars_, cvars_, opts___] := Module[
  {newpols, rels, len = Length[cvars], gb, j, k, ruls},
  rels = Flatten[Table[cvars[[j]]*cvars[[k]], {j, len}, {k, j, len}]];
  newpols = Join[polys, rels];
  gb = GroebnerBasis[newpols, Join[cvars, vars], opts];
  rul = Map[(# :> {}) &, rels];
  gb = Flatten[gb /. rul];
  Collect[gb, cvars]
  ]
  
fF = {f[1], f[2]}		(* set of ideal generators *)
vars = {x_2, x_1};	(* lexicographic ordering of variables *)

(* encode positions of ideal generators in a matrix *)
coords = Array[ee, 3];
fmat = {{fF[[1]], 1, 0}, {fF[[2]], 0, 1}};	
newfF = fmat . coords;

mgb = moduleGroebnerBasis[newfF, vars, coords];
mgb = Select[mgb, Coefficient[#1, coords[[1]]] =!= 0 &];
gG = (Coefficient[#1, coords[[1]]] &) /@ mgb	(* groebner basis *)
sS = First /@ PolynomialReduce[fF, gG, vars];

(* check with built-in groebner basis *)
gb = GroebnerBasis[fF, vars];	
gb === gG
rul = {ee[1] -> 1, ee[2] -> -fF[[1]], ee[3] -> -fF[[2]]};
Map[Expand[# /. rul] &, mgb]	(* want zeroes *)
Expand[sS . gG - fF]	(* want zeroes *)

tT = Outer[D, mgb, Rest[coords]]
Expand[gG - tT . fF]	(* want zeroes *)
\end{lstlisting}

\section{Jeffrey-Kirwan Prescription}
\label{sec:multivariate-JK-residue}

Consider the $n$-form 
\begin{equation}
	\omega = \frac{h(z) \, \dd z_1 \wedge \ldots \wedge \dd z_n}{f_1(z) \ldots f_m(z)} \,,
\end{equation}
where $m \geq n$, and $f_i$ are linear functions in $z_1,\ldots,z_n$. Following~\cite{ferro2018}, the  \index{Jefrrey-Kirwan residue} Jeffrey-Kirwan residue of $\omega$ is defined as
\begin{equation}
\label{eq:linear-JK}
	\text{JK-Res}_p (\omega) = \frac{1}{|J(p)_{f_i \in F}|} \frac{h(p)}{\prod_{f_i \not\in F} f_i} \,,
\end{equation}
where $F$ is a set of exactly $n$ factors responsible for the pole at $p$. Given a fixed charge vector $\eta$, the Jeffrey-Kirwan prescription dictates to only take the sum of residues at those poles such that the charge vectors of the responsible factors form a convex cone containing $\eta$. Moreover, the final result is independent of the choice of $\eta$.

However, this formula is only valid if all denominator factors are linear functions. Hence it is desirable to build a dictionary to evaluate Jeffrey-Kirwan residues in terms of multivariate residues. The idea is that the Jeffrey-Kirwan prescription determines the partitioning of denominator factors. 

\begin{conj}
\label{conj:multivariate-JK}
The following procedure is conjectured to compute the Jeffrey-Kirwan residue at a given contributing pole $p$.
\begin{enumerate}
\item Determine the denominator factors $f_1, \ldots, f_n$ responsible for this pole.
\item Split the denominator factors into $n$ partitions such that each responsible factor is placed in a separate partition.
\item Compute the multivariate residue of the resulting $n$-form using the transformation law~\eqref{eq:transformation-formula}.
\end{enumerate}

The result of the computation gives the individual Jeffrey-Kirwan residue up to a sign, which is determined by the ordering of the denominator factors.
\end{conj}

For two-forms, we can order the charge vectors of the denominator factors by their polar angles. This gives the correct combination of residues as in the Jeffrey-Kirwan prescription. However, for higher forms, it is yet to know how to order the charge vectors correctly.

\section{Examples}
\label{sec:example-abelian-quiver}

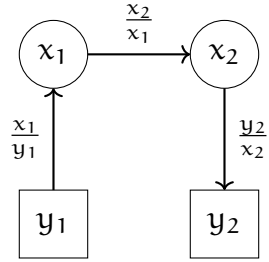
\begin{figure}[!h]
\centering
\caption{$A_2$ Linear Quiver Diagram}
\begin{tikzpicture}[scale=0.75]
	\draw (-2.1,-2.1) rectangle (-0.9,-0.9);
	\draw[->,thick] (-1.5,-0.9) -- node[left] {$\frac{x_1}{y_1}$} (-1.5,0.9);
	\draw (-1.5,1.5) circle (0.6);
	\draw[->,thick] (-0.9,1.5) -- node[above] {$\frac{x_2}{x_1}$} (0.9,1.5);
	\draw (1.5,1.5) circle (0.6);
	\draw[->,thick] (1.5,0.9) -- node[right] {$\frac{y_2}{x_2}$} (1.5,-0.9);
	\draw (2.1,-2.1) rectangle (0.9,-0.9);
	
	\node at (-1.5,-1.5) {$y_1$};
	\node at (-1.5,1.5) {$x_1$};
	\node at (1.5,1.5) {$x_2$};
	\node at (1.5,-1.5) {$y_2$};
\end{tikzpicture}
\label{fig:A2-quiver}
\end{figure}

Consider the abelian $A_2$ linear quiver gauge theory with $\mathcal{N}=4$ in three dimensions, as shown in Fig~\ref{fig:A2-quiver}. At genus $g=0$, the B-twisted index $Z_{g=0,B}^{[A_2]}$ is expected to be identical to the A-twisted index $Z_{g=0,A}^{\text{SQED}[3]}$ of the quantum electrodynamics of three hypermultiplets in the \index{mirror symmetry} mirror, and vice versa. 

\subsection{B-Twist of Abelian Linear Quiver}
First consider the mirror of the $A_2$ linear quiver, which is the supersymmetric quantum electrodynamics with three hypermultiplets. The contour integral formula for the A-twisted index is~\cite[(6.35)]{closset2016}
\begin{align}
	Z_{g=0,A}^{\text{SQED}[3]} 
	=& -(t-t^{-1})^{-1} \sum_{m} q^{m}   
	\quad \times \nonumber \\
	& \hspace{10ex}
	\frac{1}{2 \pi i} \oint_{\text{JK}} \frac{\dd x}{x}  \,  \left(\frac{x - y_1 t}{y_1 - x t}\right)^{m}
	\left(\frac{x - y_2 t}{y_2 - x t}\right)^{m}
	\left(\frac{x - y_3 t}{y_3 - x t}\right)^{m} \,.
\end{align}
Taking $\eta=1$ selects the residues at $y_1/t$, $y_2/t$, and $y_3/t$, which sum to~\cite[(6.40)]{closset2016}
\begin{equation}
Z_{g=0,A}^{\text{SQED}[3]} = \frac{t^{-1}(1-t^{-6})}{(1-t^{-2})(1-q t^{-3})(1-q^{-1} t^{-3})} \,.
\end{equation}
After the change of variables $t \mapsto t^{-1}$, and $q \mapsto \frac{y_1}{y_2}$, the A-twisted index reads
\begin{equation}
\label{eq:Z-SQED-A}
Z_{g=0,A}^{\text{SQED}[3]} = - \frac{(t+t^3+t^5) y_1 y_2}{(t^3 y_1 - y_2) (t^3 y_2 - y_1)} \,.
\end{equation}
We would like to compute the B-twisted index $Z_{g=0,B}^{[A_2]}$ of the quiver from Jeffrey-Kirwan prescription and verify if it is identical to this expression.

The contour integral formula of the index $Z_{g=0,B}^{[A_2]}$ is 
\begin{align}
	& Z_{g=0,B}^{[A_2]} \nonumber \\
	= & (t-t^{-1})^2 \sum_{m_1, m_2} q_1^{m_1} q_2^{m_2} 
	\frac{1}{(2 \pi i)^2} \oint_{\text{JK}} \frac{\dd x_1}{x_1} \frac{\dd x_2}{x_2} \, \nonumber \\ 
	& \qquad \left(\frac{x_1 - y_1 t}{y_1 - x_1 t}\right)^{m_1}
	\left(\frac{x_2 - y_2 t}{y_2 - x_2 t}\right)^{m_2}
	\left(\frac{x_1 - x_2 t}{x_2 - x_1 t}\right)^{m_1 - m_2} 
	\quad \times \nonumber \\
	& \qquad \frac{x_1 y_1 t}{(x_1 - y_1 t)(y_1 - x_1 t)}
	\frac{x_2 y_2 t}{(x_2 - y_2 t)(y_2 - x_2 t)}
	\frac{x_1 x_2 t}{(x_2 - x_1 t)(x_1 - x_2 t)} \,.
\end{align}

The only contributing sector is expected to be the zeroth sector $m_1 = 0, m_2 = 0$, which has the integrand
\begin{equation}
\omega_{0,0} = \frac{t (1-t^2)^2 y_1 y_2 x_1 x_2 \, \dd x_1 \dd x_2}{(x_1 - y_1 t)(y_1 - x_1 t)(x_2 - y_2 t)(y_2 - x_2 t)(x_2 - x_1 t)(x_1 - x_2 t)} \,.
\end{equation}
The six denominator factors have charge vectors $(Q_1, Q_2)$ listed below
\begin{equation}
\label{eq:charges}
	(-1,0), (1,0), (0,-1), (0,1), (1,-1), (-1,1) \,.
\end{equation}
Note that its Jeffrey-Kirwan residue can still be computed using~\eqref{eq:linear-JK}.

Given the charge vector $\eta = (1,1)$, the Jeffrey-Kirwan prescription picks the following three poles
$$
(y_1 / t, y_2 / t), (y_1 t, y_1 / t^2), (y_2 / t^2, y_2 / t) \,,
$$
whose charge vectors form cones containing $\eta$.

Consider first the pole at $(y_1 / t, y_2 / t)$ for the procedure in Conjecture~\ref{conj:multivariate-JK}. The responsible denominator factors are $(y_1 - x_1 t)$ and $(y_2 - x_2 t)$. We then split all denominator factors into two partitions $\{f_1, f_2\}$, each containing one responsible factor, say
$$
\{f_1, f_2\} = \{ (y_1 - x_1 t),  (y_2 - x_2 t)  (x_1 - y_1 t)(x_2 - y_2 t)(x_2 - x_1 t)(x_1 - x_2 t) \} \,.
$$

Following the transformation law~\ref{thm:transformation-law},  first we compute the Gr\"obner bases of $f_1$ and $f_2$. Taking the order $x_2 \succ x_1$ gives a basis whose first entry only depends on $x_1$, while taking the order $x_1 \succ x_2$ gives another basis whose first entry depends only on $x_2$. These two entries are then taken to be the new denominator factors $g_1$ and $g_2$ given by
\begin{align*}
	g_1 =& - y_1 + x_1 t \\
	g_2 =& \left(-1 + t^2\right) y_1 
	\quad \times \\
	& \qquad \left[-y_1^2 + \left(1 + t^2\right) y_1 x_2 - t^2 x_2^2\right]
	\quad \times \\
	& \qquad \qquad \left[y_2 x_2 + t^2 y_2 x_2 - t \left(y_2^2 + x_2^2\right)\right]
	\,.
\end{align*}
The transformation matrix $A$ taking $(f_1, f_2)$ to $(g_1, g_2)$ is found to be 
\begin{equation}
	A = 
	\begin{pmatrix}
		-1 & 0 \\
		 a_{21} & -t^2
	\end{pmatrix} \,,
\end{equation}
where 
\begin{align*}
	a_{21} =& \left[y_2 x_2 + t^2 y_2 x_2 - t \left(y_2^2 + x_2^2\right)\right] 
	\quad \times \\
	& \hspace{35pt} \qquad \big[
	-\left(\left(-1 + t^2\right) y_1^2\right) 
	\quad \times \\
	& \hspace{35pt} \qquad \qquad \left(-1 + t^2\right) y_1 \left(-t x_1 + x_2 + t^2 x_2\right)
	\quad \times \\
	&\hspace{35pt} \qquad \qquad \qquad t \left(-x_1 x_2 - t^2 x_1 x_2 + t \left(x_1^2 + x_2^2\right)\right)
	\big] \,,
\end{align*}
satisfying 
$$
A \cdot 
	\begin{pmatrix}
		f_1 \\ 
		f_2
	\end{pmatrix}
	= 
	\begin{pmatrix}
		g_1 \\ 
		g_2
	\end{pmatrix} \,.
$$
The determinant $\det(A) = t^2$, giving a transformed integrand
\begin{align}
\omega'_{0,0} &= \frac{\det(A) t (1-t^2)^2 y_1 y_2 x_1 x_2 \, \dd x_1 \dd x_2  \, \dd x_1 \dd x_2}{g_1 g_2}  \nonumber \\
&= \frac{t^3 (-1+t^2) y_2 x_1 x_2 \, \dd x_1 \dd x_2}{(y_1 - t x_1) (y_1 - x_2) (y_1 - t^2 x_2) (y_2 x_2 + t^2 y_2 x_2 - t (y_2^2 + x_2^2))} \,.
\end{align}
Now its residue can be evaluated as the product of two univariate residues.

Hence, the multivariate residue at $(y_1 / t, y_2 / t)$ with respect to this partitioning $\{f_1,f_2\}$ is computed to be 
\begin{equation}
\label{eq:Res-1}
\Res_{(y_1 / t, y_2 / t)} (\omega_{0,0}) = \frac{t y_1 y_2}{(t y_1 - y_2) (t y_2 - y_1)} \,.
\end{equation}

Note that the result flips signs if we exchange $f_1$ and $f_2$ in the partitioning. To get the correct combination of Jeffrey-Kirwan residues, we order the denominator factors such that their corresponding charge vectors are in anti-clockwise order.

Similarly, the residues at $(y_1 t, y_1 / t^2)$ and $(y_2 / t^2, y_2 / t)$ are
\begin{equation}
\label{eq:Res-2}
\Res_{(y_1 t, y_1 / t^2)} (\omega_{0,0}) = -\frac{t^2 y_1 y_2}{(y_1 - t y_2) (y_1 - t^3 y_2)}
\end{equation}
and 
\begin{equation}
\label{eq:Res-3}
\Res_{(y_2 / t^2, y_2 / t)} (\omega_{0,0}) = -\frac{t^2 y_1 y_2}{(t y_1 - y_2) (t^3 y_1 - y_2)} \,.
\end{equation}

The sum of the three residues is
\begin{equation}
Z_{g=0,A}^{\text{SQED}[3]} = \frac{(t+t^3+t^5) y_1 y_2}{(t^3 y_1 - y_2) (t^3 y_2 - y_1)} \,,
\end{equation}
which reproduces the expected result in \eqref{eq:Z-SQED-A} up to an overall minus sign. This minus sign could also have been obtained if we picked clockwise ordering for the charge vectors instead of the anti-clockwise ordering. 

It can also be verified that the sectors of $m_1 \neq 0$ or $m_2 \neq 0$ have residues summing to zero from these three poles.

Although the computation is done for the particular choice of $\eta = (1,1)$, it can be shown that the result does not depend on this choice, as long as $\eta$ is not chosen to align with any of the charges $(Q_1,Q_2)$ in~\eqref{eq:charges}. This can be explicitly seen by plotting the charges on the $Q_1$--$Q_2$ plane, and observe that any choice of $\eta$ is contained within three convex cones, each giving one of the factors in~\eqref{eq:Res-1},~\eqref{eq:Res-2},~and~\eqref{eq:Res-3}.

\subsection{A-Twist of Abelian Linear Quiver}
For the mirror theory, the contour integral formula~\cite[(6.44)]{closset2016} for the B-twisted index of quantum electrodynamics with three hypermultiplets is
\begin{align}
	Z_{g=0,B}^{\text{SQED}[3]} 
	=&  -(t-t^{-1}) \sum_{m} (-q)^{m} 
	\frac{1}{2 \pi i} \oint_{\text{JK}} \frac{\dd x}{x} \, \nonumber \\ 
	& \qquad \left(\frac{x - y_1 t}{y_1 - x t}\right)^{m}
	\left(\frac{x - y_2 t}{y_2 - x t}\right)^{m}
	\left(\frac{x - y_3 t}{y_3 - x t}\right)^{m} \quad \times \nonumber \\
	& \qquad \frac{x y_1 t}{(x - y_1 t)(y_1 - x t)}
	\frac{x y_2 t}{(x - y_2 t)(y_2 - x t)}
	\frac{x y_3 t}{(x - y_3 t)(y_3 - x t)} \,.
\end{align} 
Setting $\eta = 1$ picks the residues at $y_1/t$, $y_2/t$, and $y_3/t$, which sum to
\begin{align}
	Z_{g=0,B}^{\text{SQED}[3]} =& - \frac{t^4 y_1^2 y_2}{(1-t^2)(y_1-y_2)(t^2 y_1 - y_2)(t^2 y_1 - y_3)} \nonumber \\
	& \qquad \qquad + \frac{t^4 y_1 y_2^2}{(1-t^2)(y_1-y_2)(t^2 y_2 - y_1)(t^2 y_2 - y_3)} \nonumber \\
	& \qquad \qquad - \frac{t^2 y_1^2 y_2}{(1-t^2)(y_1-y_2)(t^2 y_2 - y_1)(t^2 y_3 - y_1)} \nonumber \\
	& \qquad \qquad - \frac{t^2 y_1 y_2^2}{(1-t^2)(y_1-y_2)(t^2 y_1 - y_2)(t^2 y_3 - y_2)} \,.
\end{align}
To map to parameters $t'=t^{-1}, q_1=y_1/y_2, q_2=y_2/y_3$ in the mirror, we implement the following change of variables
\begin{subequations}
\begin{align}
	t &\mapsto t^{-1} \,, \\
	y_1 &\mapsto q_1 q_2 y_3 \,,\\
	y_2 &\mapsto q_2 y_3 \,,
\end{align}
\end{subequations}
which results in
\begin{align}
\label{eq:Z-SQED-B}
	&Z_{g=0,B}^{\text{SQED}[3]} \nonumber \\
	=& \frac{q_1 q_2 \left[ 
	(1 + q_2) t^6 + q_1^2 q_2 (1 + q_2) t^6 + 
	q_1 \left(t^6 + q_2^2 t^6 - q_2 \left(t^2 + 2 t^4 + 2 t^8 + t^10\right)\right)\right]
	}{
	- \left(q_1 - t^2\right) (q_1 q_2 - t^2) (-q_2 + t^2) (-1 + q_1 t^2) (-1 + q_2 t^2) (-1 + q_1 q_2 t^2))} 
	\nonumber \\
	= & -\frac{
	q_1 q_2 (1 + q_1) t^2
	}{
	(q_1 - t^2) (1 - q_1 t^2)} 
	\nonumber \\
	& -\frac{
	q_1 q_2^2 \left[
	t^2 + t^4 + t^6 + q_1^2 (t^2 + t^4 + t^6) - q_1 (1 + t^8)\right] 
	}{
	t^2 (q_1 - t^2) (1 - q_1 t^2)} 
	\nonumber \\
    &  -\frac{q_1 q_2^3 
    \left[t^2 + t^4 + \ldots + t^{10} + 
    q_1^3 (t^2 + t^4 + \ldots + t^{10}) - (q_1 + q_1^2) (1 + t^{12})\right] 
    }{
    t^4 (q_1 - t^2) (1 - q_1 t^2)} 
    \nonumber \\
    &  -\frac{q_1 q_2^4 
    \left[t^2 + t^4 + \ldots + t^{14} + q_1^4 (t^2 + t^4 + \ldots + t^{14}) - (q_1+q_1^2+q_1^3) (1 + t^{16})\right] 
    }{
    t^6 (q_1 - t^2) (1 - q_1 t^2)} 
    \nonumber \\
    & + \mathcal{O}(q_2^5) \,,
\end{align}
where the last expression is the expansion in $q_2$. In this case all sectors are expected to contribute. So the computation becomes more complicated.

The contour integral formula for A-twisted index of $A_2$ quiver is
\begin{align}
	Z_{g=0,A}^{[A_2]} = & (t-t^{-1})^{-2} \sum_{m_1, m_2} q_1^{m_1} q_2^{m_2} 
	\frac{1}{(2 \pi i)^2} \oint_{\text{JK}} \frac{\dd x_1}{x_1} \frac{\dd x_2}{x_2} \, \nonumber \\ 
	& \quad \qquad \qquad \left(\frac{x_1 - y_1 t}{y_1 - x_1 t}\right)^{m_1}
	\left(\frac{x_2 - y_2 t}{y_2 - x_2 t}\right)^{m_2}
	\left(\frac{x_1 - x_2 t}{x_2 - x_1 t}\right)^{m_1 - m_2} \,.
\end{align}

Setting the charge vector $\eta=(1,1)$ again picks up the interior poles at 
$$
(y_1 / t, y_2 / t), (y_1 t, y_1 / t^2), (y_2 / t^2, y_2 / t) \,.
$$
However, the boundary poles involving $1/x_1$ and $1/x_2$ still need to be considered. In this case, we follow~\cite[(2.17)]{xu2022} to assign both $1/x_1$ and $1/x_2$ with the charge $(-\infty,-\infty)$. Because there is no Chern-Simons terms in $\mathcal{N}=4$ theory. It turns out that in this case, none of the boundary poles are selected by the Jeffrey-Kirwan prescription, as they always have a charge vector in the exact opposite direction of~$\eta$.

To compare with~\eqref{eq:Z-SQED-B}, the contour integral formula for the B-twisted index is evaluated for $m_2=0,1,2,3,4$.
\vspace{-\parskip}
\begin{itemize}
	\item The sector of $m_2=0$ does not have non-vanishing residues at the selected poles.
	\item The $m_2=1$ sector has contributions at $(y_1 / t, y_2 / t)$ and $(y_2 / t^2, y_2 / t)$ for $m_1 \geq 1$. The series of residues evaluated for each $m_1$ at individual poles do not resemble the expansion of rational functions. However, their sum gives 
$$
-\frac{q_1 q_2 (1 + q_1) t^2}{(q_1 - t^2) (1 - q_1 t^2)} \,,
$$
which is identical to the $q_2$ term in the expansion in~\eqref{eq:Z-SQED-B}. 
	\item The $m_2=2$ sector has contributions at $(y_1 / t, y_2 / t)$ and $(y_2 / t^2, y_2 / t)$ summing to 
$$
-\frac{q_1^2 q_2^2 (1+t^2)^2}{(q_1-t^2)(1-q_1 t^2)}
$$
for $m_1 \geq 2$, and contributions at $(y_1 / t, y_2 / t)$ and $(y_1 / t, y_1 / t^2)$ summing to
$$
\frac{q_1^2 q_2^2 (1+t^2+t^4)}{t^2}
$$
for $m_1 = 1$. The sum of these two factors is
$$
-\frac{q_1 q_2^2 (t^2 + t^4 + t^6 + q_1^2 (t^2 + t^4 + t^6) - q_1 (1 + t^8)) }{t^2 (q_1 - t^2) (1 - q_1 t^2)} \,,
$$
which is identical to the $q_2^2$ term in the expansion in~\eqref{eq:Z-SQED-B}. 
\item The $m_2=3$ sector has contributions at $(y_1 / t, y_2 / t)$ and $(y_2 / t^2, y_2 / t)$ summing to 
$$
-\frac{q_1^3 q_2^3 (1+t^2+t^4) (1+(1-q_1)t^2+t^4)}{t^2 (q_1-t^2)(1-q_1 t^2)}
$$
for $m_1 \geq 3$, and contributions at $(y_1 / t, y_2 / t)$ and $(y_1 / t, y_1 / t^2)$ summing to
$$
\frac{q_1 q_2^3 (1+t^2+\ldots+t^8)}{t^4} + \frac{q_1^2 q_2^3 (1+t^2)^2 (1+t^4)}{t^4}
$$
for $m_1 = 1, 2$. The total sum is
$$
-\frac{q_1 q_2^3 \left[t^2 + t^4 + \ldots + t^{10} + 
    q_1^3 (t^2 + t^4 + \ldots + t^{10}) - (q_1 + q_1^2) (1 + t^{12}) \right] }{t^4 (q_1 - t^2) (1 - q_1 t^2)} \,,
$$
which is identical to the $q_2^3$ term in the expansion in~\eqref{eq:Z-SQED-B}. 
\item The $m_2=4$ sector has contributions at $(y_1 / t, y_2 / t)$ and $(y_2 / t^2, y_2 / t)$ summing to 
$$
-\frac{q_1^4 q_2^4 (1+t^2)^2 (1-q_1 t^2+t^4) (1+t^4)}{t^4 (q_1-t^2)(1-q_1 t^2)}
$$
for $m_1 \geq 4$, and contributions at $(y_1 / t, y_2 / t)$ and $(y_1 / t, y_1 / t^2)$ summing to
\begin{align*}
	\frac{q_1 q_2^4 (1+t^2+\ldots+t^{12})}{t^6} &+ 
	\frac{q_1^2 q_2^4 (1+t^2)^2 (1+t^4+t^8)}{t^6} \\
	&+ \frac{q_1^3 q_2^4 (1+t^2+t^4) (1+t^2+\ldots+t^8)}{t^6}
\end{align*}
for $m_1 = 1, 2, 3$. The total sum is
$$
	-\frac{q_1 q_2^4 \left[t^2 + t^4 + \ldots + t^{14} + q_1^4 (t^2 + t^4 + \ldots + t^{14}) - (q_1+q_1^2+q_1^3) (1 + t^{16}) \right] }{t^6 (q_1 - t^2) (1 - q_1 t^2)} \,,
$$
which is identical to the $q_2^4$ term in the expansion in~\eqref{eq:Z-SQED-B}. 
\end{itemize}
\chapter{Physical Background}
\label{chap:prelim-physics}

This chapters reviews some of the foundations in physics for this thesis.
\vspace{-\parskip}
\begin{itemize}
	\item Section~\ref{sec:physics-gauge-theories} briefly reviews classical gauge theories following~\cite{Deligne:1999qp,nakahara2003}.
	\item Section~\ref{sec:physics-SUSY-gauge-theories} is a review of abelian $\mathcal{N} = 2$ gauge theories in three dimensions, following~\cite{aharony1997}.
\end{itemize}

\section{Gauge Theories}
\label{sec:physics-gauge-theories}

This section briefly reviews classical gauge theories following~\cite{Deligne:1999qp,nakahara2003}.

A \index{gauge theory} gauge theory~\cite{nakahara2003} is a field theory where the gauge field is the connection on a principal bundle $P \to M$, where $M$ is the spacetime manifold of dimension $n$. The structure group $G$ of $P$ is called the gauge group. 

The connection on $P$ is a $G$-invariant distribution~\cite{Deligne:1999qp} on $P$ which projects isomorphically onto the tangent space $TM$. It can be encoded in a $\mathfrak{g}$-valued one-form $A \in \Omega^1(\mathfrak{g})$, written as
\begin{equation}
	A = A_\mu \dd x^\mu
\end{equation}
in components. It induces a covariant derivative operator $\DD$ defined by
\begin{equation}
	\DD := \dd + A
\end{equation}
on any associated bundles.

The local curvature two-form $F \in \Omega^2(\mathfrak{g})$ 
\begin{equation}
	F := \dd A + [A \wedge A] = \frac{1}{2} F_{\mu \nu} \dd x^\mu \wedge \dd x^\nu
\end{equation}
is called the field strength, where the components read
\begin{equation}
	F_{\mu \nu} = \partial_\mu A_\nu - \partial_\nu A_\mu + [A_\mu, A_\nu] \,.
\end{equation}
It satisfies the Bianchi identity
\begin{equation}
	\DD F = \dd F + [A, F] = 0 \,.
\end{equation}

The lagrangian is 
\begin{equation}
	\mathcal{L} = -\frac{1}{2} \Tr_\mathfrak{g} \left(F \wedge * F\right) = - \frac{1}{4} \Tr_\mathfrak{g} \left(F_{\mu \nu} F^{\mu \nu}\right) \dd^n x \,. 
\end{equation}
The variation with respect to $A$ yields the equation of motion
\begin{equation}
	\DD * F = \DD_\mu F^{\mu \nu} = 0 \,.
\end{equation}

For example, electromagnetism \index{electromagnetism} is described by an abelian gauge theory on a manifold $M$ of dimension $n=4$. The gauge group is unitary $G = U(1)$, and the principal bundle $P$ is trivial. The gauge group elements are complex numbers $z$ of modulus $|z|=1$ on the unit circle. The elements of the corresponding Lie algebra $\mathfrak{g}= \mathfrak{u}(1) = \sqrt{-1} \mathbb{R}$ are then the imaginary phases as the tangent space to the circle. 
The connection one-form is the gauge potential $A = A_\mu \dd x^\mu$ where $A_\mu$ are imaginary numbers. The field strength is then simply $F = \dd A$ where the components are 
\begin{equation*}
	F_{\mu \nu} = \partial_\mu A_\nu - \partial_\nu A_\mu  \,.
\end{equation*}
The electric field $E$ and magnetic field $B$ can be identified as
\begin{subequations}
\begin{align}
	E_i &= -i F_{i0} \,, \\
	B_i &= -\frac{i}{2} \epsilon_{ijk} F^{jk} \,.
\end{align}
\end{subequations}
The Bianchi identity reads 
\begin{equation*}
	\partial_\mu F_{\nu \rho} + \partial_\rho F_{\mu \nu} + \partial_\nu F_{\rho \mu} = 0 \,,
\end{equation*}
which reduces to two of the Maxwell's equations
\begin{subequations}
\begin{align}
	\nabla \times E + \frac{\partial B}{\partial t} &= 0 \,, \\
	\nabla \cdot B &= 0 \,.
\end{align}
\end{subequations}
The Maxwell lagrangian is 
\begin{equation}
	\mathcal{L} = - \frac{1}{4} F_{\mu \nu} F^{\mu \nu} \dd^4 x = \frac{1}{2} (E^2 - B^2) \dd^4 x \,.
\end{equation}
The equation of motion $\partial_\mu F^{\mu\nu} = 0$ reduces to the other two Maxwell's equations
\begin{subequations}
\begin{align}
	\nabla \cdot E &= 0 \,, \\
	\nabla \times B - \frac{\partial E}{\partial t} &= 0 \,.
\end{align}
\end{subequations}

\section{Supersymmetric Gauge Theories}
\label{sec:physics-SUSY-gauge-theories}

This section is a review of abelian $\mathcal{N} = 2$ gauge theories in three dimensions, following~\cite{aharony1997}.

Consider a three-dimensional $\mathbb{N}=2$ supersymmetric quantum \index{quantum electrodynamics} electrodynamics~\cite{aharony1997} with $N_\text{f} > 0$ flavours, i.e., chiral multiplets $\{\Phi_j, \tilde{\Phi}_j\}_{j=1}^{N_\text{f}}$ of charges $\pm 1$. This theory flows to an interacting fixed point in the infra-red. In comparison, the four-dimensional theories are always infra-red free. 

The classical moduli space of vacua contain Coulomb and Higgs branches.
\vspace{-\parskip}
\begin{itemize}
	\item The \index{vacuum !Higgs branch} Higgs branch has $(2 N_\text{f} - 1)$ dimensions. It can be parametrised by gauge invariant operators $M_{ij} = \Phi_i \tilde{\Phi}_j$, subject to the constraint $M_{ij} M_{mn} = M_{mj} M_{in}$.
	\item The \index{vacuum !Coulomb branch} Coulomb branch is one-dimensional. It is parametrised by $\sigma + i \gamma$, where $\sigma$ is the vector multiplet scalar in~\eqref{eq:vector-components}, and $\gamma$ is the dual photon defined by $F_{\mu \nu} = \epsilon_{\mu \nu \rho} \partial^{\rho} \gamma$. It is cylindrical with radius proportional to the coupling $g$ in~\eqref{eq:kinetic-term-gauge-vector}. The scalar $\gamma \in S^1$ is rotated around $S^1$ by the ``magnetic'' global $U(1)_J$ symmetry, with a period of $g^2$.
\end{itemize}
\vspace{-\parskip}
The Higgs and Coulomb branches intersect at $\sigma = 0$. 

Despite that there is no instanton corrections in abelian theories, the quantum perturbative effects still change the topology of the moduli space. 
The Coulomb branch can be consistently parametrised by the vacuum expectation value of the chiral superfield $V = \exp((\sigma + i \gamma)/g^2)$ for large $\sigma$. As the metric for $\gamma$ receives quantum corrections, the topology of the moduli changes in perturbation theory. Because the Higgs branch is invariant under the $U(1)_J$ symmetry, and intersects with the Coulomb branch at $\sigma = 0$, the radius of the circle $S^1$ where $\gamma$ lives must vanish at $\sigma = 0$. 
Therefore the moduli space looks like an intersection of three cones near the origin, where the Coulomb branch splits into two distinct regions parametrised by $V_\pm \sim \exp(\pm(\sigma + i \gamma)/g^2)$. So the quantum Coulomb branch involves two unconstrained chiral superfields $V_\pm$. The Higgs branch is still parametrised by $M_{ij}$ since it does not receive corrections.

The dynamics is constrained by the global symmetries. The charges are listed in Table~\ref{table:global-charges-SQED}. The $U(1)_R$ charges of the chiral multiplets are chosen to be zero so their fermions have charge $-1$.
\begin{table}[ht]
\centering
\bgroup
\def\arraystretch{1.5}
\begin{tabular}{c|c|c|c|c|c}
     & $U(1)_R$ & $U(1)_J$ & $U(1)_A$ & $SU(N_\text{f})$ & $SU(N_\text{f})$ \\
    \hline
    $\Phi_j$ & $0$ & $0$ & $1$ & $\mathbf{N_\text{f}}$ & $\mathbf{1}$\\
    $\tilde{\Phi}_j$ & $0$ & $0$ & $1$ & $\mathbf{1}$ & $\mathbf{\widebar{N}_\text{f}}$\\
    $M_{ij}$ & $0$ & $0$ & $2$ & $\mathbf{N_\text{f}}$ & $\mathbf{\widebar{N}_\text{f}}$\\
    $V_\pm$ & $N_\text{f}$ & $\pm 1$ & $-N_\text{f}$ & $\mathbf{1}$ & $\mathbf{1}$    
\end{tabular}
\egroup
\caption{Charges of Global Symmetries}
\label{table:global-charges-SQED}
\end{table}

For $N_\text{f} = 1$, all three cones parametrised by $\{M, V_+, V_-\}$ have one complex dimension. At the origin there is a renormalisation fixed point. The same fixed point can be reached by the theory described by the fields $\{M, V_+, V_-\}$ with a superpotential $W = - M V_+ V_-$, giving the same moduli space. In general for $N_\text{f} > 1$, the corresponding theory has 
$$
	W = - N_\text{f} \left(\det(M) V_+ V_- \right)^{\sfrac{1}{N_\text{f}}} \,.
$$
This is analogous to the superpotential corresponding to four-dimensional $\mathcal{N}=1$ quantum chromodynamics~\cite{Seiberg:1994bz} when $N_\text{f} > N_\text{c} + 1$. 

\chapter{Mathematical Background}
\label{chap:prelim-mathematics}

This chapters reviews some of the mathematical foundations for this thesis.
\vspace{-\parskip}
\begin{itemize}
	\item Section~\ref{sec:maths-riemann-surfaces} briefly reviews line bundles and vector bundles on Riemann surface, following~\cite{Hitchin:1999at}.
	\item Section~\ref{sec:abelian-vortex} reviews the abelian vortex equation and the Hitchin-Kobayashi correspondence.
\end{itemize}

\section{Bundles on Riemann Surfaces}
\label{sec:maths-riemann-surfaces}

This section is a review of line bundles and vector bundles on Riemann surfaces, following~\cite{Hitchin:1999at}.

\subsection{Riemann Surfaces}
\begin{defn}
	A \index{Riemann surface} Riemann surface is a one-dimensional complex manifold with an atlas of coordinate charts $\{\phi_\alpha: U_\alpha \to \mathbb{C}\}$ such that a transition map $\phi_\beta \circ \phi_\alpha^{-1}$  is an invertible holomorphic function $\phi_\alpha(U_\alpha \cap U_\beta) \to \phi_\beta(U_\alpha \cap U_\beta)$ for all $\alpha$ and $\beta$.
\end{defn}

The most basic example is the sphere $S^2$ with the two standard stereographic charts 
\begin{subequations}
\begin{align}
	\phi_0 : U_0 \to \mathbb{C} \,, \\
	\phi_1 : U_1 \to \mathbb{C} \,,
\end{align}
\end{subequations}
where $U_0 = S^2 \setminus \{\text{N}\}$ and $U_1 = S^2 \setminus \{\text{S}\}$.
Its transition map $z'(z) = \phi_1 \circ \phi_0^{-1} (z) = z^{-1}$ is holomorphic from $\mathbb{C}^*$ to $\mathbb{C}^*$. The point $p$ where $z = \phi_0(p) = \infty$ can be understood as the north pole $\{N\}$. The two-sphere with this complex structure is known the Riemann sphere and the complex projective line, denoted by $\mathbb{CP}^1$ or $\mathbb{P}^1$.

The torus $T^2$ can be defined as $\mathbb{C}^* / \mathbb{Z}$ where the integer $n$ acts by $z \mapsto \lambda^n z$ with $|\lambda| \neq 1$. The coordinate patches can be taken as the overlapping annuli in $\mathbb{C}^*$.

\begin{defn}
	A holomorphic map $f: M \to N$ of Riemann surfaces $M$ and $N$ is a continuous map, such that for each chart $\phi_\alpha: U_\alpha \to \mathbb{C}$ on $M$ and $\tilde{\phi}_\alpha: \tilde{U}_\alpha \to \mathbb{C}$ on $N$ the representative $\phi_\alpha \circ f \circ \tilde{\phi}_\beta^{-1}$ is holomorphic. 
\end{defn}

For example, a holomorphic map from the Riemann sphere $\mathbb{CP}^1$ to itself is defined by a rational function from $\mathbb{C}$ to $\mathbb{C}$. Consider a rational function $g(z) = \phi_0 \circ f \circ \phi_0^{-1 } (z)$ of the coordinate $z = \phi_0(p)$. A point $p_0 = \phi_0^{-1}(z_0)$ where $g(z_0) = \infty$ is mapped to the north pole under $f$, which no longer belongs to $U_0$. The representative $\phi_1 \circ f \circ \phi_0^{-1} (z_0) = \frac{1}{f(z_0)} = 0$ is therefore holomorphic.

\subsection{Line Bundles}
\begin{defn}
	A \index{holomorphic line bundle} holomorphic line bundle $L$ over a Riemann surface $M$ is a two-dimensional complex manifold $L$ with a holomoprhic projection $\pi: L \to M$ such that
\vspace{-\parskip}
\begin{itemize}
	\item for each point $p \in M$, the fibre $F_p = \pi^{-1}(p) \simeq \mathbb{C}$ at $p$ is a one-dimensional vector space;
	\item each point $p \in M$ has a neighbourhood $U$ and a homeomorphism $\psi_U$ called the local trivialisation such that the diagram
	\begin{equation}
	\begin{tikzcd}
		\pi^{-1}(U) \arrow[r, "\simeq"', "\psi_U"] \arrow[dr, "\pi", swap]
		& U \times \mathbb{C} \arrow[d]\\
		& U
	\end{tikzcd}
	\end{equation}
	is commutative;
	\item the transition map $g_{VU} = \psi_V \circ \psi_U^{-1}$ is of the form
	\begin{equation}
		(p, w) \mapsto (p, f(p) w) \,,
	\end{equation}
	where $f: U \cap V \mapsto \mathbb{C}^*$ is a non-vanishing holomorphic function, which uniquely determines the transition map $g_{VU}$.
\end{itemize}
\end{defn}

For a point $p \in M$, we can construct a line bundle $L_p$, by using the coordinate $z$ on the neighbourhood $U_0$ centred at $p$ as the transition map to the patch $U_1 = M \setminus \{p\}$.  Explicitly the transition map is 
\begin{align}
	g_{01}: U_1 \times \mathbb{C} &\to U_0 \times \mathbb{C} \,, \nonumber \\
	(p, w) & \mapsto (p, z(p) w) \,.
\end{align}
The line bundle $L_p$ is holomorphic since $g_{01}(p) = z(p)$ is holomorphic and non-vanishing on the intersection $U_0 \cap U_1$. 

\begin{defn}
	A holomorphic section of a line bundle $L$ over $M$ is a holomorphic map $s: M \to L$ such that $\pi \circ s = \id_M$.
\end{defn}

In a local trivialisation $\{(U_i, \psi_i)\}$, the section is defined by a holomorphic function $s_i$ on $U_i$ via
\begin{align}
	\psi_i \circ s: U_i &\to U_i \times \mathbb{C} \,, \nonumber \\
	p & \mapsto (p, s_i(p)) \,.
\end{align}
On the overlap $U_i \cap U_j$ the holomorphic functions $s_i$, $s_j$ are then related by the transition function $s_j = g_{ji} s_i$ since
\begin{alignat}{3}
	\big(\psi_j \circ \psi_i^{-1}\big) \circ \big(\psi_i \circ s\big): 
	U_i &\to U_i \times \mathbb{C} &&\to U_j \times \mathbb{C} \,, \nonumber \\
	p & \mapsto (p, s_i(p)) &&\mapsto  (p, g_{ji} (p) s_i(p)) \,.
\end{alignat}
A section $s$ is therefore uniquely determined by a collection $\{s_i\}$ of local functions patched together by the transition functions $\{g_{ij}\}$.
Given two holomorphic sections $s$ and $t$ of $L \to M$, they can be used to construct a global meromorphic function on $M$ via
\begin{equation}
	\frac{s_i}{t_i} = \frac{g_{ij} s_j}{g_{ij} t_j} = \frac{s_j}{t_j} \,.
\end{equation}
The space of all sections of $L \to M$ form a vector space $H^0 (M,L)$.

The line bundle $L_p$ has a canonical section $s_p$, where the local functions are simply $z$ on $U_0$ and $1$ on $U_1$. It has as a single simple zero at $p$.

The \index{canonical bundle} canonical bundle $K \to M$ is the cotangent bundle $T^* M$ of holomorphic one-forms. On a chart $U_\alpha$ with coordinate $z$, the neighbourhood $\pi^{-1} (U_\alpha) \simeq U_\alpha \times \mathbb{C}$ is trivial, and we choose a section $\dd z (p)$ over $U_\alpha$ as the natural basis for each fibre $F_p$. This frame~\cite{nakahara2003} over $U_\alpha$ gives a natural map $F_p \to \mathbb{C}$ given by 
\begin{equation}
	\omega(p) = f \, \dd z (p) \mapsto f \in \mathbb{C}.
\end{equation}
The local trivialisation of the section $\omega$ is
\begin{equation}
	\psi_\alpha \circ \omega : p \mapsto (p, f(p)) \,,
\end{equation}
where the one-form coordinate $f$ is regarded as a local function.
Given two charts $(U_\alpha, \phi_\alpha)$ and $(U_\beta, \phi_\beta)$ on the base $M$ and their coordinates $z$ and $z'(z) = \phi_\beta \circ \phi_\alpha^{-1} (z)$, a one-form on the overlap can be written as $\omega = f_\alpha \, \dd z = f_\beta \, \dd z'$. The local trivialisations are then given by the one-form coordinates via
\begin{subequations}
\begin{align}
	p &\mapsto (p, f_\alpha(p)) \in U_\alpha \times \mathbb{C} \,,\\
	p &\mapsto (p, f_\beta(p)) \in U_\beta \times \mathbb{C} \,,
\end{align}
\end{subequations}
where the fibre coordinates are related by $f_\alpha = f_\beta \frac{\dd z'}{\dd z}$ by the transition function $g_{\alpha \beta} = \frac{\dd z'}{\dd z}$.

The $\mathcal{O}(n)$ bundle on $\mathbb{CP}^1$ is defined by choosing the transition function to be $g_{01} = z^n$ on the overlap $U_0 \cap U_1 \simeq \mathbb{C}^*$ of the standard patches $U_0$ and $U_1$. A holomorphic section $s$ is then given by local functions $s_0$ and $s_1$ on $\mathbb{C}$ related by
\begin{equation}
	s_0 (z) = z^n s_1(z') 
\end{equation}
on the intersection $\mathcal{C}^*$. The local functions are given by polynomials of degree less than or equal to $n$. The dimension of all sections is then
\begin{equation}
	h^0\left( \mathbb{CP}^1, \mathcal{O}(n) \right) = n + 1
\end{equation}

\begin{defn}
	For a compact Riemann surface $M$, its genus $g$ is defined to be the dimension of $H^0(M, K)$.
\end{defn}

For example, the canonical bundle $K$ on $\mathbb{CP}^1$ has sections given by
$f_0(z) \dd z$ and $f_1(z') \dd z'$ where $f_0$ and $f_1$ are holomorphic functions on $\mathbb{C}$. The transition function is given by $g_{01} = \frac{\dd z'}{\dd z} = - z^2$ piecing together the local functions via
\begin{equation}
	f_0 (z) = - z^2 f_1(z^{-1}) \,.
\end{equation}
However, expanding $f_0$ and $f_1$ shows that both must vanish. Therefore there are no non-zero global sections of the canonical bundle. The genus of $\mathbb{CP}^1$ is $g = H^0 \left( \mathbb{CP}^1, K \right) = 0$ as expected.

On the torus $T^2 = \mathbb{C}^* / \mathbb{Z}$, a one-form is given by $\frac{\dd z}{z}$ as it is invariant under the integer action, i.e., $\frac{\dd (\lambda z)}{\lambda z} = \frac{\dd z}{z}$. It defines a non-vanishing section of the canonical bundle $K \to T^2$, which means that $K$ is isomorphic to the trivial bundle $T^2 \times \mathbb{C}$. Sections of the trivial bundle are holomorphic functions which are constants on a compact manifold. Therefore the genus of $T^2$ is $g = H^0 \left( T^2, K \right) = 1$.

In terms of local objects, line bundles are given by transition functions $g_{ij}$ on $U_i \cap U_j$, and their sections are given by functions $f_i$ on $U_i$. They are examples of sheaves.

\subsubsection{Sheaves}

To classify line bundles, we need to introduce sheaf theory.
\begin{defn}
	A sheaf $\mathcal{S}$ on a topological space $X$ associates to each open set $U \subset X$ an abelian group $\mathcal{S}(U)$, and to subset $U \subset V$ a restriction map $r_{VU}: \mathcal{S}(V)\to \mathcal{S}(U)$ such that
\vspace{-\parskip}
\begin{itemize}
	\item for $U \subset V \subset W$, $r_{WU} = r_{VU} \circ r_{WV}$;
	\item if $r_{U (U \cap V)}(\sigma) = r_{V (U \cap V)}(\tau)$ for some $\sigma \in \mathcal{S}(U)$ and $\tau \in \mathcal{S}(V)$, then there exists $\rho \in \mathcal{S}(U \cup V)$ such that $r_{(U\cup V) U} (\rho) = \sigma$ and $r_{(U\cup V) V} (\rho) = \tau$;
	\item if $r_{(U\cup V) U} (\sigma) = 0$ and $r_{(U\cup V) V} (\sigma) = 0$ for some $\sigma \in \mathcal{S}(U \cup V)$, then $\sigma = 0$.
\end{itemize}
\end{defn}

Some familiar Examples of sheaves are
\vspace{-\parskip}
\begin{itemize}
	\item trivial sections, i.e., locally constant functions on $U$,
	\item holomorphic functions $\mathcal{O}(U)$ on $U$,
	\item sections $\mathcal{O}(L)(U)$ of a holomorphic line bundle $L$ over $U$,
	\item non-vanhisng holomorphic functions $\mathcal{O}^*(U)$ on $U$ \,.
\end{itemize}

\begin{defn}
	The $p$-th cohomology group of a sheaf $\mathcal{S}$ on $M$ relative to an open cover $\{U_\alpha\}$ of $M$ is the quotient group
	\begin{equation}
		H^p (M, \mathcal{S}) := \frac{\ker \delta: C^p \to C^{p+1}}{\im \delta: C^{p-1} \to C^p} \,,
	\end{equation}
	where the chain group $C^p$ is the alternating elements in $\mathcal{S}^p$ given by the sections on $p$ intersections, i.e.,
	\begin{equation}
		\mathcal{S}^p = \bigoplus_{\alpha_0 \neq \cdots \neq \alpha_p} \mathcal{S}(U_{\alpha_0} \cap \cdots \cap U_{\alpha_p}) \,,
	\end{equation}
	and the coboundary operator $\delta$ is a homomorphism of abelian groups $C^p \to C^{p+1}$ by
	\begin{equation}
		(\delta f)_{\alpha_0 \ldots \alpha_{p+1}} 
		= \sum_i (-1)^i \, 
		\left. f_{\alpha_0 \ldots \hat{\alpha}_i \ldots \alpha_{p+1}} \right|_{U_{\alpha_0} \cap \cdots \cap U_{\alpha_{p+1}}} \,.
	\end{equation}
\end{defn}

For a holomorphic line bundle $L \to M$ and the sheaf $S$ of holomorphc functions, the zero-th cohomology is 
\begin{equation}
	H^0(M, L) := H^0(M, \mathcal{L}) = \ker \delta \,,
\end{equation} 
which is the space of global holomorphic sections of $L$. The reason is that the boundary of a zero chain $f \in C^0$ is $(\delta f)_{\alpha \beta}  = f_\alpha - f_\beta$, which vanishes if and only if the local sections $f_\alpha$ piece together to give a global section.

The isomorphism classes of holomorphic line bundles are given by elements of the sheaf cohomology group $H^1(M, \mathcal{O}^*)$. This is because the transition functions $g_{\alpha \beta}$ lie in $C^1$ for the sheaf $\mathcal{O}^*$ of non-vanishing holomorphic functions, and they are in the kernel of $\delta$ since
\begin{equation}
	(\delta g)_{\alpha \beta \gamma} = g_{\alpha \beta} g_{\beta \gamma} g_{\gamma \alpha} = \id \,.
\end{equation}

\begin{thm}[Serre Duality\index{Serre duality}]
\label{thm:serre}
	If $L$ is a line bundle on a compact Riemann surface $M$, then 
	\begin{equation}
		H^1(M, L) \simeq H^0(M, K \otimes L^*)^* \,,
	\end{equation}
	where $ ^*$ denotes the dual vector space.
\end{thm}

\begin{thm}
	Given a short exact sequence 
	\begin{equation}
		0 \to \mathcal{S} \to \mathcal{T} \to \mathcal{U} \to 0
	\end{equation}		
	of sheaves on $M$, then there is a long exact sequence of cohomology groups
	\begin{align}
		0 &\to H^0(M, \mathcal{S}) \to H^0(M, \mathcal{T}) \to H^0(M, \mathcal{U}) \xrightarrow{\delta_0} H^1(M, \mathcal{S}) \to \cdots \nonumber \\
		\cdots &\to H^p(M, \mathcal{S}) \to H^p(M, \mathcal{T}) \to H^p(M, \mathcal{U}) \xrightarrow{\delta_p} H^{p+1}(M, \mathcal{S}) \to \cdots \,,
	\end{align}
	where $\{\delta_q\}$ are the coboundary operators.
\end{thm}
Consider the coboundary operator $\delta_0: H^0(M,\mathcal{U}) \to H^1(M,\mathcal{S})$. 
The elements $\{u_\alpha \in H^0(M,\mathcal{U})\}$ satisfy $u_\alpha - u_\beta = 0$. 
There exist elements $\{t_\alpha \in C^0(\mathcal{T})\}$ such that $t_\alpha \mapsto u_\alpha$. 
Then $\{(t_\alpha - t_\beta) \in C^1(\mathcal{T})\}$ are mapped to $\{u_\alpha - u_\beta = 0\}$. 
By exactness of the short exact sequence, there exists a unique $s_{\alpha \beta} \in C^1(\mathcal{S})$ such that $s_{\alpha \beta} \mapsto t_\alpha - t_\beta$. 
It satisfies $\delta s = 0$ and hence $s \in H^1(M,\mathcal{S})$. 
The coboundary operator is defined by $\delta_0 u := s$.

For example, given a line bundle $L \to M$ and the line bundle $L_p$ associated with a point $p \in M$, there is a short exact sequence
\begin{equation}
	0 \to \mathcal{O}(L L_p^{-1}) \xrightarrow{s_p} \mathcal{O}(L) \to \mathcal{O}_p(L) \to 0 \,,
\end{equation}
where $s_p$ is the canonical section of $L_p$ vanishing only at $p$. If a section $s \in \mathcal{O}(L)$ vanishes at $p$ with multiplicity $m$, then the section $s s_p^{-1} \in \mathcal{O}(L L_p^{-1})$ vanishes at $p$ with multiplicity $(m-1)$.
The sheaf $\mathcal{O}_p(L) (U)$ is the sections of $L$ over $U \cap \{p\}$. Its space of global sections is given by $\pi^{-1}(p) = \mathbb{C}$, which is one-dimensional. This gives rise to a long exact sequence
\begin{equation}
	0 \to H^0(M, L L_p^{-1}) \to H^0(M, L) \to \mathbb{C} \xrightarrow{\delta} H^1(M, L L_p^{-1}) \to \cdots \,.
\end{equation}
When the map $\delta$ is non-zero, the map $H^0(M,L) \to \mathbb{C}$ must be zero. By exactness there is an isomorphism
\begin{equation}
	H^0(M, L L_p^{-1}) \simeq H^0(M, L)
\end{equation}
via multiplication by the section $s_p$. Therefore if $\delta \neq 0$, then all global sections of $L$ must vanish at $p$.

Consider the short exact sequence of sheaves
\begin{equation}
	0 \to \mathbb{Z} \to \mathcal{O} \xrightarrow{\exp} \mathcal{O}^* \to 1 \,,
\end{equation}
where $\mathcal{O}$ is holomorphic functions, and $\mathcal{O}^*$ is non-vanishing holomorphic functions.
This gives rise to a long exact sequence
\begin{align}
	0 \to \mathbb{Z} \to \mathbb{C} \to \mathbb{C}^* &\to H^1(M,\mathbb{Z}) \to H^1(M,\mathcal{O}) \to H^1(M,\mathcal{O}^*) \nonumber \\
	&\to H^2(M,\mathbb{Z}) \to H^2(M,\mathcal{O}) \to \cdots \,.
\end{align}
The leading part of the sequence is due to the fact that holomorphic functions on compact Riemann surfaces are constants. By exactness, $H^1(M, \mathbb{Z})$ injects into $H^1(M, \mathcal{O})$ since exponentiation is surjective onto $\mathbb{C}^*$. As $H^2(M, \mathcal{O})$ must vanish and $H^2(M, \mathbb{Z}) \simeq \mathbb{Z}$, the sequence reduces to
\begin{equation}
\label{eq:exact-sequence-H1MO*}
	0 \to \frac{H^1 (M, \mathcal{O})}{H^1(M, \mathbb{Z})} \to H^1 (M, \mathcal{O}^*) \xrightarrow{\delta} \mathbb{Z} \to 0 \,.
\end{equation}

\begin{defn}
	The degree of a line bundle $L$ is
	\begin{equation}
		\deg L = c_1(L) := \delta ([L]) \,,
	\end{equation}
	which is also called the first Chern class. The degree of the $L_p$ bundle is normalised to $\deg L_p = 1$.
\end{defn}

The degree of product bundles satisfy $\deg L_1 \otimes L_2 = \deg L_1 + \deg L_2$. This implies that if a section $s \in H^0 (M, L)$ vanishes at points $\{p_i\}_{i=1}^n$ then 
$$
	\deg L = \sum_{i=1}^n m_i \,,
$$ 
where $\{m_i\}_{i=1}^n$ are the multiplicities of the zeros. This can be seen from the fact that the bundle $L L_{p_1}^{-m_1} \cdots L_{p_n}^{-m_p}$ is trivial as it has a global non-vanishing section $s s_{p_1}^{-m_1} \cdots s_{p_n}^{-m_p}$. The trivial bundle has degree zero. As a corollary, if $\deg L < 0$, then $L$ has no non-trivial holomorphic sections.

The dimension $h^1(M, \mathcal{O}) = g$ since $H^1(M, \mathcal{O}) \simeq H^0 (M, K)^*$ by Serre duality of Theorem~\ref{thm:serre}. Consider the short exact sequence,
\begin{equation}
	0 \to \mathbb{C} \to \mathcal{O} \xrightarrow{\dd} \mathcal{O}(K) \to 0 \,,
\end{equation}
where $\dd$ is the derivative operator. It gives a long exact sequence
\begin{align}
	0 \to \mathbb{C} \to \mathbb{C} &\to H^0 (M, \mathcal{O}(K)) \to H^1(M, \mathbb{C}) \to H^1(M, \mathcal{O}) \nonumber \\
	& \to H^1(M, \mathcal{O}(K)) \to H^2(M, \mathbb{C}) \to 0 \,.
\end{align}
The map $H^1(M, \mathcal{O}(K)) \to H^2(M, \mathbb{C})$ is an isomorphism since $H^1(M, \mathcal{O}(K)) \simeq H^0(M,\mathcal{O})^* \simeq \mathbb{C}$ and $H^2(M, \mathbb{C}) \simeq \mathbb{C}$. 
Hence the map $H^1(M, \mathcal{O}) \to H^1(M, \mathcal{O}(K))$ must be zero. 
Because $h^0(M, \mathcal{O}(K)) = h^1(M, \mathcal{O})$ are $g$-dimensional, we have by exactness the dimension $h^1(M,\mathbb{C}) = 2g$. 
Then $H^1(M,\mathbb{Z}) = \mathbb{Z}^{2g}$ as there is no torsion in $H^1$.
Therefore the exact sequence~\eqref{eq:exact-sequence-H1MO*} becomes
\begin{equation}
	0 \to \frac{\mathbb{C}^g}{\mathbb{Z}^{2g}} \to H^1 (M, \mathcal{O}^*) \to \mathbb{Z} \to 0 \,.
\end{equation}

The group $H^1 (M, \mathcal{O}^*)$ is called the \index{Picard group} Picard group of $M$. The space $\frac{\mathbb{C}^g}{\mathbb{Z}^{2g}}$ is topologically a $2g$-dimensional torus. 

Line bundles are classified with sheaf theory, which is essentially linear. Each line bundle has an integer invariant, its degree. The space $J^d$ of classes of line bundles of degree $d$ is a complex torus. They are isomorphic to the jacobian of the Riemann surface.

\subsection{Vector Bundles}

\begin{defn}
	A vector bundle \index{vector bundle} of rank $m$ on a Riemann surface $M$ is a complex manifold $E$ with a holomorphic projection $\pi: E \to M$ such that
\vspace{-\parskip}
\begin{itemize}
	\item for each point $p \in M$, the fibre $\pi^{-1}(p)$ at $p$ is an $m$-dimensional vector space;
	\item each point $p \in M$ has a neighbourhood $U$ and a homeomorphism $\psi_U$ called the local trivialisation such that the diagram
	\begin{equation}
	\begin{tikzcd}
		\pi^{-1}(U) \arrow[r, "\simeq"', "\psi_U"] \arrow[dr, "\pi", swap]
		& U \times \mathbb{C}^m \arrow[d]\\
		& U
	\end{tikzcd}
	\end{equation}
	is commutative;
	\item the transition map $g_{VU} = \psi_V \circ \psi_U^{-1}$ is of the form
	\begin{equation}
		(p, w) \mapsto (p, g(p) w) \,,
	\end{equation}
	where $g: U \cap V \mapsto GL(m,\mathbb{C})$ is a holomorphic map, which uniquely determines the transition map $g_{VU}$.
\end{itemize}
\end{defn}

Since the transition functions are matrices and non-commutative in general, sheaf theory cannot be used to classify vector bundles in the same way as for line bundles.

\begin{defn}
	The degree of a vector bundle $E$ of rank $m$ is
	\begin{equation}
		\deg E = c_1(E) := \deg (\det E) \,,
	\end{equation}
	where $\det E := \wedge^m E$ is the \index{determinant line bundle} determinant line bundle.
\end{defn}

Their sheaf cohomolgies can be related by the Riemann-Roch theorem.
\begin{thm}[Riemann-Roch]
	If $E$ is a vector bundle on a compact Riemann surface of genus $g$, then
	\begin{equation}
		h^0(M,E) - h^1(M,E) = \deg E + (1-g) \rank E  \,.
	\end{equation}
\end{thm}

\section{Abelian Vortex Equations}
\label{sec:abelian-vortex}
This section briefly reviews the abelian vortex equations, and their moduli space of solutions. The analytic description can be translated to an algebraic description by the \index{Hitchin-Kobayashi correspondence} Hitchin-Kobayashi correspondence~\cite{Jaffe:1980mj,Garcia-Prada:1993usn,Alvarez-Consul:2001mqd,manton_sutcliffe_2004}.

Consider a holomorphic line bundle $L$ of degree $d$ with a hermitian metric on a Riemann surface $\Sigma$. Let $A$ and $\phi$ be a smooth connection and a smooth section respectively. 
The space $\mathfrak{F}_d$ of pairs $(A, \phi)$ is an infinite-dimensional K\"ahler manifold. It inherits from the metric on $\Sigma$ and the hermitian metric on $L$ a flat metric
\begin{equation}
	g = \frac{1}{4 \pi} \int_\Sigma 
	\left(
	\frac{1}{e^2} \delta A \wedge * \delta A + * |\delta \phi|^2
	\right)
	\, \dd \Sigma \,.
\end{equation}

The abelian vortex equations are
\begin{subequations}
\label{eq:abelian-vortex}
\begin{align}
	\frac{1}{e^2} * F_A +  |\phi|^2 - \tau &= 0 \,,  \label{eq:abelian-vortex-1}\\
	\bar{\partial}_A \phi_i &=0 \label{eq:abelian-vortex-2}\,, 
\end{align}
\end{subequations}
where $F_A$ is the curvature and $\bar{\partial}_A$ is the holomorphic structure inherited from $\dd_A$ and the complex structure on $\Sigma$. 
The moduli space of vortices \index{moduli space !of vortices} is the quotient
\begin{equation}
	\mathfrak{M}_d := \mathfrak{N}_d / G \,,
\end{equation}
where $\mathfrak{N}_d \subset \mathfrak{F}_d$ is the space of solutions $(A,\phi)$, and $G: \Sigma \to U(1)$ is the gauge group.

The moduli space $\mathfrak{M}_d$ can be interpreted as an infinite-dimensional K\"ahler quotient. 
The second constraint~\eqref{eq:abelian-vortex-2} defines a K\"ahler submanifold $\mathfrak{E}_d \subset \mathfrak{F}_d$ where the gauge group $G$ acts with a moment map
\begin{equation}
	\frac{1}{e^2} * F_A + \mu(\phi)
\end{equation}
with $\mu(\phi) = |\phi|^2$.
Therefore the vortex moduli space can be written as a  K\"ahler quotient
\begin{equation}
	\mathfrak{M}_d = \mathfrak{E}_d \sslash G \,.
\end{equation}

By the Hitchin-Kobayashi correspondence, the moduli space $\mathfrak{M}_d$ can be parametrised by pairs $(L,\phi)$, where $L$ is a holomorphic line bundle of degree $d$ and $\phi$ is a non-vanishing holomorphic section of $L$. There exists a map $\mathfrak{M}_d \to \sym{d}{\Sigma}$ to the symmetric product parametrising degree $d$ divisors on $\Sigma$, which is given by simply taking the divisor
\begin{equation}
	D = p_1 + \cdots + p_d 
\end{equation}
of zeroes of $\phi$.
The points $\{p_1, \ldots, p_d\}$ are the centres of the vortices. 
The hermitian line bundle can be recovered by the map
\begin{align}
	j: \sym{d}{\Sigma} &\to \pic{d}{\Sigma} \simeq J_\Sigma \nonumber \\
	\{D\} & \mapsto \mathcal{O}_\Sigma(D) \,,
\end{align}
from which the connection $A$ can be defined uniquely.
Therefore the moduli space $\mathfrak{M}_d$ can be described as
\begin{equation}
	\mathfrak{M}_d \simeq \sym{d}{\Sigma} \,.
\end{equation}

\setlength{\parskip}{0pt}
\phantomsection
\addcontentsline{toc}{chapter}{Bibliography}
\bibliography{thesis}{}

\newcommand{\etalchar}[1]{$^{#1}$}
\begin{thebibliography}{AHCCK10}

\bibitem[ACGP03]{Alvarez-Consul:2001mqd}
L.~Alvarez-Consul and O.~Garcia-Prada, \textsl{ {Hitchin-Kobayashi
  Correspondence, Quivers, and Vortices}},
\newblock Commun. Math. Phys. \textbf{ 238}, 1--33 (2003),
  \href{https://arxiv.org/abs/math/0112161}{math/0112161}.

\bibitem[AGW84]{Alvarez-Gaume:1983ihn}
L.~Alvarez-Gaume and E.~Witten, \textsl{ {Gravitational Anomalies}},
\newblock Nucl. Phys. B \textbf{ 234}, 269 (1984).

\bibitem[AHCCK10]{arkani-hamed2009}
N.~Arkani-Hamed, F.~Cachazo, C.~Cheung and J.~Kaplan, \textsl{ {A Duality for
  the S Matrix}},
\newblock JHEP \textbf{ 03}, 020 (2010),
  \href{https://arxiv.org/abs/0907.5418}{0907.5418}.

\bibitem[AHI{\etalchar{+}}97]{aharony1997}
O.~Aharony, A.~Hanany, K.~Intriligator, N.~Seiberg and M.~J. Strassler,
  \textsl{ {Aspects of $\mathcal{N}=2$ Supersymmetric Gauge Theories in Three
  Dimensions}},
\newblock Nucl. Phys. B \textbf{ 499}, 67--99 (1997),
  \href{https://arxiv.org/abs/hep-th/9703110}{hep-th/9703110}.

\bibitem[Arb85]{arbarello1985}
E.~Arbarello,
\newblock \textsl{ {Geometry of Algebraic Curves}},
\newblock Grundlehren der Mathematischen Wissenschaften, Springer, New York,
  1985.

\bibitem[ARW17]{aharony2018}
O.~Aharony, S.~S. Razamat and B.~Willett, \textsl{ {From 3D Duality to 2D
  Duality}},
\newblock JHEP \textbf{ 11}, 090 (2017),
  \href{https://arxiv.org/abs/1710.00926}{1710.00926}.

\bibitem[Bat94]{Batyrev:1993oya}
V.~V. Batyrev, \textsl{ {Dual polyhedra and mirror symmetry for Calabi-Yau
  hypersurfaces in toric varieties}},
\newblock J. Alg. Geom. \textbf{ 3}, 493--545 (1994),
  \href{https://arxiv.org/abs/alg-geom/9310003}{alg-geom/9310003}.

\bibitem[BC15]{benini2015a}
F.~Benini and S.~Cremonesi, \textsl{ {Partition Functions of $\mathcal{N} =
  (2,2)$ Gauge Theories on $S^2$ and Vortices}},
\newblock Commun. Math. Phys. \textbf{ 334}(3), 1483--1527 (2015),
  \href{https://arxiv.org/abs/1206.2356}{1206.2356}.

\bibitem[BDG{\etalchar{+}}18]{Bullimore:2016hdc}
M.~Bullimore, T.~Dimofte, D.~Gaiotto, J.~Hilburn and H.~Kim, \textsl{ {Vortices
  and Vermas}},
\newblock Adv. Theor. Math. Phys. \textbf{ 22}, 803--917 (2018),
  \href{https://arxiv.org/abs/1609.04406}{1609.04406}.

\bibitem[BFK19]{bullimore2019a}
M.~Bullimore, A.~E.~V. Ferrari and H.~Kim, \textsl{ {Twisted Indices of 3D
  $\mathcal{N} = 4$ Gauge Theories and Enumerative Geometry of Quasi-Maps}},
\newblock JHEP \textbf{ 07}, 014 (2019),
  \href{https://arxiv.org/abs/1812.05567}{1812.05567}.

\bibitem[BFK22]{bullimore2019}
M.~Bullimore, A.~E.~V. Ferrari and H.~Kim, \textsl{ {The 3D Twisted Index and
  Wall-Crossing}},
\newblock SciPost Phys. \textbf{ 12}(6), 186 (2022),
  \href{https://arxiv.org/abs/1912.09591}{1912.09591}.

\bibitem[BFKX22]{xu2022}
M.~Bullimore, A.~E.~V. Ferrari, H.~Kim and G.~Xu, \textsl{ {The Twisted Index
  and Topological Saddles}},
\newblock JHEP \textbf{ 05}, 116 (2022),
  \href{https://arxiv.org/abs/2007.11603}{2007.11603}.

\bibitem[BJG{\etalchar{+}}71]{grothendieck1971}
P.~Berthelot, O.~Jussila, A.~Grothendieck, M.~Raynaud, S.~Kleiman, L.~Illusie
  and P.~Berthelot,
\newblock \textsl{ {Th\'eorie des Intersections et Théorème de
  Riemann-Roch}},
\newblock Lecture Notes in Mathematics, Springer, Berlin, 1971.

\bibitem[BMO10]{braverman2009}
A.~Braverman, D.~Maulik and A.~Okounkov, \textsl{ {Quantum Cohomology of the
  Springer Resolution}},
\newblock (2010), \href{https://arxiv.org/abs/1001.0056}{1001.0056}.

\bibitem[BP14]{benini2014}
F.~Benini and W.~Peelaers, \textsl{ {Higgs Branch Localization in Three
  Dimensions}},
\newblock JHEP \textbf{ 05}, 030 (2014),
  \href{https://arxiv.org/abs/1312.6078}{1312.6078}.

\bibitem[BX22]{xu2022a}
M.~Bullimore and G.~Xu, \textsl{ {Chern-Simons Terms and Quantum K-Theory}},
\newblock In~Preparation  (2022).

\bibitem[BZ15]{benini2015}
F.~Benini and A.~Zaffaroni, \textsl{ {A Topologically Twisted Index for
  Three-Dimensional Supersymmetric Theories}},
\newblock JHEP \textbf{ 07}, 127 (2015),
  \href{https://arxiv.org/abs/1504.03698}{1504.03698}.

\bibitem[BZ17]{benini2017}
F.~Benini and A.~Zaffaroni, \textsl{ {Supersymmetric Partition Functions on
  Riemann Surfaces}},
\newblock Proc. Symp. Pure Math. \textbf{ 96}, 13--46 (2017),
  \href{https://arxiv.org/abs/1605.06120}{1605.06120}.

\bibitem[CCP15]{closset2015}
C.~Closset, S.~Cremonesi and D.~S. Park, \textsl{ {The Equivariant A-Twist and
  Gauged Linear Sigma Models on Two-Sphere}},
\newblock JHEP \textbf{ 06}, 076 (2015),
  \href{https://arxiv.org/abs/1504.06308}{1504.06308}.

\bibitem[CK16]{closset2016}
C.~Closset and H.~Kim, \textsl{ {Comments on Twisted Indices in 3D
  Supersymmetric Gauge Theories}},
\newblock JHEP \textbf{ 08}, 059 (2016),
  \href{https://arxiv.org/abs/1605.06531}{1605.06531}.

\bibitem[CKW17]{Closset:2017zgf}
C.~Closset, H.~Kim and B.~Willett, \textsl{ {Supersymmetric Partition Functions
  and the Three-Dimensional A-Twist}},
\newblock JHEP \textbf{ 03}, 074 (2017),
  \href{https://arxiv.org/abs/1701.03171}{1701.03171}.

\bibitem[CT08]{collie2008}
B.~Collie and D.~Tong, \textsl{ {The Dynamics of Chern-Simons Vortices}},
\newblock Phys. Rev. D \textbf{ 78}, 065013 (2008),
  \href{https://arxiv.org/abs/0805.0602}{0805.0602}.

\bibitem[DEF{\etalchar{+}}99]{Deligne:1999qp}
P.~Deligne, P.~Etingof, D.~S. Freed, L.~C. Jeffrey, D.~Kazhdan, J.~W. Morgan,
  D.~R. Morrison and E.~Witten, editors,
\newblock \textsl{ {Quantum Fields and Strings: A Course for Mathematicians.
  Vol. 1, 2}},
\newblock AMS, Providence, 1999.

\bibitem[DGLFL13]{doround2013}
N.~Dorond, J.~Gomis, B.~Le~Floch and S.~Lee, \textsl{ {Exact Results in $D=2$
  Supersymmetric Gauge Theories}},
\newblock JHEP \textbf{ 05}, 093 (2013),
  \href{https://arxiv.org/abs/1206.2606}{1206.2606}.

\bibitem[Don96]{donaldson1996}
S.~K. Donaldson, \textsl{ {The Seiberg-Witten Equations and $4$-Manifold
  Topology}},
\newblock Bulletin of the American Mathematical Society \textbf{ 33}, 45--70
  (1996).

\bibitem[DT00]{dorey2000}
N.~Dorey and D.~Tong, \textsl{ {Mirror Symmetry and Toric Geometry in Three
  Dimensional Gauge Theories}},
\newblock JHEP \textbf{ 05}, 018 (2000),
  \href{https://arxiv.org/abs/hep-th/9911094}{hep-th/9911094}.

\bibitem[FHY14]{fujitsuka2014}
M.~Fujitsuka, M.~Honda and Y.~Yoshida, \textsl{ {Higgs Branch Localization of
  3D $\mathcal{N} = 2$ Theories}},
\newblock PTEP \textbf{ 2014}(12), 123B02 (2014),
  \href{https://arxiv.org/abs/1312.3627}{1312.3627}.

\bibitem[F{\L}M19]{ferro2018}
L.~Ferro, T.~{\L}ukowski and P.~Matteo, \textsl{ {Amplituhedron meets
  Jeffrey-Kirwan Residue}},
\newblock J. Phys. A \textbf{ 52}(4), 045201 (2019),
  \href{https://arxiv.org/abs/1805.01301}{1805.01301}.

\bibitem[GH78]{griffiths1978}
P.~Griffiths and J.~Harris,
\newblock \textsl{ {Principles of Algebraic Geometry}},
\newblock John Wiley {\&} Sons, New York, 1978.

\bibitem[Giv00]{givental2000}
A.~B. Givental, \textsl{ {On the WDVV-Equation in Quantum K-Theory}},
\newblock Michigan Mathematical Journal \textbf{ 48}, 295--304 (2000).

\bibitem[Giv15]{givental2015}
A.~Givental, \textsl{ {Permutation-Equivariant Quantum K-Theory I-XI}},
\newblock (2015), \href{https://arxiv.org/abs/1508.02690}{1508.02690}.

\bibitem[GL03]{givental2001}
A.~B. Givental and Y.~P. Lee, \textsl{ {Quantum K-Theory on Flag Manifolds,
  Finite-Difference Toda Lattices and Quantum Groups}},
\newblock Invent. Math. \textbf{ 151}, 193--219 (2003),
  \href{https://arxiv.org/abs/math/0108105}{math/0108105}.

\bibitem[GP93]{Garcia-Prada:1993usn}
O.~Garcia-Prada, \textsl{ {Invariant Connections and Vortices}},
\newblock Commun. Math. Phys. \textbf{ 156}, 527--546 (1993).

\bibitem[HHL11a]{hama2011a}
N.~Hama, K.~Hosomichi and S.~Lee, \textsl{ {Notes on SUSY Gauge Theories on
  Three-Spheres}},
\newblock JHEP \textbf{ 03}, 127 (2011),
  \href{https://arxiv.org/abs/1012.3512}{1012.3512}.

\bibitem[HHL11b]{hama2011}
N.~Hama, K.~Hosomichi and S.~Lee, \textsl{ {SUSY Gauge Theories on Squashed
  Three-Spheres}},
\newblock JHEP \textbf{ 05}, 014 (2011),
  \href{https://arxiv.org/abs/1102.4716}{1102.4716}.

\bibitem[HKK{\etalchar{+}}03]{hori2003}
K.~Hori, S.~Katz, A.~Klemm, R.~Pandharipande, R.~Thomas, C.~Vafa, R.~Vakil and
  E.~Zaslow,
\newblock \textsl{ {Mirror Symmetry}}, volume~1 of \textsl{ Clay Mathematics
  Monographs},
\newblock AMS, Providence, 2003.

\bibitem[HSW99]{Hitchin:1999at}
N.~J. Hitchin, G.~B. Segal and R.~S. Ward,
\newblock \textsl{ {Integrable Systems: Twistors, Loop groups, and Riemann
  surfaces}},
\newblock Clarendon Press, 1999.

\bibitem[IS13]{intriligator2013}
K.~Intriligator and N.~Seiberg, \textsl{ {Aspects of $3D$ $\mathcal{N}=2$
  Chern-Simons-Matter Theories}},
\newblock JHEP \textbf{ 07}, 079 (2013),
  \href{https://arxiv.org/abs/1305.1633}{1305.1633}.

\bibitem[JM19]{jockers2019}
H.~Jockers and P.~Mayr, \textsl{ {Quantum K-Theory of Calabi-Yau Manifolds}},
\newblock JHEP \textbf{ 11}, 011 (2019),
  \href{https://arxiv.org/abs/1905.03548}{1905.03548}.

\bibitem[JM20]{jockers2018}
H.~Jockers and P.~Mayr, \textsl{ {A 3D Gauge Theory/Quantum K-Theory
  Correspondence}},
\newblock Adv. Theor. Math. Phys. \textbf{ 24}(2), 327--457 (2020),
  \href{https://arxiv.org/abs/1808.02040}{1808.02040}.

\bibitem[JMNT20]{jockers2019a}
H.~Jockers, P.~Mayr, U.~Ninad and A.~Tabler, \textsl{ {Wilson Loop Algebras and
  Quantum K-Theory for Grassmannians}},
\newblock JHEP \textbf{ 10}, 036 (2020),
  \href{https://arxiv.org/abs/1911.13286}{1911.13286}.

\bibitem[JT80]{Jaffe:1980mj}
A.~M. Jaffe and C.~H. Taubes,
\newblock \textsl{ {Vortices and Monopoles: Structure of Static Gauge
  Theories}},
\newblock Birkh\"auser, Basel, 1980.

\bibitem[KWY10]{kapustin2010}
A.~Kapustin, B.~Willett and I.~Yaakov, \textsl{ {Exact Results for Wilson Loops
  in Superconformal Chern-Simons Theories with Matter}},
\newblock JHEP \textbf{ 03}, 089 (2010),
  \href{https://arxiv.org/abs/0909.4559}{0909.4559}.

\bibitem[Lee04]{lee2001}
Y.~P. Lee, \textsl{ {Quantum K-Theory I: Foundations}},
\newblock Duke Math. J. \textbf{ 121}, 389--424 (2004),
  \href{https://arxiv.org/abs/math/0105014}{math/0105014}.

\bibitem[Lib07]{libine2010}
M.~Libine, \textsl{ {Lecture Notes on Equivariant Cohomology}},
\newblock (2007), \href{https://arxiv.org/abs/0709.3615}{0709.3615}.

\bibitem[Lic14]{lichtblau2014}
D.~Lichtblau, \textsl{ {Practical Computations with Gr\"obner Bases}},
\newblock (2014).

\bibitem[LR18]{larsen2019}
K.~J. Larsen and R.~Rietkerk, \textsl{ {MultivariateResidues: A Mathematica
  Package for Computing Multivariate Residues}},
\newblock Comput. Phys. Commun. \textbf{ 222}, 250--262 (2018),
  \href{https://arxiv.org/abs/1701.01040}{1701.01040}.

\bibitem[LV98]{Leung:1997tw}
N.~C. Leung and C.~Vafa, \textsl{ {Branes and toric geometry}},
\newblock Adv. Theor. Math. Phys. \textbf{ 2}, 91--118 (1998),
  \href{https://arxiv.org/abs/hep-th/9711013}{hep-th/9711013}.

\bibitem[Mac62]{macdonald1962}
I.~G. MacDonald, \textsl{ {Symmetric Products of an Algebraic Curve}},
\newblock Topology , 319--343 (1962).

\bibitem[Man82]{Manton:1981mp}
N.~S. Manton, \textsl{ {A Remark on the Scattering of BPS Monopoles}},
\newblock Phys. Lett. B \textbf{ 110}, 54--56 (1982).

\bibitem[MO12]{maulik2012}
D.~Maulik and A.~Okounkov, \textsl{ {Quantum Groups and Quantum Cohomology}},
\newblock (2012), \href{https://arxiv.org/abs/1211.1287}{1211.1287}.

\bibitem[MO15]{okounkov2015}
D.~Maulik and A.~Okounkov, \textsl{ {Lectures on K-Theoretic Computations in
  Enumerative Geometry}},
\newblock (2015), \href{https://arxiv.org/abs/1512.07363}{1512.07363}.

\bibitem[MS04]{manton_sutcliffe_2004}
N.~Manton and P.~Sutcliffe,
\newblock \textsl{ Topological Solitons},
\newblock Cambridge Monographs on Mathematical Physics, Cambridge University
  Press, Cambridge, 2004.

\bibitem[Nak03]{nakahara2003}
M.~Nakahara,
\newblock \textsl{ {Geometry, Topology and Physics}},
\newblock IoP, London, 2003.

\bibitem[NS09]{Nekrasov:2009ui}
N.~A. Nekrasov and S.~L. Shatashvili, \textsl{ {Quantum Integrability and
  Supersymmetric Vacua}},
\newblock Prog. Theor. Phys. Suppl. \textbf{ 177}, 105--119 (2009),
  \href{https://arxiv.org/abs/0901.4748}{0901.4748}.

\bibitem[NS15]{Nekrasov:2014xaa}
N.~A. Nekrasov and S.~L. Shatashvili, \textsl{ {Bethe/Gauge Correspondence on
  Curved Spaces}},
\newblock JHEP \textbf{ 01}, 100 (2015),
  \href{https://arxiv.org/abs/1405.6046}{1405.6046}.

\bibitem[NW10]{Nekrasov:2010ka}
N.~Nekrasov and E.~Witten, \textsl{ {The Omega Deformation, Branes,
  Integrability, and Liouville Theory}},
\newblock JHEP \textbf{ 09}, 092 (2010),
  \href{https://arxiv.org/abs/1002.0888}{1002.0888}.

\bibitem[Red84a]{Redlich:1983kn}
A.~N. Redlich, \textsl{ {Gauge Noninvariance and Parity Violation of
  Three-Dimensional Fermions}},
\newblock Phys. Rev. Lett. \textbf{ 52}, 18 (1984).

\bibitem[Red84b]{Redlich:1983dv}
A.~N. Redlich, \textsl{ {Parity Violation and Gauge Noninvariance of the
  Effective Gauge Field Action in Three-Dimensions}},
\newblock Phys. Rev. D \textbf{ 29}, 2366--2374 (1984).

\bibitem[RSV20]{Robbins:2020msp}
D.~Robbins, E.~Sharpe and T.~Vandermeulen, \textsl{ {A Generalization of
  Decomposition in Orbifolds}},
\newblock JHEP \textbf{ 21}, 134 (2020),
  \href{https://arxiv.org/abs/2101.11619}{2101.11619}.

\bibitem[RWZ20]{ruan2020}
Y.~Ruan, Y.~Wen and Z.~Zhou, \textsl{ {Quantum K-Theory of Toric Varieties,
  Level Structures, and 3D Mirror Symmetry}},
\newblock (2020), \href{https://arxiv.org/abs/2011.07519}{2011.07519}.

\bibitem[RZ18]{ruan2019}
Y.~Ruan and M.~Zhang, \textsl{ {The Level Structure in Quantum K-Theory and
  Mock Theta Functions}},
\newblock (2018), \href{https://arxiv.org/abs/1804.06552}{1804.06552}.

\bibitem[Sei94]{Seiberg:1994bz}
N.~Seiberg, \textsl{ {Exact Results on the Space of Vacua of Four-Dimensional
  SUSY Gauge Theories}},
\newblock Phys. Rev. D \textbf{ 49}, 6857--6863 (1994),
  \href{https://arxiv.org/abs/hep-th/9402044}{hep-th/9402044}.

\bibitem[SW94]{seiberg1994}
N.~Seiberg and E.~Witten, \textsl{ {Electric-Magnetic Duality, Monopole
  Condensation, and Confinement in $\mathcal{N} = 2$ Supersymmetric Yang-Mills
  Theory}},
\newblock Nucl. Phys. B \textbf{ 426}, 19--52 (1994),
  \href{https://arxiv.org/abs/hep-th/9407087}{hep-th/9407087}.

\bibitem[Tha92]{thaddeus1992}
M.~Thaddeus, \textsl{ {Stable Pairs, Linear Systems and the Verlinde Formula}},
\newblock (1992),
  \href{https://arxiv.org/abs/alg-geom/9210007}{alg-geom/9210007}.

\bibitem[UY20]{Ueda:2019qhg}
K.~Ueda and Y.~Yoshida, \textsl{ {3D $ \mathcal{N} $ = 2 Chern-Simons-Matter
  Theory, Bethe ansatz, and Quantum $K$-Theory of Grassmannians}},
\newblock JHEP \textbf{ 08}, 157 (2020),
  \href{https://arxiv.org/abs/1912.03792}{1912.03792}.

\bibitem[WB92]{Wess:1992cp}
J.~Wess and J.~Bagger,
\newblock \textsl{ {Supersymmetry and Supergravity}},
\newblock Princeton University Press, Princeton, 1992.

\bibitem[Wil17]{willet2017}
B.~Willett, \textsl{ {Localization on Three-Dimensional Manifolds}},
\newblock J. Phys. A \textbf{ 50}(44), 443006 (2017),
  \href{https://arxiv.org/abs/1608.02958}{1608.02958}.

\bibitem[Wit82]{witten1982}
E.~Witten, \textsl{ {Constraints on Supersymmetry Breaking}},
\newblock Nucl. Phys. B \textbf{ 202}, 253 (1982).

\bibitem[Wit98]{witten1991}
E.~Witten, \textsl{ {Mirror Manifolds and Topological Field Theory}},
\newblock AMS/IP Stud. Adv. Math. \textbf{ 9}, 121--160 (1998),
  \href{https://arxiv.org/abs/hep-th/9112056}{hep-th/9112056}.

\bibitem[ZZ20]{Zhang:2020rou}
M.~Zhang and Y.~Zhou, \textsl{ {K-Theoretic Quasimap Wall-Crossing}},
\newblock (2020), \href{https://arxiv.org/abs/2012.01401}{2012.01401}.

\end{thebibliography}
\bibliographystyle{hep}

\cleardoublepage
\phantomsection
\addcontentsline{toc}{chapter}{List of Symbols}
\printindex[sym]

\cleardoublepage
\phantomsection
\addcontentsline{toc}{chapter}{Index}
\printindex

\end{document}